%% file: ms.tex
\documentclass[12pt, oneside, table, xcdraw]{book}

\usepackage[utf8]{inputenc}
\usepackage[T1]{fontenc}
\usepackage{lmodern}
\usepackage[
a4paper,
margin=3cm,
bindingoffset=1cm,
lines=25,
includeheadfoot
]{geometry}
\usepackage{graphicx}

\usepackage[hidelinks]{hyperref}
\hypersetup{
	pdftitle={Statistical Analysis to Support CSI-Based Sensing Methods},
	pdfpagemode=FullScreen,
	pdfauthor={Elena Tonini}
}

\usepackage{setspace}
\usepackage{amsmath}
\usepackage{amsfonts}
\usepackage[binary-units,per-mode=symbol]{siunitx}
\usepackage{tikz}
\usepackage{url}
\usepackage[american, italian]{babel}
\usepackage[capitalise]{cleveref}
\crefname{appendix}{App.}{App.}
\crefname{section}{Sect.}{Sect.}
\crefname{figure}{Fig.}{Fig.}
\crefname{table}{Tab.}{Tab.}
\crefname{equation}{Eq.}{Eq.}
\Crefname{table}{Table}{Tables}
\crefname{lstlisting}{Lst.}{Lst.}
\Crefname{lstlisting}{Listing}{Listings}
\Crefname{equation}{Equation}{Equations}
\Crefname{appendix}{Appendix}{Appendices}
\def\appname{\csname cref@appendix@name\endcsname\xspace}
\def\figname{\csname cref@figure@name\endcsname\xspace}
\def\tabname{\csname cref@table@name\endcsname\xspace}
\def\secname{\csname cref@section@name\endcsname\xspace}
\def\eqname{\csname cref@equation@name\endcsname\xspace}
\def\lstname{\csname cref@lstlisting@name\endcsname\xspace}
\def\eqpname{\csname cref@equation@name@plural\endcsname\xspace}
\usepackage{xspace}
\usepackage{multirow}
\usepackage{hyphenat}
\usepackage{listings}
\usepackage{bytefield}
\usepackage{subcaption}

\usepackage{xcolor}
\usepackage{color,soul}

\definecolor{light-gray}{gray}{0.97} 
\definecolor{codegreen}{rgb}{0,0.6,0}

\usepackage[backend=biber,style=ieee,doi=false,isbn=false]{biblatex}
\DeclareFieldFormat{sentencecase}{#1} 
\DeclareFieldFormat{titlecase}{#1} 
\usepackage{xpatch}
\xpatchbibmacro{textcite}{\addspace}{\addnbspace}{}{}
\xpatchbibmacro{Textcite}{\addspace}{\addnbspace}{}{}
\DefineBibliographyStrings{english}{
	andothers = et~al\adddot\addspace
}

\graphicspath{{./images/}}
\usepackage{todonotes}
\presetkeys{todonotes}{inline}{}

\DeclareSIUnit \belm{Bm}
\DeclareSIUnit \beli{Bi}

\RequirePackage{xstring}
\RequirePackage{xparse}
\RequirePackage[]{acro}   
\NewDocumentCommand\acrodef{mO{#1}mG{}}{\DeclareAcronym{#1}{short={#2}, long={#3}, #4}}

\acrodef{3GPP}{3rd Generation Partnership Project}
\acrodef{4G}{4$^{\text{th}}$ Generation}
\acrodef{5G}{5$^{\text{th}}$ Generation}
\acrodef{5GV2X}[5G-V2X]{5$^{\text{th}}$ Generation Cellular V2X}
\acrodef{ACI}{Adjacent Channel Interference}
\acrodef{AGC}{Automatic Gain Control}
\acrodef{AI}[AI]{Artificial Intelligence}
\acrodef{AoI}{Age of Information}
\acrodef{AP}{Access Point}
\acrodef{API}{Application Programming Interface}
\acrodef{AWGN}{Additive White Gaussian Noise}
\acrodef{BER}{Bit Error Probability}
\acrodef{BPSK}{Binary Phase Shift Keying}
\acrodef{CSI}{Channel State Information}
\acrodef{CSMA}{Carrier-Sense Multiple Access}
\acrodef{CV2X}[C-V2X]{Cellular V2X}
\acrodef{D2D}{Device-to-Device}
\acrodef{DES}{Discrete Event Simulation}
\acrodef{eNB}{eNodeB}
\acrodef{FCC}{Federal Communications Commission}
\acrodef{FDMA}{Frequency-Division Multiple Access}
\acrodef{FSM}{Finite State Machine}
\acrodef{gNB}{gNodeB}
\acrodef{GUI}{Graphical User Interface}
\acrodef{HARQ}{Hybrid Automatic Repeat reQuest}
\acrodef{IC}{In Coverage}
\acrodef{JCAS}{Joint Communication and Sensing}
\acrodef{WLAN}{Wireless Local Area Network}
\acrodef{LOS}[LoS]{Line of Sight}
\acrodef{LTE}{Long Term Evolution}
\acrodef{LTEV2X}[LTE-V2X]{Cellular V2X}
\acrodef{MAC}{Medium Access Control}
\acrodef{MEC}{Mobile Edge Computing}
\acrodef{MI}{Mutual Information}
\acrodef{MIMO}{Multiple Input Multiple Output}
\acrodef{ML}[ML]{Machine Learning}
\acrodef{MLO}{Multi-Link Operation}
\acrodef{MUMIMO}[MU-MIMO]{Multi-User \ac{MIMO}}
\acrodef{MPC}{Model Predictive Control}{short-indefinite={an}}
\acrodef{NR}{New Radio}
\acrodef{OFDM}{Orthogonal Frequency-Division Multiplexing}
\acrodef{OFDMA}{Orthogonal Frequency-Division Multiple Access}
\acrodef{OoC}{Out of Coverage}
\acrodef{PC5}{PC5}
\acrodef{PDR}{Packet Delivery Ratio}
\acrodef{PDF}{Probability Density Function}
\acrodef{PHY}{Physical Layer}
\acrodef{PMF}{Probability Mass Function}
\acrodef{PSK}{Phase Shift Keying}
\acrodef{QAM}{Quadrature Amplitude Modulation}
\acrodef{QOS}[QoS]{Quality of Service}
\acrodef{QBPSK}{Quadrature Binary Phase Shift Keying}
\acrodef{QPSK}{Quadrature Phase Shift Keying}
\acrodef{RU}{Resource Unit}
\acrodef{SINR}{Signal to Interference plus Noise Ratio}{short-indefinite={an}}
\acrodef{SIR}{Signal to Interference Ratio}
\acrodef{SNR}{Signal to Noise Ratio}
\acrodef{STA}{Station}
\acrodef{TDMA}{Time-Division Multiple Access}
\acrodef{UE}{User Equipment}
\acrodef{URLLC}{Ultra-Reliable Low-Latency Communication}  
\acrodef{Uu}{User to Network Interface}
\acrodef{V2V}{Vehicle to Vehicle communication}
\acrodef{V2X}{Vehicle to Everything}
\acrodef{VLC}{Visible Light Communication}
\acrodef{VRU}{Vulnerable Road Users}
\acrodef{WHD}{Weighted Hamming Distance}
\acrodef{ZOH}{Zero Order Hold}

\newcommand{\sensing}{ambient sensing\xspace}
\newcommand{\wifi}{Wi-Fi\xspace}
\newcommand{\csi}{\ensuremath{\mathbf{C}}\xspace}

\newcommand{\csia}{\ensuremath{\mathbf{A}}\xspace}
\newcommand{\csib}{\ensuremath{\mathbf{B}}\xspace}

\newcommand{\acsi}{\ensuremath{A_{\csi}}\xspace}
\newcommand{\racsi}{\ensuremath{A^{\star}_{\csi}}\xspace}

\newcommand{\iacsi}{\ensuremath{\delta_{\csi}}\xspace}
\newcommand{\iacsih}{\ensuremath{\delta{h}_{\csi}}\xspace}
\newcommand{\iacsimax}{\ensuremath{\delta^{\star}}\xspace}
\newcommand{\tin}{\ensuremath{k}\xspace}
\newcommand{\subc}{\ensuremath{n}\xspace}
\newcommand{\nsamp}{\ensuremath{M_{\csi}}\xspace}
\newcommand{\nsampa}{\ensuremath{M_{\csia}}\xspace}
\newcommand{\nsampb}{\ensuremath{M_{\csib}}\xspace}

\newcommand{\ncomb}{\ensuremath{M_{\csi{T}}}\xspace}
\newcommand{\nsc}{\ensuremath{N_{\text{\tiny SC}}}\xspace}
\newcommand{\dt}{\ensuremath{\delta t}\xspace}
\newcommand{\rtm}{\ensuremath{t_r}\xspace}
\newcommand{\nan}{\ensuremath{\text{NaN}}\xspace}
\newcommand{\ndis}{\ensuremath{\mathcal{N}}\xspace}
\newcommand{\qndis}{\ensuremath{\mathcal{N'}}\xspace}
\newcommand{\nbits}{\ensuremath{q_{\text{\tiny inc}}}\xspace}
\newcommand{\acsibits}{\ensuremath{q_{\text{\tiny amp}}}\xspace}
\newcommand{\mint}{\ensuremath{\mathcal{I}\xspace}}
\newcommand{\mext}{\ensuremath{\mathcal{E}\xspace}}
\newcommand{\whd}{\ensuremath{\text{WHD}}\xspace}

\DeclareSIUnit \km {\kilo\meter}
\DeclareSIUnit \kmh {\kilo\meter\per\hour}
\DeclareSIUnit \mps {\meter\per\second}
\DeclareSIUnit \mpsq {\meter\per\second\squared}
\DeclareSIUnit \mpsc {\meter\per\cubic\second}
\DeclareSIUnit \mhz {\mega\hertz}

\def\BibTeX{{\rm B\kern-.05em{\subc i\kern-.025em b}\kern-.08em
		T\kern-.1667em\lower.7ex\hbox{E}\kern-.125emX}}

\usepackage{titlesec}
\titleformat{\chapter}{\normalfont\fontsize{18}{21}\bfseries}{\thechapter}{1.5em}{}
\titlespacing*{\chapter}{0pt}{5.5ex plus 1ex minus .2ex}{5.3ex plus .2ex}

\titleformat{\section}{\normalfont\fontsize{16}{18}\bfseries}{\thesection}{1em}{}
\titlespacing*{\section}{0pt}{2.5ex plus 1ex minus .2ex}{2.3ex plus .2ex}

\titleformat{\subsection}{\normalfont\fontsize{13}{15}\bfseries}{\thesubsection}{1em}{}
\titlespacing*{\subsection}{0pt}{1.5ex plus 1ex minus .2ex}{1.3ex plus .2ex}

\titleformat{\subsubsection}{\normalfont\fontsize{12}{13}\itshape\bfseries}{\thesubsubsection}{1em}{}
\titlespacing*{\subsection}{0pt}{1.5ex plus 1ex minus .2ex}{1.3ex plus .2ex}

\usepackage{fancyhdr}

\fancypagestyle{plain}{%
	\fancyhf{}
	\fancyfoot[C]{
		\if\relax\thepage\else%
		--- {\thepage} ---
		\fi
	}

}
\begin{document}
	\pagenumbering{gobble}
	\include{frontespice}
	\cleardoublepage
	\pagenumbering{Roman}
	\setstretch{1.5}
	
	\include{sommario}
	
\newpage
	\include{summary}
	
	\tableofcontents
	\cleardoublepage
	
	\pagenumbering{arabic}

	\include{introduction}
	\include{wifi_fundamentals}

	\include{prev_work}
	\include{experimental_setup}
	\include{notation}
	\include{norm_quant}
	\include{mutual_information}
	\include{hamming}
	\include{elab_workflow}
	\include{norm_res}
	\include{hamming_res}
	\include{future_work}
	
	\section*{Acknowledgments}
	\label{sec:acks}
	The work of this thesis has been carried out within the framework of PNRR-funded research projects granted to the Department of Information Engineering at the University of Brescia, PI the supervisor of this thesis.
	In particular, the topic of \ac{CSI} analysis lies at the base of these two projects: 
	\begin{itemize}
		\item Joint Communication and Sensing: CSI-Based Sensing for Future Wireless Networks (CSI-Future), PRIN 2022 PNRR Prot. P2022FP9W3 (CUP D53D23016040001);  
		\item  RESearch and innovation on future Telecommunications systems and networks, to make Italy more smART (RESTART -- PE00000001) funded by the European Union (EU) and the Italian Ministry for Universities and Research (MUR), National Recovery and Resilience Plan (NRRP), Spoke 4, Structural Project SUPER, Cascade Call Project ``Architettura, PROgetto, ottimizzazione e valutazione di sistemi di percezione collaborativa distribuiti per Smart Driving Spaces'' (PROSDS -- CUP C89J24000270004).
	\end{itemize}
	
	\addcontentsline{toc}{chapter}{Bibliography}
	\printbibliography
	
	\appendix 
	\include{appendixA}
	\include{appendixB}

	\include{acknowledgements}
	\clearpage
	
\end{document}

%% file: frontespice.tex
\newgeometry{left=20mm,right=20mm,top=25mm,bottom=20mm} 
\begin{titlepage}
	\begin{center}
		
		\begin{figure}[t]
			\centering
			\includegraphics[height=30mm]{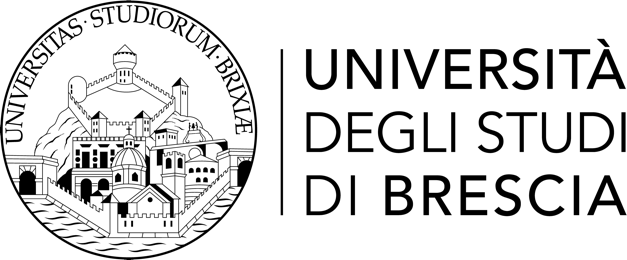}    
		\end{figure}
		\vspace*{4mm}
		{\fontsize{17}{12mm}\fontfamily{lmss}\selectfont
			DIPARTIMENTO DI INGEGNERIA DELL'INFORMAZIONE\\
			Corso di Laurea Magistrale\\
			\vspace*{1mm}
			in Ingegneria Informatica
		}
		\vfill
		
		{\fontsize{20}{20}\fontfamily{lmss}\selectfont
			Tesi di Laurea\\
			\vspace*{5mm}
			\textbf{Statistical Analysis to Support\\\vspace*{2mm}CSI-Based Sensing Methods}\\
			\vspace*{8mm}
			\textbf{Analisi Statistica a Supporto\\\vspace*{2mm}di Metodi di Misura Basati su CSI}
		}
		
	\end{center}
	
	\vfill
	
	\begin{minipage}{.75\textwidth}
		\fontsize{16}{16}\fontfamily{lmss}\selectfont 
		\textbf{Relatore:} Chiar.mo~Prof.~Renato~Lo~Cigno
	\end{minipage}
	
	\vspace*{10mm}
	
	\begin{flushright}
		{\fontsize{17}{17}\fontfamily{lmss}\selectfont 
			Laureanda:\\
			Elena~Tonini\\
			Matricola n. 727382\\
		}
		
	\end{flushright}

	\vspace*{10mm}
	\begin{center}
		{\fontsize{17}{17}\fontfamily{lmss}\selectfont 
			Anno Accademico 2023/24
		}
	\end{center}
\end{titlepage}
\restoregeometry 

%% file: sommario.tex
\selectlanguage{italian}
\section*{Sommario}
\label{sec:sommario}
Prendendo spunto dal lavoro della Tesi di Laurea Triennale intitolata ``Analysis and Characterization of \wifi Channel State Information'', questa tesi amplia e approfondisce la ricerca conducendo un'analisi dettagliata delle CSI, offrendo nuovi approcci che si spingono oltre i risultati dello studio originale. 
L'obiettivo del lavoro è estendere la rappresentazione matematica e statistica di un canale wireless attraverso lo studio del comportamento e dell'evoluzione nel tempo e nella frequenza delle CSI.

Le CSI forniscono una descrizione ad alto livello del comportamento di un segnale che si propaga da un trasmettitore a un ricevitore, rappresentando così la struttura dell'ambiente che il segnale attraversa. 
Questa conoscenza può essere utilizzata per effettuare \textit{\sensing}, una tecnica che permette di estrarre informazioni rilevanti sull'ambiente di propagazione in funzione delle proprietà che il segnale presenta al ricevitore, dopo aver interagito con le superfici degli oggetti presenti nello spazio analizzato. 
L'\textit{\sensing} svolge già un ruolo essenziale nelle nuove reti wireless come 5G e Beyond 5G; il suo impiego nelle applicazioni di \textit{Joint Communication and Sensing} e per l'ottimizzazione della propagazione del segnale tramite \textit{beamforming} potrebbe supportare \textit{\sensing} cooperativo efficiente anche nelle reti veicolari, consentendo la \textit{Cooperative Perception} e aumentando di conseguenza la sicurezza stradale.

A causa della mancanza di ricerca sulla caratterizzazione delle CSI, l'attuale studio intraprende un'analisi della struttura delle CSI raccolte in un ambiente sperimentale controllato, al fine di descriverne le proprietà statistiche. 
I risultati potrebbero fornire un approccio matematico di supporto alle attività di \textit{environment classification} e di \textit{movement recognition} che attualmente sono eseguite solo tramite approcci basati su Machine Learning, introducendo invece efficienti algoritmi dedicati.
\selectlanguage{american}

%% file: summary.tex
\section*{Summary}
\label{sec:summary}

Building upon the foundational work of the Bachelor's Degree Thesis titled ``Analysis and Characterization of \wifi Channel State Information'', this thesis significantly advances the research by conducting an in-depth analysis of CSIs, offering new insights that extend well beyond the original study.
The goal of this work is to broaden the mathematical and statistical representation of a wireless channel through the study of CSI behavior and evolution over time and frequency.

CSI provides a high-level description of the behavior of a signal propagating from a transmitter to a receiver, thereby representing the structure of the environment where the signal propagates.
This knowledge can be used to perform \textit{\sensing}, a technique that extracts relevant information about the surroundings of the receiver from the properties of the received signal, which are affected by interactions with the surfaces of the objects within the analyzed environment.
Ambient sensing already plays an essential role in new wireless networks such as 5G and Beyond 5G; its use in \textit{Joint Communication and Sensing} applications and for the optimization of signal propagation through \textit{beamforming} could also enable the implementation of efficient cooperative \sensing in vehicular networks, facilitating \textit{Cooperative Perception} and, consequently, increasing road safety.

Due to the lack of research on CSI characterization, this study aims to begin analyzing the structure of CSI traces collected in a controlled experimental environment and to describe their statistical properties.
The results of such characterization could provide mathematical support for environment classification and movement recognition tasks that are currently performed only through Machine Learning techniques, introducing instead efficient, dedicated algorithms.

%% file: introduction.tex
\chapter{Introduction}
\label{ch:intro}
With the ever-increasing applications of wireless telecommunication networks in all aspects of everyday life, ensuring users' security and privacy has become an increasingly delicate field of study requiring dedicated research.

As users across the world are becoming accustomed to approaching \wifi as a means of quickly transferring their own data, the ongoing development of this technology is both guaranteeing more users access to network coverage and high bit rates and raising awareness about previously unforeseen threats to users' security \cite{lamers2021securing}. 
Aside from the challenges of \wifi managed through the introduction of cryptographic protocols employed to ensure users' security when accessing the Internet \cite{wpa3}, some features of \wifi can still be exploited by attackers to violate users' privacy. 
Specifically, it may become more straightforward in the near future to perform attacks based on \wifi \ac{CSI} \cite{cai18,WangLiu2014,Stud18_Access-CC}. 

\acp{CSI} are pieces of information associated with packets transmitted on a \wifi channel and whose structure allows for the description of the behavior of a signal propagating from a transmitter to a receiver. 
Essentially, they provide a numerical representation of how the signal bounces off the surfaces it meets during its propagation by including information about the signal's phase shift and attenuation \cite{ma2019wifi}. 
\acp{CSI} do not intrinsically qualify as tools that can be exploited to perform attacks on wireless networks, but rather as features that should be used to improve the quality of telecommunications over \wifi.
Newly developed technologies benefit from the use of \acp{CSI} when implementing \ac{MIMO} techniques and improving channel equalization.

In fact, \acp{CSI} can also be used to perform \textit{\sensing}, a technique that extracts spatial information about the environment in which a signal propagates. 
Depending on the reflection, scattering, and absorption of the signal by the surroundings of both transmitter and receiver, the content of a \ac{CSI} is altered and it can be interpreted as a representation of the environment itself.
The content of a \ac{CSI} becomes a useful descriptor of both the static and dynamic structure of an environment, while also allowing to locate electronic devices within it.
Moreover, it is not necessary for a person to be carrying a communication device to be correctly located within the environment through the analysis of \ac{CSI} content, as the propagating signal will interact with the person's body regardless of the presence of any other electronic device \cite{ShiSig18,CoGri21_MCN}. 
This allows to both identify the position of the person and give an idea about their movements around the environment based solely on the properties the signal displays once it is received and its associated \acp{CSI} are extracted \cite{basiri2017indoor}.

Of course, this property of \acp{CSI} may be exploited by attackers to locate users within a given environment, violating their privacy without giving them the chance to defend themselves from sensing-based attacks \cite{felix20}. 
Research conducted in this field has identified signal jamming and information obfuscation --- which do not interfere with the quality and understandability of the transmitted content at the receiver --- as possible countermeasures to prevent attackers from obtaining sensing information directly from extracted \acp{CSI} \cite{CoGri22_ComCom, locobfusc}. 

The feature that makes sensing attacks apparently easy to carry out is that to effectively perform \sensing, the only requirements are that a fixed transmitter be placed within the analyzed environment and that a sensing receiver --- which should also be in a fixed position so as not to externally alter the \ac{CSI} content --- be used to capture and analyze the extracted \acp{CSI}. 

The feasibility of \sensing, for the time being, has only been tested in indoor environments using \wifi-based technologies \cite{WangLiu2014,WangWu2017}, but multiple applications could benefit from its implementation in outdoor locations and from the use of different technologies (e.g., cellular networks). 
Specifically, a useful extension to \sensing as we know it would be \ac{JCAS}, an approach that allows multiple parties to share information alongside the more ``traditional'' sensing activity. 

\ac{JCAS} is expected to play a significant role in \ac{5G} New Radio and Beyond \ac{5G} networks, where the concept of ``sharing \sensing information'' becomes more relevant. 
As communications rely on increasingly higher frequencies (over 20-30 GHz), difficulties may arise when \ac{LOS} between transmitter and receiver becomes strictly required for communications to effectively take place, as omnidirectional antennae no longer provide sufficient power to support data exchange at such frequencies.
Communications at high frequencies undergo significant signal attenuation during outdoor propagation, requiring the implementation of \textit{beamforming} to increase the directionality of a transmitter radiation pattern: this approach ensures that the pattern covers only the area where the targeted receiver is expected to be, significantly reducing power waste compared to omnidirectional antennae and increasing efficiency in communication through an increment in power density in the direction of the receiver. 
Without \textit{beamforming}, guaranteeing that all receiving devices have \ac{LOS} with an omnidirectional transmitter would be infeasible \cite{cox1987robust}.

As implementing \textit{beamforming} remains technologically challenging, the introduction of \sensing may help identify obstacles along the signal propagation path and automatically steer beams or move transmission to a device that guarantees better \ac{QOS} when operating in a mesh-like network topology.

Other fields of research may draw advantage from the implementation of \ac{CSI}-based \ac{JCAS}, specifically when high data rates are required.
Above all others, autonomous vehicle networks may see the implementation of \ac{JCAS} as a tool to improve the quality of shared sensing information and to make the process of sharing such data more efficient. 
The main requirement for autonomous vehicles to perform cooperative \sensing is the availability of high data rates, as each vehicle should ultimately be able to share tens of gigabytes of information per second with all surrounding vehicles \cite{velasco2020autonomous}. 
Cooperative \sensing allows all vehicles participating in the activity to build a full virtual representation of the surrounding real world, deriving information on static obstacles, \ac{VRU}, other vehicles, etc. from what has been sensed and shared by the others through \ac{V2V} \cite{tha2020coop}. 
This approach, albeit currently infeasible on a large scale given the available technologies and supported data rates, would greatly improve the performance of autonomous driving applications, allowing vehicles to identify obstacles that are hidden from their own sensors through what has been detected and shared by surrounding road users \cite{hobert2015enhancements}.

Implementing a network whose users are allowed to share gigabytes of data per second (with each transmission possibly being similar to previous ones, as sensor data may not change drastically from one second to another, especially when travelling at low speed) while simultaneously granting a minimum \ac{QOS} in a safety-critical application is not a simple task; nonetheless, it may benefit from the introduction of \ac{CSI}-based \sensing to reduce the necessity for an autonomous vehicle to share raw sensor data with all surrounding vehicles, by instead only sending the extracted \acp{CSI} as already-parsed information about the surrounding environment.

Studies are already being conducted on the possibility of exploiting shared frequencies and hardware when performing \ac{JCAS} to improve spectrum efficiency and reduce hardware cost: this could result in larger applicability of \ac{JCAS}, even in contexts where it is currently infeasible \cite{9148935, 9705498}, with cheaper implementations on a larger scale from which also applications in autonomous driving could greatly benefit. 

State-of-the-art mechanisms to perform \sensing mainly consist of Artificial Intelligence and Machine Learning applications \cite{arnold2018deep}, but they often require more computational resources and resolution time than are available, especially when working with safety-critical or real-time applications.
Moreover, understanding the mathematical characterization of the electromagnetic channel supporting the transmission may result in efficient dedicated algorithms to extract \acp{CSI} and gain useful information to make \ac{JCAS} more efficient.

This work serves as a continuation of the introductory study proposed in the Bachelor's Degree Thesis titled ``Analysis and Characterization of \wifi Channel State Information'' \cite{bscthesis}.
The goal of this work is to study the statistical properties of a \wifi channel through the analysis of \ac{CSI} behaviour and evolution in time and frequency. 
This analytical approach aims to help identify and describe some channel characteristics that can be used by AI and ML techniques to classify and use \acp{CSI} to perform movement recognition.

%% file: wifi_fundamentals.tex
\chapter{\wifi Fundamentals}
\label{ch:wififun}
\wifi is a trademarked brand name indicating one of the most widespread means of wireless connection used by manufacturers to certify interoperability.
It is commonly associated with the IEEE 802.11 standard, a family of standards --- strictly linked to the Ethernet 802.3 standard --- that defines rules to implement wireless communication between \wifi-enabled devices. 

IEEE 802.11 is the standard for \acp{WLAN} and multiple versions exist, each one supporting different radio technologies and therefore allowing different radio frequencies, maximum ranges, and achievable speeds.
\wifi most commonly uses the 2.4 GHz and 5 GHz frequency bands, but the latest versions of the standard (802.11ax and 802.11be, associated with \wifi 6/6E and \wifi 7 respectively) also support communication on the 6 GHz band. 
Both spectra are divided into channels, each of them identified by its own center frequency, whose number varies depending on the supported channel bandwidth: initially, all channels had a 20 MHz bandwidth, whereas now bandwidths of 40, 80, 160, 240, and 320 MHz are supported. 

The 2.4 GHz frequency band by default is made of 14 overlapping 22 MHz channels, with the possibility of modifying channel bandwidth to either 20 or 40 MHz when using OFDM modulation technique.

The 5 and 6 GHz frequency bands are subject to different regulations depending on the Country, meaning that their channel partition may be different from one Nation to another and that their use may be allowed for different activities in different regions.

Each version of the 802.11 standard implements different modulation techniques by building on the same \ac{MAC} and \ac{PHY} specifications for \acp{WLAN}.

The \acp{CSI} commented and analyzed in this study were collected using the 802.11ax standard on channel 157 at 5 GHz with 20-40-80 MHz bandwidths.

\section{Modulation Techniques}
\label{sec:modulation}
Modulation is a procedure that allows the mapping of information on a physical dimension.
The most straightforward technique is \textit{amplitude modulation}, which consists in mapping the information on the amplitude of a selected dimension. 
An implementable example could be to map binary values onto voltage values, such that values below a selected threshold are mapped onto 0 and values over such threshold are mapped onto 1. 

Amplitude modulation only requires working with one dimension, but as the amount of information to represent grows, the number of dimensions to map such information onto may increase as well. 
From simple amplitude modulation, it is possible to switch to \textit{phase modulation} (known as \ac{PSK}), whose logic is based on the representation of complex numbers, as it represents information exploiting the phase of the exponential used to represent the complex value.
Phase modulation can be obtained through the combination of two non-interfering orthogonal dimensions, which define the signal space as a Cartesian plane as shown in Fig. 2.1. 

\begin{figure}[h!]
	\centering
	\begin{tikzpicture}
		\draw (0,0) circle (2);
		\draw[->] (-2.5,0) -- (2.5,0) node[right] {$sin()$};
		\draw[->] (0,-2.5) -- (0,2.5) node[above] {$cos()$};
		\filldraw[fill=black] (1.414,1.414) circle (0.15) node[above right] { 11};
		\filldraw[fill=black] (1.414,-1.414) circle (0.15) node[below right] { 10};
		\filldraw[fill=black] (-1.414,-1.414) circle (0.15) node[below left] { 00};
		\filldraw[fill=black] (-1.414,1.414) circle (0.15) node[above left] { 01};
	\end{tikzpicture}
	\caption{Example of the signal space defined for \ac{PSK} modulation.}
	\label{fig:psk}
\end{figure}
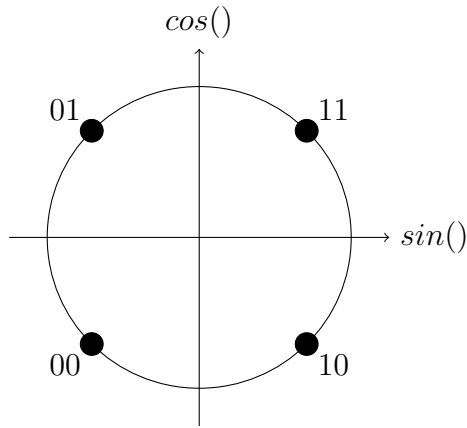

Mapping onto more than two linearly independent dimensions is possible, albeit more complex.
However, one of the most widely employed modulation techniques is called \ac{QAM}, which allows for the transmission of large quantities of data with a relatively small number of symbols. 
\ac{QAM} consists in a representation of information through the combination of two amplitude-modulated signals into a single channel. 
This is achieved by modulating the amplitude of two carrier waves, one cosine (in-phase, indicated as $I$) and the other sine (quadrature-phase, indicated as $Q$), which must be 90 degrees out of phase with each other. 

The in-phase component $I$ represents the $x$ axis of the signal space, while the quadrature component $Q$ represents the $y$ axis. Their combination originates the \ac{QAM} signal. 

A traditional representation of \ac{QAM} modulation relies on the `constellation diagram', which displays a set of points, each one corresponding to a unique combination of amplitude and phase. 
Depending on the number of points making up the diagram, the amount of transmitted information varies; for instance, the diagram for 16-\ac{QAM} shown in Fig.~2.2 consists of 16 points, allowing 4 bits per symbol. 

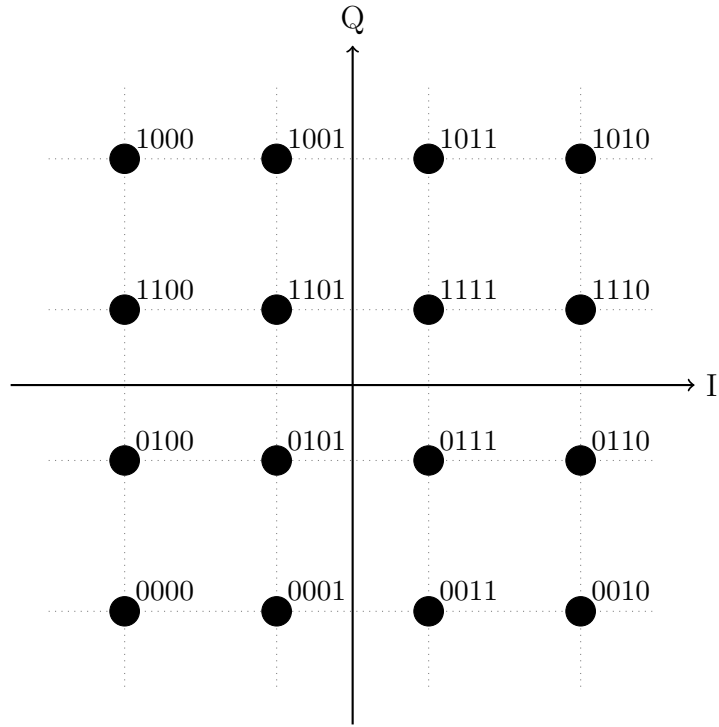
\begin{figure}[h!]
	\centering
	\begin{tikzpicture}
		\foreach \x in {-3,-1,1,3}
		\draw[gray, thin, dotted] (\x, -4) -- (\x, 4);
		\foreach \y in {-3,-1,1,3}
		\draw[gray, thin, dotted] (-4, \y) -- (4, \y);
		
		\draw[thick, ->] (-4.5, 0) -- (4.5, 0) node[right] {I};
		\draw[thick, ->] (0, -4.5) -- (0, 4.5) node[above] {Q};
		
		\foreach \x in {-3,-1,1,3}
		\foreach \y in {-3,-1,1,3}
		\node[draw, circle, fill=black] at (\x, \y) {};
		
		\foreach \x/\y/\label in {-3/-3/0000, -1/-3/0001, 1/-3/0011, 3/-3/0010, 
			-3/-1/0100, -1/-1/0101, 1/-1/0111, 3/-1/0110,
			-3/1/1100, -1/1/1101, 1/1/1111, 3/1/1110,
			-3/3/1000, -1/3/1001, 1/3/1011, 3/3/1010}
		\node[above right] at (\x, \y) {\small \label};
	\end{tikzpicture}
	\label{fig:qam}
	\caption{Example of the constellation diagram defined for \ac{QAM} modulation.}
\end{figure}

\section{\ac{OFDM}}
\ac{OFDM} is a multi-carrier modulation and multiplexing system that transmits data streams as multiple orthogonal narrowband signals named \textit{sub-carriers} \cite{booktlc}, each subject to one of multiple available modulation schemes, such as \ac{QAM}, \ac{BPSK}, \ac{QBPSK}, etc. 
The \ac{OFDM} symbol is given by a combination of all sub-carriers, meaning that each symbol can correspond to more than one bit of information.

\begin{figure}
	\centering
	\includegraphics[width=\textwidth]{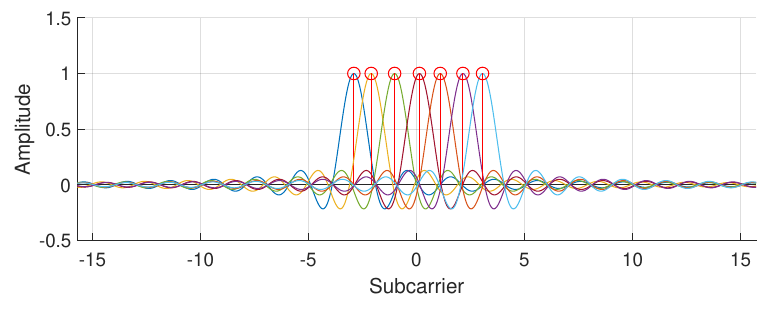}
	\caption{\ac{OFDM} sub-carriers orthogonality.}
	\label{fig:ofdm}
\end{figure}

Given a transmission period $T$, sub-carriers are linearly independent if they are spaced by $\frac{k}{T}$ for $k \in \mathbf{N}$. 
If this constraint is satisfied, their combination shows sub-carrier nulls in correspondence to peaks of adjacent sub-carriers, as shown in \cref{fig:ofdm}. 

One of the main advantages introduced by \ac{OFDM} is the scalability of the rate of transmission: by increasing the transmission period by one symbol, the sub-carriers `widen', causing the bandwidth to increase; vice versa, it decreases by reducing the transmission period. 
Partially overlapping adjacent sub-carriers can contribute to increasing the bandwidth; this is only feasible because sub-carriers are mathematically orthogonal, hence they do not require an interposed guard band that guarantees non-interference.
Moreover, due to sub-carrier orthogonality, possible disturbing interference, noise, and fading phenomena only affect a portion of the sub-carriers, allowing the others to continue their transmission unhindered.

\section{802.11ax Standard Version}
IEEE 802.11ax is associated with \wifi 6 and it operates on the 2.4 and 5 GHz bands, with an additional 6 GHz band in \wifi 6E \cite{802.11ax, 802.11ax_structure}. 
Compared to previous versions of the protocol, 802.11ax uses the frequency spectrum more efficiently, thus increasing the overall network throughput and the per-user performance. 

The improved performances derive from the implementation --- for multi-user communications --- of \ac{OFDMA}, which was already in use in cellular networks since \ac{3GPP} \ac{LTE} but comes as a new approach in \wifi. 
\ac{OFDMA} relies on the same structure as \ac{OFDM}: the available channel is divided into sub-channels, each having its assigned sub-carriers. 
The user can send their data, split into packets, on a sub-channel for a specific amount of time (or frames)\footnote{The \ac{LTE} implementation of \ac{OFDMA} is time-based, meaning that a \ac{RU} is allocated to a single user for each specific amount of time. The implementation in \wifi is frame-based, meaning that a \ac{RU} contains data belonging to different users, thus becoming a Multi-User resource.}, using one or more \acp{RU}, each one consisting of a set of 26 sub-carriers. 
An \ac{AP} can dynamically choose the best \ac{RU} for each \ac{STA} it is communicating with, resulting in higher \ac{SINR} and throughput; \ac{OFDMA} is also more efficient when the quantity of shared data is limited, as the number of selected \acp{RU} can vary depending on the sender's needs.

Compared to previous versions of 802.11, 802.11ax also greatly improves the \ac{QOS} in crowded environments thanks to Uplink Multi-User MIMO. 

The structure of a generic 802.11ax frame respects the following model when implemented in Single-User mode \cite{802.11ax_structure}:

\begin{center}
	\begin{bytefield}[bitheight=1.5\baselineskip, bitwidth=0.3\baselineskip]{60}
		\bitbox{20}{\texttt{Legacy preamble}} 
		& \bitbox{4}{\texttt{RL-\\SIG}} 
		& \bitbox{8}{\texttt{HE-SIG}} 
		& \bitbox{4}{\texttt{HE-\\STF}}
		& \bitbox{4}{\texttt{HE-\\LTF}}
		& \bitbox{4}{\texttt{HE-\\Data}}
		& \bitbox{16}{\texttt{Packet \\Extension}}
	\end{bytefield}
\end{center}

The \texttt{Legacy preamble} guarantees backwards compatibility with previous versions of the 802.11 protocol. The preamble contains information that allows time and frequency synchronization and channel estimation, together with some data regarding payload length and rate of the transmission. 

The \texttt{RL-SIG} (\texttt{Repeated Legacy Signal}) field is used to repeat the content of the \texttt{SIGNAL} field of the \texttt{Legacy preamble}.

The rest of the preamble consists of fields that can only be decoded by 802.11ax devices and whose names start with \texttt{HE} (\texttt{High Efficiency}) to distinguish them from the homonymous parameters of the previous versions of the standard. \texttt{HE-SIG} is used to signal the parameters that are needed to correctly decode the rest of the frame (e.g. bandwidth, number of spatial streams, etc.) while \texttt{HE-STF} and \texttt{HE-LTF} are training fields (respectively short and long) used to perform frequency tuning and channel response estimation. 
The \texttt{HE-Data} field contains the actual user's data and is followed by a \texttt{Packet Extension} field.

When used in Multi-User mode, the packet structure changes slightly: the \texttt{HE-SIG} field is split into two fields (\texttt{HE-SIG-A} and \texttt{HE-SIG-B}) used to set up and tune \ac{MUMIMO} transmission.

\section{802.11be Standard Version}
The updated standard is associated with \wifi 7 --- released in January 2024, final approval expected by the end of 2024 \cite{wifi7schedule, wifi7taskgroup,ehtphy2020} ---, whose key features include \cite{wifi7}:
\begin{itemize}
	\item \acp{MLO};
	\item Support for 320 MHz-wide channels;
	\item 4096-QAM modulation scheme;
	\item Allocation of multiple \acp{RU} to a single \ac{STA};
	\item Uplink and Downlink single user and multi-user \ac{OFDMA} and \ac{MIMO} with up to sixteen spatial streams.
\end{itemize}
The standard aims to enhance \ac{QOS} and reduce latency in transmission. 

The doubling in the channel's maximum bandwidth is supported in all Countries that allow the use of \wifi on the 6 GHz band, granting speed in the order of gigabits and higher throughput compared to previous versions of the standard.
Moreover, the channel bandwidth can be obtained through the juxtaposition of contiguous and non-contiguous 160+160 MHz bands; an additional bandwidth of 240/160+80 MHz is made available.

The 4096-\ac{QAM} modulation scheme achieves 20\% higher transmission rates than the previously employed 1024-\ac{QAM}; this improvement contributes to the enhancement of the \ac{QOS}, combined with the possibility of allocating multiple \acp{RU} to one \ac{STA}, which enhances spectral efficiency. 

The increased throughput obtained through wider channels, higher order modulation, and \ac{MUMIMO} allows the transmission rate to reach up to 46 Gbps while maintaining backwards compatibility with previous \wifi standards. 
An overview of the main differences between 802.11be and 802.11ax is provided in \cref{tab:80211axbe}.

\begin{table}[]
	\centering
	\resizebox{\textwidth}{!}{%
		\begin{tabular}{|
				>{\columncolor[HTML]{C1C1C1}}c |
				>{\columncolor[HTML]{FFFFFF}}c |
				>{\columncolor[HTML]{FFFFFF}}c |}
			\hline
			& \cellcolor[HTML]{C1C1C1}\textbf{\begin{tabular}[c]{@{}c@{}}\wifi 7\\ 802.11be\end{tabular}} & \cellcolor[HTML]{C1C1C1}\textbf{\begin{tabular}[c]{@{}c@{}}\wifi 6E\\ 802.11ax\end{tabular}} \\ \hline
			\textbf{Launch year}        & 2024                                                                                        & \cellcolor[HTML]{FFFFFF}2021                                                                 \\ \hline
			\textbf{Maximum Throughput} & 46 Gbps                                                                                     & \cellcolor[HTML]{FFFFFF}9.6 Gbps                                                             \\ \hline
			\textbf{Frequency Bands}    & 2.4 GHz, 5 GHz, 6 GHz                                                                       & 2.4 GHz, 5 GHz, 6 GHz                                                                        \\ \hline
			\textbf{Supported Channels} & \begin{tabular}[c]{@{}c@{}}Up to 320/160+160 MHz, \\ 240/160+80 MHz\end{tabular}            & \cellcolor[HTML]{FFFFFF}20, 40, 80, 80+80, 160 MHz                                           \\ \hline
			\textbf{Modulation Scheme}  & 4096-\ac{QAM}                                                                               & 1024-\ac{QAM}                                                                                \\ \hline
			\textbf{\ac{MIMO}}          & $16\times 16$ UL/DL \ac{MUMIMO}                                                              & $8\times 8$ UL/DL \ac{MUMIMO}                                                                 \\ \hline
			\textbf{\ac{RU}}            & Multi-\acp{RU}                                                                              & \ac{RU}                                                                                      \\ \hline
		\end{tabular}%
	}
	\caption{Main differences between 802.11ax and 802.11be standards \cite{diffwifi7}.}
	\label{tab:80211axbe}
\end{table}

The \wifi standard 802.11be was not used during the experiments carried out in this study; nonetheless, it was deemed important to highlight its main features, as its imminent introduction to the market will soon impact studies on \ac{CSI} characterization. 
Analysis of the behaviour of channels up to 160 MHz wide is going to contribute to the study of the 240 and 320 MHz channels newly introduced by 802.11be.

\section{\ac{CSI} Structure}
\acp{CSI} can be represented mathematically as a complex number, according to the following formula \cite{Wiar-dataset19}: 
\begin{equation}\label{eq:expCSIold}
	\csi(n)=||\csi(n)||e^{j\angle \csi(n)}
\end{equation}
In this expression, $\csi(n)$ is a \ac{CSI} of the $n$-th sub-carrier, $||\csi(n)||$ corresponds to its amplitude and $\angle \csi(n)$ to its phase.
To maintain consistency with the notation that will be introduced further on, \cref{eq:expCSIold} can be rewritten as:

\begin{equation}\label{eq:expCSI}
	\begin{split}
		\csi(\tin, \subc) & = \acsi(\tin,\subc)\cdot e^{j\angle\csi(\tin,\subc)}\\
						  & = \sqrt{\Re(\csi(\tin,\subc))^2 + 		\Im(\csi(\tin,\subc))^2} \cdot e^{j \tan^{-1} \left( \frac{\Im(\csi(\tin,\subc))}{\Re(\csi(\tin,\subc))} \right)}
	\end{split}
\end{equation}
where \csi(\tin, \subc) represents the \tin-th \ac{CSI} of an experiment on the \subc-th sub-carrier and \acsi indicates its amplitude.

Amplitude and phase take on different values depending on the properties of the signal at the receiver, according to scattering, reflection, and attenuation of the transmitted signal. 

\begin{figure}[h!]
	\centering
	\includegraphics[width=.75\textwidth]{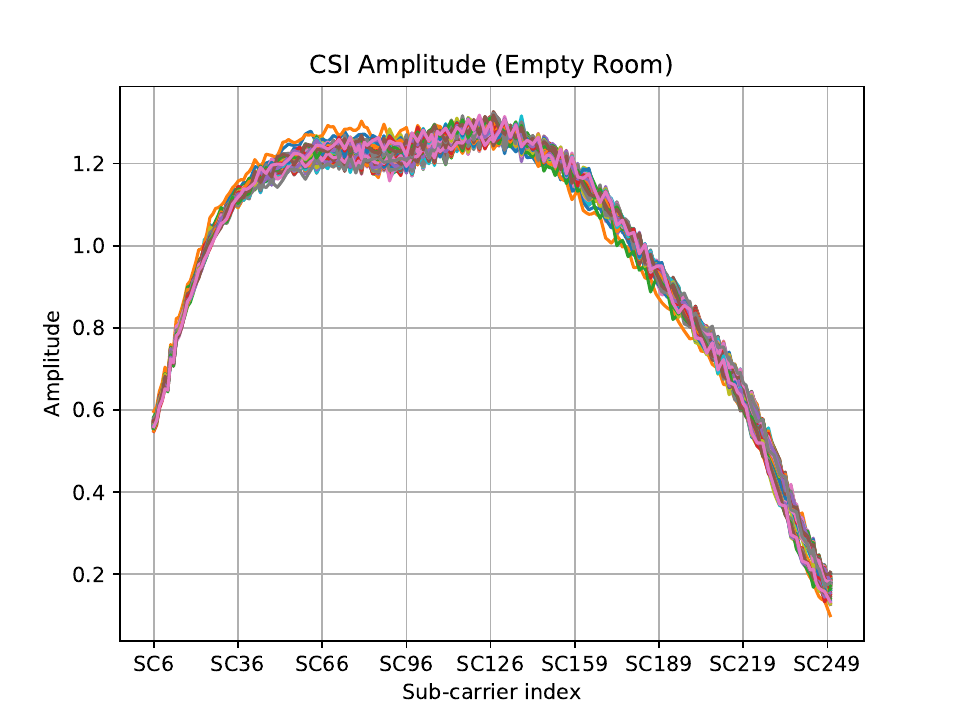}
	\caption{Amplitude \acp{CSI} collected in an empty room.}
	\label{fig:csiempty}
\end{figure}

\begin{figure}[h!]
	\centering
	\includegraphics[width=.75\textwidth]{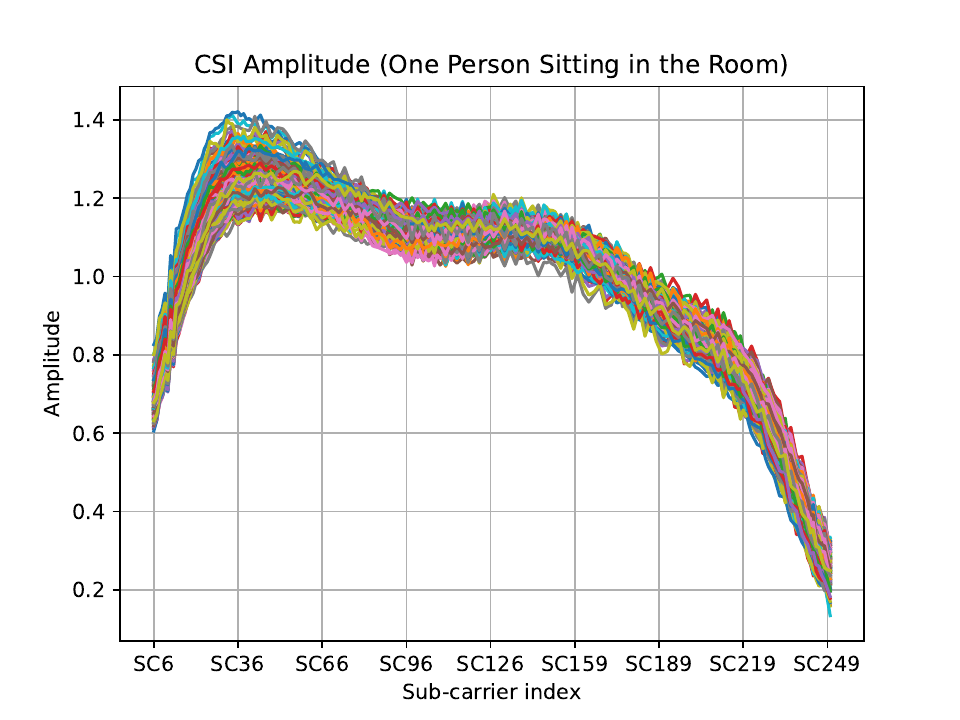}
	\caption{Amplitude of \acp{CSI} collected in a room with one person sitting at a desk.}
	\label{fig:csioneperson}
\end{figure}

This property of \acp{CSI} is already evident in a comparison between two basic scenarios, the first (\cref{fig:csiempty}) representing \acp{CSI} collected in an empty room, the second (\cref{fig:csioneperson}) in the same room with one person sitting at a desk.
All \acp{CSI} plotted in the two figures were collected on channel 157 with 20 MHz bandwidth using 802.11ax; the two experiments were performed at different times of the day, but they both consist of about 18000 \acp{CSI} collected during 10-minute-long captures. 
The effect of the \ac{AGC} was removed from both datasets before plotting.

\begin{figure}[h!]
	\centering
	\includegraphics[width=.75\textwidth]{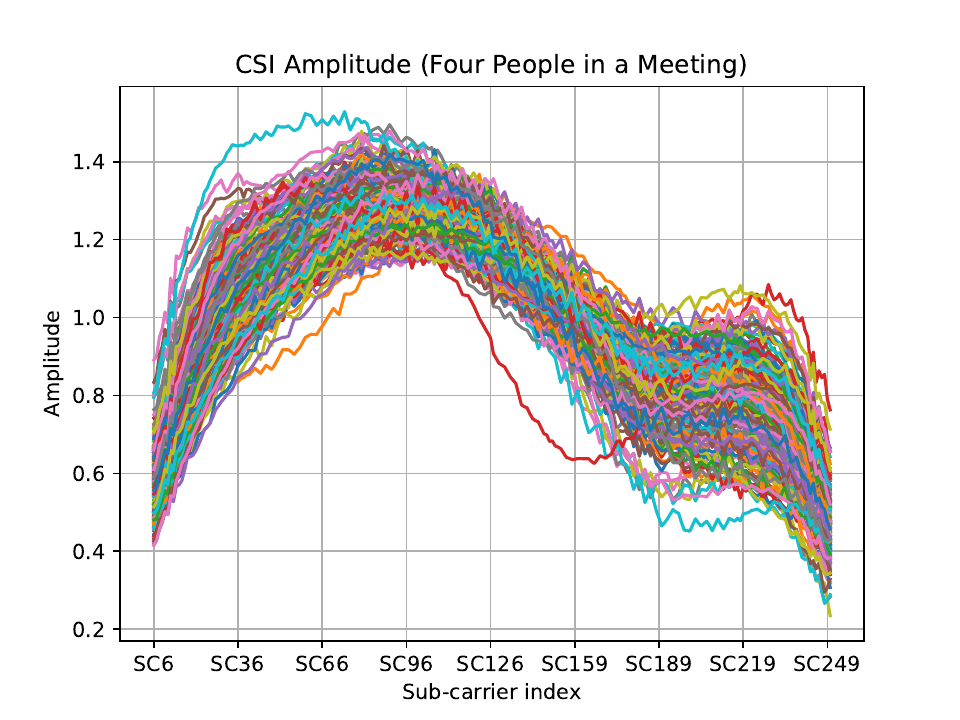}
	\caption{Amplitude of \acp{CSI} collected in a room with four people during a meeting.}
	\label{fig:csifourppl}
\end{figure}

It immediately comes to the eye that the two plots have different trends, but, more importantly, that the \acp{CSI} collected in the empty room are more similar to each other compared to those collected in the room with one person, which have a more visible variability. 
This consideration highlights how the presence of a person --- even though they are not moving around --- can be detected based on the dispersion of the amplitudes of the \acp{CSI}. 
Since the mere presence of a person affects the behavior of the traces, we can expect --- and indeed observe in \cref{fig:csifourppl} --- that the more modifications the environment undergoes, the more variable the corresponding \acp{CSI} become, reflecting people's presence and movements in their amplitudes.

This graphical representation of \acp{CSI} helps to identify the distinction between traces collected in various environments. 
Since \acp{CSI} coming from distinct scenarios clearly display different characteristics, we can assume that environment identification based on the collected \acp{CSI} is feasible. 
This hypothesis will be thoroughly justified in the discussion carried out in the following chapters.

%% file: prev_work.tex
\chapter{Background and Previous Results}
\label{ch:prevres}
During the BSc Thesis \cite{bscthesis}, the analysis of \ac{CSI} traces was mainly focused on the identification of a probability distribution that could be used to describe the increments between consecutive \acp{CSI}. 
This chapter serves as a contextualization for the study that is carried out in the following chapters, to provide a uniform background.
The results commented in this chapter, as well as some considerations that were already discussed in the previous study, are reported solely for a better understanding of the current work and to make this research self-consistent and comprehensive of all results.

\section{Data Collection}
\begin{figure}
	\centering
	\includegraphics[width=0.6\textwidth]{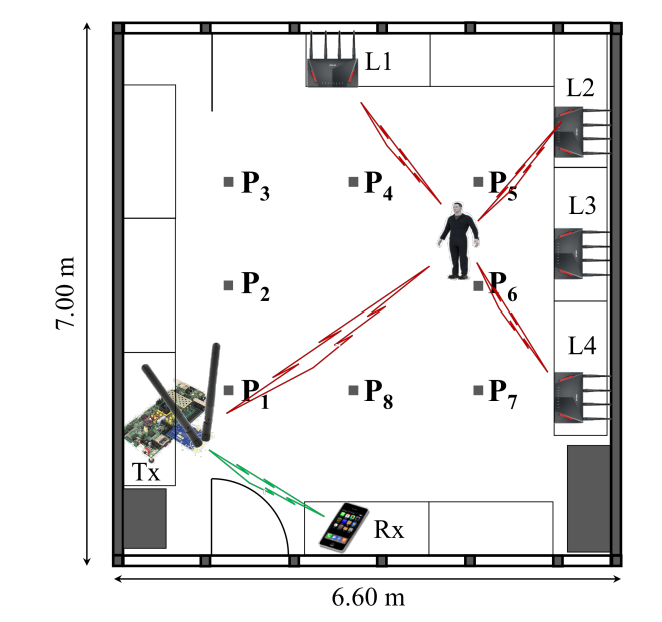}
	\caption{Plan of the Telecommunications Laboratory where \acp{CSI} were collected. Figure taken from \cite{locobfusc} with permission by the authors.}
	\label{fig:tlclabold}
\end{figure}

A relevant difference from the current study is that the data analyzed in \cite{bscthesis} consisted of shorter experiments than those performed for this work; specifically, albeit the number of \ac{CSI} is elevated, the experiments consisted of collections of bursts of \acp{CSI} with a limited duration (i.e. in the order of tens of seconds) performed in the Telecommunications Laboratory within the Department of Information Engineering at the University of Brescia. 
Each capture was collected while one person was standing in one of eight fixed spots within the room, with the transmitter and receivers placed along the walls of the laboratory, as can be seen in \cref{fig:tlclabold}.
Moreover, being a preliminary analysis, data categorization was not yet done as described in \cref{ch:exp_setup}, therefore the configuration files of the experimental setup are not available. 

No experiments in an empty laboratory were available; nevertheless, comparison between traces collected in different environments and with a varying number of people in the room has only gained relevance in this study, therefore its absence in previous work does not have a meaningful impact.

It must be noted that the impact of \ac{AGC} was not initially eliminated from the amplitudes, as its removal was introduced in this work, together with normalization and quantization of both increments and amplitudes (see \cref{ch:normquant}). 

\section{Amplitude Evolution in Time}
\begin{figure}
	\centering
	\includegraphics[width=0.7\textwidth]{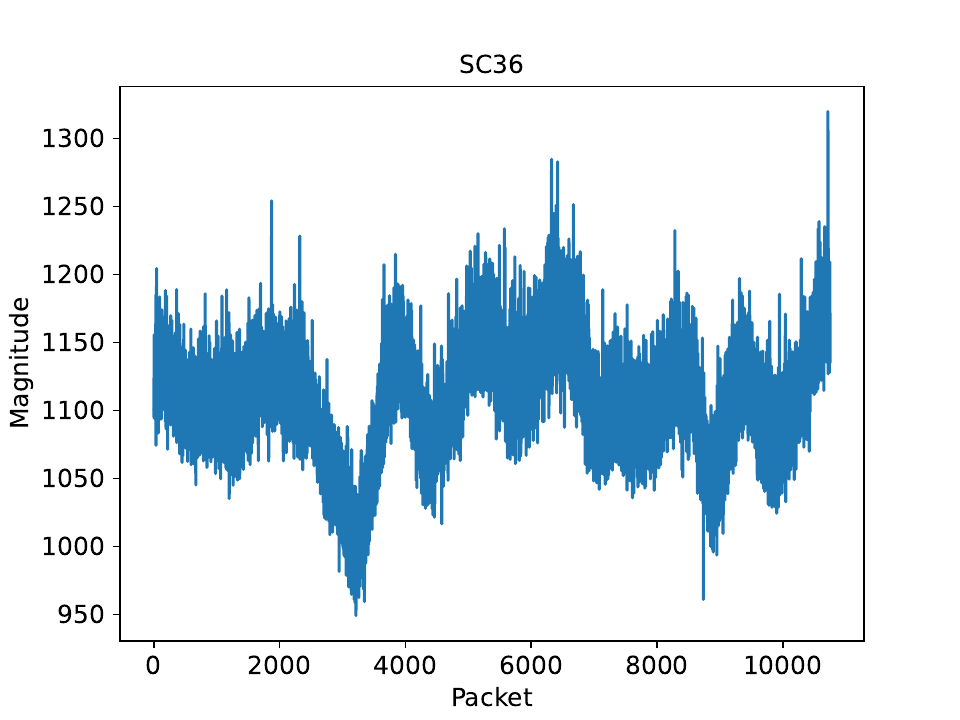}
	\caption{Example of amplitude evolution in time on sub-carrier 36. \acp{CSI} were collected on a 20 MHz bandwidth channel. Image taken from \cite{bscthesis}.}
	\label{fig:timeevol}
\end{figure}

The initial goal of the study was to identify the presence of correlation in time between the amplitudes on the same sub-carrier. 
As the first step in the analysis, graphs showing the amplitude evolution in time were presented, with discrete time being the variable on the $x$ axis and amplitude on the $y$ axis.
An example of such plots is shown in \cref{fig:timeevol}.

The fluctuating trend of the \ac{CSI} is mainly due to the \ac{AGC}, which undermines considerations about the stationary nature of the process.

It is relevant to specify that multiple features can be identified in various plots showing similar trends on adjacent sub-carriers, which may imply the presence of frequency correlation between the amplitudes.

\section{Amplitude Relative Frequency Observation}
\begin{figure}
	\centering
	\includegraphics[width=0.7\textwidth]{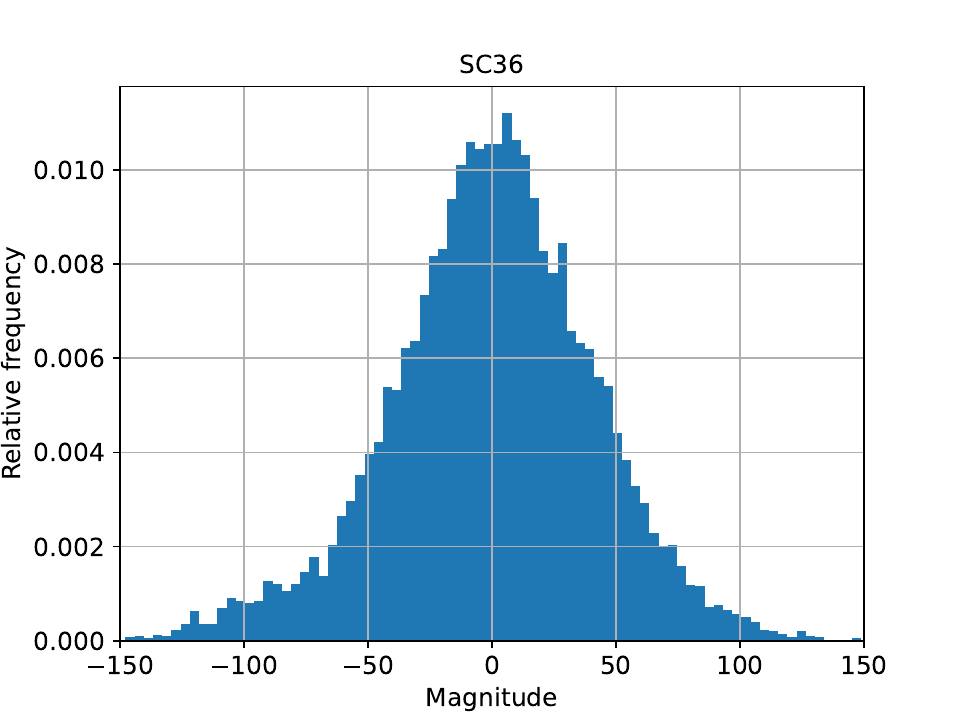}
	\caption{Example of amplitude relative frequency on sub-carrier 36. \acp{CSI} were collected on a 20 MHz bandwidth channel. Image taken from \cite{bscthesis}.}
	\label{fig:amplrel}
\end{figure}

The \ac{CSI} amplitudes on the different sub-carriers were also shown using histograms having normalized amplitude on the $x$ axis and its relative frequency on the $y$ axis. 
\cref{fig:amplrel} is an example of the analysis that was carried out.
This approach allowed to make an initial hypothesis about the family of probability distributions that could be used to describe the process of the amplitudes. Nonetheless, the true process that we wanted to characterize was that of the increments, which are the next main topic in the previous research.

\section{Amplitude Increments and Auto-Correlation}
\begin{figure}[]
	\centering
	\includegraphics[width=0.7\textwidth]{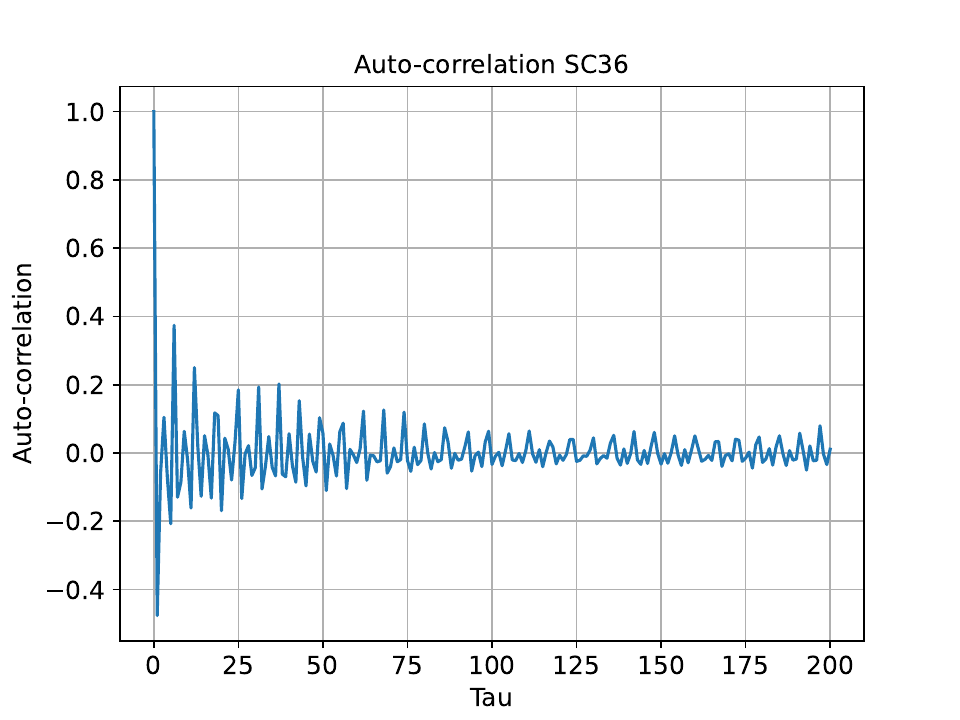}
	\caption{Example of increments auto-correlation on sub-carrier 36. \acp{CSI} were collected on a 20 MHz bandwidth channel. Image taken from \cite{bscthesis}.}
	\label{fig:autocorr}
\end{figure}

As analysis of the amplitude itself was not deemed sufficient to characterize the evolution of \acp{CSI}, increments were then taken into account as well by computing their auto-correlation over time on each separate sub-carrier.
Their values were assumed to belong to a Markovian process --- hence memoryless ---, which means that the increments auto-correlation should appear to be noise-like around value 0. 
This assumption was tested through the empirical evaluation of the auto-correlation of the increments, as shown in \cref{fig:autocorr}, which displays a rapid reduction of the values towards zero, as expected. 
Whether the process can actually be described as Markovian remains to be explored by observing longer experiments performed in different contexts. 

\section{Amplitude Increments Analysis}
\begin{figure}[]
	\centering
	\includegraphics[width=0.7\textwidth]{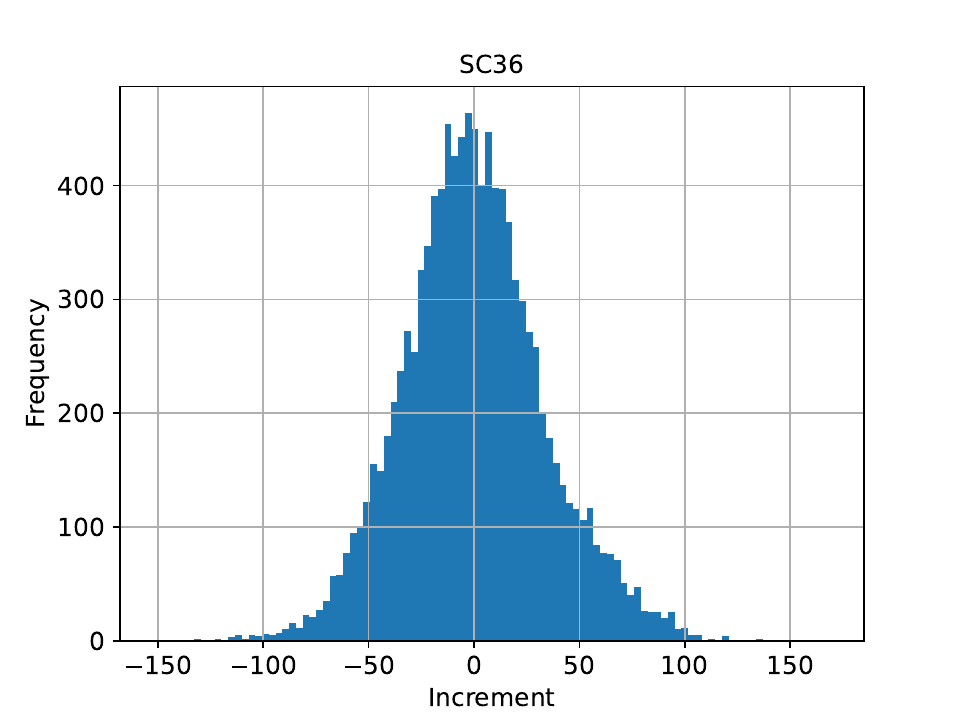}
	\caption{Example of increments distribution on sub-carrier 36. \acp{CSI} were collected on a 20 MHz bandwidth channel. Image taken from \cite{bscthesis}.}
	\label{fig:incrdistold}
\end{figure}

The distributions of the increments --- see \cref{fig:incrdistold} for an example --- were compared to a set of known distributions; the Gaussian distribution turned out to be the best-fitting one and was hence chosen as the best approximation and the final proposed model.

%% file: experimental_setup.tex
\chapter{Experimental Setup}
\label{ch:exp_setup}
The collection of all \acp{CSI} used in the analysis proposed in this thesis was built through multiple separate experiments, with different possible configurations. 
Two datasets have been analyzed for this work, as described in the following sections.

\section{Collected data}
\label{sec:colldata}
\begin{figure}
	\centering
	\includegraphics[width=0.7\textwidth]{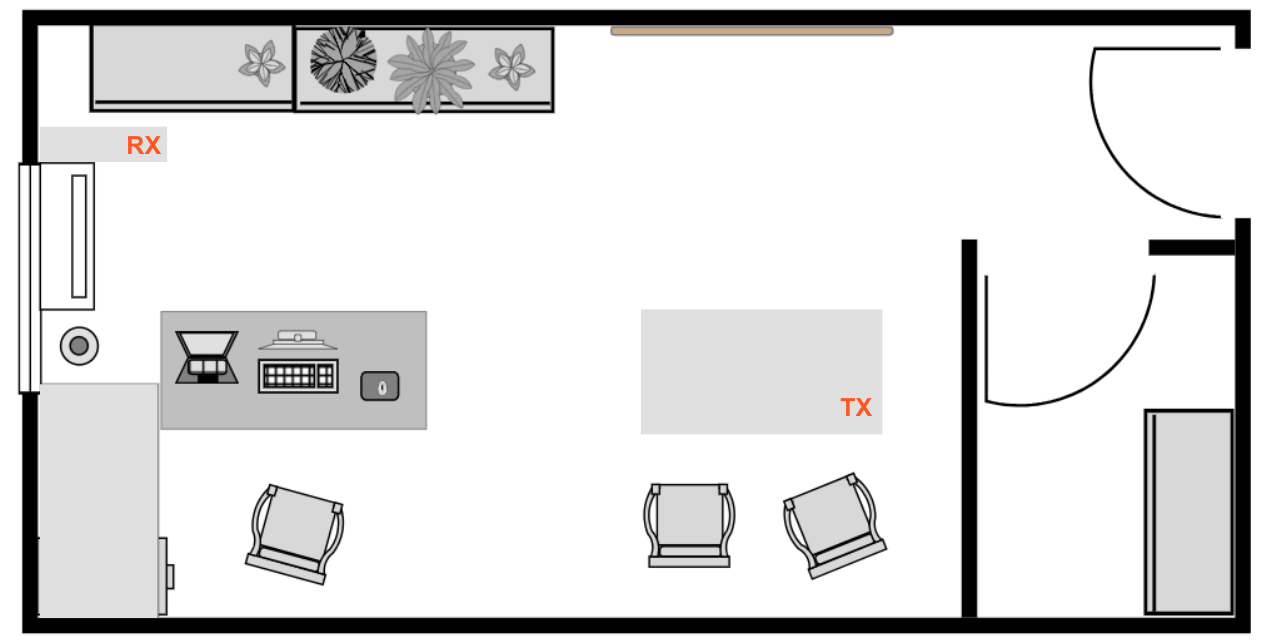}
	\caption{Layout of the office where all \acp{CSI} were collected.}
	\label{fig:office}
\end{figure}

The main dataset employed in this study was collected within the same office in the Department of Information Engineering at the University of Brescia by Elena Tonini. 
An approximate layout of the office is provided in \cref{fig:office}, which shows the locations of the transmitter and the receiver alongside the main working stations used during office hours.

\ac{CSI} captures were performed in three distinct scenarios: 
\begin{itemize}
	\item Empty Scenario: empty office;
	\item Static Scenario: one person sitting in the office and working at the desk;
	\item Fully Dynamic Scenario: multiple people moving around the office.
\end{itemize}
An additional setup called `Dynamic Scenario' (i.e., one person moving around the office) has been defined and will be taken into account in the future to compare its results with those obtained for the Static and Fully Dynamic scenarios, as it could be considered an intermediate setup between these two. 

By associating a \texttt{json} file to each capture, all experiments are categorized according to their own scenario.
The file also contains other mandatory fields used to keep track of configuration parameters needed by the hardware itself to set up the data exchange from which \acp{CSI} are collected; other complementary fields provide corollary information that can be used to fully characterize the experiment. 
An example of the \texttt{json} configuration file is shown in \cref{lst:expconfig}, while a thorough description of the metadata it contains is provided in \cref{app:experiments}.

\begin{lstlisting}[caption={Example of configuration file.},label={lst:expconfig}]
	{	
		"date": {
			"day": 12,
			"month": 12,
			"year": 2023
		},
		"locationID": "U004",
		"experiment": "capture",
		"adHocTransmission": true,
		"usleep": 10000,
		"avgDuration": 600,
		"bandwidth": 20,
		"modulation": "ax",
		"numRx": 1,
		"numTx": 1,
		"numAntennasTx": 1,
		"numAntennasRx": 1,
		"numSpatialStreams": 1,
		"people":{
			"present": true,
			"num": 2,
			"moving": false,
			"names": ["John Smith", "Jane Doe"]
		},
		"notes": "JS sitting at the main desk, JD facing him."	
	}
\end{lstlisting}

All \ac{CSI} traces are extracted from \ac{OFDM}-modulated \wifi frames transmitted over a channel regulated by the 802.11ax protocol. 
The used channel is number 157 (whose center frequency is 5785 MHz) within the 5 GHz frequency band with 20, 40, and 80 MHz bandwidth. 

The traces were extracted using Nexmon Channel State Information Extractor \cite{2019:freecsi, NexmonProject}. 
The analyzed traffic is generated by a board communicating with a receiving device: looking at \cref{fig:office}, the transmitter was placed on the bottom right corner of the rightmost desk, whereas the receiver was placed on a rigid support close to the closet on the top left of the room.

The collected data are summarized in \cref{tab:expsummary}.

\begin{table}[]
	\centering
	\resizebox{\textwidth}{!}{%
		\begin{tabular}{|
				>{\columncolor[HTML]{C1C1C1}}c |cccc|cccc|ccccccccc|}
			\hline
			\textbf{SCENARIO} &
			\multicolumn{4}{c|}{\cellcolor[HTML]{C1C1C1}\textbf{Empty}} &
			\multicolumn{4}{c|}{\cellcolor[HTML]{C1C1C1}\textbf{Static}} &
			\multicolumn{9}{c|}{\cellcolor[HTML]{C1C1C1}\textbf{Fully Dynamic}} \\ \hline
			\textbf{802.11} &
			\multicolumn{4}{c|}{ax} &
			\multicolumn{4}{c|}{ax} &
			\multicolumn{9}{c|}{ax} \\ \hline
			\textbf{BW (MHz)} &
			\multicolumn{2}{c|}{20} &
			\multicolumn{1}{c|}{40} &
			80 &
			\multicolumn{2}{c|}{20} &
			\multicolumn{1}{c|}{40} &
			80 &
			\multicolumn{1}{c|}{20} &
			\multicolumn{4}{c|}{40} &
			\multicolumn{4}{c|}{80} \\ \hline
			\textbf{\begin{tabular}[c]{@{}c@{}}\# Spatial\\ Streams\end{tabular}} &
			\multicolumn{1}{c|}{1} &
			\multicolumn{1}{c|}{1} &
			\multicolumn{1}{c|}{1} &
			1 &
			\multicolumn{1}{c|}{1} &
			\multicolumn{1}{c|}{1} &
			\multicolumn{1}{c|}{1} &
			1 &
			\multicolumn{1}{c|}{1} &
			\multicolumn{1}{c|}{1} &
			\multicolumn{1}{c|}{1} &
			\multicolumn{1}{c|}{1} &
			\multicolumn{1}{c|}{1} &
			\multicolumn{1}{c|}{1} &
			\multicolumn{1}{c|}{1} &
			\multicolumn{1}{c|}{1} &
			1 \\ \hline
			\textbf{\# Experiments} &
			\multicolumn{1}{c|}{4} &
			\multicolumn{1}{c|}{5} &
			\multicolumn{1}{c|}{3} &
			1 &
			\multicolumn{1}{c|}{4} &
			\multicolumn{1}{c|}{5} &
			\multicolumn{1}{c|}{1} &
			1 &
			\multicolumn{1}{c|}{4} &
			\multicolumn{1}{c|}{2} &
			\multicolumn{1}{c|}{1} &
			\multicolumn{1}{c|}{1} &
			\multicolumn{1}{c|}{2} &
			\multicolumn{1}{c|}{2} &
			\multicolumn{1}{c|}{1} &
			\multicolumn{1}{c|}{1} &
			1 \\ \hline
			\textbf{\begin{tabular}[c]{@{}c@{}}Avg. Duration\\ (sec.)\end{tabular}} &
			\multicolumn{1}{c|}{600} &
			\multicolumn{1}{c|}{10} &
			\multicolumn{1}{c|}{600} &
			600 &
			\multicolumn{1}{c|}{600} &
			\multicolumn{1}{c|}{10} &
			\multicolumn{1}{c|}{600} &
			600 &
			\multicolumn{1}{c|}{600} &
			\multicolumn{1}{c|}{600} &
			\multicolumn{1}{c|}{600} &
			\multicolumn{1}{c|}{600} &
			\multicolumn{1}{c|}{600} &
			\multicolumn{1}{c|}{600} &
			\multicolumn{1}{c|}{600} &
			\multicolumn{1}{c|}{600} &
			60 \\ \hline
			\textbf{\begin{tabular}[c]{@{}c@{}}\#\ac{CSI}/exp.\\ (approx.)\end{tabular}} &
			\multicolumn{1}{c|}{18k} &
			\multicolumn{1}{c|}{300} &
			\multicolumn{1}{c|}{18k} &
			18k &
			\multicolumn{1}{c|}{18k} &
			\multicolumn{1}{c|}{300} &
			\multicolumn{1}{c|}{18k} &
			18k &
			\multicolumn{1}{c|}{18k} &
			\multicolumn{1}{c|}{18k} &
			\multicolumn{1}{c|}{18k} &
			\multicolumn{1}{c|}{18k} &
			\multicolumn{1}{c|}{18k} &
			\multicolumn{1}{c|}{18k} &
			\multicolumn{1}{c|}{18k} &
			\multicolumn{1}{c|}{18k} &
			2.6k \\ \hline
			\textbf{\# People} &
			\multicolumn{1}{c|}{0} &
			\multicolumn{1}{c|}{0} &
			\multicolumn{1}{c|}{0} &
			0 &
			\multicolumn{1}{c|}{1} &
			\multicolumn{1}{c|}{1} &
			\multicolumn{1}{c|}{1} &
			1 &
			\multicolumn{1}{c|}{4} &
			\multicolumn{1}{c|}{2} &
			\multicolumn{1}{c|}{3} &
			\multicolumn{1}{c|}{4} &
			\multicolumn{1}{c|}{5} &
			\multicolumn{1}{c|}{2} &
			\multicolumn{1}{c|}{3} &
			\multicolumn{1}{c|}{4} &
			5 \\ \hline
		\end{tabular}%
	}
	\caption{Summary of the experiments performed for this study. Collected \acp{CSI} are classified primarily by scenario, additional parameters are then specified as they may vary from one experiment to another.}
	\label{tab:expsummary}
\end{table}

\section{Additional available dataset}
\label{sec:altdata}
Some additional collections of \acp{CSI} have been made available by the authors of \cite{CoGri22_ComCom}. 
In this work, the data is analyzed to implement \ac{CSI} obfuscation against unauthorized \wifi sensing, therefore most of it consists of obfuscated traces. 
Nonetheless, some `clean' collections are available --- i.e., retrieved without activating the obfuscator --- which are the ones that have been used in this thesis.
The goal of studying the channel characterization using data that originates from a different work is to support the sensing techniques implemented in other studies with an innovative approach, leveraging the quantification of the information content of a \ac{CSI} instead of \ac{ML} alone.  

\begin{figure}[h!]
	\centering
	\includegraphics[width=0.4\textwidth]{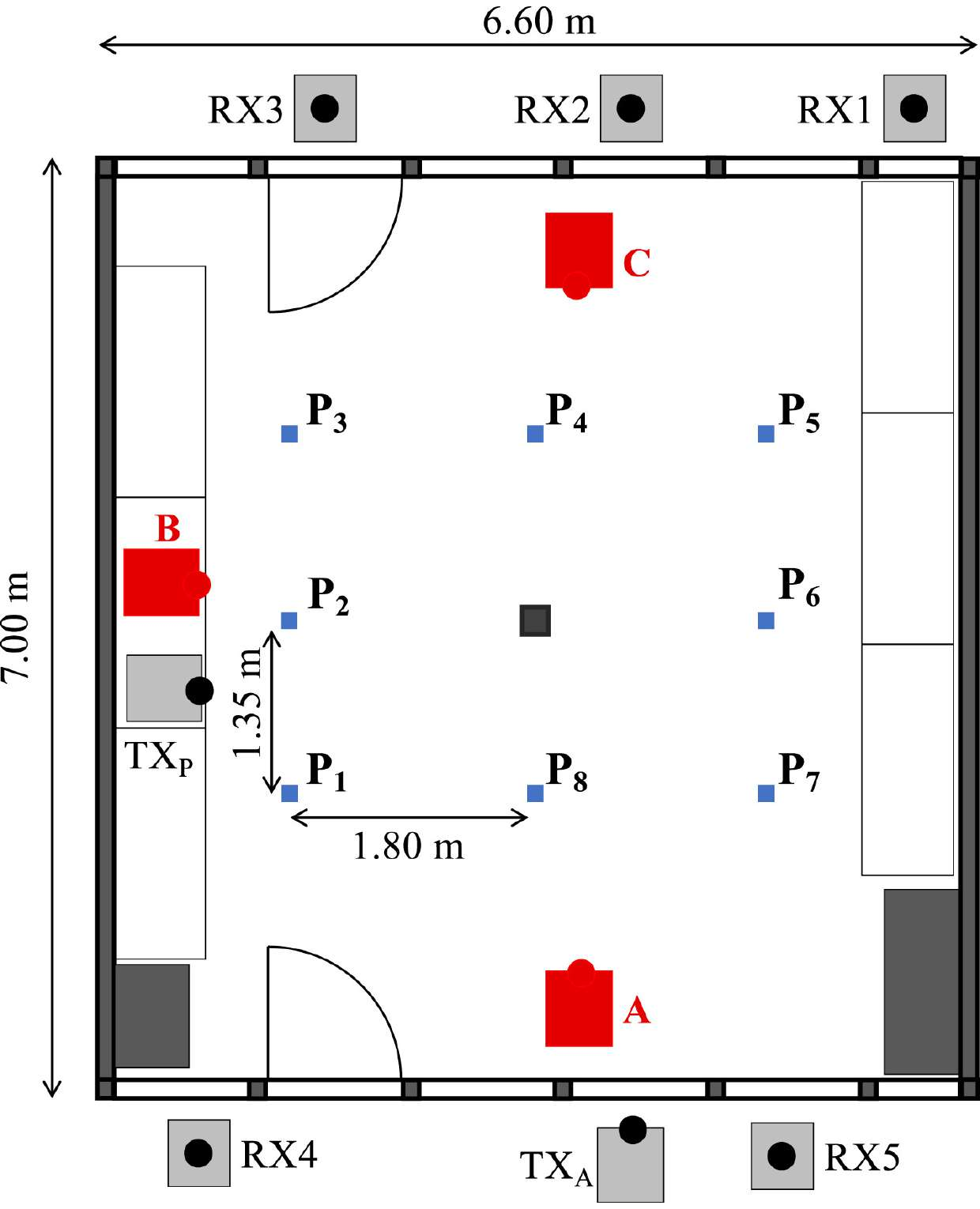}
	\caption{Plan of the lab where the AntiSense dataset was collected. The blue squares represent the eight possible locations of the person within the room; the red boxes labelled A, B, and C are the locations of the obfuscator in different scenarios: for the scope of this work, their effect is irrelevant; the transmitter inside the room is used for passive attacks, the one outside is used for active ones. Figure taken from \cite{CoGri22_ComCom} with permission by the authors.}
	\label{fig:tlclab}
\end{figure}

This dataset was collected on an 80 MHz 802.11ac channel in the Telecommunications Laboratory of the Department of Information Engineering at the University of Brescia in August 2021 by Dr.~Marco Cominelli, Prof.~Francesco Gringoli, and Prof.~Renato Lo Cigno, and it has been used to study device-free localization and test the performance of different obfuscation systems in \cite{CoGri22_ComCom}.
From this point onward, the dataset will be referred to as the `AntiSense dataset'.
Its content is relative to \acp{CSI} captured with the same person standing in one of 8 pre-determined target positions and with a receiver placed in one of 5 fixed spots just outside the perimeter of the laboratory, as displayed in \cref{fig:tlclab}. 

The transmitter was placed either on the left wall of the laboratory or just outside the door at the bottom of \cref{fig:tlclab}, depending on the experiment. 
In our study, we will only focus on the \acp{CSI} collected when the active transmitter was that outside of the laboratory. 
The technology used to extract the \acp{CSI} is the same as that described in \cref{sec:colldata}.

For each of the two positions of the transmitter, the dataset has then been partitioned into a training, a testing, and a validation dataset, each consisting of eight captures for each position the receiver was placed in.
For the scope of this work, only the training and testing datasets will be analyzed.
As a whole, the dataset employed in this study consists of 120 captures, 40 of which are discarded (the validation partition), divided as shown in \cref{tab:antisensesummary}.

\begin{table}[h!]
	\centering
	\resizebox{\textwidth}{!}{%
		\begin{tabular}{|cc|ccccc|ccccc|ccccc|}
			\hline
			\rowcolor[HTML]{C1C1C1} 
			\multicolumn{2}{|c|}{\cellcolor[HTML]{C1C1C1}\textbf{PARTITION}} &
			\multicolumn{5}{c|}{\cellcolor[HTML]{C1C1C1}\textbf{TRAINING}} &
			\multicolumn{5}{c|}{\cellcolor[HTML]{C1C1C1}\textbf{TESTING}} &
			\multicolumn{5}{c|}{\cellcolor[HTML]{C1C1C1}\textbf{VALIDATION}} \\ \hline
			\rowcolor[HTML]{FFFFFF} 
			\multicolumn{2}{|c|}{\cellcolor[HTML]{C1C1C1}\textbf{RX POS.}} &
			\multicolumn{1}{c|}{\cellcolor[HTML]{FFFFFF}\textbf{rx1}} &
			\multicolumn{1}{c|}{\cellcolor[HTML]{FFFFFF}\textbf{rx2}} &
			\multicolumn{1}{c|}{\cellcolor[HTML]{FFFFFF}\textbf{rx3}} &
			\multicolumn{1}{c|}{\cellcolor[HTML]{FFFFFF}\textbf{rx4}} &
			\cellcolor[HTML]{FFFFFF}\textbf{rx5} &
			\multicolumn{1}{c|}{\cellcolor[HTML]{FFFFFF}\textbf{rx1}} &
			\multicolumn{1}{c|}{\cellcolor[HTML]{FFFFFF}\textbf{rx2}} &
			\multicolumn{1}{c|}{\cellcolor[HTML]{FFFFFF}\textbf{rx3}} &
			\multicolumn{1}{c|}{\cellcolor[HTML]{FFFFFF}\textbf{rx4}} &
			\textbf{rx5} &
			\multicolumn{1}{c|}{\cellcolor[HTML]{FFFFFF}\textbf{rx1}} &
			\multicolumn{1}{c|}{\cellcolor[HTML]{FFFFFF}\textbf{rx2}} &
			\multicolumn{1}{c|}{\cellcolor[HTML]{FFFFFF}\textbf{rx3}} &
			\multicolumn{1}{c|}{\cellcolor[HTML]{FFFFFF}\textbf{rx4}} &
			\textbf{rx5} \\ \hline
			\rowcolor[HTML]{FFFFFF} 
			\multicolumn{1}{|c|}{\cellcolor[HTML]{C1C1C1}} &
			\cellcolor[HTML]{C1C1C1}\textbf{P1} &
			\multicolumn{1}{c|}{\cellcolor[HTML]{FFFFFF}1000} &
			\multicolumn{1}{c|}{\cellcolor[HTML]{FFFFFF}1000} &
			\multicolumn{1}{c|}{\cellcolor[HTML]{FFFFFF}1000} &
			\multicolumn{1}{c|}{\cellcolor[HTML]{FFFFFF}1000} &
			1000 &
			\multicolumn{1}{c|}{\cellcolor[HTML]{FFFFFF}1000} &
			\multicolumn{1}{c|}{\cellcolor[HTML]{FFFFFF}1000} &
			\multicolumn{1}{c|}{\cellcolor[HTML]{FFFFFF}1000} &
			\multicolumn{1}{c|}{\cellcolor[HTML]{FFFFFF}1000} &
			1000 &
			\multicolumn{1}{c|}{\cellcolor[HTML]{FFFFFF}200} &
			\multicolumn{1}{c|}{\cellcolor[HTML]{FFFFFF}200} &
			\multicolumn{1}{c|}{\cellcolor[HTML]{FFFFFF}200} &
			\multicolumn{1}{c|}{\cellcolor[HTML]{FFFFFF}200} &
			200 \\ \cline{2-17} 
			\rowcolor[HTML]{FFFFFF} 
			\multicolumn{1}{|c|}{\cellcolor[HTML]{C1C1C1}} &
			\cellcolor[HTML]{C1C1C1}\textbf{P2} &
			\multicolumn{1}{c|}{\cellcolor[HTML]{FFFFFF}1000} &
			\multicolumn{1}{c|}{\cellcolor[HTML]{FFFFFF}1000} &
			\multicolumn{1}{c|}{\cellcolor[HTML]{FFFFFF}1000} &
			\multicolumn{1}{c|}{\cellcolor[HTML]{FFFFFF}1000} &
			1000 &
			\multicolumn{1}{c|}{\cellcolor[HTML]{FFFFFF}1000} &
			\multicolumn{1}{c|}{\cellcolor[HTML]{FFFFFF}1000} &
			\multicolumn{1}{c|}{\cellcolor[HTML]{FFFFFF}1000} &
			\multicolumn{1}{c|}{\cellcolor[HTML]{FFFFFF}1000} &
			1000 &
			\multicolumn{1}{c|}{\cellcolor[HTML]{FFFFFF}200} &
			\multicolumn{1}{c|}{\cellcolor[HTML]{FFFFFF}200} &
			\multicolumn{1}{c|}{\cellcolor[HTML]{FFFFFF}200} &
			\multicolumn{1}{c|}{\cellcolor[HTML]{FFFFFF}200} &
			200 \\ \cline{2-17} 
			\rowcolor[HTML]{FFFFFF} 
			\multicolumn{1}{|c|}{\cellcolor[HTML]{C1C1C1}} &
			\cellcolor[HTML]{C1C1C1}\textbf{P3} &
			\multicolumn{1}{c|}{\cellcolor[HTML]{FFFFFF}1000} &
			\multicolumn{1}{c|}{\cellcolor[HTML]{FFFFFF}1000} &
			\multicolumn{1}{c|}{\cellcolor[HTML]{FFFFFF}1000} &
			\multicolumn{1}{c|}{\cellcolor[HTML]{FFFFFF}1000} &
			1000 &
			\multicolumn{1}{c|}{\cellcolor[HTML]{FFFFFF}1000} &
			\multicolumn{1}{c|}{\cellcolor[HTML]{FFFFFF}1000} &
			\multicolumn{1}{c|}{\cellcolor[HTML]{FFFFFF}1000} &
			\multicolumn{1}{c|}{\cellcolor[HTML]{FFFFFF}1000} &
			1000 &
			\multicolumn{1}{c|}{\cellcolor[HTML]{FFFFFF}200} &
			\multicolumn{1}{c|}{\cellcolor[HTML]{FFFFFF}200} &
			\multicolumn{1}{c|}{\cellcolor[HTML]{FFFFFF}200} &
			\multicolumn{1}{c|}{\cellcolor[HTML]{FFFFFF}200} &
			200 \\ \cline{2-17} 
			\rowcolor[HTML]{FFFFFF} 
			\multicolumn{1}{|c|}{\cellcolor[HTML]{C1C1C1}} &
			\cellcolor[HTML]{C1C1C1}\textbf{P4} &
			\multicolumn{1}{c|}{\cellcolor[HTML]{FFFFFF}1000} &
			\multicolumn{1}{c|}{\cellcolor[HTML]{FFFFFF}1000} &
			\multicolumn{1}{c|}{\cellcolor[HTML]{FFFFFF}1000} &
			\multicolumn{1}{c|}{\cellcolor[HTML]{FFFFFF}1000} &
			1000 &
			\multicolumn{1}{c|}{\cellcolor[HTML]{FFFFFF}1000} &
			\multicolumn{1}{c|}{\cellcolor[HTML]{FFFFFF}1000} &
			\multicolumn{1}{c|}{\cellcolor[HTML]{FFFFFF}1000} &
			\multicolumn{1}{c|}{\cellcolor[HTML]{FFFFFF}1000} &
			1000 &
			\multicolumn{1}{c|}{\cellcolor[HTML]{FFFFFF}200} &
			\multicolumn{1}{c|}{\cellcolor[HTML]{FFFFFF}200} &
			\multicolumn{1}{c|}{\cellcolor[HTML]{FFFFFF}200} &
			\multicolumn{1}{c|}{\cellcolor[HTML]{FFFFFF}200} &
			200 \\ \cline{2-17} 
			\rowcolor[HTML]{FFFFFF} 
			\multicolumn{1}{|c|}{\cellcolor[HTML]{C1C1C1}} &
			\cellcolor[HTML]{C1C1C1}\textbf{P5} &
			\multicolumn{1}{c|}{\cellcolor[HTML]{FFFFFF}1000} &
			\multicolumn{1}{c|}{\cellcolor[HTML]{FFFFFF}1000} &
			\multicolumn{1}{c|}{\cellcolor[HTML]{FFFFFF}1000} &
			\multicolumn{1}{c|}{\cellcolor[HTML]{FFFFFF}1000} &
			1000 &
			\multicolumn{1}{c|}{\cellcolor[HTML]{FFFFFF}1000} &
			\multicolumn{1}{c|}{\cellcolor[HTML]{FFFFFF}1000} &
			\multicolumn{1}{c|}{\cellcolor[HTML]{FFFFFF}1000} &
			\multicolumn{1}{c|}{\cellcolor[HTML]{FFFFFF}1000} &
			1000 &
			\multicolumn{1}{c|}{\cellcolor[HTML]{FFFFFF}200} &
			\multicolumn{1}{c|}{\cellcolor[HTML]{FFFFFF}200} &
			\multicolumn{1}{c|}{\cellcolor[HTML]{FFFFFF}200} &
			\multicolumn{1}{c|}{\cellcolor[HTML]{FFFFFF}200} &
			200 \\ \cline{2-17} 
			\rowcolor[HTML]{FFFFFF} 
			\multicolumn{1}{|c|}{\cellcolor[HTML]{C1C1C1}} &
			\cellcolor[HTML]{C1C1C1}\textbf{P6} &
			\multicolumn{1}{c|}{\cellcolor[HTML]{FFFFFF}1000} &
			\multicolumn{1}{c|}{\cellcolor[HTML]{FFFFFF}1000} &
			\multicolumn{1}{c|}{\cellcolor[HTML]{FFFFFF}1000} &
			\multicolumn{1}{c|}{\cellcolor[HTML]{FFFFFF}1000} &
			1000 &
			\multicolumn{1}{c|}{\cellcolor[HTML]{FFFFFF}1000} &
			\multicolumn{1}{c|}{\cellcolor[HTML]{FFFFFF}1000} &
			\multicolumn{1}{c|}{\cellcolor[HTML]{FFFFFF}1000} &
			\multicolumn{1}{c|}{\cellcolor[HTML]{FFFFFF}1000} &
			1000 &
			\multicolumn{1}{c|}{\cellcolor[HTML]{FFFFFF}200} &
			\multicolumn{1}{c|}{\cellcolor[HTML]{FFFFFF}200} &
			\multicolumn{1}{c|}{\cellcolor[HTML]{FFFFFF}200} &
			\multicolumn{1}{c|}{\cellcolor[HTML]{FFFFFF}200} &
			200 \\ \cline{2-17} 
			\rowcolor[HTML]{FFFFFF} 
			\multicolumn{1}{|c|}{\cellcolor[HTML]{C1C1C1}} &
			\cellcolor[HTML]{C1C1C1}\textbf{P7} &
			\multicolumn{1}{c|}{\cellcolor[HTML]{FFFFFF}1000} &
			\multicolumn{1}{c|}{\cellcolor[HTML]{FFFFFF}1000} &
			\multicolumn{1}{c|}{\cellcolor[HTML]{FFFFFF}1000} &
			\multicolumn{1}{c|}{\cellcolor[HTML]{FFFFFF}1000} &
			1000 &
			\multicolumn{1}{c|}{\cellcolor[HTML]{FFFFFF}1000} &
			\multicolumn{1}{c|}{\cellcolor[HTML]{FFFFFF}1000} &
			\multicolumn{1}{c|}{\cellcolor[HTML]{FFFFFF}1000} &
			\multicolumn{1}{c|}{\cellcolor[HTML]{FFFFFF}1000} &
			1000 &
			\multicolumn{1}{c|}{\cellcolor[HTML]{FFFFFF}200} &
			\multicolumn{1}{c|}{\cellcolor[HTML]{FFFFFF}200} &
			\multicolumn{1}{c|}{\cellcolor[HTML]{FFFFFF}200} &
			\multicolumn{1}{c|}{\cellcolor[HTML]{FFFFFF}200} &
			200 \\ \cline{2-17} 
			\rowcolor[HTML]{FFFFFF} 
			\multicolumn{1}{|c|}{\multirow{-8}{*}{\rotatebox[origin=c]{90}{\textbf{\begin{tabular}[c]{@{}c@{}}\cellcolor[HTML]{C1C1C1}POS. OF PERSON \\ \cellcolor[HTML]{C1C1C1}IN THE ROOM\end{tabular}}\cellcolor[HTML]{C1C1C1}}}} &
			\cellcolor[HTML]{C1C1C1}\textbf{P8} &
			\multicolumn{1}{c|}{\cellcolor[HTML]{FFFFFF}1000} &
			\multicolumn{1}{c|}{\cellcolor[HTML]{FFFFFF}1000} &
			\multicolumn{1}{c|}{\cellcolor[HTML]{FFFFFF}1000} &
			\multicolumn{1}{c|}{\cellcolor[HTML]{FFFFFF}1000} &
			1000 &
			\multicolumn{1}{c|}{\cellcolor[HTML]{FFFFFF}1000} &
			\multicolumn{1}{c|}{\cellcolor[HTML]{FFFFFF}1000} &
			\multicolumn{1}{c|}{\cellcolor[HTML]{FFFFFF}1000} &
			\multicolumn{1}{c|}{\cellcolor[HTML]{FFFFFF}1000} &
			1000 &
			\multicolumn{1}{c|}{\cellcolor[HTML]{FFFFFF}200} &
			\multicolumn{1}{c|}{\cellcolor[HTML]{FFFFFF}200} &
			\multicolumn{1}{c|}{\cellcolor[HTML]{FFFFFF}200} &
			\multicolumn{1}{c|}{\cellcolor[HTML]{FFFFFF}200} &
			200 \\ \hline
			\rowcolor[HTML]{C1C1C1} 
			\multicolumn{2}{|c|}{\cellcolor[HTML]{C1C1C1}\textbf{TOT. CSI}} &
			\multicolumn{1}{c|}{\cellcolor[HTML]{C1C1C1}8000} &
			\multicolumn{1}{c|}{\cellcolor[HTML]{C1C1C1}8000} &
			\multicolumn{1}{c|}{\cellcolor[HTML]{C1C1C1}8000} &
			\multicolumn{1}{c|}{\cellcolor[HTML]{C1C1C1}8000} &
			8000 &
			\multicolumn{1}{c|}{\cellcolor[HTML]{C1C1C1}8000} &
			\multicolumn{1}{c|}{\cellcolor[HTML]{C1C1C1}8000} &
			\multicolumn{1}{c|}{\cellcolor[HTML]{C1C1C1}8000} &
			\multicolumn{1}{c|}{\cellcolor[HTML]{C1C1C1}8000} &
			8000 &
			\multicolumn{1}{c|}{\cellcolor[HTML]{C1C1C1}1600} &
			\multicolumn{1}{c|}{\cellcolor[HTML]{C1C1C1}1600} &
			\multicolumn{1}{c|}{\cellcolor[HTML]{C1C1C1}1600} &
			\multicolumn{1}{c|}{\cellcolor[HTML]{C1C1C1}1600} &
			1600 \\ \hline
		\end{tabular}%
	}
	\caption{Summary of the experiments making up the AntiSense dataset. Collected \acp{CSI} are classified according to the dataset partition they belong to, the position of the receiver and that of the person in the room. P1 through P8 reference the positions displayed in \cref{fig:tlclab}.}
	\label{tab:antisensesummary}
\end{table}

%% file: notation.tex
\chapter{Notation}
\label{ch:notation}

Let $\csi(\tin,\subc)$ be the \ac{CSI} collected during a generic experiment; $\tin \in [1,\nsamp]$ is the sequence number (ordering) of the collection, which consists of \nsamp samples, and  $\subc \in [1,\nsc]$ is the index of the sub-carrier. 
\nsc is the number of useful sub-carriers, i.e., those that are not suppressed in transmission and can be usefully employed to estimate the \ac{CSI}.
$\csi(\tin,\subc)$ is a bi-dimensional vector containing the I/Q samples of the \ac{CSI}, represented as a complex number with real and imaginary parts, so that  
\[
\acsi(\tin,\subc) = \sqrt{\Re(\csi(\tin,\subc))^2 + \Im(\csi(\tin,\subc))^2}
\]
is the amplitude of $\csi(\tin,\subc)$. 

The total collection of the samples of an experiment $\csi(\cdot,\cdot)$ can be (and normally is) annotated with additional data such as the descriptor of the experiment, the sub-carrier spacing, and so forth, as described in \cref{ch:exp_setup}, while each sample $\csi(\tin,\cdot)$ is annotated at least with the absolute time $\dt_{\tin} = \rtm[\csi(\tin,\cdot)] - \rtm[\csi(\tin-1,\cdot)]$, where $\rtm(\cdot)$ is a function measuring the actual reception time of the frame with the collected \ac{CSI}. 
Clearly, $\dt_0=\nan$ is undefined and irrelevant. 

\cref{tab:notation} summarizes relevant symbols, including those that have not been introduced yet in this chapter as they will be encountered further on in the discussion. 
\begin{table}[]
	\centering
	\resizebox{\textwidth}{!}{%
		\begin{tabular}{|c|l|}
			\hline
			\rowcolor[HTML]{C1C1C1} 
			\textbf{SYMBOL}                 & \multicolumn{1}{c|}{\cellcolor[HTML]{C1C1C1}\textbf{DESCRIPTION}} \\ \hline
			\rowcolor[HTML]{FFFFFF} 
			$\csi(\tin,\subc)$              & \ac{CSI} collected during a generic experiment                    \\ \hline
			\rowcolor[HTML]{FFFFFF} 
			$\tin \in [1,\nsamp]$           & Sequence number of a \ac{CSI} within the collection               \\ \hline
			\rowcolor[HTML]{FFFFFF} 
			\nsamp                          & Number of samples of the collection                               \\ \hline
			\rowcolor[HTML]{FFFFFF} 
			$\subc \in [1,\nsc]$            & Index of the sub-carrier                                          \\ \hline
			\rowcolor[HTML]{FFFFFF} 
			\nsc                            & Number of useful sub-carriers                                     \\ \hline
			\rowcolor[HTML]{FFFFFF} 
			$\acsi(\tin,\subc)$             & Amplitude of $\csi(\tin, \subc)$                                  \\ \hline
			\rowcolor[HTML]{FFFFFF} 
			$\dt_{\tin}$                    & Absolute time of $\csi(\tin, \subc)$                              \\ \hline
			\rowcolor[HTML]{FFFFFF} 
			$\rtm(\cdot)$                   & Reception time of the frame with the collected \ac{CSI}           \\ \hline
			\rowcolor[HTML]{FFFFFF} 
			$\overline{\acsi}$              & Energy of a \ac{CSI}                                              \\ \hline
			\rowcolor[HTML]{FFFFFF} 
			$\acsi^{\min}$                  & Minimum amplitude value across all \acp{CSI}                      \\ \hline
			\rowcolor[HTML]{FFFFFF} 
			$\acsi^{\max}$                  & Maximum amplitude value across all \acp{CSI}                      \\ \hline
			\rowcolor[HTML]{FFFFFF} 
			$\racsi(\subc)$                 & Reference \ac{CSI} computed on each experiment                    \\ \hline
			\rowcolor[HTML]{FFFFFF} 
			$\iacsi(\tin, \subc)$              & Increment between two \acp{CSI} on the same sub-carrier \subc \\ \hline
			\rowcolor[HTML]{FFFFFF} 
			$\iacsi^{\min}$                 & Minimum increment value across all \acp{CSI}                      \\ \hline
			\rowcolor[HTML]{FFFFFF} 
			$\iacsi^{\max}$                 & Maximum increment value across all \acp{CSI}                      \\ \hline
			\rowcolor[HTML]{FFFFFF} 
			$\ndis(\sigma)$                 & Gaussian distribution with standard deviation $\sigma$            \\ \hline
			\rowcolor[HTML]{FFFFFF} 
			$\qndis(\sigma)$                & Quantized Gaussian distribution with standard deviation $\sigma$  \\ \hline
			\rowcolor[HTML]{FFFFFF} 
			$\nbits$                        & Number of bits used to quantize the increments                    \\ \hline
			\rowcolor[HTML]{FFFFFF} 
			$\acsibits$                     & Number of bits used to quantize the amplitude                     \\ \hline
			\rowcolor[HTML]{FFFFFF} 
			$P_w$                           & Probability weight of the tails of $\ndis$                        \\ \hline
			\rowcolor[HTML]{FFFFFF} 
			$\iacsimax$                     & Value of the increments after which tails are discarded            \\ \hline
			\rowcolor[HTML]{FFFFFF} 
			$I(X;Y)$                        & Mutual information between random variables $X$ and $Y$           \\ \hline
			\rowcolor[HTML]{FFFFFF} 
			$\mint_{A}$                     & Internal \ac{MI} for experiment $A$                               \\ \hline
			\rowcolor[HTML]{FFFFFF} 
			$\mext_{A,B}$                   & External \ac{MI} between experiment $A$ and $B$                   \\ \hline
			\rowcolor[HTML]{FFFFFF} 
			$\whd(\racsi, \acsi(\tin, \cdot))$ & Weighted Hamming Distance between \racsi and \acsi            \\ \hline
			\rowcolor[HTML]{FFFFFF} 
			$\overline{\whd(\racsi,\acsi)}$ & Average \whd between \racsi and \acsi                          \\ \hline
		\end{tabular}%
	}
	\caption{Summary of the used symbols, in order of appearance.}
	\label{tab:notation}
\end{table}

%% file: norm_quant.tex
\chapter{Normalization and Quantization}
\label{ch:normquant}
 
Before delving into the processing and analysis of the collected data, it is necessary to introduce a standard representation of $\acsi(\tin, \subc)$ to ensure the feasibility of the comparisons between different experiments. 
The following processing is done separately on each experiment.

The first step in the conditioning of the collected data is the normalization of the \ac{CSI} amplitude in the assumption that the transmitted energy is constant, as it should be, and variations in the collected data are due only to different gains of the \ac{AGC} at the receiver\footnote{Note that whenever the same variable appears on both sides of an equation, the equal sign should be interpreted as an assignment rather than a comparison between left and right sides.}:
\begin{equation}
	\overline{\acsi} = \frac{1}{\nsc} \sum_{\subc=1}^{\nsc} \acsi(\tin,\subc);\quad 
	\acsi(\tin,\subc) = \frac{\acsi(\tin,\subc)}{\overline{\acsi}} \forall  \subc \in [1,\nsc]
	\label{eq:normenergy}
\end{equation} 

Next, all values are mapped in the $[0,1]$ interval as follows. 
First, the minimum amplitude value is computed and subtracted from all values across all \acp{CSI} and sub-carriers:  
\begin{equation}
	\acsi^{\min} = \min_{\tin \in [1,\nsamp],\subc \in [1,\nsc]} \acsi(\tin,\subc) 
	\label{eq:csimin}
\end{equation} 
\begin{equation}
	\acsi(\tin,\subc) = \acsi(\tin,\subc) - \acsi^{\min},\, \tin \in [1,\nsamp], \subc \in [1,\nsc] 
	\label{eq:submin}
\end{equation} 
Next, the maximum is computed; this is in practice the maximum difference between the minimum and the maximum of the original sequence. Its value is employed to normalize the amplitude to one:
\begin{equation}
	\acsi^{\max} = \max_{\tin \in [1,\nsamp],\subc \in [1,\nsc]} \acsi(\tin,\subc) 
	\label{eq:csimax}
\end{equation} 
\begin{equation}
	\acsi(\tin,\subc) = \frac{\acsi(\tin,\subc)}{\acsi^{\max}},\, \tin \in [1,\nsamp], \subc \in [1,\nsc] 
	\label{eq:normuno}
\end{equation} 
Since the minimum and maximum amplitude values are computed over the whole experiment rather than referring to a single trace, it is possible for some \acp{CSI} to not fully cover the interval from 0 to 1. 
Hence, some traces may not reach the limits of the normalization interval at all, but, over the entire experiment, there will be \textit{at least} one \ac{CSI} that is equal to 0 --- and, similarly, to 1 --- on \textit{at least} one sub-carrier. 
The \acp{CSI} taking on these two values may be distinct traces.

Finally, a reference \ac{CSI} amplitude is calculated for each experiment as the average over $\tin \in [1,\nsamp]$ of all the \acp{CSI} collected during the experiment: 
\begin{equation}
	\racsi(\subc) = \frac{1}{\nsamp} \sum_{\tin=1}^{\nsamp} \acsi(\tin,\subc)
	\label{eq:csiav}
\end{equation} 
This reference \ac{CSI} is taken as the representative of the experiment to estimate the information content embedded in the \ac{CSI} by the propagation environment in the different experiments.

\section{Estimate of the Increments Process}
\label{sec:increment}

Once the amplitude of the \acp{CSI} is properly normalized, the process of the increments can be estimated. 
An increment in amplitude is defined as the difference between the values of the amplitude of two different (not necessarily consecutive) \acp{CSI} on the same sub-carrier. 
Mathematically:
\begin{equation}
	\iacsi(\tin, \subc) = \acsi(\tin+\dt, \subc) - \acsi(\tin, \subc)
	\label{eq:increment}
\end{equation}

This topic has been analyzed more in-depth in \cite{bscthesis}, whose goal is to provide a simple mathematical model that can be used to approximate the process of the increments on each sub-carrier. 
The work suggests that --- at least in an initial approach to the process estimation --- it is possible and sufficient to use a Normal distribution to approximate the process on each sub-carrier. 
Through the properties of Gaussian distributions, it is possible to combine the Normal distribution of each sub-carrier to create a single Gaussian distribution $\ndis(\sigma)$ that can be used to represent the entire process of the increments across all sub-carriers used during transmission. 
In this process, the average is zero by construction, but it must be zero also because an increment process with non-zero mean implies a non-stationary process on the one hand, and, on the other hand, either a diverging received --- and transmitted --- power or a vanishing signal, and both cases are not meaningful in this work.

In other words, this work assumes that all increments of each element $\acsi(\tin,\subc)$ of the \ac{CSI} are i.i.d.. 

Given these assumptions, a stochastic model of the \ac{CSI} amplitude evolution is:
\begin{equation}
	\acsi(\tin,\subc) = \acsi(\tin-1,\subc) + \ndis(\sigma),\, \tin \in [2,\nsamp], \subc \in [1,\nsc],
	\label{eq:increments}
\end{equation} 
thus there is only the need to estimate $\sigma$ given all the available increment samples $\iacsi(\tin,\subc) = \acsi(\tin,\subc) - \acsi(\tin-1,\subc),\, \tin \in [2,\nsamp], \subc \in [1,\nsc]$. 
To avoid cluttering the notation, $\sigma$ and its estimate are indicated with the same symbol: 
\begin{equation}
	\sigma = \frac{1}{\nsc} \sum_{\subc=1}^{\nsc} 
	\left[ \frac{1}{\nsamp-1} \sqrt{\sum_{\tin=2}^{\nsamp} \iacsi^2(\tin,\subc)} \right] 
	\label{eq:sigma}
\end{equation} 

Indeed, $\sigma$ can be estimated differently, as the evolution of $\acsi(\tin,\subc)$ has memory. 
In a process with memory, \cref{eq:sigma} correctly estimates the one-step increment marginal distribution, but may not represent the n-step increment marginal distribution correctly, as well known, as memory may even eventually make processes self-similar. 
Although this discussion will not be brought on further, note that $\sigma$ can be estimated as: 
\begin{equation}
	\sigma = \frac{1}{\nsc} \sum_{\subc=1}^{\nsc} 
	\left[ \frac{1}{\ncomb - 1} \sqrt{\sum_{h}^{\ncomb} \iacsih^2(\tin,\subc)} \right] 
	\label{eq:sigmalt}
\end{equation} 
where 
\[
\ncomb = (\nsamp-1)+(\nsamp-2)+\cdots +(\nsamp-\left\lfloor \frac{\nsamp}{2}\right\rfloor)
\] 
and $\iacsih = \acsi(\tin,\subc) - \acsi(\tin-h,\subc),\, \tin \in [2,\nsamp-h+1], \subc \in [1,\nsc]$. 
The n-step increments are limited to $\left\lfloor \frac{\nsamp}{2}\right\rfloor$ to have enough samples for each increment gap.
This latter estimate should yield a larger variance of the process as normally $\iacsih > \iacsi$ for $h \geq 2$. 
Which estimate is better and hence chosen will be decided based on the effectiveness of the modelling in predicting the information content of experiments. 

\section{Quantization and Mapping}
\label{sec:qmap}
Amplitude $\acsi(\cdot,\cdot)$ needs to be quantized for two reasons. 
First and foremost, working with real numbers makes estimating information content (in the sense of Shannon theory) of \acp{CSI} extremely difficult.
Second, indeed, the measure of the \ac{CSI} itself is already quantized by the hardware that collects it but, unfortunately, access to the low-level measures is not given. 
The hardware exports \acsi values in floating point format, so knowing the exact representation of $\csi(\cdot,\cdot)$ is impossible and, in any case, the pre-processing described so far is best done using floating point. 
Since both $\acsi$ and $\iacsi$ values need to be quantized to provide a correct and comprehensive representation of the collected data, this section will start with the approach to quantization of $\iacsi$ values.

Before quantizing the increments, it is necessary to apply the same procedure used on the amplitudes to ensure that $\iacsi(\tin, \subc) \in [0,1]$. 
Firstly, we compute the minimum value of the increments:
\begin{equation}
	\iacsi^{\min} = \min_{\tin \in [1,\nsamp],\subc \in [1,\nsc]} \iacsi(\tin,\subc) 
	\label{eq:iacsimin}
\end{equation} 
\begin{equation}
	\iacsi(\tin,\subc) = \iacsi(\tin,\subc) - \iacsi^{\min},\, \tin \in [1,\nsamp], \subc \in [1,\nsc] 
	\label{eq:isubmin}
\end{equation} 
Next, compute the maximum and normalize the increments to one:
\begin{equation}
	\iacsi^{\max} = \max_{\tin \in [1,\nsamp],\subc \in [1,\nsc]} \iacsi(\tin,\subc) 
	\label{eq:iacsimax}
\end{equation} 
\begin{equation}
	\iacsi(\tin,\subc) = \frac{\iacsi(\tin,\subc)}{\iacsi^{\max}},\, \tin \in [1,\nsamp], \subc \in [1,\nsc] 
	\label{eq:inormuno}
\end{equation} 
The approach to increment normalization is clearly the same as that used with amplitudes, as shown by \cref{eq:csimin} to \cref{eq:normuno}.

For the reasons mentioned at the beginning of this section, the quantization process of the increments is based on some simple reasoning: the number of used bits should be the smallest possible to represent the increments reasonably accurately; in other words, the \ac{PDF} of $\ndis(\sigma)$ should be reasonably approximated by the \ac{PMF} of  $\qndis(\sigma)$, where $\qndis(\sigma)$ is the quantized version of  $\ndis(\sigma)$; notice that there is no need to quantize $\sigma$, but only the output of the distribution (whether it is used as a random generator of synthetic $\iacsi$ values or empirically built on experimental data). 

There are many ways of defining a good approximation, both in terms of residual errors and in probabilistic terms.
Let us, for the time being, neglect this specific step and assume that \nbits bits are used to represent $\qndis(\sigma)$, or, equivalently, the quantized version of $\iacsi(\tin,\subc)$. 

First of all, a maximum (and minimum) value of $\ndis(\sigma)$ needs to be set. This helps to define a symmetrical and finite interval of values that $\ndis$ can be defined on, essentially cutting off the tails of the distribution that would make its domain infinite and hard to work with.    
This is easily done by defining the probability weight $P_w$ of the tails that are thus discarded and selecting an appropriate value. 
The reference equation is:
\begin{equation}
	P_w = \frac{2\sigma}{\sqrt{\pi}} \int_{\iacsimax}^{\infty} \ndis(\sigma) d\delta  
	\label{eq:ptails}
\end{equation}
To properly select \iacsimax given a desired $P_w$, error function tables or calculators can be used. 
To maintain discussion and implementation simple, \iacsimax is set to a value that is an integer multiple of the standard deviation of the Normal distribution; specifically, $\iacsimax = n \sigma$ with $n$ integer such that \cref{eq:ptails} is smaller than the desired probability. 
This probability can be selected simply by observing that, given a certain number of collected samples, probabilities smaller than the inverse of the number itself cannot be estimated. Therefore, in this case, selecting $\frac{1}{\nsc} < P_w < \frac{10}{\nsc}$ is appropriate. 

For reasons that will become clear later in the discussion, \nbits can be selected such that $\qndis(\sigma)$ is centred around zero (obvious) and its support is over 7, 15, or 31 values only. 
As \nbits bits are used to represent the entire interval $[0,1]$ with uniform quantization, \nbits selection as a function of \iacsimax and the cardinality of $\qndis(\sigma)$ support is straightforward.  
Indeed, there are boundary conditions to be fixed in the numerical computation as $\iacsimax$ may not be coincident with any sampling interval and $\qndis(\sigma)$ must be normalized to be a proper distribution, i.e., the weight $P_w$ must be accounted for. 

A simple and effective way of fixing the boundary conditions is to approximate $\iacsimax$ with the nearest sampling interval larger than $\iacsimax$ and accumulate $P_w$ on the boundary intervals. 
This is a good approximation method as long as the probabilities that are accumulated on the boundary intervals do not alter the structure of the Normal distribution, i.e. as long as they do not increase the probability of the outermost intervals to the point that the resulting distribution no longer resembles a Normal distribution. 

By leveraging the boundary conditions --- and therefore limiting the domain of $\qndis(\sigma)$ to $[-\iacsimax, \iacsimax]$ --- symmetry of \qndis around zero is ensured. 

Up to this point, only quantization of the distribution of the increments has been taken into account, but \acsi needs to be quantized as well. 
This will allow to finally switch to work with integer, finite numbers representing both the absolute amplitude values and the increments of the \acp{CSI} coherently. 
Coherent representation means that the quantization interval of both \ndis and \acsi must be the same, which implies that the number of bits used to quantize \ndis is smaller than the number of bits used to represent \acsi.  

Switching to this representation, a \ac{CSI} is a vector of integer positive numbers of dimension \nsc. 
The generation of a synthetic trace of \acsi values --- that is, a new \ac{CSI} --- is obtained by adding a vector of increments (\iacsi) with the same dimension to generate a new \ac{CSI}; iterating the procedure produces new \acp{CSI}. 
To be able to add \iacsi to $\acsi(\tin,\subc)$ we have to make sure that the quantization process uses the same quantization interval for increments and amplitudes, which may require the introduction of some tricks to obtain a coherent and consistent result. 

Forcing the quantization intervals of the increments and the amplitudes to be exactly the same is not easy because \iacsimax is not necessarily equal to $k\cdot2^{-n}$ with $k, n\in\mathbb{N}$. 
Therefore, we first have to set the number of quantization bits of \acsi to  
\begin{equation}
	\acsibits =  \left\lceil \log_2\left( \frac{1}{\iacsimax} \times (2^{\nbits}+1)\right) \right\rceil 
	\label{eq:acsibits}
\end{equation} 
where \nbits is the number of bits used to quantize \iacsi. 
Next, we have to `tune' $\iacsimax$ on the first sampling interval boundary larger than $\iacsimax$ and re-sample the increments. 
From now on $\iacsimax$ refers to the tuned version, so that \acsi and \iacsi are sampled with exactly the same sampling interval and each value of $\qndis(\sigma)$ has the appropriate probability value. 
It is important to note that in a generative process all of this will be done before generating increments. 

With this approach, the study is not strictly bound to use Normal distributions and it is also possible to compute a quantized empirical distribution starting from measured values. 

This approach ensures that both quantities are defined and quantized over the same interval and using the appropriate numbers of bits, allowing easy and correct summation of \acsi and \iacsi values. 
Since a \ac{CSI} normalized between 0 and 1 may not reach the ends of the normalization interval, as said in the paragraph following \cref{eq:normuno}, the quantized \acp{CSI} are naturally subject to the same consideration. 
Therefore, the generic quantized \ac{CSI} may not be equal to 0 or $2^{\acsibits}-1$ on any sub-carrier, but at least one \ac{CSI} within each experiment will reach such values on at least one sub-carrier.

Again, to avoid cluttering the notation, from now on we will assume that all the quantities have been correctly quantized and mapped; the same symbols (\acsi, \iacsi, \ndis, \ldots) introduced so far will be used to represent the quantized version of the variables. 

\section{Visualization of the Normalization and Quantization Processes}
\label{sec:visualnq}
\begin{figure}[p]
	\centering
	\begin{subfigure}{0.41\linewidth}
		\includegraphics[width=\linewidth]{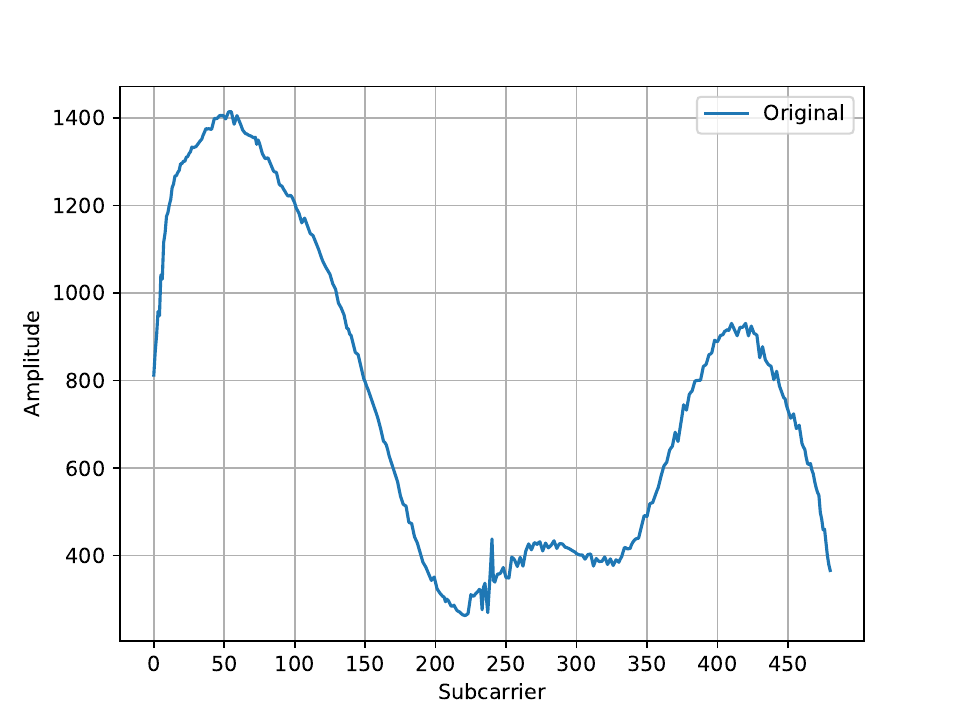}
		\caption{Original \ac{CSI} amplitude}
		\label{fig:subfigA}
	\end{subfigure}
	\begin{subfigure}{0.41\linewidth}
		\includegraphics[width=\linewidth]{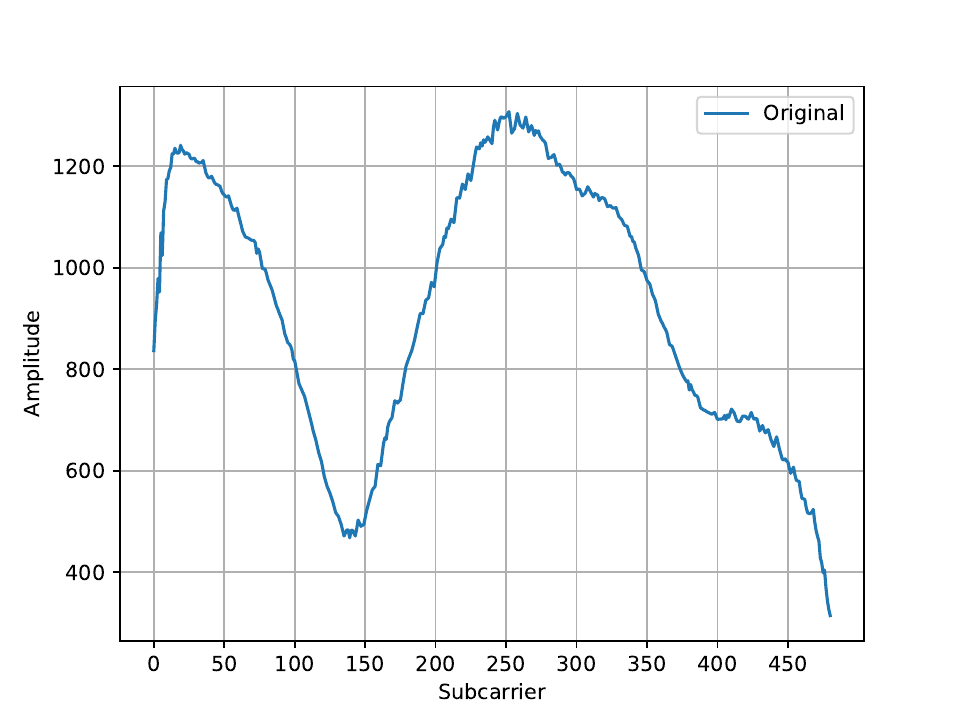}
		\caption{Original \ac{CSI} amplitude}
		\label{fig:subfigB}
	\end{subfigure}\\
	\begin{subfigure}{0.41\linewidth}
		\includegraphics[width=\textwidth]{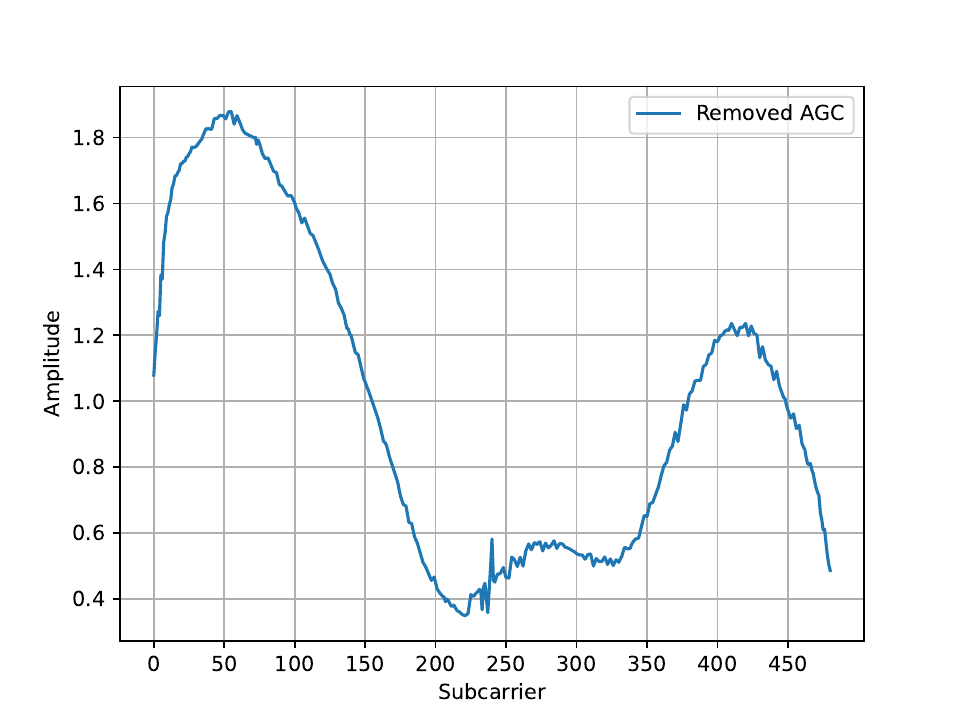}
		\caption{Removed impact of \ac{AGC} as per \cref{eq:normenergy}}
		\label{fig:subfigC}
	\end{subfigure}
	\begin{subfigure}{0.41\linewidth}
		\includegraphics[width=\textwidth]{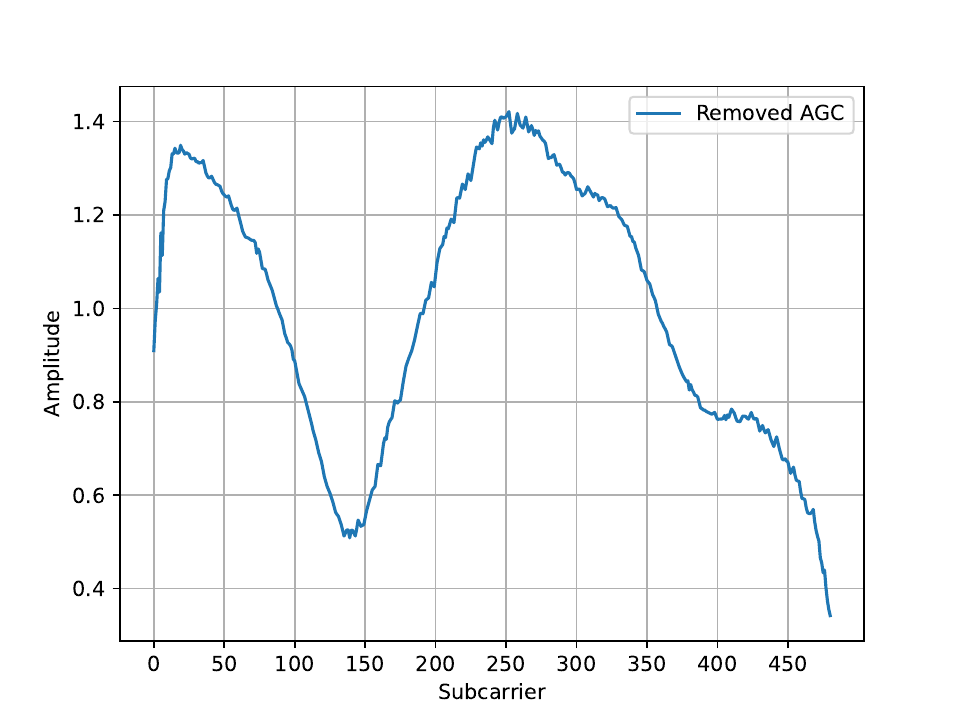}
		\caption{Removed impact of \ac{AGC} as per \cref{eq:normenergy}}
		\label{fig:subfigD}
	\end{subfigure}\\
	\begin{subfigure}{0.41\linewidth}
		\includegraphics[width=\textwidth]{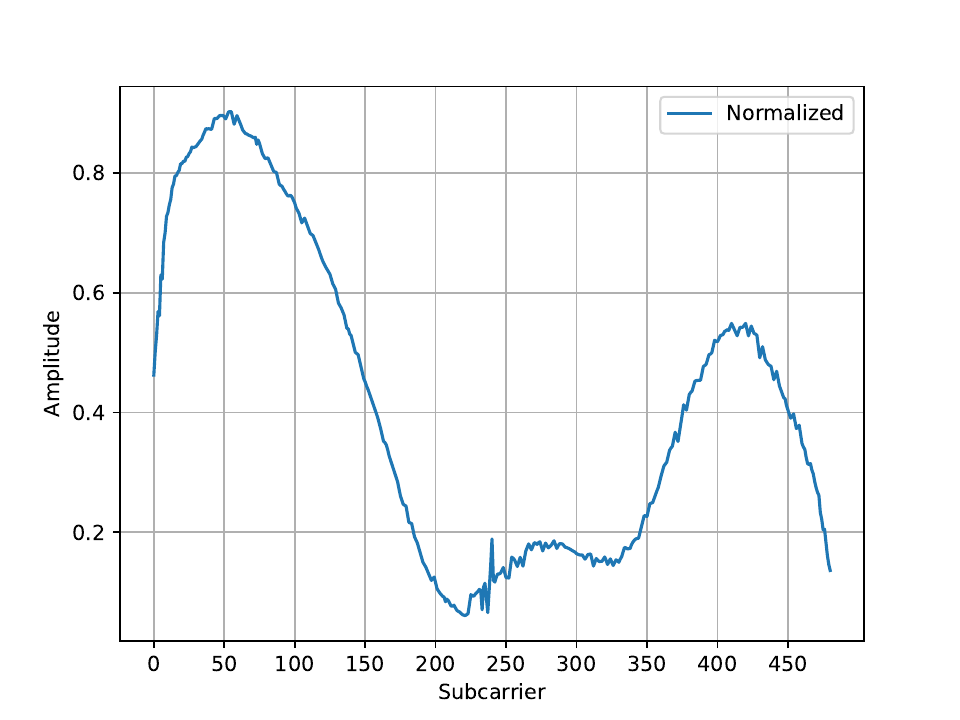}
		\caption{Normalized}
		\label{fig:subfigE}
	\end{subfigure}
	\begin{subfigure}{0.41\linewidth}
		\includegraphics[width=\textwidth]{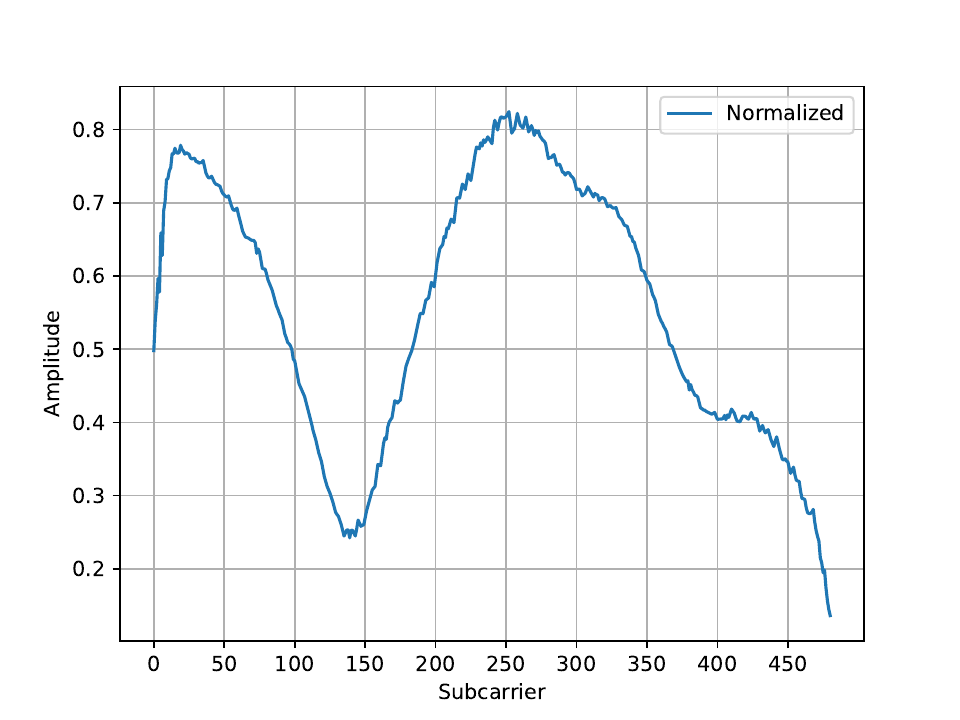}
		\caption{Normalized}
		\label{fig:subfigF}
	\end{subfigure}\\
	\begin{subfigure}{0.41\linewidth}
		\includegraphics[width=\textwidth]{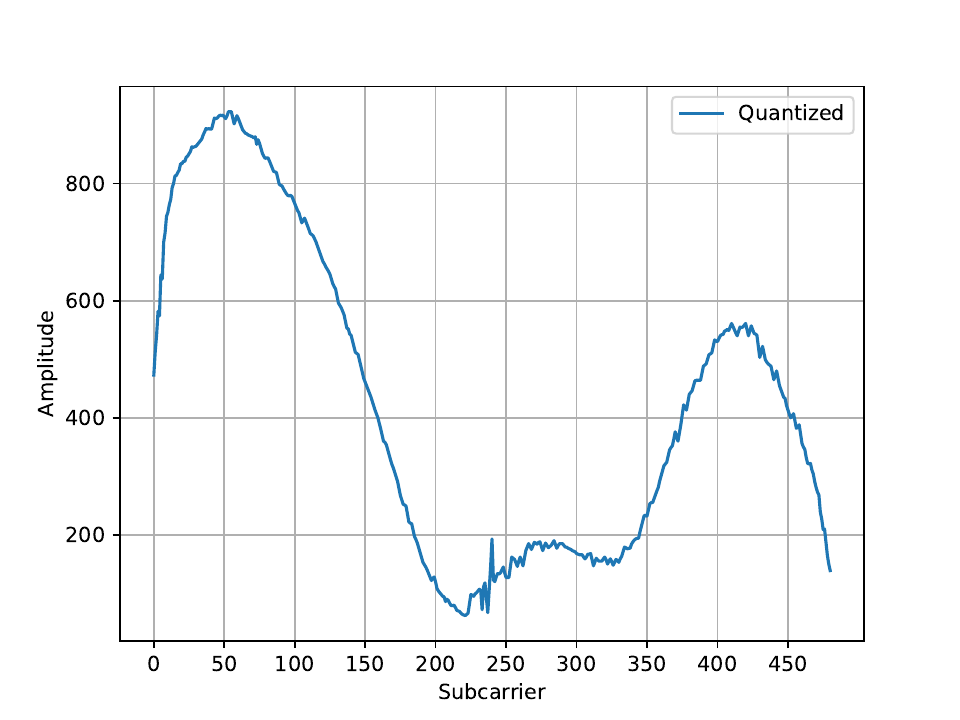}
		\caption{Quantized}
		\label{fig:subfigG}
	\end{subfigure}
	\begin{subfigure}{0.41\linewidth}
		\includegraphics[width=\textwidth]{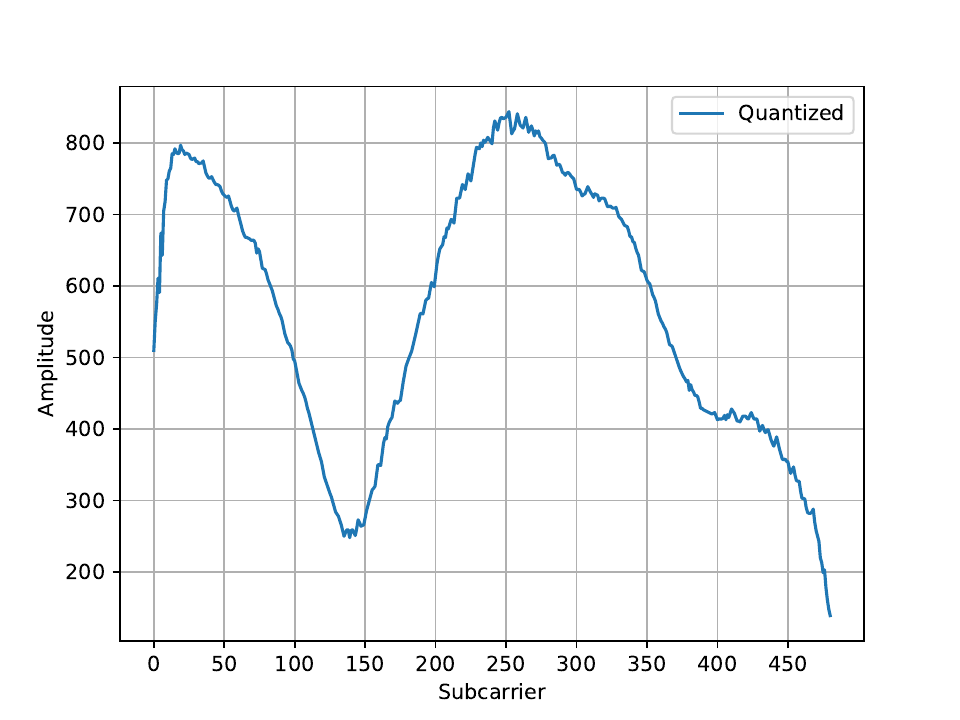}
		\caption{Quantized}
		\label{fig:subfigH}
	\end{subfigure}
	\caption{Visualization of the numerical processing on the \acp{CSI} in the Empty (left) and Static (right) Scenarios. Data collected on a 40 MHz bandwidth channel using 802.11ax.}
	\label{fig:normquant}
\end{figure}

To support the understanding of the normalization and quantization processes, a visual representation of the variations that amplitudes undergo is shown in the following paragraphs. 
Two examples of randomly chosen \acp{CSI} are selected: the first, displayed on the left column of \cref{fig:normquant}, belongs to a ten-minute-long experiment performed in the Empty Scenario, and the second (right column) comes from an equally long experiment performed in the Static Scenario. 
Both setups are described in \cref{ch:exp_setup}.
The original amplitude values indicated on the $y$-axis in \cref{fig:subfigA,,fig:subfigB} are arbitrary values detected by the receiver, therefore the measurement has no reference scale. 
This consideration is at the base of the whole normalization and quantization processes, as without such processing the comparison of different \acp{CSI} would be harder to carry out.

The relevant feature that comes across by looking at \cref{fig:normquant} is that the structure of the \acp{CSI} remains unaltered after each step of elaboration. 
The changing element of the displayed plots is the scale on the $y$ axis, as the value of the amplitude is scaled on different intervals. 
Specifically, the mitigation of the effects of \ac{AGC} through the normalization with respect to the energy of the \ac{CSI} (\cref{eq:normenergy}) brings the amplitude values closer to 1, which is then set as the maximum value by the normalization described through \cref{eq:csimin} to \cref{eq:normuno}. 

As already highlighted in the comments to \cref{eq:normuno}, it is possible that some \acp{CSI} across the experiment do not reach the values 0 and 1 (i.e., both ends of the normalization range) because the normalization is done using the maximum amplitude reached during the whole experiment, rather than that of the individual \ac{CSI}. 
The traces displayed in \cref{fig:normquant} are an example of this behavior. 

The amplitudes in the $[0,1]$ interval are then mapped onto the $[0, 2^{\acsibits}-1]$ interval through quantization.
Moreover, it is also evident that the \ac{CSI} collected in the Static scenario (i.e., with one person in the room sitting at the desk while working on a laptop) differs from the one collected in the empty room, highlighting how \acp{CSI} directly reflect the properties of the environment in the changes of their amplitude structure.

%% file: mutual_information.tex
\chapter{Mutual Shannon Information}
\label{ch:mi}
The \ac{MI} between two random variables is a measure of the mutual dependence of the two variables. In terms of \acp{PMF} for discrete distributions, the \ac{MI} between two discrete random variables $X$ and $Y$ is computed as a double sum:
\begin{equation}
	\begin{split}
		I(X;Y) & = \sum_{y\in \mathcal{Y}}\sum_{x\in \mathcal{X}} P_{(X,Y)}(x,y)\log\left (\frac{P_{(X,Y)}(x,y)}{P_{X}(x)P_{Y}(y)}\right)\\
		& =\sum_{y\in \mathcal{Y}}\sum_{x\in \mathcal{X}} P(x|y)\cdot P_{Y}(y) \log\left (\frac{P(x|y)\cdot P_{Y}(y)}{P_{X}(x)P_{Y}(y)}\right)\\
		& = \sum_{y\in \mathcal{Y}}\sum_{x\in \mathcal{X}} P(x|y)\cdot P_{Y}(y) \log\left (\frac{P(x|y)}{P_{X}(x)}\right)
	\end{split}
	\label{eq:mi}
\end{equation}
where $P_{(X,Y)}$ is the joint probability mass function of $X$ and $Y$ and $P_X$ and $P_Y$ are the marginal probability functions of $X$ and $Y$ respectively.
In terms of \acp{PDF} for continuous distributions, the sums in the formula are exchanged for integrals, allowing integration in $dx$ and $dy$ respectively.

\ac{MI} essentially measures how knowledge of the probability of an event impacts knowledge about the other. 
In the analysis of \acp{CSI}, \ac{MI} represents the amount of information that a reference \ac{CSI} \racsi provides about another \ac{CSI} \acsi or vice versa.
If $X$ and $Y$ are two disjoint discrete random variables, knowing anything about either of them provides no additional information about the other variable. 
Contrarily, if the value of $X$ can be deterministically calculated based on that of $Y$, the \ac{MI} is the same as the uncertainty about either of the two variables' values (i.e., the entropy of $X$ or $Y$).

Some relevant properties of the \ac{MI} are:
\begin{itemize}
	\item $I(X;Y)=0 \Leftrightarrow X$ and $Y$ are independent random variables. This is due to the fact that $p_{(X,Y)}(x,y)=p_X(x)\cdot p_Y(y)$, causing the content of the logarithm function to be equal to 1, meaning that $\log\left ( \frac{p_{(X,Y)}(x,y)}{p_{X}(x)p_{Y}(y)}\right) = \log(1) = 0$;
	\item Non-negativity: $I(X;Y) \geq 0$;
	\item Symmetry: $I(X;Y) = I(Y;X)$.
\end{itemize}
Note that the non-negativity holds when $\left (\frac{P(x|y)\cdot P_{Y}(y)}{P_{X}(x)P_{Y}(y)}\right) = 0$ and $log(0)$ is undefined by leveraging the properties of infinitesimal calculus: in such condition, in fact, $P(x|y)$ is what causes the argument of the logarithm to be zero, but this value also multiplies the logarithm, making it unnecessary to compute the product between their finite values as it will always be equal to zero regardless of the resulting logarithm. 

\ac{MI} can alternatively be computed as a function of entropy and conditional entropy: 
\begin{equation}
	\begin{split}
		I(X;Y)& \equiv H(X)-H(X|Y)\\
		&	\equiv H(Y)-H(Y|X)\\
		&	\equiv H(X)+H(Y)-H(X,Y)\\
		&	\equiv H(X,Y)-H(X|Y)-H(Y|X)
	\end{split}
	\label{eq:mi_entropy}
\end{equation}
where $H(X)$ represents the entropy of $X$, $H(X|Y)$ represents the conditional entropy of $X$ given the knowledge about $Y$, and $H(X,Y)$ is the joint entropy of $X$ and $Y$.

The application of the \ac{MI} equation in this study works as a quantitative measurement to determine whether two \acp{CSI} belong to the same experiment, assuming that two \acp{CSI} coming from different captures (i.e., different locations, number of people in the room, etc.) bear little additional information about each other, whereas two samples belonging to the same experiment have a higher \ac{MI} value. 
Numerically, we assume that samples belonging to experiments performed in distinct environments have a \ac{MI} value closer to (or equal to, in case of complete independence) zero, whereas samples coming from experiments performed with the same setup have a value asymptotically growing to infinity. 
To represent an infinite value using a finite set of numbers, an upper limit is set to the value of the \ac{MI}. 

The analysis can start by computing the \ac{MI} between the value taken by the average \ac{CSI} $\racsi$ --- which is representative of the whole experiment --- and that of another \ac{CSI} $\acsi(\tin, \subc)$ on a chosen sub-carrier $\subc\in [0,\nsc]$, with any $\tin\in [1, \nsamp]$.
To derive each $\acsi(\tin,\subc)$, an increment $\iacsi$ is added to $\acsi(\tin-1,\subc)$, with $\iacsi$ belonging to a known discrete probability distribution that can be modelled as a quantized Gaussian distribution (according to the quantization process described in \cref{ch:normquant}). 
This characterization of the increments as belonging to a Normal distribution simplifies the computation of the probabilities of an increment $\iacsi$ being added to $\acsi(\tin-1,\subc)$ and that of $\iacsi$ occurring at all. 
From now on we will consider $\acsi(\tin-1,\subc)=\racsi(\subc)$ to compute the \ac{MI} between the reference \ac{CSI} and another one from the same capture. 

Once $\tin\in [1,\nsamp]$ is defined as the index of the \ac{CSI} to consider within the experiment and $\subc \in [0,\nsc]$ is chosen as the analyzed sub-carrier, the computation of the \ac{MI} requires knowing some probability values, such as:
\begin{itemize}
	\item $P(\acsi(\tin, \subc)| \racsi(\subc))$: it can be computed as the probability of drawing a specific $\iacsi$ value from the quantized Normal distribution and obtaining $\acsi(\tin, \subc)$ by adding the increment $\iacsi$ to $\racsi(\subc)$. Essentially, it is equal to $P[\iacsi : \racsi(\subc)+\iacsi==\acsi(\tin, \subc)]$;
	\item $P(\acsi(\tin, \subc))$
	\item $P(\racsi(\subc))$
\end{itemize}
Computing these probabilities allows to calculate the \ac{MI} between two amplitude values at consecutive time steps on a fixed sub-carrier. 

Given that the goal is to compute the \ac{MI} between \acp{CSI} as a whole and not on each sub-carrier by itself, a value to represent the probability of an entire \ac{CSI} $\csi(\tin)$ happening in an experiment is also needed.
Assuming, as a simplification, that all sub-carriers are independent, this is given by:
\begin{equation}
	P(\csi(\tin)) = \prod_{\subc\in [0,\nsc]} P(\acsi(\tin, \subc))
	\label{eq:pcsi}
\end{equation}

Considering that an analysis that only looks at \ac{MI} sub-carrier by sub-carrier would be too limited and that it would not return the actual \ac{MI} between \acp{CSI}, it becomes necessary to translate what has been described in this chapter up to this point to work with \acp{CSI} as a whole rather than splitting them $\nsc$ times.

We can, at this point, consider the amplitudes of the \ac{CSI} $\acsi$ across the $\nsc$ sub-carriers as symbols of an alphabet. 
The alphabet is very large, but finite, having $2^{(\nsc \cdot \acsibits)}$ symbols, hence \ac{MI} is always finite and numerical evaluations can proceed, albeit with care to avoid numerical problems in case of very large (or very small) numbers. 

First of all, \cref{eq:pcsi} can be extended as follows:
\begin{equation}
	\begin{split}
		P(\csi(\tin)) & = \prod_{\subc\in [0,\nsc]} P(\acsi(\tin, \subc))\\
		& = \prod_{\subc\in [0,\nsc]} \frac{1}{2^{\acsibits}}\\
		& = \frac{1}{2^{\nsc\cdot\acsibits}}
	\end{split}
	\label{eq:pcsiext}
\end{equation}
This implies that --- ignoring cross-sub-carrier dependence --- any \ac{CSI} $\csi(\tin)$ has the same probability of happening, given the available alphabet. 
Unfortunately, it is clear that, given any reasonable \nsc and \acsibits, $\frac{1}{2^{\nsc\cdot\acsibits}}$ is way too small to allow any numerical evaluation of \cref{eq:mi} or derivations thereof without further manipulation or approximation. 

One possible, quick solution is to use a polynomial expansion of the logarithm and exploit the fact that $P(\csi(\tin))$ is constant. 	
One possibility is to use the bilinear expansion:
\begin{equation}
	\log(x) = 2\left[  \left( \frac{x-1}{x+1} \right) + \frac{1}{3} \left( \frac{x-1}{x+1} \right)^3 + \frac{1}{5} \left( \frac{x-1}{x+1} \right)^5 + \ldots \right]
	\label{eq:bilin}
\end{equation}
An alternative method to approximate the logarithm could be the following: it is known that
\begin{equation}
	\log(x) - \log(1) = (x-1)-\frac{(x-1)^2}{2}+\frac{(x-1)^3}{3}-\frac{(x-1)^4}{4}+\ldots
	\label{eq:expcentred1}
\end{equation}
and that the logarithm of a fraction can be computed as the difference of two logarithms: \[\log\left( \frac{x}{y}\right) = \log(x)-\log(y) \]
In the computations presented up to this point, it has been stated that $P(\csi(\tin))$ is constant, therefore $P(\acsi)=P(\racsi)$ and $\log\left( \frac{P(\acsi| \racsi)}{P(\acsi)}\right) $ can be expanded as $\log\left( \frac{P(\acsi| \racsi)}{a}\right) $ with $a$ constant. 
\begin{equation}
	\log(x) - \log(a) = \frac{1}{a}(x-a)-\frac{(x-a)^2}{2a^2}+\frac{(x-a)^3}{3a^3}-\frac{(x-a)^4}{4a^4}+\ldots
	\label{eq:explogxa}
\end{equation}
This simplifies \cref{eq:mi} to:
\begin{equation}
	\begin{split}
		I(X;Y) & = \sum\sum P(\acsi|\racsi)\cdot P(\racsi) \log\left (\frac{P(\acsi|\racsi)}{P(\acsi)}\right)\\
		& = \sum\sum P(\acsi|\racsi)\cdot P(\racsi)\cdot \\& \qquad \left[ \log(P(\acsi)) + \frac{P(\acsi|\racsi)}{P(\acsi)} - \frac{P(\acsi)}{P(\acsi)} - \log(P(\acsi))\right ]\\
		& = \sum\sum P(\acsi|\racsi)\cdot P(\racsi) \left [\frac{P(\acsi|\racsi)}{P(\acsi)} - 1\right ]\\
		& = \sum\sum P(\acsi|\racsi)\cdot[P(\acsi|\racsi) - P(\acsi)]
	\end{split}
	\label{eq:miextended}
\end{equation}
Where $P(\acsi)$ and $P(\racsi)$ can be simplified with each other because they are equal.
The \ac{MI} of any two \acp{CSI} can be evaluated exploiting \cref{eq:miextended} and the probability model derived in \cref{ch:normquant}.
 
Once again, unfortunately, the probability values that are needed to compute the \ac{MI} are infinitesimal, resulting in calculations that are not only difficult to carry out but also hardly significant. 
Nonetheless, it is deemed appropriate to complete the mathematical reasoning behind the computation of the \ac{MI}, as it still maintains theoretical relevance.  

In particular, once a solution to the numerical representation of infinitely small numbers has been found, it would be possible to estimate the average \ac{MI} of \acp{CSI} collected in the same experiment using the experimental distribution of increments; it still remains feasible to compute the theoretical \ac{MI} based on the Gaussian approximation performed in \cref{ch:normquant}\footnote{Note that any other distribution can be used rather than Gaussian, so additional investigation may lead to other, better approximations.}. 
For the time being, as these final considerations are merely theoretical, no distinction is assumed between the two distributions and a little overloaded notation is used. 

Let $\mint_{A}$ be the \textit{internal} \ac{MI} for an experiment $A$ 
\begin{equation}
	\mint_{A} = \sum_{i}^{\nsampa} I(\csia^*, \csia(i))
	\label{eq:miint}
\end{equation}
and similarly for any other experiment $B, C, D, \ldots$. 

A larger internal \ac{MI} would identify experiments that are intrinsically more variable, which does not necessarily imply noisier, as for instance experiments performed with people moving inside the room have an obviously larger variability. 

It would also be interesting to compute a pair of \textit{external} \ac{MI} values between any two experiments $A,B$, using the increment process estimated either in $A$ or $B$, according to which experiment the average \ac{CSI} belongs to:
\begin{equation}
	\mext_{A,B} = \sum_{i}^{\nsampb} I(\csia^*, \csib(i))
	\label{eq:miextab}
\end{equation}
and
\begin{equation}
	\mext_{B,A} = \sum_{i}^{\nsampa} I(\csib^*, \csia(i))
	\label{eq:miextba}
\end{equation}
The two will be different because the process of the increments is distinct in any experiment. 

\section{Future Research Directions}
Further investigation is needed to identify alternative solutions to the quantitative representation of \ac{MI}, as its theoretical analysis only becomes more significant after it is correlated with empirical results. 
For the time being, the hypothesis of using \ac{MI} as a measurement of the mutual additional information content is set aside and other options are analyzed to compute the distance between \acp{CSI} belonging to either the same or a different experiment. 

%% file: hamming.tex
\chapter{Weighted Hamming Distance}
\label{ch:whd}
As the computation of the \ac{MI} has been proven, for the time being, infeasible, the characterization of \ac{CSI} amplitude requires the introduction of a new unit of measurement to quantify the information carried by each trace. 
The task of associating a \ac{CSI} to a specific scenario can now be reformulated as follows: after computing the distance between a \ac{CSI} \acsi and the reference \ac{CSI} \racsi of a selected experiment, the more similar \acsi is to \racsi, the shorter the distance between the two \acp{CSI}. Consequently, the shorter the distance, the more likely \acsi is to belong to the same experiment as \racsi. 
The choice of unit of measurement to fulfill this goal has fallen on the Hamming Distance. 

By definition, the Hamming distance between two equal-length strings of symbols is the number of positions at which the corresponding symbols are different. 
Contextualizing the use of the Hamming distance in this work, we can see it as a tool to measure the difference between two equally long strings \textit{of bits}. 
Whether the comparison starts from the most or least significant bit of the string is irrelevant when computing the standard Hamming distance, as it does not account for the position of the differing symbols but rather looks at their difference itself. 
For binary strings $a$ and $b$, the Hamming distance is equal to the number of ones in the result of the $a \oplus b$ operation.

An intuitive example of its computation is provided below:
\begin{center}
	1\color{red}0\color{black}01\color{red}1\color{black}0\color{red}1\color{black}1\\
	1\color{red}1\color{black}01\color{red}0\color{black}0\color{red}0\color{black}1\\
\end{center}
Given the two bytes above, the Hamming distance between them is $3$, as the mismatched bits highlighted in red indicate. 

Directly implementing the computation of the Hamming distance, albeit straightforward, bypasses some necessary logical assumptions. 
Its implementation would be used to quantify the difference in the information contents of two \acp{CSI}.
In particular, the standard Hamming distance as-is would only be capable of representing the existence of a difference between the \acp{CSI} but it would not show \textit{how} they differ. 
Specifically, two \acp{CSI} --- represented as binary strings after quantization --- differing by the most significant bit would have the same Hamming distance as two \acp{CSI} differing by the least significant bit. 
Of course, this would result in inconsistent interpretations of the experimental results because the positions of the differing bits would not be accounted for. 
The mismatch in the most significant bits should be weighed differently than that in the least significant ones, as the information content brought along by the discrepancies of the strings in the two cases is different.

These considerations lead to the need for the identification of a \ac{WHD} as a more appropriate metric to compute the information content linked to the differences between two \acp{CSI}.
We propose that such a metric associates a larger weight to differences in the more significant bits of the compared strings. 
To do so, we need to introduce a list of weights that is as long as the strings of bits being considered. 
Such weights should be set by default and left unaltered within the same experiment regardless of the compared strings to ensure that all measures belonging to the same experiment are consistent with one another (provided that the strings of bits belonging to the same experiment all have the same length, which is also compatible with the length of the list of weights). 
The list of weights should be configured so that it gives an arbitrarily larger or smaller weight to differences in more significant bits; in this study, the choice was made to assign a larger weight to differences in more significant bits, while mismatches in less significant bits will have a smaller impact on the value of the metric. 

Let's assume that we have a dataset of 8-bit strings to compute the \whd on. The list of weights can be represented as an array of integer values, such as:
\begin{center}
	$w$ = [8 7 6 5 4 3 2 1]
\end{center}
This array allows for the computation of the \whd between string $a$ and string $b$ as:
\begin{equation}
	\whd = \sum_{i} \mathopen|a[i]-b[i]\mathclose| \cdot w[i]
	\label{eq:whd}
\end{equation}
\cref{eq:whd} implies that $ 0 \leq \whd \leq \sum_{i} w[i]$, where $\whd = 0$ when $a$ and $b$ are equal and $\whd = \sum_{i} w[i]$ when $a$ and $b$ are one's complements of each other. 
For example, 
\begin{center}
	$a = 10110010$\\
	$b = 11101100$\\
	$w$ = [8 7 6 5 4 3 2 1]\\
	$\whd = 7+5+4+3+2 = 21$ 
\end{center}
Given the suggested characterization of the \whd, the closer the value of the measure to its maximum reachable value, the more likely it is that more significant bits are different in the considered strings. 

In this study, after quantization of \ac{CSI} amplitudes, we do not work directly with strings of bits but rather with their representation in base 10. 
This implies that the weight that has to be given to mismatching bits in different positions along the strings is implicitly accounted for in the binary-to-decimal conversion. 
Therefore, the array of weights can be left out of \cref{eq:whd} as all its items will be equal to 1 in the base 10 representation of the compared strings. 

As an initial characterization of the experiments, we compute the \whd between the reference \ac{CSI} \racsi of each experiment and each \ac{CSI} $\tin\in\nsamp$ of the experiment. 
The formula presented in \cref{eq:whd} becomes:
\begin{equation}
	\whd(\racsi, \acsi(\tin, \cdot)) = \sum_{\subc=0}^{\nsc} \mathopen|\acsi(\tin, \subc)-\racsi(\tin,\subc)\mathclose|
	\label{eq:whdacsi}
\end{equation}
This equation is used to compute the `internal' \whd of an experiment, as well as the `external' distance between two different experiments.
The `internal' \whd is defined as the average distance between the reference \ac{CSI} \racsi and all \acp{CSI} of the experiment that \racsi is computed on. 
Contrarily, the `external' distance is defined as the average distance between \racsi and all \acp{CSI} of an experiment different than the one \racsi is computed on but belonging to the same experimental setup. 
Moreover, `cross-setup' distance (also called `cross distance') is defined as a variation of the external distance such that the \racsi and the \acp{CSI} used to compute the \whd belong to experiments with different experimental setups, e.g. \racsi is computed on data collected within the Empty Scenario and it is compared to data collected in the Static Scenario.

The expected results of these computations are that the `internal' and `external' distances take on significantly lower values than the `cross-setup' distance, with the `internal' distance possibly remaining lower than the `external', albeit with less substantial variation.
Such results would provide a basic tool to support environment identification: given a \ac{CSI} extracted from an unknown environment, the closer it is to correctly classified reference \acp{CSI}, the more likely it is that it was collected within the same scenario. 

%% file: elab_workflow.tex
\chapter{CSI Processing}
\label{ch:elab}

Before proceeding with the analysis of the results derived from the elaboration of the collected \acp{CSI}, we provide an overview of the process that was followed to obtain them. 

Upon extraction, \ac{CSI} traces are represented as non-null complex numbers within which amplitude and phase can be identified and separated. 
All \acp{CSI} belonging to the same experiment are saved in a \texttt{csv} file, with each row corresponding to a different \ac{CSI}. 
Each \ac{CSI} is composed of one complex number for each sub-carrier; all traces belonging to the same capture are made of the same number of values, as the number of sub-carriers \nsc obviously remains unaltered throughout the experiment.
Depending on the used bandwidth, the number of sub-carriers changes as displayed in \cref{lst:sccount}.

\begin{lstlisting}[caption={Computation of the number of sub-carriers as a function of bandwidth (20, 40, 80 MHz) and 802.11 standard.}, label={lst:sccount}]
	# if working with 802.11ac
	nsc = 3.2 * BW
	if STD == 'ax': # if working with 802.11ax
		nsc = nsc * 4
\end{lstlisting}

Some sub-carriers are suppressed during transmission and therefore the corresponding \ac{CSI} values are set to $0i+0$. 
Such sub-carriers are identified and removed from each sample, as they do not carry information about the environment where the trace was captured. 

At this point, only \ac{CSI} amplitudes are kept into account, while phase values are discarded, as they are not analyzed within this thesis. 
Since \acp{CSI} are subject to the effect of \ac{AGC}, its impact is removed before further processing is carried out. 

Then, \acp{CSI} are normalized and quantized, according to what has been described in \cref{ch:normquant}. 
All remaining elaboration is performed on the quantized version of both \ac{CSI} increments and amplitude values. 

\begin{figure}[h]
	\centering
	\includegraphics[height=0.5\textheight]{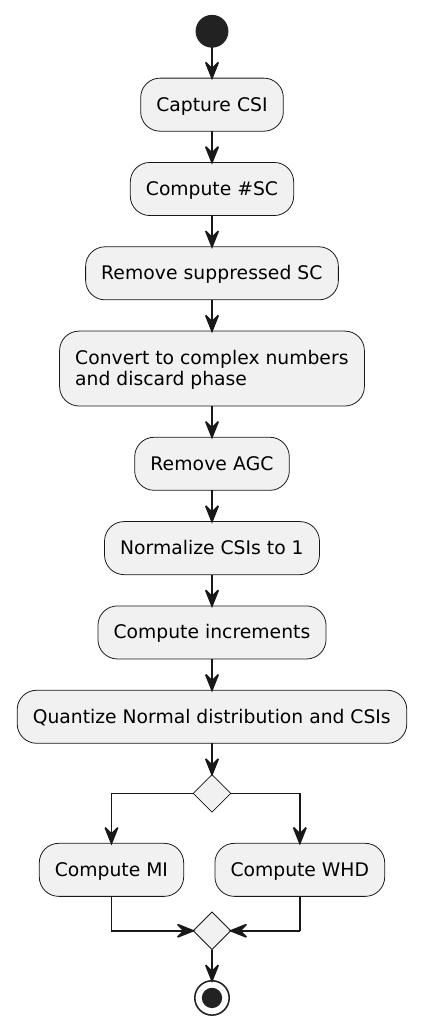}
	\caption{Overview of the workflow followed during \ac{CSI} processing.}
	\label{fig:workflow}
\end{figure}

\cref{fig:workflow} depicts a summarized overview of the followed workflow.

\cref{ch:prevres} provides a summary of the results produced in the previous work, to improve contextualization of this analysis.
The current version of the code maintains backwards compatibility with the processing carried out throughout the BSc Thesis.
To support this statement, we reproduce the results showcased in \cref{ch:prevres} on the new dataset. 

\begin{figure}[h]
	\centering
	\begin{subfigure}[t]{0.49\linewidth}
		\includegraphics[width=\textwidth]{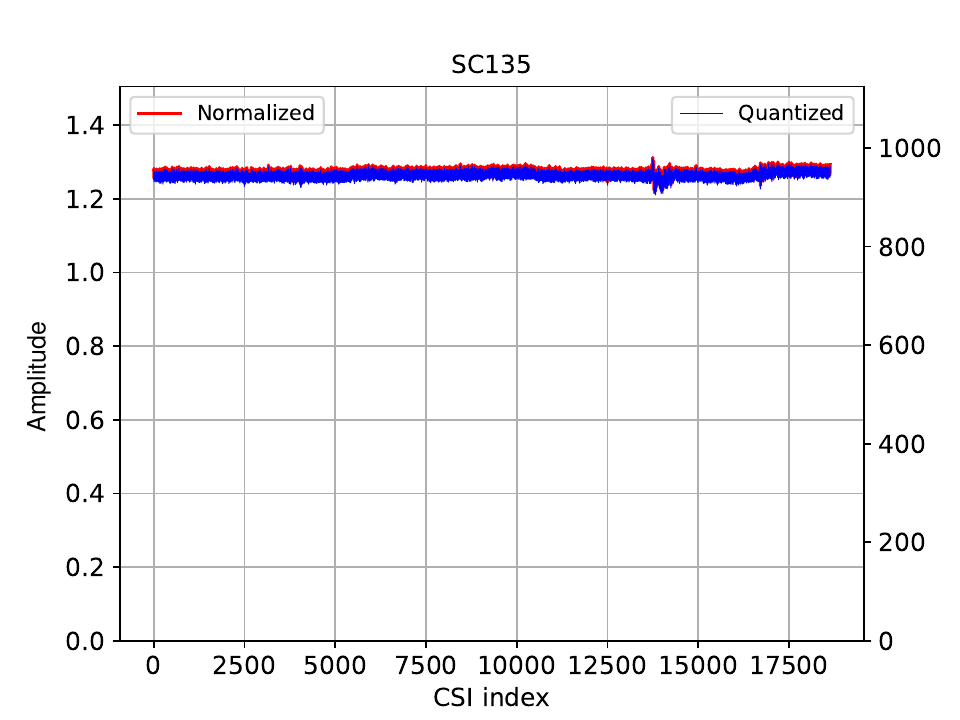}
		\caption{\acp{CSI} collected on a 20 MHz channel using 802.11ax in an empty room.}
		\label{fig:timeevolnewempty}
	\end{subfigure}
	\begin{subfigure}[t]{0.49\linewidth}
		\includegraphics[width=\textwidth]{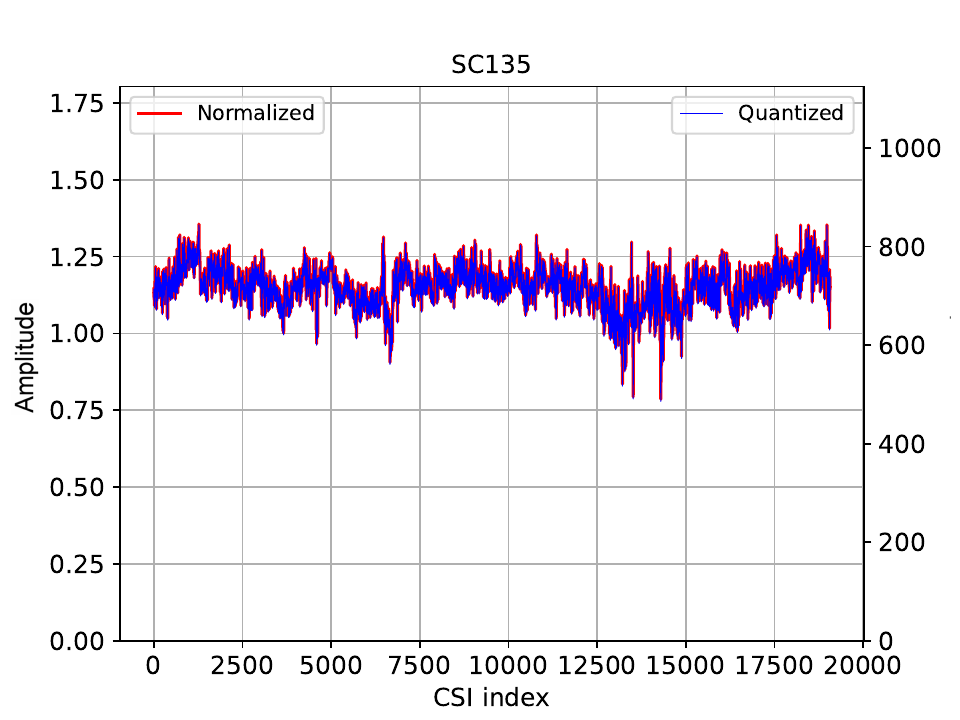}
		\caption{\acp{CSI} collected on a 20 MHz channel using 802.11ax in a room with four people in it.}
		\label{fig:timeevolnewfd}
	\end{subfigure}\\
	\caption{Example of amplitude evolution in time on sub-carrier 135 in the Empty and Fully Dynamic Scenarios.}
	\label{fig:timeevolnew}
\end{figure}

\cref{fig:timeevolnew} displays the evolution in time of the amplitude of \acp{CSI} collected on a randomly chosen sub-carrier in two different scenarios. 
Each plot represents two distinct yet visibly superimposable graphs: in red, referencing the left $y$-axis, the normalized amplitude is displayed, whereas in blue, referencing the right $y$-axis, the quantized version is plotted. 
Note that the line widths used to represent the two processes have been set to different values to allow distinction of the two series that are otherwise almost exactly superimposed within each of the two scenarios.
By observing these figures, we come to the conclusion that \ac{CSI} amplitude before and after quantization remains structurally unaltered, regardless of the scenario the traces were collected in. 
As one can expect, the \acp{CSI} representing a more dynamic scenario (\cref{fig:timeevolnewfd}) display higher variability in their evolution in time, which highlights how the amplitude indeed reflects the structure of the environment.
It must be noted that the removal of the effects of the \ac{AGC} positively contributes to enhancing the `true' behavior of the \acp{CSI}, mitigating the fluctuations that their amplitudes undergo and that were more evident in the results commented in \cite{bscthesis}. 

\begin{figure}[h]
	\centering
	\includegraphics[width=0.8\textwidth]{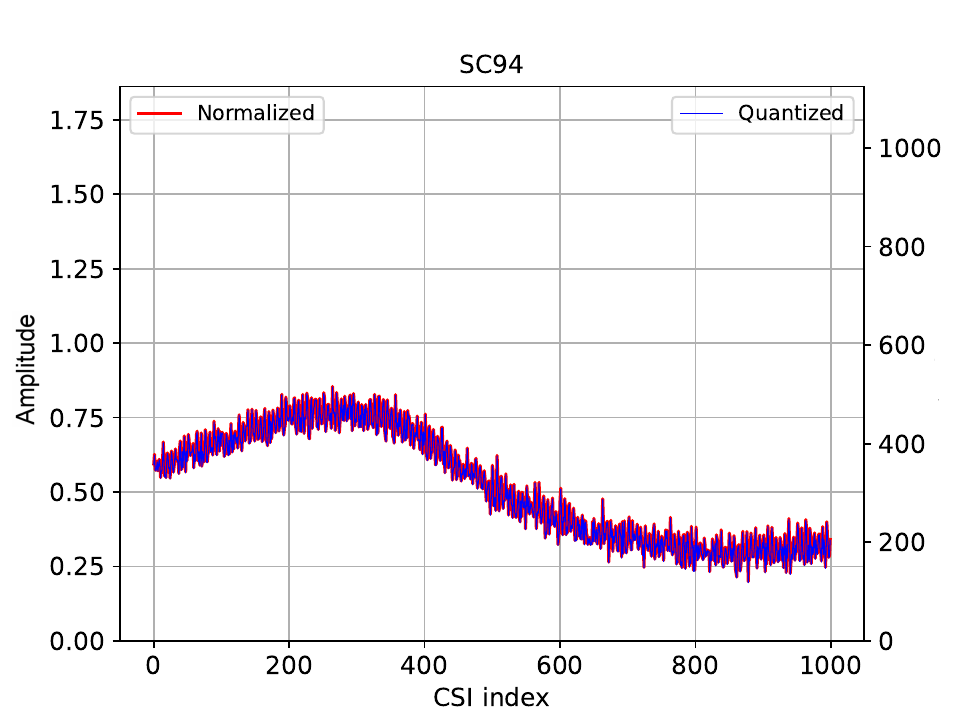}
	\caption{Example of amplitude evolution in time on sub-carrier 94. \acp{CSI} belong to the AntiSense dataset.}
	\label{fig:timeevolnew2}
\end{figure}

To provide a complete evaluation of the available captures, the code developed for \cite{bscthesis} was also tested against the AntiSense dataset; an example of the results is showcased in \cref{fig:timeevolnew2}.

\begin{figure}[h]
	\centering
	\includegraphics[width=0.8\textwidth]{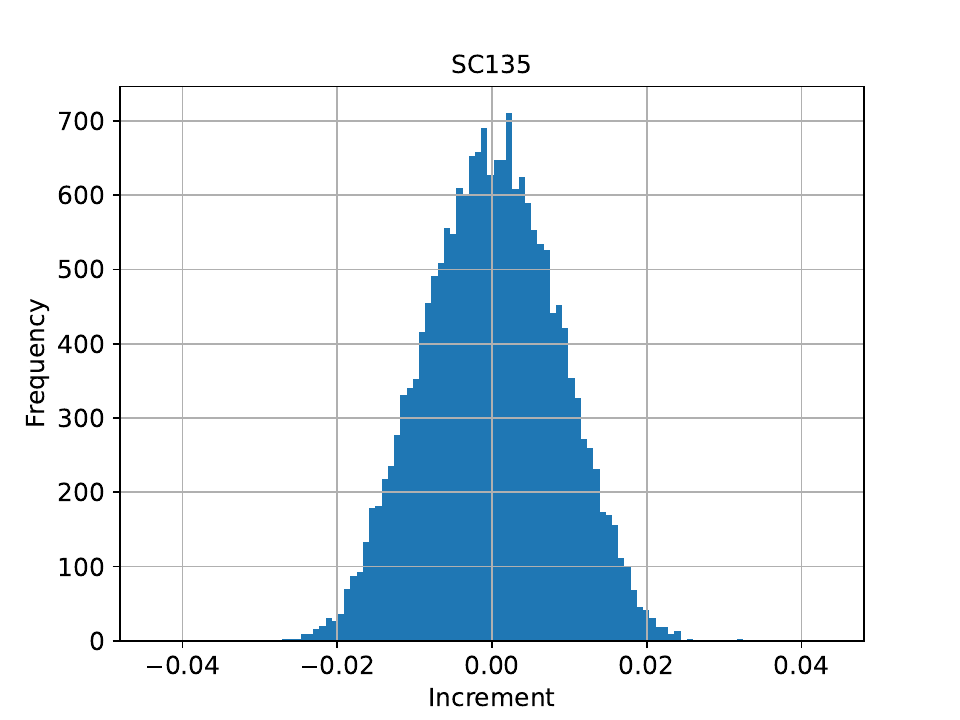}
	\caption{Example of increments distribution on sub-carrier 135. \acp{CSI} were collected on a 20 MHz channel using 802.11ax in an empty room.}
	\label{fig:incrdistrnew}
\end{figure}

\cref{fig:incrdistrnew} displays the distribution of the amplitude increments measured on sub-carrier 135: it can be observed that the histogram resembles a Gaussian distribution, which is coherent with the model proposed in \cite{bscthesis}.
\begin{figure}[h]
	\centering
	\begin{subfigure}[t]{0.49\linewidth}
		\includegraphics[width=\textwidth]{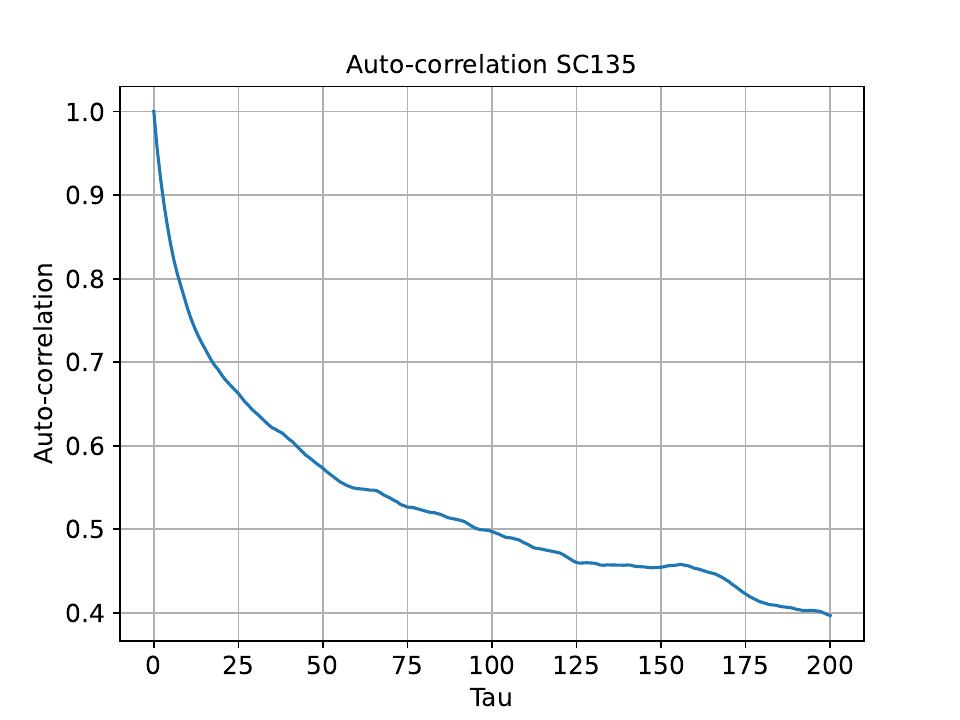}
		\caption{Amplitudes auto-correlation.}
		\label{fig:autocorrampl}
	\end{subfigure}
	\begin{subfigure}[t]{0.49\linewidth}
		\includegraphics[width=\textwidth]{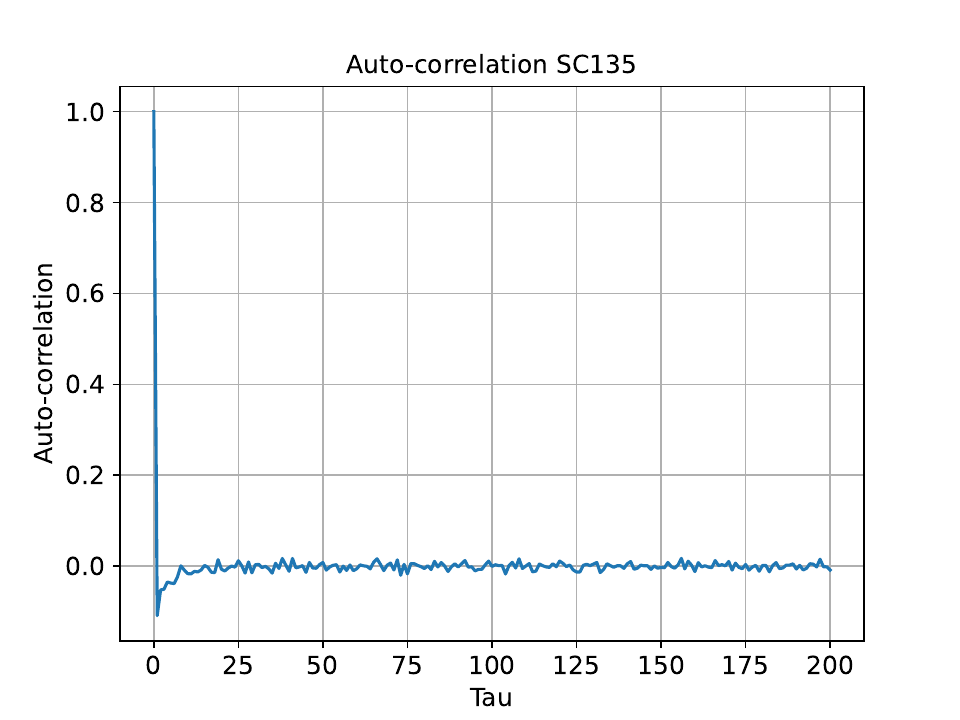}
		\caption{Increments auto-correlation.}
		\label{fig:autocorrincr}
	\end{subfigure}\\
	\caption{Examples of amplitudes and increments auto-correlation on sub-carrier 135. \acp{CSI} collected on a 20 MHz channel using 802.11ax in a room with four people.}
	\label{fig:autocorrnew}
\end{figure}

Finally, by examining the results of the auto-correlation function computed on the amplitudes (\cref{fig:autocorrampl}), we observe that the process indeed has memory. However, when looking at the auto-correlation of the increments (\cref{fig:autocorrincr}), we find that the function returns noise-like values, which are consistent with the results expected from a Markovian process. 
Whether such mathematical description could accurately represent the behavior of the increments will be the subject to future further analysis.

%% file: norm_res.tex
\chapter{Results of the Normalization and Quantization Processes}
\label{ch:normres}

Following the theoretical considerations carried out in \cref{sec:qmap}, we provide a sketch of the implementation of the quantization of $\ndis(\sigma)$ to obtain $\qndis(\sigma)$. 

The pseudo-code for the quantization process is briefly displayed in the following snippet.

\begin{lstlisting}[caption={Pseudo-code of the algorithm used to normalize and quantize \ac{CSI} and increments.}, label={lst:normquant}]
	csi = csi - min(csi)
	csi = csi / max(csi) # CSI ranges from 0 to 1
	a = 0
	b = 2 ** (2 * q_amp) - 1
	csi_quant = round(csi * (b - a) + a) # quantize CSI 
	
	incr = csi.diff() # computes increments using normalized CSI
	mu, sigma = norm.fit(incr)
	dstar = 3 * sigma
	sample = numpy.random.normal(loc=mu, scale=sigma, size=incr.size)
	sample = filterTails(sample, dstar) # fix boundary conditions
	# apply the same logic used to quantize amplitudes on increments
	incr_quant = int(round((sample - min(sample)) / (max(sample) - min(sample)) * (2 ** q_inc - 2) - (2 ** (q_inc - 1) - 1))) 
\end{lstlisting}

By running this code on the collected data, we create a quantized version of the Normal distribution that is used to approximate the empirical distribution of the increments. 
This is evident in the presented pseudo-code, as the values of the \texttt{sample} array are randomly selected from a Normal distribution with the same mean and standard deviation as the distribution of the increments. 
Should the approximation prove ineffective in correctly representing the empirical increments, the overall logic of the code would remain unaltered and all computations would be carried out on the original \texttt{incr} array instead of \texttt{sample}.
As stated at the end of \cref{sec:qmap}, we will assume that the Gaussian distribution correctly approximates the increments distribution.

\begin{figure}[h!]
	\centering
	\includegraphics[width=0.8\textwidth]{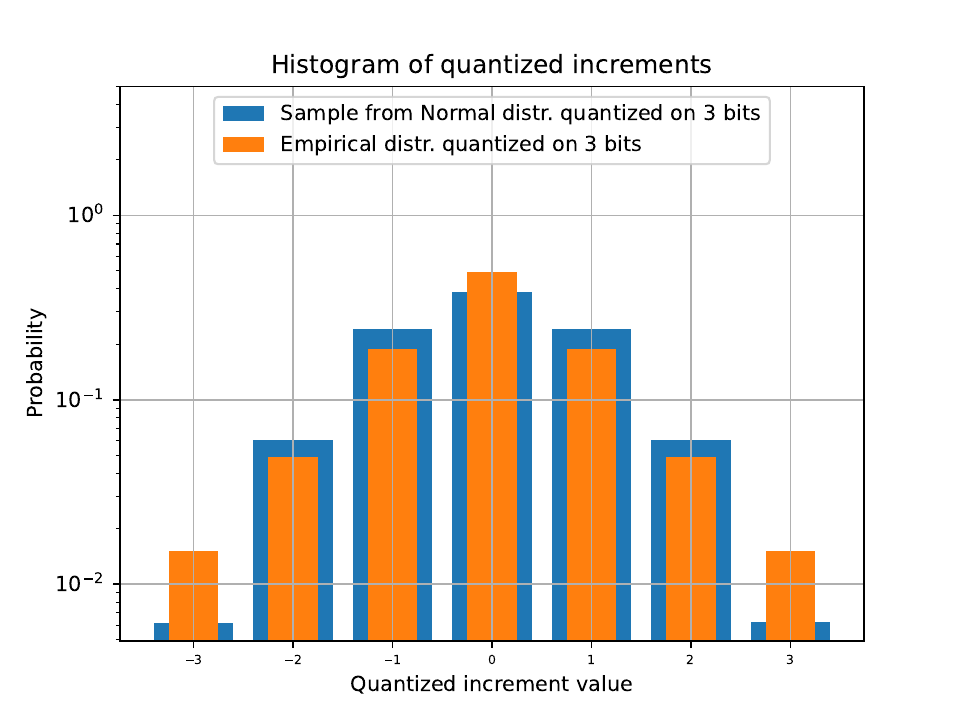}
	\caption{Distribution of increments VS sample from Gaussian distribution after quantization over 3 bits. Increments are computed on an experiment performed in the Empty Scenario at 20 MHz.}
	\label{fig:quantVSsample3}
\end{figure}

\begin{figure}[h]
	\centering
	\includegraphics[width=0.8\textwidth]{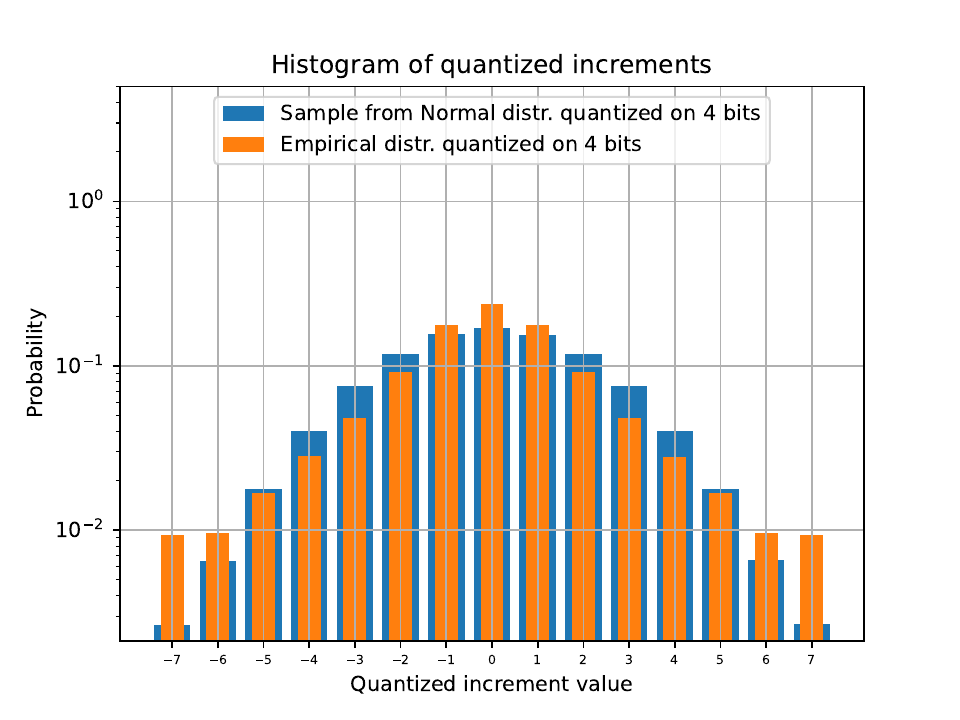}
	\caption{Distribution of increments VS sample from Gaussian distribution after quantization over 4 bits. Increments are computed on an experiment performed in the Empty Scenario at 20 MHz.}
	\label{fig:quantVSsample4}
\end{figure}

\begin{figure}[h]
	\centering
	\includegraphics[width=0.8\textwidth]{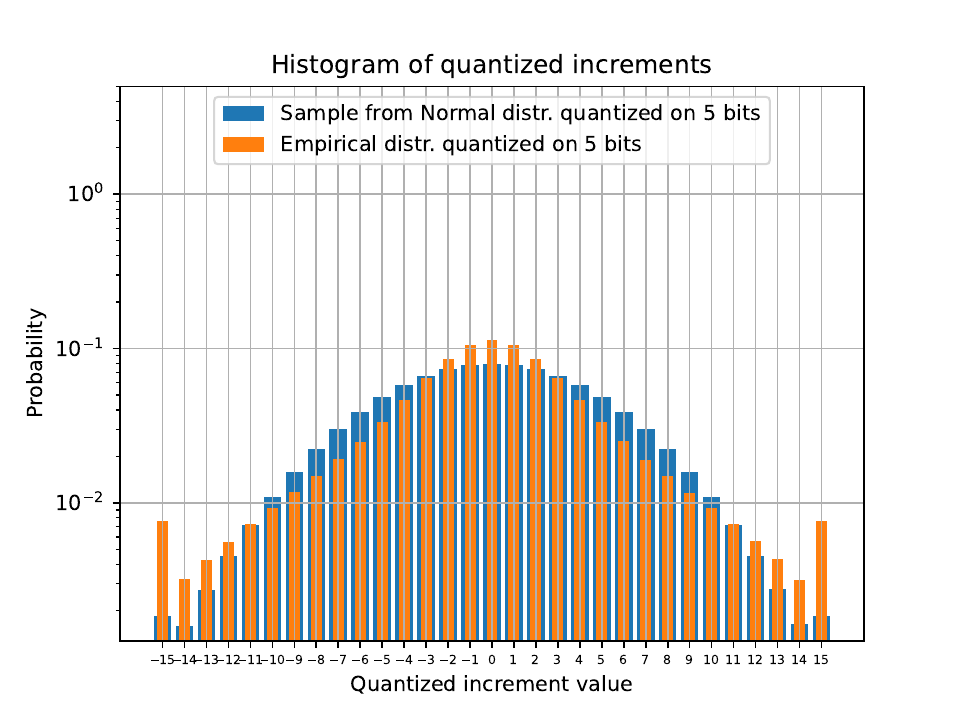}
	\caption{Distribution of increments VS sample from Gaussian distribution after quantization over 5 bits. Increments are computed on an experiment performed in the Empty Scenario at 20 MHz.}
	\label{fig:quantVSsample5}
\end{figure}

To choose a suitable $\nbits$ value to quantize the increments and to compute the correct $\acsibits$, values $\nbits=3,4,5$ have been selected for evaluation. 
To support the final choice of $\nbits=4$, we present three histograms comparing the increments obtained from the collected data and an equally large sample of values randomly extracted from the Gaussian distribution, as described in \cref{lst:normquant}.
The three plots displayed in \cref{fig:quantVSsample3,,fig:quantVSsample4,,fig:quantVSsample5} compare increments and sampled values after quantization over 3, 4, and 5 bits respectively. 
All three histograms have been normalized with respect to the integral of the distribution and use a logarithmic scale on the $y$-axis to simplify data comparison.

It is easily observable that the measured increments follow a slightly more peaked trend, with the central and the outermost values (i.e. the tails of the quantized distribution) appearing more often than they do in the sampled version of the increments. 
In contrast, the intermediate values appear less frequently than in the sampled increments.

Choosing 3 bits to quantize the Gaussian distribution is limiting in terms of information that can be represented. 
The quantization over 4 bits, even though the resulting empirical distribution has a higher probability of the external intervals compared to the values sampled from the Gaussian distribution, results in a sufficiently informative representation of the increments. 
Moreover, the tails of the Normal distribution are correctly quantized and do not alter the distribution itself. 
We will use the Gaussian distribution with 4-bit quantization to approximate the increments, as it allows a sufficiently informative representation of the \iacsi values, without misrepresenting the process due to the use of too many quantization bits. 
Contrarily, quantization over 5 bits using $3\cdot\sigma$ to fix boundary conditions results in evident excessive accumulation of the tails of both the Normal and empirical distribution on the boundary quantization intervals; this excludes the possibility of using 5 bits to quantize the Gaussian distribution, as the behavior of its tails is altered.

\begin{figure}[h!]
	\centering
	\includegraphics[width=0.8\textwidth]{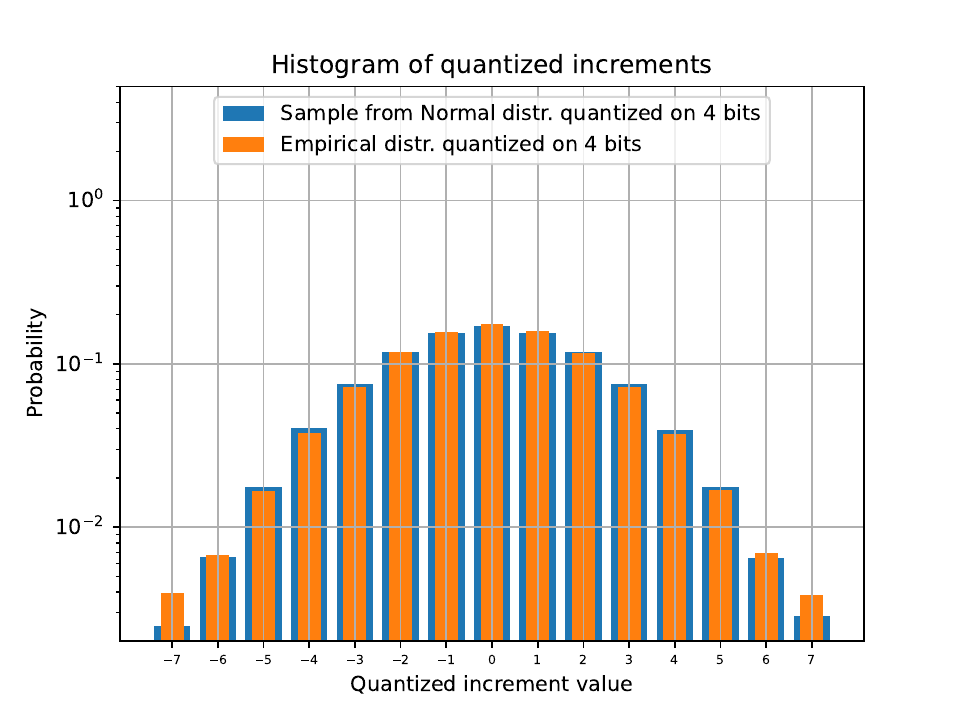}
	\caption{Distribution of increments VS sample from Gaussian distribution after quantization over 4 bits. Increments are computed on an experiment belonging to the AntiSense dataset.}
	\label{fig:ASquantVSsample4}
\end{figure}

\begin{figure}[h!]
	\centering
	\includegraphics[width=0.8\textwidth]{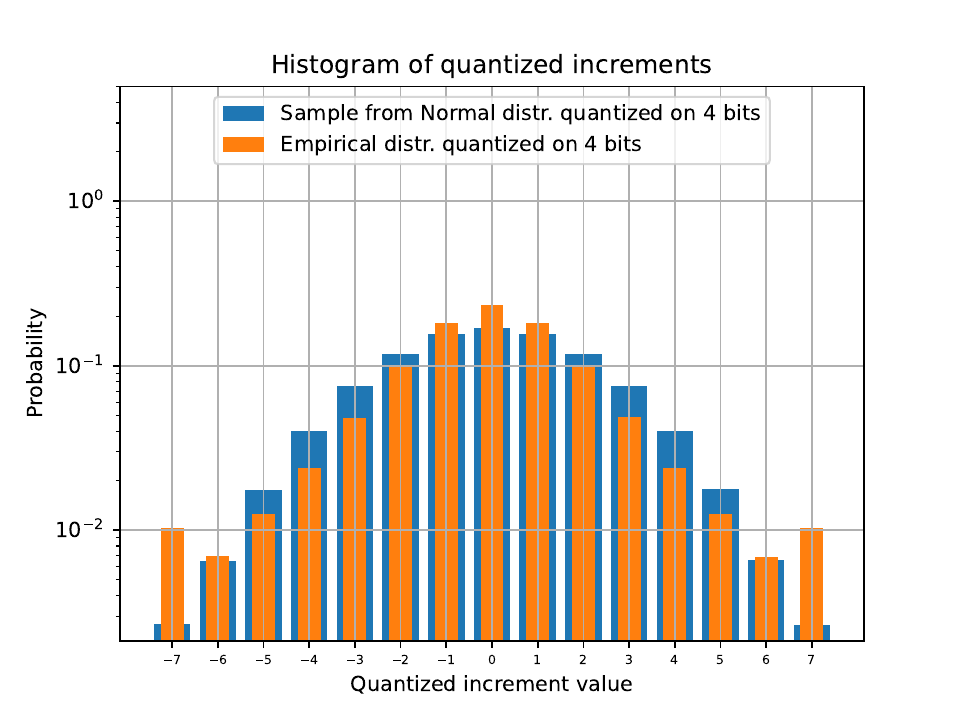}	\caption{Distribution of increments VS sample from Gaussian distribution after quantization over 4 bits. Increments are computed on an experiment performed in the Empty Scenario at 40 MHz.}
	\label{fig:qvss40}
\end{figure}

\begin{figure}[h!]
	\centering
	\includegraphics[width=0.8\textwidth]{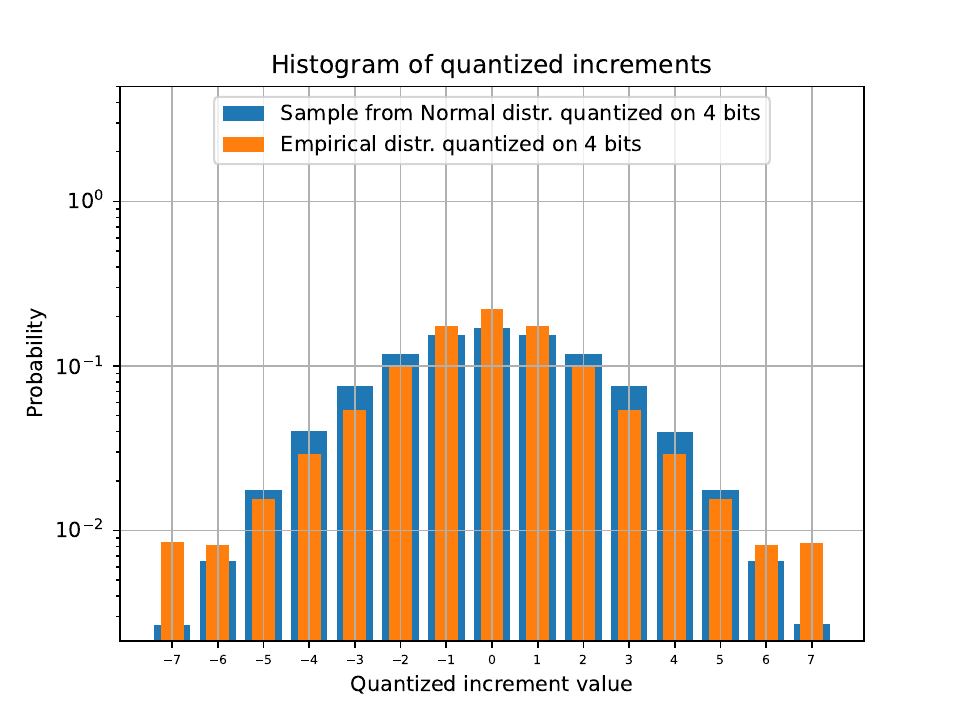}	\caption{Distribution of increments VS sample from Gaussian distribution after quantization over 4 bits. Increments are computed on an experiment performed in the Empty Scenario at 80 MHz.}
	\label{fig:qvss80}
\end{figure}

To further support the choice of $\nbits=4$, we provide an example of the distribution of the quantized increments of \acp{CSI} belonging to the AntiSense dataset in \cref{fig:ASquantVSsample4}. 
In this case, the increments behave almost exactly like the Gaussian distribution, without facing any distortion of the values of the boundary intervals.
Regardless of the number of quantization bits and the technology used to capture the \acp{CSI}, the original distribution of the increments is symmetric around zero, as can also be seen in \cref{fig:qvss40,,fig:qvss80} where the results of the quantization over 4 bits of the increments computed on traces collected at 40 and 80 MHz are displayed. 
This consideration remains valid even in those cases where the empirical distribution no longer resembles the Gaussian curve. 

\begin{figure}[h!]
	\centering
	\includegraphics[width=\textwidth]{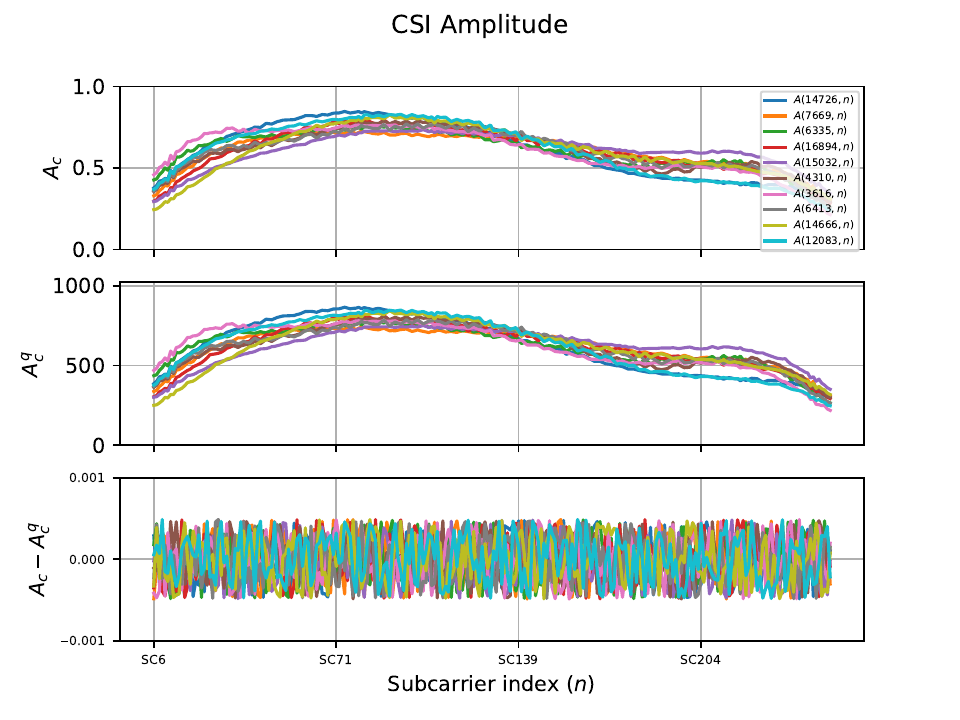}
	\caption{Comparison of the structure of \acsi before and after quantization. The third plot shows the difference between the original normalized \acsi and that obtained after reversing the quantization process and normalizing the result between $0$ and $1$. The represented \acp{CSI} are randomly selected from an experiment performed at 20 MHz in the Fully Dynamic Scenario, with 4 people in the room.}
	\label{fig:ac_acquant_diff20}
\end{figure}

\begin{figure}[h!]
	\centering
	\includegraphics[width=\textwidth]{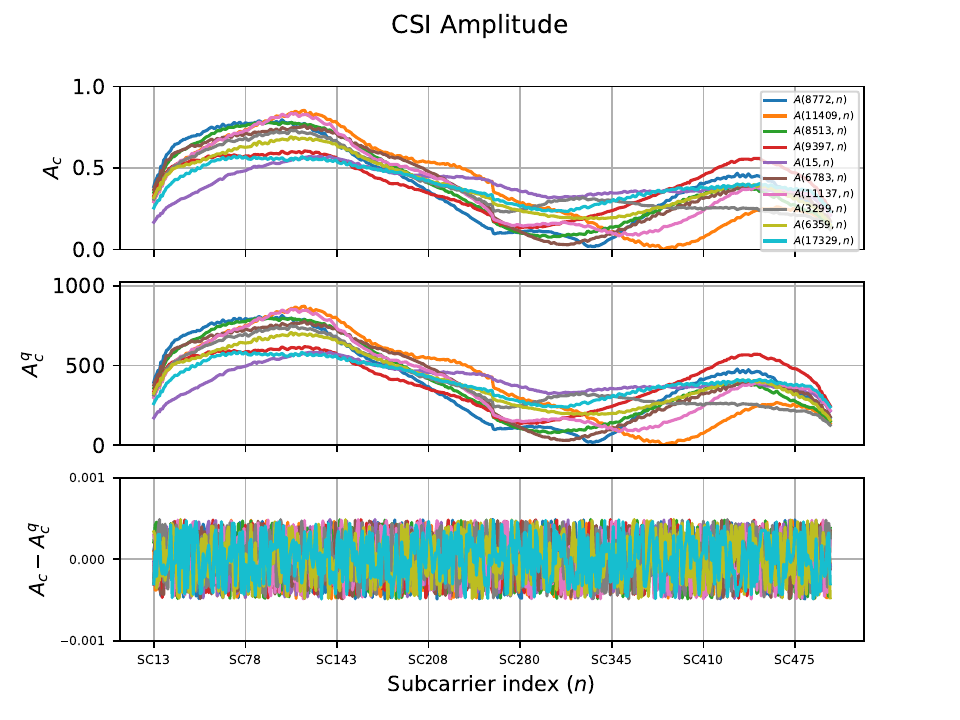}
	\caption{Comparison of the structure of \acsi before and after quantization. The third plot shows the difference between the original normalized \acsi and that obtained after reversing the quantization process and normalizing the result between $0$ and $1$. The represented \acp{CSI} are randomly selected from an experiment performed at 40 MHz in the Fully Dynamic Scenario, with 5 people in the room.}
	\label{fig:ac_acquant_diff40}
\end{figure}

\begin{figure}[h!]
	\centering
	\includegraphics[width=\textwidth]{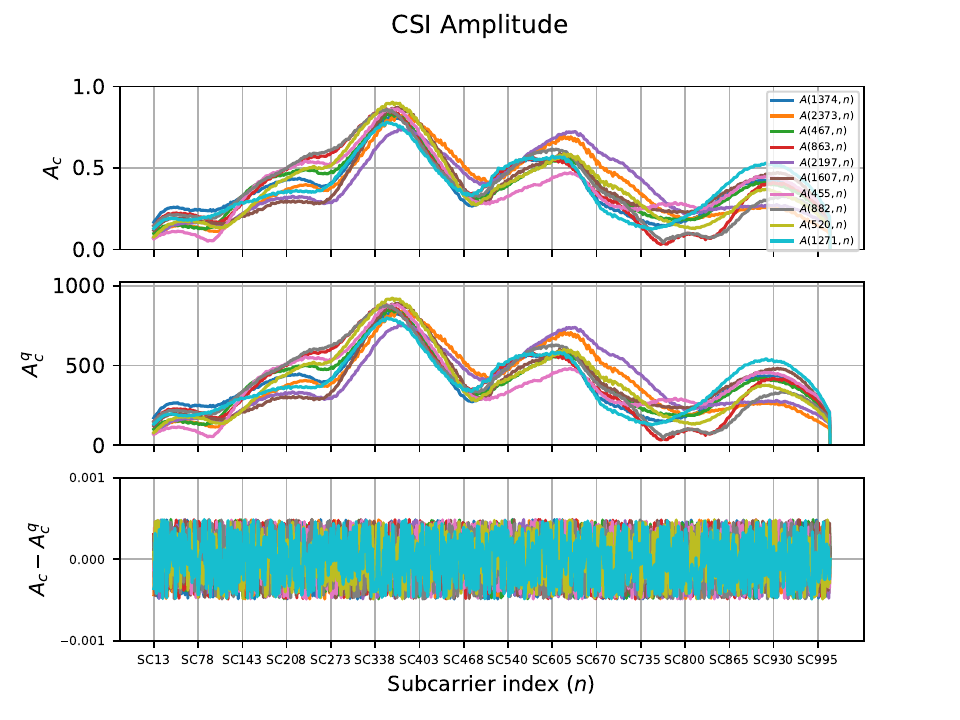}
	\caption{Comparison of the structure of \acsi before and after quantization. The third plot shows the difference between the original normalized \acsi and that obtained after reversing the quantization process and normalizing the result between $0$ and $1$. The represented \acp{CSI} are randomly selected from an experiment performed at 80 MHz in the Fully Dynamic Scenario, with 5 people in the room.}
	\label{fig:ac_acquant_diff80}
\end{figure}

\cref{fig:ac_acquant_diff20,,fig:ac_acquant_diff40,,fig:ac_acquant_diff80} offer a straightforward comparison between \acsi values before and after undergoing quantization. The trends followed by the two processes can be superimposed with an irrelevant mismatch, as highlighted in the third plot of each figure. 
The third plot is obtained by displaying the difference between normalized $\acsi(\tin, \subc)$ and the value computed after reversing the quantization process and re-normalizing the resulting values.
Regardless of the bandwidth the collection was obtained on, $\acsibits=10$ has been chosen as a function of $\nbits$, as per \cref{eq:acsibits}: this choice should provide a semantically equal representation of the amplitudes across scenarios and experiments, facilitating comparison of the results obtained through different technologies and setups. 


%% file: hamming_res.tex
\chapter{Results of the Analysis of the Weighted Hamming Distance}
\label{ch:whdres}
In \cref{ch:whd}, the following different characterizations of the \whd have been introduced:
\begin{itemize}
	\item Internal: $\whd(\racsi, \acsi(\tin,\cdot))$ with \racsi belonging to the same experiment as all \acsi;
	\item External: $\whd(\racsi, \acsi(\tin,\cdot))$ with \racsi belonging to the same scenario as all \acsi but to a distinct experiment;
	\item Cross-setup: $\whd(\racsi, \acsi(\tin,\cdot))$ with \racsi belonging to a different scenario than all \acsi.
\end{itemize}
Considering that a 10-minute-long experiment consists of nearly twenty thousand \acp{CSI}, albeit feasible, computing all distances between each and every \ac{CSI} and \racsi would make the discussion of the results ineffective and impossible to compact into a limited yet meaningful amount of data.
Therefore, the discussion will initially revolve around the average distances between \racsi and all \acsi of an experiment. 
The computation of the average \whd will be performed as follows:
\begin{equation}\label{eq:avgwhd}
	\overline{\whd(\racsi,\acsi)} = \frac{1}{\nsamp}\sum_{\tin=1}^{\nsamp}\whd(\racsi, \acsi(\tin,\cdot))
\end{equation}
where $\overline{\whd(\racsi,\acsi)}$ employs the symbol $\acsi$ to indicate  the whole considered experiment through its generic \ac{CSI}. 
The classification of the average \acp{WHD} as internal, external or cross-setup remains the same as that described at the beginning of this chapter.

Together with the average \whd, the standard deviation of the \whd is also computed, to provide a numerical evaluation of the dispersion of the distance values.

Initially, we compute the internal and external $\overline{\whd}$ under the assumption that the average distance between an experiment and its mean \ac{CSI} \racsi should be slightly lower than the external distance of an experiment. 
Such a difference may not be extremely noticeable (i.e. they may not differ by orders of magnitude) as the compared experiments belong to the same scenario. 
Data collected in the same scenario is expected to display similar behavior, especially in static environments: the Empty and Static Scenarios, given the minor modifications that the propagation environments undergoes by nature, should produce \acp{CSI} that are --- to some extent --- similar to each other.

Since the average distances are computed as floating point values that do not have a reference scale, comparison of the results obtained across different scenarios is ineffective. 

To correctly compare the measurements, we normalize the distances dividing each value by the maximum achievable distance.
According to the quantization process that each \ac{CSI} undergoes, the minimum and maximum values that they can take on each sub-carrier are $0$ and $2^{\acsibits}-1$ respectively. 
Therefore, the minimum distance between two \acp{CSI} is obviously zero --- in case the \acp{CSI} take the same value on all sub-carriers ---, whereas the maximum distance is reached when one \ac{CSI} is `null' (i.e. zero on all sub-carriers) and the other is equal to $2^{\acsibits}-1$ on all sub-carriers.
Hence,
\begin{equation}
	0 \leq \frac{\overline{\whd(\racsi,\acsi)}}{\nsc\cdot(2^{\acsibits}-1)} \leq 1
	\label{eq:normwhd}
\end{equation}
Using this normalization of $\overline{\whd(\racsi,\acsi)}$, the values obtained from different experiments across distinct setups become comparable, as they reference a scale going from 0 to 1. 
From this point on, $\overline{\whd(\racsi,\acsi)}$ will be used to refer to the \textit{normalized} average \whd, to avoid overloading the notation.

\section{Results on the Collected Dataset}
To answer the question of whether it is easily understandable if a given \ac{CSI} belongs to a specific experimental setup, \cref{tab:avg20ax} presents the normalized average \whd obtained from computing the internal, external and cross-distance between \racsi and the \acp{CSI} collected at 20 MHz in the Empty, Static, and Fully Dynamic Scenarios. 
Values highlighted in yellow represent the internal and external \whd for the Empty, Static and Fully Dynamic Scenarios.
Specifically, on the diagonal of the matrix, the values of the internal distances of the experiments are displayed. 
Values highlighted in green, orange, and blue represent the `cross-setup' distances between each couple of scenarios. 

\begin{table}[]
	\centering
	\resizebox{\textwidth}{!}{%
		\begin{tabular}{|
				>{\columncolor[HTML]{C1C1C1}}c |c|c|c|c|c|c|c|c|c|c|c|c|}
			\hline
			& \cellcolor[HTML]{C1C1C1}\textbf{E0} & \cellcolor[HTML]{C1C1C1}\textbf{E1} & \cellcolor[HTML]{C1C1C1}\textbf{E2} & \cellcolor[HTML]{C1C1C1}\textbf{E3} & \cellcolor[HTML]{C1C1C1}\textbf{S0}                & \cellcolor[HTML]{C1C1C1}\textbf{S1}                & \cellcolor[HTML]{C1C1C1}\textbf{S2}                & \cellcolor[HTML]{C1C1C1}\textbf{S3}                & \cellcolor[HTML]{C1C1C1}\textbf{FD0}               & \cellcolor[HTML]{C1C1C1}\textbf{FD1}               & \cellcolor[HTML]{C1C1C1}\textbf{FD2}               & \cellcolor[HTML]{C1C1C1}\textbf{FD3}               \\ \hline
			\cellcolor[HTML]{C1C1C1}\textbf{\racsi E0} & \cellcolor[HTML]{FFFFC7}0.008       & \cellcolor[HTML]{FFFFC7}0.021       & \cellcolor[HTML]{FFFFC7}0.021       & \cellcolor[HTML]{FFFFC7}0.016       & \multicolumn{1}{l|}{\cellcolor[HTML]{C4FF99}0.083} & \multicolumn{1}{l|}{\cellcolor[HTML]{C4FF99}0.069} & \multicolumn{1}{l|}{\cellcolor[HTML]{C4FF99}0.073} & \multicolumn{1}{l|}{\cellcolor[HTML]{C4FF99}0.086} & \multicolumn{1}{l|}{\cellcolor[HTML]{FFD098}0.145} & \multicolumn{1}{l|}{\cellcolor[HTML]{FFD098}0.217} & \multicolumn{1}{l|}{\cellcolor[HTML]{FFD098}0.191} & \multicolumn{1}{l|}{\cellcolor[HTML]{FFD098}0.108} \\ \hline
			\cellcolor[HTML]{C1C1C1}\textbf{\racsi E1} & \cellcolor[HTML]{FFFFC7}0.021       & \cellcolor[HTML]{FFFFC7}0.007       & \cellcolor[HTML]{FFFFC7}0.007       & \cellcolor[HTML]{FFFFC7}0.012       & \multicolumn{1}{l|}{\cellcolor[HTML]{C4FF99}0.102} & \multicolumn{1}{l|}{\cellcolor[HTML]{C4FF99}0.089} & \multicolumn{1}{l|}{\cellcolor[HTML]{C4FF99}0.091} & \multicolumn{1}{l|}{\cellcolor[HTML]{C4FF99}0.105} & \multicolumn{1}{l|}{\cellcolor[HTML]{FFD098}0.163} & \multicolumn{1}{l|}{\cellcolor[HTML]{FFD098}0.231} & \multicolumn{1}{l|}{\cellcolor[HTML]{FFD098}0.204} & \multicolumn{1}{l|}{\cellcolor[HTML]{FFD098}0.125} \\ \hline
			\cellcolor[HTML]{C1C1C1}\textbf{\racsi E2} & \cellcolor[HTML]{FFFFC7}0.021       & \cellcolor[HTML]{FFFFC7}0.007       & \cellcolor[HTML]{FFFFC7}0.006       & \cellcolor[HTML]{FFFFC7}0.012       & \multicolumn{1}{l|}{\cellcolor[HTML]{C4FF99}0.102} & \multicolumn{1}{l|}{\cellcolor[HTML]{C4FF99}0.089} & \multicolumn{1}{l|}{\cellcolor[HTML]{C4FF99}0.091} & \multicolumn{1}{l|}{\cellcolor[HTML]{C4FF99}0.105} & \multicolumn{1}{l|}{\cellcolor[HTML]{FFD098}0.163} & \multicolumn{1}{l|}{\cellcolor[HTML]{FFD098}0.231} & \multicolumn{1}{l|}{\cellcolor[HTML]{FFD098}0.204} & \multicolumn{1}{l|}{\cellcolor[HTML]{FFD098}0.125} \\ \hline
			\cellcolor[HTML]{C1C1C1}\textbf{\racsi E3} & \cellcolor[HTML]{FFFFC7}0.017       & \cellcolor[HTML]{FFFFC7}0.012       & \cellcolor[HTML]{FFFFC7}0.012       & \cellcolor[HTML]{FFFFC7}0.006       & \multicolumn{1}{l|}{\cellcolor[HTML]{C4FF99}0.095} & \multicolumn{1}{l|}{\cellcolor[HTML]{C4FF99}0.082} & \multicolumn{1}{l|}{\cellcolor[HTML]{C4FF99}0.086} & \multicolumn{1}{l|}{\cellcolor[HTML]{C4FF99}0.098} & \multicolumn{1}{l|}{\cellcolor[HTML]{FFD098}0.153} & \multicolumn{1}{l|}{\cellcolor[HTML]{FFD098}0.220} & \multicolumn{1}{l|}{\cellcolor[HTML]{FFD098}0.193} & \multicolumn{1}{l|}{\cellcolor[HTML]{FFD098}0.115} \\ \hline
			\textbf{\racsi S0}                         & \cellcolor[HTML]{C4FF99}0.081       & \cellcolor[HTML]{C4FF99}0.101       & \cellcolor[HTML]{C4FF99}0.101       & \cellcolor[HTML]{C4FF99}0.094       & \cellcolor[HTML]{FFFFC7}0.018                      & \cellcolor[HTML]{FFFFC7}0.025                      & \cellcolor[HTML]{FFFFC7}0.032                      & \cellcolor[HTML]{FFFFC7}0.020                      & \cellcolor[HTML]{BCFFFB}0.085                      & \cellcolor[HTML]{BCFFFB}0.155                      & \cellcolor[HTML]{BCFFFB}0.136                      & \cellcolor[HTML]{BCFFFB}0.059                      \\ \hline
			\textbf{\racsi S1}                         & \cellcolor[HTML]{C4FF99}0.067       & \cellcolor[HTML]{C4FF99}0.087       & \cellcolor[HTML]{C4FF99}0.087       & \cellcolor[HTML]{C4FF99}0.080       & \cellcolor[HTML]{FFFFC7}0.025                      & \cellcolor[HTML]{FFFFC7}0.017                      & \cellcolor[HTML]{FFFFC7}0.030                      & \cellcolor[HTML]{FFFFC7}0.029                      & \cellcolor[HTML]{BCFFFB}0.093                      & \cellcolor[HTML]{BCFFFB}0.168                      & \cellcolor[HTML]{BCFFFB}0.147                      & \cellcolor[HTML]{BCFFFB}0.065                      \\ \hline
			\textbf{\racsi S2}                         & \cellcolor[HTML]{C4FF99}0.070       & \cellcolor[HTML]{C4FF99}0.089       & \cellcolor[HTML]{C4FF99}0.088       & \cellcolor[HTML]{C4FF99}0.083       & \cellcolor[HTML]{FFFFC7}0.030                      & \cellcolor[HTML]{FFFFC7}0.026                      & \cellcolor[HTML]{FFFFC7}0.022                      & \cellcolor[HTML]{FFFFC7}0.028                      & \cellcolor[HTML]{BCFFFB}0.102                      & \cellcolor[HTML]{BCFFFB}0.179                      & \cellcolor[HTML]{BCFFFB}0.156                      & \cellcolor[HTML]{BCFFFB}0.072                      \\ \hline
			\textbf{\racsi S3}                         & \cellcolor[HTML]{C4FF99}0.084       & \cellcolor[HTML]{C4FF99}0.104       & \cellcolor[HTML]{C4FF99}0.103       & \cellcolor[HTML]{C4FF99}0.097       & \cellcolor[HTML]{FFFFC7}0.020                      & \cellcolor[HTML]{FFFFC7}0.029                      & \cellcolor[HTML]{FFFFC7}0.030                      & \cellcolor[HTML]{FFFFC7}0.019                      & \cellcolor[HTML]{BCFFFB}0.087                      & \cellcolor[HTML]{BCFFFB}0.159                      & \cellcolor[HTML]{BCFFFB}0.139                      & \cellcolor[HTML]{BCFFFB}0.061                      \\ \hline
			\textbf{\racsi FD0}                        & \cellcolor[HTML]{FFD098}0.142       & \cellcolor[HTML]{FFD098}0.159       & \cellcolor[HTML]{FFD098}0.161       & \cellcolor[HTML]{FFD098}0.150       & \cellcolor[HTML]{BCFFFB}0.079                      & \cellcolor[HTML]{BCFFFB}0.087                      & \cellcolor[HTML]{BCFFFB}0.097                      & \cellcolor[HTML]{BCFFFB}0.081                      & \cellcolor[HTML]{FFFFC7}0.041                      & \cellcolor[HTML]{FFFFC7}0.108                      & \cellcolor[HTML]{FFFFC7}0.090                      & \cellcolor[HTML]{FFFFC7}0.050                      \\ \hline
			\textbf{\racsi FD1}                        & \cellcolor[HTML]{FFD098}0.212       & \cellcolor[HTML]{FFD098}0.226       & \cellcolor[HTML]{FFD098}0.227       & \cellcolor[HTML]{FFD098}0.216       & \cellcolor[HTML]{BCFFFB}0.149                      & \cellcolor[HTML]{BCFFFB}0.163                      & \cellcolor[HTML]{BCFFFB}0.174                      & \cellcolor[HTML]{BCFFFB}0.153                      & \cellcolor[HTML]{FFFFC7}0.104                      & \cellcolor[HTML]{FFFFC7}0.049                      & \cellcolor[HTML]{FFFFC7}0.062                      & \cellcolor[HTML]{FFFFC7}0.115                      \\ \hline
			\textbf{\racsi FD2}                        & \cellcolor[HTML]{FFD098}0.187       & \cellcolor[HTML]{FFD098}0.201       & \cellcolor[HTML]{FFD098}0.201       & \cellcolor[HTML]{FFD098}0.190       & \cellcolor[HTML]{BCFFFB}0.128                      & \cellcolor[HTML]{BCFFFB}0.140                      & \cellcolor[HTML]{BCFFFB}0.151                      & \cellcolor[HTML]{BCFFFB}0.132                      & \cellcolor[HTML]{FFFFC7}0.087                      & \cellcolor[HTML]{FFFFC7}0.063                      & \cellcolor[HTML]{FFFFC7}0.047                      & \cellcolor[HTML]{FFFFC7}0.092                      \\ \hline
			\textbf{\racsi FD3}                        & \cellcolor[HTML]{FFD098}0.106       & \cellcolor[HTML]{FFD098}0.123       & \cellcolor[HTML]{FFD098}0.123       & \cellcolor[HTML]{FFD098}0.113       & \cellcolor[HTML]{BCFFFB}0.056                      & \cellcolor[HTML]{BCFFFB}0.062                      & \cellcolor[HTML]{BCFFFB}0.070                      & \cellcolor[HTML]{BCFFFB}0.057                      & \cellcolor[HTML]{FFFFC7}0.057                      & \cellcolor[HTML]{FFFFC7}0.122                      & \cellcolor[HTML]{FFFFC7}0.097                      & \cellcolor[HTML]{FFFFC7}0.030                      \\ \hline
		\end{tabular}%
	}
	\caption{Normalized average \whd between each experiment performed in the Empty, Static, and Fully Dynamic Scenarios and the reference average \ac{CSI} computed on the same experiments. Data collected on a 20 MHz channel using 802.11ax. The FD Scenario consisted of four people in the room. }
	\label{tab:avg20ax}
\end{table}

It is evident that the internal \whd always takes the lowest value compared to all other distances --- i.e., the lowest values of \cref{tab:avg20ax} are on the diagonal of the matrix.
Similarly, the highest values of \whd can be found in the blue and orange sub-matrices: this is representative of the fact that the Fully Dynamic Scenario is a more changing environment, which results in more varying \ac{CSI} traces. 
Such variability of the environment is reflected in higher distance values.

These considerations can be extended to data collected on the 40 and 80 MHz bandwidths, as can be seen in \cref{tab:avg40ax,,tab:avg80ax}. 
The color coding of these two tables remains the same as that of \cref{tab:avg20ax}. 
Once again, the lowest values can be found on the diagonals of the matrices, corresponding to the internal \whd. 
Contrarily, the largest differences can be found in the orange sections of the tables, as they contain the distances between the Fully Dynamic and the Empty Scenario: the great variability of the Fully Dynamic Scenario --- where multiple people are present in the room and possibly moving around --- is compared to the extremely static nature of the Empty Scenario, originating the highest \whd values. 
This corroborates the assumption that both the presence and the movements of people within an environment significantly affect the behavior of the signal traveling from the transmitter to the receiver.  

\begin{table}[h!]
	\centering
	\resizebox{\textwidth}{!}{%
		\begin{tabular}{|
				>{\columncolor[HTML]{C1C1C1}}c |
				>{\columncolor[HTML]{C1C1C1}}c |
				>{\columncolor[HTML]{FFD098}}c 
				>{\columncolor[HTML]{FFD098}}c 
				>{\columncolor[HTML]{FFD098}}c |
				>{\columncolor[HTML]{BCFFFB}}c |
				>{\columncolor[HTML]{FFFFC7}}c 
				>{\columncolor[HTML]{FFFFC7}}c 
				>{\columncolor[HTML]{FFFFC7}}c 
				>{\columncolor[HTML]{FFFFC7}}c 
				>{\columncolor[HTML]{FFFFC7}}c 
				>{\columncolor[HTML]{FFFFC7}}c |}
			\hline
			\textbf{}                                                                                                           & \textbf{}                          & \multicolumn{3}{c|}{\cellcolor[HTML]{C1C1C1}\textbf{EMPTY}}                                                                                            & \cellcolor[HTML]{C1C1C1}\textbf{STATIC} & \multicolumn{6}{c|}{\cellcolor[HTML]{C1C1C1}\textbf{FULLY DYNAMIC}}                                                                                                                                                                                                                                                                  \\ \hline
			\textbf{}                                                                                                           & \textbf{\# PPL}                     & \multicolumn{1}{c|}{\cellcolor[HTML]{C1C1C1}\textbf{0}} & \multicolumn{1}{c|}{\cellcolor[HTML]{C1C1C1}\textbf{0}} & \cellcolor[HTML]{C1C1C1}\textbf{0} & \cellcolor[HTML]{C1C1C1}\textbf{1}      & \multicolumn{1}{c|}{\cellcolor[HTML]{C1C1C1}\textbf{2}} & \multicolumn{1}{c|}{\cellcolor[HTML]{C1C1C1}\textbf{3}} & \multicolumn{1}{c|}{\cellcolor[HTML]{C1C1C1}\textbf{4}} & \multicolumn{1}{c|}{\cellcolor[HTML]{C1C1C1}\textbf{4}} & \multicolumn{1}{c|}{\cellcolor[HTML]{C1C1C1}\textbf{5}} & \cellcolor[HTML]{C1C1C1}\textbf{5} \\ \hline
			\cellcolor[HTML]{C1C1C1}                                                                                            & \cellcolor[HTML]{C1C1C1}\textbf{0} & \multicolumn{1}{c|}{\cellcolor[HTML]{FFFFC7}0.004}      & \multicolumn{1}{c|}{\cellcolor[HTML]{FFFFC7}0.247}      & \cellcolor[HTML]{FFFFC7}0.244      & \cellcolor[HTML]{C4FF99}0.232           & \multicolumn{1}{c|}{\cellcolor[HTML]{FFD098}0.164}      & \multicolumn{1}{c|}{\cellcolor[HTML]{FFD098}0.135}      & \multicolumn{1}{c|}{\cellcolor[HTML]{FFD098}0.169}      & \multicolumn{1}{c|}{\cellcolor[HTML]{FFD098}0.157}      & \multicolumn{1}{c|}{\cellcolor[HTML]{FFD098}0.146}      & \cellcolor[HTML]{FFD098}0.169      \\ \cline{2-12} 
			\cellcolor[HTML]{C1C1C1}                                                                                            & \cellcolor[HTML]{C1C1C1}\textbf{0} & \multicolumn{1}{c|}{\cellcolor[HTML]{FFFFC7}0.246}      & \multicolumn{1}{c|}{\cellcolor[HTML]{FFFFC7}0.007}      & \cellcolor[HTML]{FFFFC7}0.008      & \cellcolor[HTML]{C4FF99}0.344           & \multicolumn{1}{c|}{\cellcolor[HTML]{FFD098}0.292}      & \multicolumn{1}{c|}{\cellcolor[HTML]{FFD098}0.184}      & \multicolumn{1}{c|}{\cellcolor[HTML]{FFD098}0.119}      & \multicolumn{1}{c|}{\cellcolor[HTML]{FFD098}0.131}      & \multicolumn{1}{c|}{\cellcolor[HTML]{FFD098}0.175}      & \cellcolor[HTML]{FFD098}0.210      \\ \cline{2-12} 
			\multirow{-3}{*}{\cellcolor[HTML]{C1C1C1}\textbf{\begin{tabular}[c]{@{}c@{}}EMPTY\\ \racsi\end{tabular}}}           & \cellcolor[HTML]{C1C1C1}\textbf{0} & \multicolumn{1}{c|}{\cellcolor[HTML]{FFFFC7}0.243}      & \multicolumn{1}{c|}{\cellcolor[HTML]{FFFFC7}0.008}      & \cellcolor[HTML]{FFFFC7}0.005      & \cellcolor[HTML]{C4FF99}0.341           & \multicolumn{1}{c|}{\cellcolor[HTML]{FFD098}0.289}      & \multicolumn{1}{c|}{\cellcolor[HTML]{FFD098}0.182}      & \multicolumn{1}{c|}{\cellcolor[HTML]{FFD098}0.117}      & \multicolumn{1}{c|}{\cellcolor[HTML]{FFD098}0.129}      & \multicolumn{1}{c|}{\cellcolor[HTML]{FFD098}0.173}      & \cellcolor[HTML]{FFD098}0.208      \\ \hline
			\textbf{\begin{tabular}[c]{@{}c@{}}STATIC\\ \racsi\end{tabular}}                                                    & \cellcolor[HTML]{C1C1C1}\textbf{1} & \multicolumn{1}{c|}{\cellcolor[HTML]{C4FF99}0.230}      & \multicolumn{1}{c|}{\cellcolor[HTML]{C4FF99}0.344}      & \cellcolor[HTML]{C4FF99}0.341      & \cellcolor[HTML]{FFFFC7}0.031           & \multicolumn{1}{c|}{\cellcolor[HTML]{BCFFFB}0.182}      & \multicolumn{1}{c|}{\cellcolor[HTML]{BCFFFB}0.264}      & \multicolumn{1}{c|}{\cellcolor[HTML]{BCFFFB}0.267}      & \multicolumn{1}{c|}{\cellcolor[HTML]{BCFFFB}0.259}      & \multicolumn{1}{c|}{\cellcolor[HTML]{BCFFFB}0.258}      & \cellcolor[HTML]{BCFFFB}0.245      \\ \hline
			\cellcolor[HTML]{C1C1C1}                                                                                            & \textbf{2}                         & \multicolumn{1}{c|}{\cellcolor[HTML]{FFD098}0.150}      & \multicolumn{1}{c|}{\cellcolor[HTML]{FFD098}0.278}      & 0.275                              & 0.168                                   & \multicolumn{1}{c|}{\cellcolor[HTML]{FFFFC7}0.060}      & \multicolumn{1}{c|}{\cellcolor[HTML]{FFFFC7}0.223}      & \multicolumn{1}{c|}{\cellcolor[HTML]{FFFFC7}0.201}      & \multicolumn{1}{c|}{\cellcolor[HTML]{FFFFC7}0.178}      & \multicolumn{1}{c|}{\cellcolor[HTML]{FFFFC7}0.210}      & 0.182                              \\ \cline{2-12} 
			\cellcolor[HTML]{C1C1C1}                                                                                            & \textbf{3}                         & \multicolumn{1}{c|}{\cellcolor[HTML]{FFD098}0.127}      & \multicolumn{1}{c|}{\cellcolor[HTML]{FFD098}0.180}      & 0.179                              & 0.263                                   & \multicolumn{1}{c|}{\cellcolor[HTML]{FFFFC7}0.233}      & \multicolumn{1}{c|}{\cellcolor[HTML]{FFFFC7}0.044}      & \multicolumn{1}{c|}{\cellcolor[HTML]{FFFFC7}0.130}      & \multicolumn{1}{c|}{\cellcolor[HTML]{FFFFC7}0.162}      & \multicolumn{1}{c|}{\cellcolor[HTML]{FFFFC7}0.096}      & 0.166                              \\ \cline{2-12} 
			\cellcolor[HTML]{C1C1C1}                                                                                            & \textbf{4}                         & \multicolumn{1}{c|}{\cellcolor[HTML]{FFD098}0.154}      & \multicolumn{1}{c|}{\cellcolor[HTML]{FFD098}0.099}      & 0.097                              & 0.257                                   & \multicolumn{1}{c|}{\cellcolor[HTML]{FFFFC7}0.211}      & \multicolumn{1}{c|}{\cellcolor[HTML]{FFFFC7}0.119}      & \multicolumn{1}{c|}{\cellcolor[HTML]{FFFFC7}0.074}      & \multicolumn{1}{c|}{\cellcolor[HTML]{FFFFC7}0.094}      & \multicolumn{1}{c|}{\cellcolor[HTML]{FFFFC7}0.109}      & 0.154                              \\ \cline{2-12} 
			\cellcolor[HTML]{C1C1C1}                                                                                            & \textbf{4}                         & \multicolumn{1}{c|}{\cellcolor[HTML]{FFD098}0.134}      & \multicolumn{1}{c|}{\cellcolor[HTML]{FFD098}0.113}      & 0.110                              & 0.245                                   & \multicolumn{1}{c|}{\cellcolor[HTML]{FFFFC7}0.189}      & \multicolumn{1}{c|}{\cellcolor[HTML]{FFFFC7}0.149}      & \multicolumn{1}{c|}{\cellcolor[HTML]{FFFFC7}0.087}      & \multicolumn{1}{c|}{\cellcolor[HTML]{FFFFC7}0.080}      & \multicolumn{1}{c|}{\cellcolor[HTML]{FFFFC7}0.128}      & 0.154                              \\ \cline{2-12} 
			\cellcolor[HTML]{C1C1C1}                                                                                            & \textbf{5}                         & \multicolumn{1}{c|}{\cellcolor[HTML]{FFD098}0.132}      & \multicolumn{1}{c|}{\cellcolor[HTML]{FFD098}0.162}      & 0.160                              & 0.248                                   & \multicolumn{1}{c|}{\cellcolor[HTML]{FFFFC7}0.219}      & \multicolumn{1}{c|}{\cellcolor[HTML]{FFFFC7}0.082}      & \multicolumn{1}{c|}{\cellcolor[HTML]{FFFFC7}0.107}      & \multicolumn{1}{c|}{\cellcolor[HTML]{FFFFC7}0.132}      & \multicolumn{1}{c|}{\cellcolor[HTML]{FFFFC7}0.067}      & 0.129                              \\ \cline{2-12} 
			\multirow{-6}{*}{\cellcolor[HTML]{C1C1C1}\textbf{\begin{tabular}[c]{@{}c@{}}FULLY\\ DYNAMIC\\ \racsi\end{tabular}}} & \textbf{5}                         & \multicolumn{1}{c|}{\cellcolor[HTML]{FFD098}0.117}      & \multicolumn{1}{c|}{\cellcolor[HTML]{FFD098}0.183}      & 0.180                              & 0.213                                   & \multicolumn{1}{c|}{\cellcolor[HTML]{FFFFC7}0.176}      & \multicolumn{1}{c|}{\cellcolor[HTML]{FFFFC7}0.129}      & \multicolumn{1}{c|}{\cellcolor[HTML]{FFFFC7}0.119}      & \multicolumn{1}{c|}{\cellcolor[HTML]{FFFFC7}0.124}      & \multicolumn{1}{c|}{\cellcolor[HTML]{FFFFC7}0.094}      & 0.116                              \\ \hline
		\end{tabular}%
	}
	\caption{Normalized average \whd between each experiment performed in the Empty, Static, and Fully Dynamic Scenarios and the reference average \ac{CSI} computed on the same experiments. Data collected on a 40 MHz channel using 802.11ax. }
	\label{tab:avg40ax}
\end{table}

\begin{table}[]
	\centering
	\resizebox{\textwidth}{!}{%
		\begin{tabular}{|
				>{\columncolor[HTML]{C1C1C1}}c |
				>{\columncolor[HTML]{C1C1C1}}c |
				>{\columncolor[HTML]{FFD098}}c |
				>{\columncolor[HTML]{BCFFFB}}c |
				>{\columncolor[HTML]{FFFFC7}}c 
				>{\columncolor[HTML]{FFFFC7}}l 
				>{\columncolor[HTML]{FFFFC7}}c 
				>{\columncolor[HTML]{FFFFC7}}c 
				>{\columncolor[HTML]{FFFFC7}}c |}
			\hline
			\textbf{} &
			\textbf{} &
			\cellcolor[HTML]{C1C1C1}\textbf{EMPTY} &
			\cellcolor[HTML]{C1C1C1}\textbf{STATIC} &
			\multicolumn{5}{c|}{\cellcolor[HTML]{C1C1C1}\textbf{FULLY DYNAMIC}} \\ \hline
			\textbf{} &
			\textbf{\# PPL} &
			\cellcolor[HTML]{C1C1C1}\textbf{0} &
			\cellcolor[HTML]{C1C1C1}\textbf{1} &
			\multicolumn{1}{c|}{\cellcolor[HTML]{C1C1C1}\textbf{2}} &
			\multicolumn{1}{c|}{\cellcolor[HTML]{C1C1C1}\textbf{2}} &
			\multicolumn{1}{c|}{\cellcolor[HTML]{C1C1C1}\textbf{3}} &
			\multicolumn{1}{c|}{\cellcolor[HTML]{C1C1C1}\textbf{4}} &
			\cellcolor[HTML]{C1C1C1}\textbf{5} \\ \hline
			\textbf{\begin{tabular}[c]{@{}c@{}}EMPTY\\ \racsi\end{tabular}} &
			\cellcolor[HTML]{C1C1C1}\textbf{0} &
			\cellcolor[HTML]{FFFFC7}0.005 &
			\cellcolor[HTML]{C4FF99}0.075 &
			\multicolumn{1}{c|}{\cellcolor[HTML]{FFD098}0.181} &
			\multicolumn{1}{l|}{\cellcolor[HTML]{FFD098}0.163} &
			\multicolumn{1}{c|}{\cellcolor[HTML]{FFD098}0.231} &
			\multicolumn{1}{c|}{\cellcolor[HTML]{FFD098}0.232} &
			\cellcolor[HTML]{FFD098}0.221 \\ \hline
			\textbf{\begin{tabular}[c]{@{}c@{}}STATIC\\ \racsi\end{tabular}} &
			\cellcolor[HTML]{C1C1C1}\textbf{1} &
			\cellcolor[HTML]{C4FF99}0.072 &
			\cellcolor[HTML]{FFFFC7}0.027 &
			\multicolumn{1}{c|}{\cellcolor[HTML]{BCFFFB}0.143} &
			\multicolumn{1}{l|}{\cellcolor[HTML]{BCFFFB}0.128} &
			\multicolumn{1}{c|}{\cellcolor[HTML]{BCFFFB}0.208} &
			\multicolumn{1}{c|}{\cellcolor[HTML]{BCFFFB}0.206} &
			\cellcolor[HTML]{BCFFFB}0.188 \\ \hline
			\cellcolor[HTML]{C1C1C1} &
			\textbf{2} &
			0.176 &
			0.137 &
			\multicolumn{1}{c|}{\cellcolor[HTML]{FFFFC7}0.047} &
			\multicolumn{1}{l|}{\cellcolor[HTML]{FFFFC7}0.082} &
			\multicolumn{1}{c|}{\cellcolor[HTML]{FFFFC7}0.149} &
			\multicolumn{1}{c|}{\cellcolor[HTML]{FFFFC7}0.142} &
			0.152 \\ \cline{2-9} 
			\cellcolor[HTML]{C1C1C1} &
			\textbf{2} &
			0.156 &
			0.122 &
			\multicolumn{1}{c|}{\cellcolor[HTML]{FFFFC7}0.080} &
			\multicolumn{1}{c|}{\cellcolor[HTML]{FFFFC7}0.045} &
			\multicolumn{1}{c|}{\cellcolor[HTML]{FFFFC7}0.160} &
			\multicolumn{1}{c|}{\cellcolor[HTML]{FFFFC7}0.154} &
			0.139 \\ \cline{2-9} 
			\cellcolor[HTML]{C1C1C1} &
			\textbf{3} &
			0.228 &
			0.207 &
			\multicolumn{1}{c|}{\cellcolor[HTML]{FFFFC7}0.152} &
			\multicolumn{1}{l|}{\cellcolor[HTML]{FFFFC7}0.161} &
			\multicolumn{1}{c|}{\cellcolor[HTML]{FFFFC7}0.034} &
			\multicolumn{1}{c|}{\cellcolor[HTML]{FFFFC7}0.085} &
			0.111 \\ \cline{2-9} 
			\cellcolor[HTML]{C1C1C1} &
			\textbf{4} &
			0.224 &
			0.200 &
			\multicolumn{1}{c|}{\cellcolor[HTML]{FFFFC7}0.139} &
			\multicolumn{1}{l|}{\cellcolor[HTML]{FFFFC7}0.149} &
			\multicolumn{1}{c|}{\cellcolor[HTML]{FFFFC7}0.074} &
			\multicolumn{1}{c|}{\cellcolor[HTML]{FFFFC7}0.055} &
			0.114 \\ \cline{2-9} 
			\multirow{-5}{*}{\cellcolor[HTML]{C1C1C1}\textbf{\begin{tabular}[c]{@{}c@{}}FULLY\\ DYNAMIC\\ \racsi\end{tabular}}} &
			\textbf{5} &
			0.210 &
			0.176 &
			\multicolumn{1}{c|}{\cellcolor[HTML]{FFFFC7}0.147} &
			\multicolumn{1}{l|}{\cellcolor[HTML]{FFFFC7}0.133} &
			\multicolumn{1}{c|}{\cellcolor[HTML]{FFFFC7}0.101} &
			\multicolumn{1}{c|}{\cellcolor[HTML]{FFFFC7}0.112} &
			0.063 \\ \hline
		\end{tabular}%
	}
	\caption{Normalized average \whd between each experiment performed in the Empty, Static, and Fully Dynamic Scenarios and the reference average \ac{CSI} computed on the same experiments. Data collected on an 80 MHz channel using 802.11ax. }
	\label{tab:avg80ax}
\end{table}

It is important to highlight that, since the maximum distance between two \acp{CSI} is an edge case that is extremely unlikely to happen, if not impossible to reach at all, values of the \whd around $0.2$ can be considered large, as they identify substantially different environments. 
This is supported by the fact that the distance between each experiment and its own \racsi is in the order of $10^{-2}$ and lower, which means that most collections of \acp{CSI} are extremely close to their representative \ac{CSI} \racsi.
It is important to note that the Fully Dynamic Scenario intrinsically has such higher variability compared to other experimental setups that the computation of the distances on this scenario may be affected by the significant variations of the amplitudes in the captures.

\cref{tab:avg20ax,,tab:avg40ax,,tab:avg80ax} all display the interesting feature of being \textit{almost} symmetrical. 
Semantically, symmetry means that comparing an experiment $A$ with the average \ac{CSI} of experiment $B$ produces the same result that can be obtained by comparing $B$ with the average \ac{CSI} of $A$. 
In this case, comparing experiments consists in computing the normalized $\overline{\whd}$ between them, as per \cref{eq:avgwhd,,eq:normwhd}.
Especially in the Empty Scenario, the modifications that the environment undergoes are microscopic, therefore the symmetry of the matrix is more accentuated. 
Contrarily, more dynamic scenarios are subject to more impactful and macroscopic alterations, which implies the possibility of less symmetric distances. 

Exact symmetry of the external or cross-setup $\overline{\whd}$ between experiments is difficult to obtain: the $\overline{\whd}$ is computed as the \textit{average} distance between each \ac{CSI} of an experiment $A$ and the \textit{average} \ac{CSI} of another experiment $B$. 
Since the average \ac{CSI} is a summarized visualization of a whole experiment, its use implies that some information about the experiment is discarded or lost, decreasing the accuracy of the representation. 
Nonetheless, comparing all experiments \ac{CSI} by \ac{CSI} would result in an unmanageable amount of distances, therefore it is necessary to merge such information into a single significant number for each couple of experiments.
Computing, once again, an average implies losing some additional information, which can result in minimal --- or more significant, depending on the case --- asymmetries of the $\overline{\whd}$ values.
The average \whd is deemed an initial effective approach to the computation of the `distance' between \ac{CSI} captures and is used as a preliminary metric to evaluate the behavior of the collected data. 
Should a more effective metric be identified, the $\overline{\whd}$ could still be used as an approximated indicator of the similarity between captures, while for a more detailed description of the experiments the new metric would be used.

\begin{figure}[h!]
	\centering
	\begin{subfigure}{\linewidth}
		\centering
		\includegraphics[width=\linewidth]{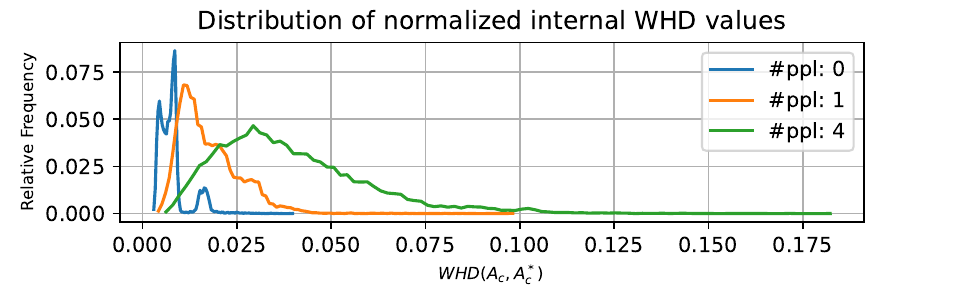}
		\caption{20 MHz bandwidth.}
		\label{fig:20MHz}
	\end{subfigure}\\
	\begin{subfigure}{\linewidth}
		\centering
		\includegraphics[width=\linewidth]{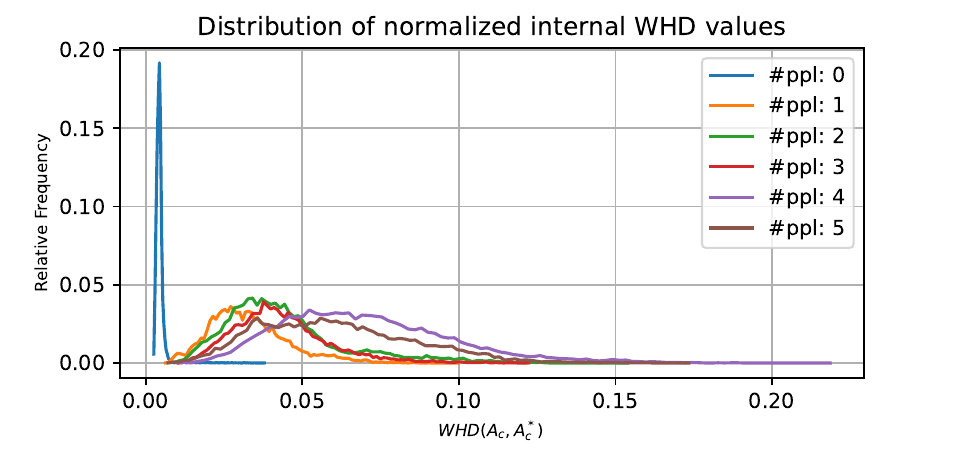}
		\caption{40 MHz bandwidth.}
		\label{fig:40MHz}
	\end{subfigure}\\
	\begin{subfigure}{\linewidth}
		\centering
		\includegraphics[width=\linewidth]{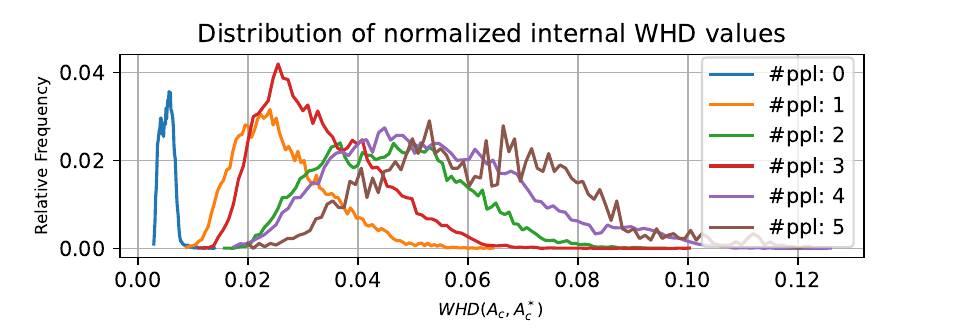}
		\caption{80 MHz bandwidth.}
		\label{fig:80MHz}
	\end{subfigure}
	\caption{Distribution of the normalized internal \whd across different experiments, each characterized by a distinct number of people in the environment.}
	\label{fig:whddistr}
\end{figure}

To provide a more compact visualization of the average internal distances depending on the experiment, we focus the categorization of the collections of \ac{CSI} on the number of people that were in the room when the data was captured. 
This way, we can summarize the values of the internal distance in distributions of the \whd, which allow us to avoid the computation of the \textit{average} \whd. 
Such an approach simplifies comparison between the different experimental setups, allowing to infer how the \acp{CSI} are modified according to the variability of the environment: before looking at the resulting distributions, we can assume that the Empty and Static Scenarios will undergo fewer modifications than the Fully Dynamic one, as the latter also encompasses movements of people in the room, whereas the former are only affected by the furniture and appliances and, possibly, the presence of a person sitting almost still. 
Therefore, the distributions of distances relative to an Empty or Static Scenario can be expected to be much less spread out than those describing a Fully Dynamic Scenario with four or five people moving around the room.
These considerations are supported by figure \cref{fig:whddistr}.

In \cref{fig:20MHz} we can observe that the distribution of the internal \whd derived from data collected in an empty room is significantly more peaked compared to the distributions relative to the Static and Fully Dynamic Scenarios. 
By juxtaposing the three distributions, we can notice that the increasing number of people altering the structure of the experimental setup with their movements impacts the spread-out of the distributions. 

\cref{fig:40MHz} highlights how easily distinguishable the Empty Scenario is compared to any other scenario. 
This means that the Empty Scenario is extremely self-similar, with internal distances being the closest to 0 across all scenarios. 
Looking at the other distributions, they maintain the expected behavior of an increasing standard deviation as the number of people in the room grows, albeit they are often overlapped.

The same considerations can be made about the distributions in \cref{fig:80MHz}, with the Empty Scenario clearly distinct from all others and the remaining experiments showing an increasingly higher dispersion of the distances. 
It can be noted that, since the 5-people Fully Dynamic Scenario consists of one-minute-long experiments --- compared to the 10 minutes of the other captures ---, the behavior of the corresponding distribution is slightly more fluctuating than the other ones due to the fewer available \acp{CSI} to compute the \whd on. 

It is necessary to state that analyzing the dispersion of the distances is insufficient to determine the number of people within the room: while the Empty Scenario is easily distinguishable from the others, telling Dynamic Scenarios apart based on the \whd distributions alone is a tough task, due to the distributions being significantly overlapped. 
A more in-depth and sophisticated analysis is required to determine a tool or metric to perform this task. 

\begin{figure}[h!]
	\centering
	\includegraphics[width=\textwidth]{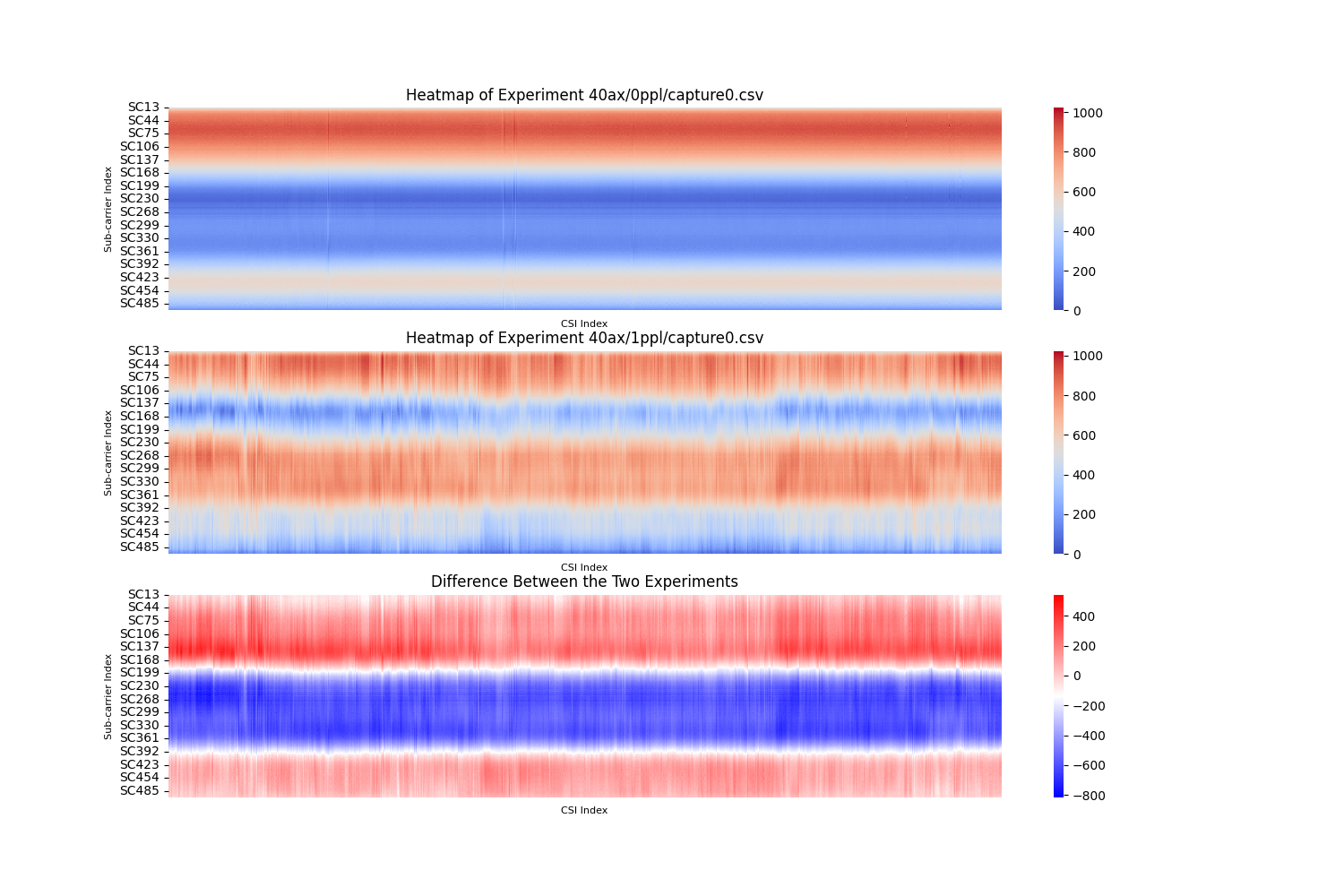}
	\caption{Difference in quantized \ac{CSI} amplitude values between an experiment performed in the Empty Scenario and one in the Static Scenario. The third heatmap depicts their difference. Data collected on a 40 MHz bandwidth channel using 802.11ax.}
	\label{fig:0vs1}
\end{figure}

\begin{figure}[h!]
	\centering
	\includegraphics[width=\textwidth]{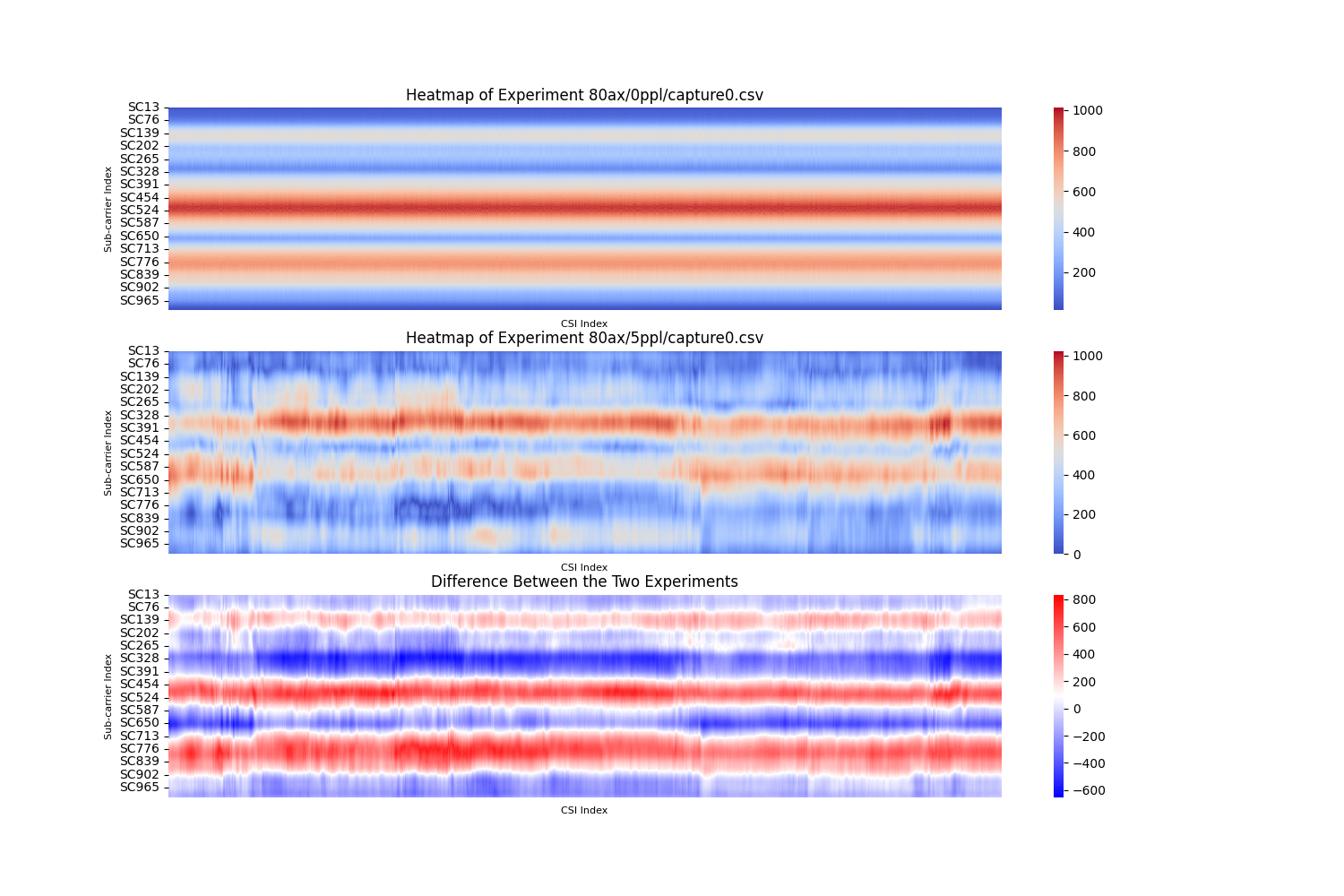}
	\caption{Difference in quantized \ac{CSI} amplitude values between an experiment performed in the Empty Scenario and one in the Fully Dynamic Scenario with 5 people in the room. The third heatmap depicts their difference. Data collected on an 80 MHz bandwidth channel using 802.11ax.}
	\label{fig:0vs5}
\end{figure}

\begin{figure}[h!]
	\centering
	\includegraphics[width=\textwidth]{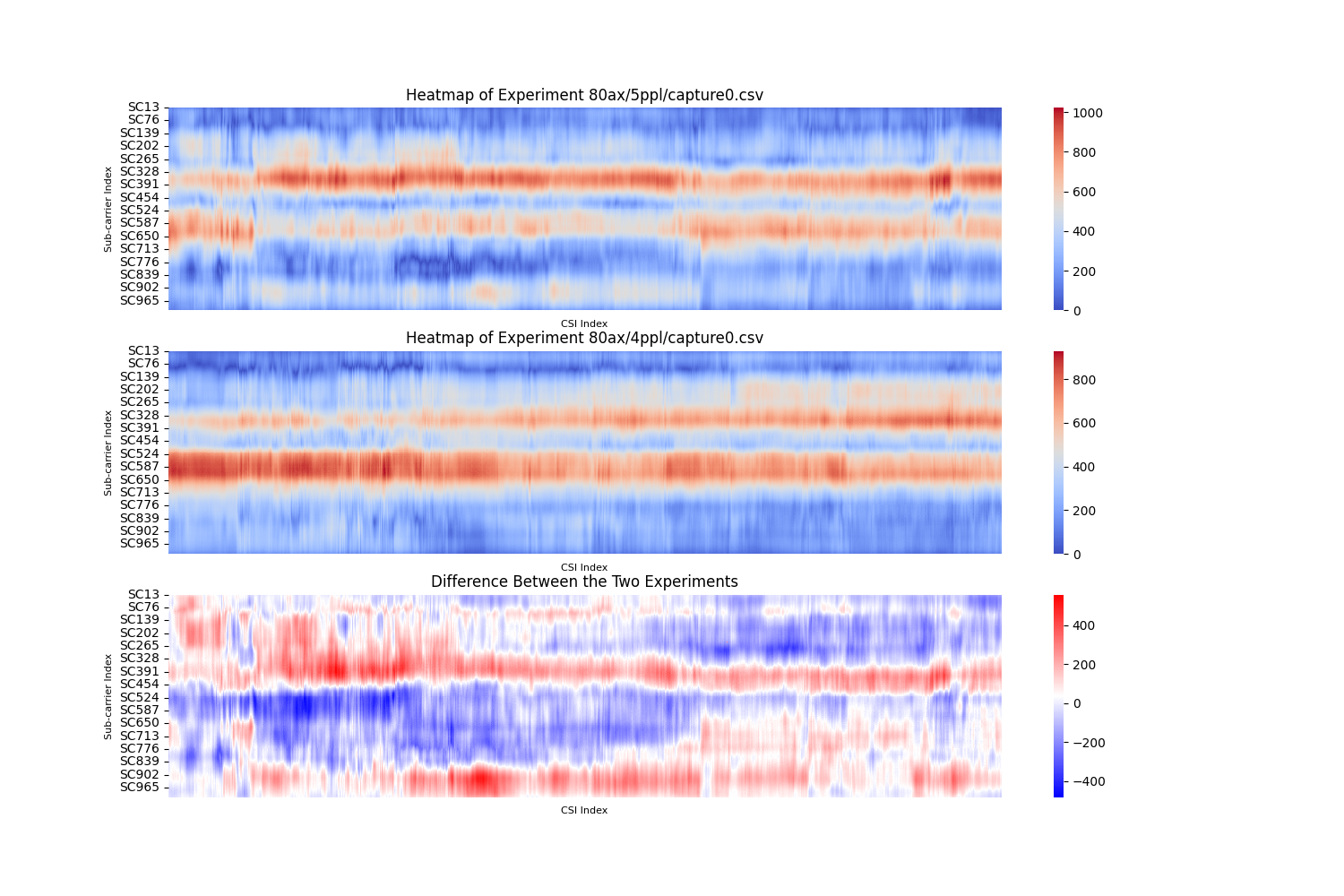}
	\caption{Difference in quantized \ac{CSI} amplitude values between two experiments performed in the fully Dynamic Scenario with 5 and 4 people in the room. The third heatmap depicts their difference. Data collected on an 80 MHz bandwidth channel using 802.11ax.}
	\label{fig:5vs4}
\end{figure}

\begin{figure}[h!]
	\centering
	\includegraphics[width=\textwidth]{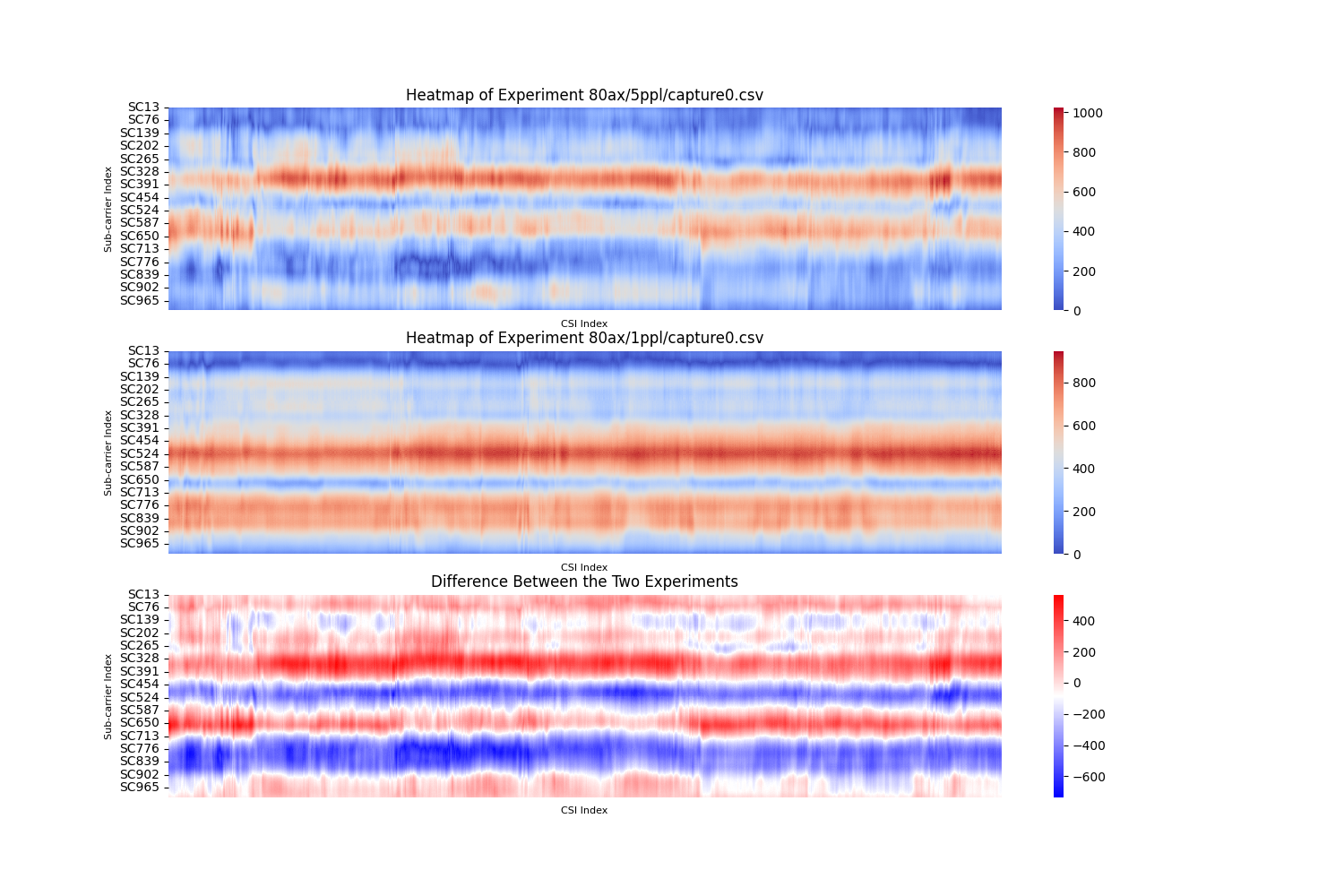}
	\caption{Difference in quantized \ac{CSI} amplitude values between an experiment performed in the Fully Dynamic Scenario with 5 people in the room and one in the Static Scenario. The third heatmap depicts their difference. Data collected on an 80 MHz bandwidth channel using 802.11ax.}
	\label{fig:5vs1}
\end{figure}

Nevertheless, an intuitive representation of the differences between captures collected with a varying number of people in the room is provided by \cref{fig:0vs1,,fig:0vs5,,fig:5vs4,,fig:5vs1}. 
Each figure represents the comparison between two experiments performed on either the 40 or the 80 MHz bandwidth. 
The first two plots of each figure represent the evolution in time of the quantized \ac{CSI} amplitude for the selected experiments: the $y$-axis corresponds to the sub-carriers (excluding those suppressed in transmission) and the $x$-axis represents the \ac{CSI} index within the experiments. 
The third plot depicts the difference between the first two (specifically, the first minus the second).
The color bar indicates that lower amplitude --- or difference --- values correspond to the color blue and that higher ones are red, with $0$ being white. 

These figures allow us to easily identify the overall structure of an experiment: for instance, in \cref{fig:0vs1} we can observe that the first heatmap, describing an experiment performed in the Empty Scenario, maintains a static behavior over time on each sub-carrier, with higher amplitude values in the first 150 sub-carriers and lower values in the central band. 
Similarly, the amplitude trend of the central heatmap --- relative to an experiment performed in the Static Scenario --- keeps a stationary behavior with lower absolute values. 
As can be seen in the third plot, the difference between the two scenarios is very distinct, as the three main `bands' that can be identified in the heatmap take on values that are far from $0$.
This simple consideration allows for the hypothesis that the two scenarios will be easily distinguishable from one another, as their difference is hardly ever close to zero. 

Similar considerations can be made on the \acp{CSI} represented in \cref{fig:0vs5}: the Empty Scenario is characterized by an extremely static behavior in time, whereas the Fully Dynamic Scenario with five people in the room has more variable amplitudes. 
In this second experimental setup, by looking at each sub-carrier by itself, we can observe that they all undergo modifications in time, necessarily due to the alterations of the environment due to the presence of multiple people. 
This implies that the higher variability of the Fully Dynamic Scenario makes it easily distinguishable from the Empty one.

\cref{fig:5vs4} indicates the opposite situation: the two analyzed experiments were both performed in the Fully Dynamic Scenario with 5 and 4 people in the room respectively. 
As can be noticed in the third heatmap, the difference between the two collections is more limited compared to the previous examples, especially in the second half of the experiment. 
It is also less definite in its structure, as the colored bands are much more variable in height and color intensity, indicating that distinguishing between the two scenarios may be less straightforward.

Contrarily, comparing the Fully Dynamic Scenario (5 people) with the Static one (see \cref{fig:5vs1}) results in more marked differences that set the two environments apart. 

Since these results and figures provide a more qualitative insight into the differences between two experiments, a metric to quantify such distance needs to be introduced. 
The goal would be to produce a distribution of \acp{CSI} around the average \ac{CSI} of each experiment; the main difficulty in achieving this characterization of the captures consists in finding a one-dimensional representation of a \ac{CSI}, accounting for the amplitude values on all sub-carriers simultaneously. 

\section{Results on the AntiSense dataset}
\label{sec:resantisense}
Considering that the goal of this work is to provide a mathematical background to \ac{ML}-based positioning and localization algorithms, testing the code written to carry out the current analysis against \ac{CSI} traces that have already been classified through \ac{ML} techniques could provide quantitative results that support such classifications.

Similarly to what has been done for the 20, 40, and 80 MHz bandwidth, we provide a summary of the normalized $\overline{\whd}$ computed on the data within the AntiSense dataset.

The results showcased in \cref{tab:rx1whd,,tab:rx5whd} highlight that each experiment is significantly self-similar, as the normalized $\overline{\whd}$ between each capture and its \racsi is at most in the order of $10^{-2}$. 
This initial consideration can be observed on the main diagonal of the two matrices. 

The green and blue sub-matrices represent the distance between the training and testing partitions of the dataset; their contents allow to highlight the correspondences between the values of the \whd and the performances of the neural network used in \cite{CoGri22_ComCom} to carry out the positioning task.
By observing the green\footnote{The same reasoning can be done on the blue sub-matrix, comparing it to the yellow one on its right.} matrix in \cref{tab:rx1whd}, we can see that some values on its diagonal significantly differ from those in the yellow matrix on its left, especially for experiments 3, 5, 6, 7, and 8. 
This result should correspond to degraded performances of the neural network: the larger the distance between testing and training, the more likely a neural network is to misinterpret the corresponding data, classifying a set of \acp{CSI} as belonging to the wrong experimental setup. 
If the positioning results obtained from evaluating the \whd are consistent with those produced by the neural network in \cite{CoGri22_ComCom}, the \whd would gain significance and reliability. 
\begin{table}[]
	\centering
	\resizebox{\textwidth}{!}{%
		\begin{tabular}{|
				>{\columncolor[HTML]{C1C1C1}}c |
				>{\columncolor[HTML]{C1C1C1}}c |cccccccc|cccccccc|}
			\hline
			\textbf{} &
			\textbf{} &
			\multicolumn{8}{c|}{\cellcolor[HTML]{C1C1C1}\textbf{TRAINING}} &
			\multicolumn{8}{c|}{\cellcolor[HTML]{C1C1C1}\textbf{TESTING}} \\ \hline
			\textbf{} &
			\textbf{POS} &
			\multicolumn{1}{c|}{\cellcolor[HTML]{C1C1C1}\textbf{1}} &
			\multicolumn{1}{c|}{\cellcolor[HTML]{C1C1C1}\textbf{2}} &
			\multicolumn{1}{c|}{\cellcolor[HTML]{C1C1C1}\textbf{3}} &
			\multicolumn{1}{c|}{\cellcolor[HTML]{C1C1C1}\textbf{4}} &
			\multicolumn{1}{c|}{\cellcolor[HTML]{C1C1C1}\textbf{5}} &
			\multicolumn{1}{c|}{\cellcolor[HTML]{C1C1C1}\textbf{6}} &
			\multicolumn{1}{c|}{\cellcolor[HTML]{C1C1C1}\textbf{7}} &
			\cellcolor[HTML]{C1C1C1}\textbf{8} &
			\multicolumn{1}{c|}{\cellcolor[HTML]{C1C1C1}\textbf{1}} &
			\multicolumn{1}{c|}{\cellcolor[HTML]{C1C1C1}\textbf{2}} &
			\multicolumn{1}{c|}{\cellcolor[HTML]{C1C1C1}\textbf{3}} &
			\multicolumn{1}{c|}{\cellcolor[HTML]{C1C1C1}\textbf{4}} &
			\multicolumn{1}{c|}{\cellcolor[HTML]{C1C1C1}\textbf{5}} &
			\multicolumn{1}{c|}{\cellcolor[HTML]{C1C1C1}\textbf{6}} &
			\multicolumn{1}{c|}{\cellcolor[HTML]{C1C1C1}\textbf{7}} &
			\cellcolor[HTML]{C1C1C1}\textbf{8} \\ \hline
			\cellcolor[HTML]{C1C1C1} &
			\textbf{1} &
			\multicolumn{1}{c|}{\cellcolor[HTML]{FFFFC7}\textbf{006}} &
			\multicolumn{1}{c|}{\cellcolor[HTML]{FFFFC7}232} &
			\multicolumn{1}{c|}{\cellcolor[HTML]{FFFFC7}019} &
			\multicolumn{1}{c|}{\cellcolor[HTML]{FFFFC7}033} &
			\multicolumn{1}{c|}{\cellcolor[HTML]{FFFFC7}179} &
			\multicolumn{1}{c|}{\cellcolor[HTML]{FFFFC7}019} &
			\multicolumn{1}{c|}{\cellcolor[HTML]{FFFFC7}054} &
			\cellcolor[HTML]{FFFFC7}047 &
			\multicolumn{1}{c|}{\cellcolor[HTML]{C4FF99}066} &
			\multicolumn{1}{c|}{\cellcolor[HTML]{C4FF99}256} &
			\multicolumn{1}{c|}{\cellcolor[HTML]{C4FF99}223} &
			\multicolumn{1}{c|}{\cellcolor[HTML]{C4FF99}036} &
			\multicolumn{1}{c|}{\cellcolor[HTML]{C4FF99}047} &
			\multicolumn{1}{c|}{\cellcolor[HTML]{C4FF99}263} &
			\multicolumn{1}{c|}{\cellcolor[HTML]{C4FF99}259} &
			\cellcolor[HTML]{C4FF99}232 \\ \cline{2-18} 
			\cellcolor[HTML]{C1C1C1} &
			\textbf{2} &
			\multicolumn{1}{c|}{\cellcolor[HTML]{FFFFC7}231} &
			\multicolumn{1}{c|}{\cellcolor[HTML]{FFFFC7}\textbf{019}} &
			\multicolumn{1}{c|}{\cellcolor[HTML]{FFFFC7}242} &
			\multicolumn{1}{c|}{\cellcolor[HTML]{FFFFC7}210} &
			\multicolumn{1}{c|}{\cellcolor[HTML]{FFFFC7}102} &
			\multicolumn{1}{c|}{\cellcolor[HTML]{FFFFC7}243} &
			\multicolumn{1}{c|}{\cellcolor[HTML]{FFFFC7}180} &
			\cellcolor[HTML]{FFFFC7}189 &
			\multicolumn{1}{c|}{\cellcolor[HTML]{C4FF99}167} &
			\multicolumn{1}{c|}{\cellcolor[HTML]{C4FF99}034} &
			\multicolumn{1}{c|}{\cellcolor[HTML]{C4FF99}037} &
			\multicolumn{1}{c|}{\cellcolor[HTML]{C4FF99}201} &
			\multicolumn{1}{c|}{\cellcolor[HTML]{C4FF99}235} &
			\multicolumn{1}{c|}{\cellcolor[HTML]{C4FF99}061} &
			\multicolumn{1}{c|}{\cellcolor[HTML]{C4FF99}052} &
			\cellcolor[HTML]{C4FF99}041 \\ \cline{2-18} 
			\cellcolor[HTML]{C1C1C1} &
			\textbf{3} &
			\multicolumn{1}{c|}{\cellcolor[HTML]{FFFFC7}016} &
			\multicolumn{1}{c|}{\cellcolor[HTML]{FFFFC7}242} &
			\multicolumn{1}{c|}{\cellcolor[HTML]{FFFFC7}\textbf{009}} &
			\multicolumn{1}{c|}{\cellcolor[HTML]{FFFFC7}036} &
			\multicolumn{1}{c|}{\cellcolor[HTML]{FFFFC7}189} &
			\multicolumn{1}{c|}{\cellcolor[HTML]{FFFFC7}020} &
			\multicolumn{1}{c|}{\cellcolor[HTML]{FFFFC7}065} &
			\cellcolor[HTML]{FFFFC7}055 &
			\multicolumn{1}{c|}{\cellcolor[HTML]{C4FF99}076} &
			\multicolumn{1}{c|}{\cellcolor[HTML]{C4FF99}266} &
			\multicolumn{1}{c|}{\cellcolor[HTML]{C4FF99}233} &
			\multicolumn{1}{c|}{\cellcolor[HTML]{C4FF99}045} &
			\multicolumn{1}{c|}{\cellcolor[HTML]{C4FF99}051} &
			\multicolumn{1}{c|}{\cellcolor[HTML]{C4FF99}274} &
			\multicolumn{1}{c|}{\cellcolor[HTML]{C4FF99}269} &
			\cellcolor[HTML]{C4FF99}241 \\ \cline{2-18} 
			\cellcolor[HTML]{C1C1C1} &
			\textbf{4} &
			\multicolumn{1}{c|}{\cellcolor[HTML]{FFFFC7}033} &
			\multicolumn{1}{c|}{\cellcolor[HTML]{FFFFC7}211} &
			\multicolumn{1}{c|}{\cellcolor[HTML]{FFFFC7}038} &
			\multicolumn{1}{c|}{\cellcolor[HTML]{FFFFC7}\textbf{006}} &
			\multicolumn{1}{c|}{\cellcolor[HTML]{FFFFC7}160} &
			\multicolumn{1}{c|}{\cellcolor[HTML]{FFFFC7}039} &
			\multicolumn{1}{c|}{\cellcolor[HTML]{FFFFC7}040} &
			\cellcolor[HTML]{FFFFC7}036 &
			\multicolumn{1}{c|}{\cellcolor[HTML]{C4FF99}050} &
			\multicolumn{1}{c|}{\cellcolor[HTML]{C4FF99}236} &
			\multicolumn{1}{c|}{\cellcolor[HTML]{C4FF99}201} &
			\multicolumn{1}{c|}{\cellcolor[HTML]{C4FF99}017} &
			\multicolumn{1}{c|}{\cellcolor[HTML]{C4FF99}047} &
			\multicolumn{1}{c|}{\cellcolor[HTML]{C4FF99}243} &
			\multicolumn{1}{c|}{\cellcolor[HTML]{C4FF99}238} &
			\cellcolor[HTML]{C4FF99}210 \\ \cline{2-18} 
			\cellcolor[HTML]{C1C1C1} &
			\textbf{5} &
			\multicolumn{1}{c|}{\cellcolor[HTML]{FFFFC7}178} &
			\multicolumn{1}{c|}{\cellcolor[HTML]{FFFFC7}103} &
			\multicolumn{1}{c|}{\cellcolor[HTML]{FFFFC7}189} &
			\multicolumn{1}{c|}{\cellcolor[HTML]{FFFFC7}159} &
			\multicolumn{1}{c|}{\cellcolor[HTML]{FFFFC7}\textbf{018}} &
			\multicolumn{1}{c|}{\cellcolor[HTML]{FFFFC7}183} &
			\multicolumn{1}{c|}{\cellcolor[HTML]{FFFFC7}158} &
			\cellcolor[HTML]{FFFFC7}160 &
			\multicolumn{1}{c|}{\cellcolor[HTML]{C4FF99}144} &
			\multicolumn{1}{c|}{\cellcolor[HTML]{C4FF99}112} &
			\multicolumn{1}{c|}{\cellcolor[HTML]{C4FF99}102} &
			\multicolumn{1}{c|}{\cellcolor[HTML]{C4FF99}153} &
			\multicolumn{1}{c|}{\cellcolor[HTML]{C4FF99}172} &
			\multicolumn{1}{c|}{\cellcolor[HTML]{C4FF99}113} &
			\multicolumn{1}{c|}{\cellcolor[HTML]{C4FF99}112} &
			\cellcolor[HTML]{C4FF99}109 \\ \cline{2-18} 
			\cellcolor[HTML]{C1C1C1} &
			\textbf{6} &
			\multicolumn{1}{c|}{\cellcolor[HTML]{FFFFC7}019} &
			\multicolumn{1}{c|}{\cellcolor[HTML]{FFFFC7}243} &
			\multicolumn{1}{c|}{\cellcolor[HTML]{FFFFC7}023} &
			\multicolumn{1}{c|}{\cellcolor[HTML]{FFFFC7}039} &
			\multicolumn{1}{c|}{\cellcolor[HTML]{FFFFC7}018} &
			\multicolumn{1}{c|}{\cellcolor[HTML]{FFFFC7}\textbf{007}} &
			\multicolumn{1}{c|}{\cellcolor[HTML]{FFFFC7}065} &
			\cellcolor[HTML]{FFFFC7}057 &
			\multicolumn{1}{c|}{\cellcolor[HTML]{C4FF99}078} &
			\multicolumn{1}{c|}{\cellcolor[HTML]{C4FF99}266} &
			\multicolumn{1}{c|}{\cellcolor[HTML]{C4FF99}235} &
			\multicolumn{1}{c|}{\cellcolor[HTML]{C4FF99}046} &
			\multicolumn{1}{c|}{\cellcolor[HTML]{C4FF99}045} &
			\multicolumn{1}{c|}{\cellcolor[HTML]{C4FF99}273} &
			\multicolumn{1}{c|}{\cellcolor[HTML]{C4FF99}269} &
			\cellcolor[HTML]{C4FF99}243 \\ \cline{2-18} 
			\cellcolor[HTML]{C1C1C1} &
			\textbf{7} &
			\multicolumn{1}{c|}{\cellcolor[HTML]{FFFFC7}054} &
			\multicolumn{1}{c|}{\cellcolor[HTML]{FFFFC7}180} &
			\multicolumn{1}{c|}{\cellcolor[HTML]{FFFFC7}066} &
			\multicolumn{1}{c|}{\cellcolor[HTML]{FFFFC7}039} &
			\multicolumn{1}{c|}{\cellcolor[HTML]{FFFFC7}158} &
			\multicolumn{1}{c|}{\cellcolor[HTML]{FFFFC7}065} &
			\multicolumn{1}{c|}{\cellcolor[HTML]{FFFFC7}\textbf{009}} &
			\cellcolor[HTML]{FFFFC7}024 &
			\multicolumn{1}{c|}{\cellcolor[HTML]{C4FF99}034} &
			\multicolumn{1}{c|}{\cellcolor[HTML]{C4FF99}205} &
			\multicolumn{1}{c|}{\cellcolor[HTML]{C4FF99}173} &
			\multicolumn{1}{c|}{\cellcolor[HTML]{C4FF99}033} &
			\multicolumn{1}{c|}{\cellcolor[HTML]{C4FF99}062} &
			\multicolumn{1}{c|}{\cellcolor[HTML]{C4FF99}213} &
			\multicolumn{1}{c|}{\cellcolor[HTML]{C4FF99}207} &
			\cellcolor[HTML]{C4FF99}182 \\ \cline{2-18} 
			\multirow{-8}{*}{\rotatebox[origin=c]{90}{\cellcolor[HTML]{C1C1C1}\textbf{\racsi TRAINING}}} &
			\textbf{8} &
			\multicolumn{1}{c|}{\cellcolor[HTML]{FFFFC7}047} &
			\multicolumn{1}{c|}{\cellcolor[HTML]{FFFFC7}190} &
			\multicolumn{1}{c|}{\cellcolor[HTML]{FFFFC7}057} &
			\multicolumn{1}{c|}{\cellcolor[HTML]{FFFFC7}035} &
			\multicolumn{1}{c|}{\cellcolor[HTML]{FFFFC7}161} &
			\multicolumn{1}{c|}{\cellcolor[HTML]{FFFFC7}057} &
			\multicolumn{1}{c|}{\cellcolor[HTML]{FFFFC7}023} &
			\cellcolor[HTML]{FFFFC7}\textbf{010} &
			\multicolumn{1}{c|}{\cellcolor[HTML]{C4FF99}031} &
			\multicolumn{1}{c|}{\cellcolor[HTML]{C4FF99}216} &
			\multicolumn{1}{c|}{\cellcolor[HTML]{C4FF99}180} &
			\multicolumn{1}{c|}{\cellcolor[HTML]{C4FF99}029} &
			\multicolumn{1}{c|}{\cellcolor[HTML]{C4FF99}061} &
			\multicolumn{1}{c|}{\cellcolor[HTML]{C4FF99}224} &
			\multicolumn{1}{c|}{\cellcolor[HTML]{C4FF99}218} &
			\cellcolor[HTML]{C4FF99}187 \\ \hline
			\cellcolor[HTML]{C1C1C1} &
			\textbf{1} &
			\multicolumn{1}{c|}{\cellcolor[HTML]{BCFFFB}066} &
			\multicolumn{1}{c|}{\cellcolor[HTML]{BCFFFB}168} &
			\multicolumn{1}{c|}{\cellcolor[HTML]{BCFFFB}077} &
			\multicolumn{1}{c|}{\cellcolor[HTML]{BCFFFB}050} &
			\multicolumn{1}{c|}{\cellcolor[HTML]{BCFFFB}145} &
			\multicolumn{1}{c|}{\cellcolor[HTML]{BCFFFB}077} &
			\multicolumn{1}{c|}{\cellcolor[HTML]{BCFFFB}034} &
			\cellcolor[HTML]{BCFFFB}031 &
			\multicolumn{1}{c|}{\cellcolor[HTML]{FFFFC7}\textbf{007}} &
			\multicolumn{1}{c|}{\cellcolor[HTML]{FFFFC7}193} &
			\multicolumn{1}{c|}{\cellcolor[HTML]{FFFFC7}158} &
			\multicolumn{1}{c|}{\cellcolor[HTML]{FFFFC7}042} &
			\multicolumn{1}{c|}{\cellcolor[HTML]{FFFFC7}074} &
			\multicolumn{1}{c|}{\cellcolor[HTML]{FFFFC7}202} &
			\multicolumn{1}{c|}{\cellcolor[HTML]{FFFFC7}196} &
			\cellcolor[HTML]{FFFFC7}167 \\ \cline{2-18} 
			\cellcolor[HTML]{C1C1C1} &
			\textbf{2} &
			\multicolumn{1}{c|}{\cellcolor[HTML]{BCFFFB}255} &
			\multicolumn{1}{c|}{\cellcolor[HTML]{BCFFFB}036} &
			\multicolumn{1}{c|}{\cellcolor[HTML]{BCFFFB}267} &
			\multicolumn{1}{c|}{\cellcolor[HTML]{BCFFFB}236} &
			\multicolumn{1}{c|}{\cellcolor[HTML]{BCFFFB}113} &
			\multicolumn{1}{c|}{\cellcolor[HTML]{BCFFFB}266} &
			\multicolumn{1}{c|}{\cellcolor[HTML]{BCFFFB}205} &
			\cellcolor[HTML]{BCFFFB}216 &
			\multicolumn{1}{c|}{\cellcolor[HTML]{FFFFC7}192} &
			\multicolumn{1}{c|}{\cellcolor[HTML]{FFFFC7}\textbf{014}} &
			\multicolumn{1}{c|}{\cellcolor[HTML]{FFFFC7}048} &
			\multicolumn{1}{c|}{\cellcolor[HTML]{FFFFC7}227} &
			\multicolumn{1}{c|}{\cellcolor[HTML]{FFFFC7}259} &
			\multicolumn{1}{c|}{\cellcolor[HTML]{FFFFC7}049} &
			\multicolumn{1}{c|}{\cellcolor[HTML]{FFFFC7}055} &
			\cellcolor[HTML]{FFFFC7}053 \\ \cline{2-18} 
			\cellcolor[HTML]{C1C1C1} &
			\textbf{3} &
			\multicolumn{1}{c|}{\cellcolor[HTML]{BCFFFB}223} &
			\multicolumn{1}{c|}{\cellcolor[HTML]{BCFFFB}040} &
			\multicolumn{1}{c|}{\cellcolor[HTML]{BCFFFB}234} &
			\multicolumn{1}{c|}{\cellcolor[HTML]{BCFFFB}201} &
			\multicolumn{1}{c|}{\cellcolor[HTML]{BCFFFB}102} &
			\multicolumn{1}{c|}{\cellcolor[HTML]{BCFFFB}235} &
			\multicolumn{1}{c|}{\cellcolor[HTML]{BCFFFB}173} &
			\cellcolor[HTML]{BCFFFB}180 &
			\multicolumn{1}{c|}{\cellcolor[HTML]{FFFFC7}158} &
			\multicolumn{1}{c|}{\cellcolor[HTML]{FFFFC7}048} &
			\multicolumn{1}{c|}{\cellcolor[HTML]{FFFFC7}\textbf{016}} &
			\multicolumn{1}{c|}{\cellcolor[HTML]{FFFFC7}192} &
			\multicolumn{1}{c|}{\cellcolor[HTML]{FFFFC7}227} &
			\multicolumn{1}{c|}{\cellcolor[HTML]{FFFFC7}062} &
			\multicolumn{1}{c|}{\cellcolor[HTML]{FFFFC7}063} &
			\cellcolor[HTML]{FFFFC7}041 \\ \cline{2-18} 
			\cellcolor[HTML]{C1C1C1} &
			\textbf{4} &
			\multicolumn{1}{c|}{\cellcolor[HTML]{BCFFFB}034} &
			\multicolumn{1}{c|}{\cellcolor[HTML]{BCFFFB}202} &
			\multicolumn{1}{c|}{\cellcolor[HTML]{BCFFFB}045} &
			\multicolumn{1}{c|}{\cellcolor[HTML]{BCFFFB}016} &
			\multicolumn{1}{c|}{\cellcolor[HTML]{BCFFFB}154} &
			\multicolumn{1}{c|}{\cellcolor[HTML]{BCFFFB}046} &
			\multicolumn{1}{c|}{\cellcolor[HTML]{BCFFFB}032} &
			\cellcolor[HTML]{BCFFFB}029 &
			\multicolumn{1}{c|}{\cellcolor[HTML]{FFFFC7}042} &
			\multicolumn{1}{c|}{\cellcolor[HTML]{FFFFC7}227} &
			\multicolumn{1}{c|}{\cellcolor[HTML]{FFFFC7}192} &
			\multicolumn{1}{c|}{\cellcolor[HTML]{FFFFC7}\textbf{009}} &
			\multicolumn{1}{c|}{\cellcolor[HTML]{FFFFC7}048} &
			\multicolumn{1}{c|}{\cellcolor[HTML]{FFFFC7}234} &
			\multicolumn{1}{c|}{\cellcolor[HTML]{FFFFC7}229} &
			\cellcolor[HTML]{FFFFC7}201 \\ \cline{2-18} 
			\cellcolor[HTML]{C1C1C1} &
			\textbf{5} &
			\multicolumn{1}{c|}{\cellcolor[HTML]{BCFFFB}040} &
			\multicolumn{1}{c|}{\cellcolor[HTML]{BCFFFB}233} &
			\multicolumn{1}{c|}{\cellcolor[HTML]{BCFFFB}045} &
			\multicolumn{1}{c|}{\cellcolor[HTML]{BCFFFB}039} &
			\multicolumn{1}{c|}{\cellcolor[HTML]{BCFFFB}170} &
			\multicolumn{1}{c|}{\cellcolor[HTML]{BCFFFB}037} &
			\multicolumn{1}{c|}{\cellcolor[HTML]{BCFFFB}056} &
			\cellcolor[HTML]{BCFFFB}057 &
			\multicolumn{1}{c|}{\cellcolor[HTML]{FFFFC7}069} &
			\multicolumn{1}{c|}{\cellcolor[HTML]{FFFFC7}257} &
			\multicolumn{1}{c|}{\cellcolor[HTML]{FFFFC7}224} &
			\multicolumn{1}{c|}{\cellcolor[HTML]{FFFFC7}043} &
			\multicolumn{1}{c|}{\cellcolor[HTML]{FFFFC7}\textbf{017}} &
			\multicolumn{1}{c|}{\cellcolor[HTML]{FFFFC7}264} &
			\multicolumn{1}{c|}{\cellcolor[HTML]{FFFFC7}259} &
			\cellcolor[HTML]{FFFFC7}233 \\ \cline{2-18} 
			\cellcolor[HTML]{C1C1C1} &
			\textbf{6} &
			\multicolumn{1}{c|}{\cellcolor[HTML]{BCFFFB}263} &
			\multicolumn{1}{c|}{\cellcolor[HTML]{BCFFFB}062} &
			\multicolumn{1}{c|}{\cellcolor[HTML]{BCFFFB}275} &
			\multicolumn{1}{c|}{\cellcolor[HTML]{BCFFFB}243} &
			\multicolumn{1}{c|}{\cellcolor[HTML]{BCFFFB}113} &
			\multicolumn{1}{l|}{\cellcolor[HTML]{BCFFFB}273} &
			\multicolumn{1}{c|}{\cellcolor[HTML]{BCFFFB}213} &
			\cellcolor[HTML]{BCFFFB}223 &
			\multicolumn{1}{c|}{\cellcolor[HTML]{FFFFC7}202} &
			\multicolumn{1}{c|}{\cellcolor[HTML]{FFFFC7}049} &
			\multicolumn{1}{c|}{\cellcolor[HTML]{FFFFC7}061} &
			\multicolumn{1}{c|}{\cellcolor[HTML]{FFFFC7}234} &
			\multicolumn{1}{c|}{\cellcolor[HTML]{FFFFC7}265} &
			\multicolumn{1}{l|}{\cellcolor[HTML]{FFFFC7}\textbf{017}} &
			\multicolumn{1}{c|}{\cellcolor[HTML]{FFFFC7}059} &
			\cellcolor[HTML]{FFFFC7}075 \\ \cline{2-18} 
			\cellcolor[HTML]{C1C1C1} &
			\textbf{7} &
			\multicolumn{1}{c|}{\cellcolor[HTML]{BCFFFB}258} &
			\multicolumn{1}{c|}{\cellcolor[HTML]{BCFFFB}051} &
			\multicolumn{1}{c|}{\cellcolor[HTML]{BCFFFB}270} &
			\multicolumn{1}{c|}{\cellcolor[HTML]{BCFFFB}238} &
			\multicolumn{1}{c|}{\cellcolor[HTML]{BCFFFB}111} &
			\multicolumn{1}{l|}{\cellcolor[HTML]{BCFFFB}269} &
			\multicolumn{1}{c|}{\cellcolor[HTML]{BCFFFB}206} &
			\cellcolor[HTML]{BCFFFB}218 &
			\multicolumn{1}{c|}{\cellcolor[HTML]{FFFFC7}196} &
			\multicolumn{1}{c|}{\cellcolor[HTML]{FFFFC7}052} &
			\multicolumn{1}{c|}{\cellcolor[HTML]{FFFFC7}061} &
			\multicolumn{1}{c|}{\cellcolor[HTML]{FFFFC7}229} &
			\multicolumn{1}{c|}{\cellcolor[HTML]{FFFFC7}260} &
			\multicolumn{1}{c|}{\cellcolor[HTML]{FFFFC7}059} &
			\multicolumn{1}{c|}{\cellcolor[HTML]{FFFFC7}\textbf{021}} &
			\cellcolor[HTML]{FFFFC7}054 \\ \cline{2-18} 
			\multirow{-8}{*}{\rotatebox[origin=c]{90}{\cellcolor[HTML]{C1C1C1}\textbf{\racsi TESTING}}} &
			\textbf{8} &
			\multicolumn{1}{c|}{\cellcolor[HTML]{BCFFFB}232} &
			\multicolumn{1}{c|}{\cellcolor[HTML]{BCFFFB}042} &
			\multicolumn{1}{c|}{\cellcolor[HTML]{BCFFFB}242} &
			\multicolumn{1}{c|}{\cellcolor[HTML]{BCFFFB}209} &
			\multicolumn{1}{c|}{\cellcolor[HTML]{BCFFFB}109} &
			\multicolumn{1}{l|}{\cellcolor[HTML]{BCFFFB}243} &
			\multicolumn{1}{c|}{\cellcolor[HTML]{BCFFFB}182} &
			\cellcolor[HTML]{BCFFFB}187 &
			\multicolumn{1}{c|}{\cellcolor[HTML]{FFFFC7}167} &
			\multicolumn{1}{c|}{\cellcolor[HTML]{FFFFC7}052} &
			\multicolumn{1}{c|}{\cellcolor[HTML]{FFFFC7}039} &
			\multicolumn{1}{c|}{\cellcolor[HTML]{FFFFC7}201} &
			\multicolumn{1}{c|}{\cellcolor[HTML]{FFFFC7}235} &
			\multicolumn{1}{c|}{\cellcolor[HTML]{FFFFC7}075} &
			\multicolumn{1}{c|}{\cellcolor[HTML]{FFFFC7}055} &
			\cellcolor[HTML]{FFFFC7}\textbf{017} \\ \hline
		\end{tabular}%
	}
	\caption{Normalized average \whd computed on the partitions of the AntiSense dataset dedicated to training and testing with the receiver located in position 1 (rx1). The integer values are the first three digits after the comma, rounded to the nearest value. The POS parameter indicates the position of the person standing still within the experimental environment.}
	\label{tab:rx1whd}
\end{table}

By looking at the results of \cite{CoGri22_ComCom}, many of these larger \whd values are consistent with misclassifications by the neural network: over 1000 samples, the neural network has a success rate of over 90\% (except for `rx5', which will be discussed later) but the classifications relative to `rx1' show higher failure rates for experiments 3, 6, 7, and 8, which is coherent with the content of \cref{tab:rx1whd}.
Unfortunately, the \textit{average} \whd is too `summarizing' to be used for classification while expecting the same level of precision and detail as a neural network. 

The results relative to the set of experiments tagged `rx5' under `active attack' were not analyzed in depth in \cite{CoGri22_ComCom}, as the classification performed by the neural network was close to a random guess. 
\cref{tab:rx5whd} contains the $\overline{\whd}$ values relative to the experiments performed in such setup of the laboratory, where the receiver was placed extremely close to the transmitter; this caused the \acp{CSI} to be dominated by the transmitted signal itself rather than reflecting the modifications it undergoes after propagating through the environment. 
Whether we observe the training or testing partition of the dataset, the distances contained in \cref{tab:rx5whd} are all similar to each other, regardless of the compared experiments. 
This behaviour directly impacted the performance of the neural network used in the cited work, as values that are similar across all experiments make it harder to correctly classify a collection of \acp{CSI} as belonging to a specific experimental setup.
\begin{table}[]
	\centering
	\resizebox{\textwidth}{!}{%
		\begin{tabular}{|
				>{\columncolor[HTML]{C1C1C1}}c |
				>{\columncolor[HTML]{C1C1C1}}c |cccccccc|cccccccc|}
			\hline
			\textbf{} &
			\textbf{} &
			\multicolumn{8}{c|}{\cellcolor[HTML]{C1C1C1}\textbf{TRAINING}} &
			\multicolumn{8}{c|}{\cellcolor[HTML]{C1C1C1}\textbf{TESTING}} \\ \hline
			\textbf{} &
			\textbf{POS} &
			\multicolumn{1}{c|}{\cellcolor[HTML]{C1C1C1}\textbf{1}} &
			\multicolumn{1}{c|}{\cellcolor[HTML]{C1C1C1}\textbf{2}} &
			\multicolumn{1}{c|}{\cellcolor[HTML]{C1C1C1}\textbf{3}} &
			\multicolumn{1}{c|}{\cellcolor[HTML]{C1C1C1}\textbf{4}} &
			\multicolumn{1}{c|}{\cellcolor[HTML]{C1C1C1}\textbf{5}} &
			\multicolumn{1}{c|}{\cellcolor[HTML]{C1C1C1}\textbf{6}} &
			\multicolumn{1}{c|}{\cellcolor[HTML]{C1C1C1}\textbf{7}} &
			\cellcolor[HTML]{C1C1C1}\textbf{8} &
			\multicolumn{1}{c|}{\cellcolor[HTML]{C1C1C1}\textbf{1}} &
			\multicolumn{1}{c|}{\cellcolor[HTML]{C1C1C1}\textbf{2}} &
			\multicolumn{1}{c|}{\cellcolor[HTML]{C1C1C1}\textbf{3}} &
			\multicolumn{1}{c|}{\cellcolor[HTML]{C1C1C1}\textbf{4}} &
			\multicolumn{1}{c|}{\cellcolor[HTML]{C1C1C1}\textbf{5}} &
			\multicolumn{1}{c|}{\cellcolor[HTML]{C1C1C1}\textbf{6}} &
			\multicolumn{1}{c|}{\cellcolor[HTML]{C1C1C1}\textbf{7}} &
			\cellcolor[HTML]{C1C1C1}\textbf{8} \\ \hline
			\cellcolor[HTML]{C1C1C1} &
			\textbf{1} &
			\multicolumn{1}{c|}{\cellcolor[HTML]{FFFFC7}\textbf{006}} &
			\multicolumn{1}{c|}{\cellcolor[HTML]{FFFFC7}012} &
			\multicolumn{1}{c|}{\cellcolor[HTML]{FFFFC7}010} &
			\multicolumn{1}{c|}{\cellcolor[HTML]{FFFFC7}007} &
			\multicolumn{1}{c|}{\cellcolor[HTML]{FFFFC7}013} &
			\multicolumn{1}{c|}{\cellcolor[HTML]{FFFFC7}021} &
			\multicolumn{1}{c|}{\cellcolor[HTML]{FFFFC7}011} &
			\cellcolor[HTML]{FFFFC7}018 &
			\multicolumn{1}{c|}{\cellcolor[HTML]{C4FF99}017} &
			\multicolumn{1}{c|}{\cellcolor[HTML]{C4FF99}017} &
			\multicolumn{1}{c|}{\cellcolor[HTML]{C4FF99}010} &
			\multicolumn{1}{c|}{\cellcolor[HTML]{C4FF99}008} &
			\multicolumn{1}{c|}{\cellcolor[HTML]{C4FF99}011} &
			\multicolumn{1}{c|}{\cellcolor[HTML]{C4FF99}010} &
			\multicolumn{1}{c|}{\cellcolor[HTML]{C4FF99}022} &
			\cellcolor[HTML]{C4FF99}035 \\ \cline{2-18} 
			\cellcolor[HTML]{C1C1C1} &
			\textbf{2} &
			\multicolumn{1}{c|}{\cellcolor[HTML]{FFFFC7}012} &
			\multicolumn{1}{c|}{\cellcolor[HTML]{FFFFC7}\textbf{006}} &
			\multicolumn{1}{c|}{\cellcolor[HTML]{FFFFC7}007} &
			\multicolumn{1}{c|}{\cellcolor[HTML]{FFFFC7}011} &
			\multicolumn{1}{c|}{\cellcolor[HTML]{FFFFC7}008} &
			\multicolumn{1}{c|}{\cellcolor[HTML]{FFFFC7}012} &
			\multicolumn{1}{c|}{\cellcolor[HTML]{FFFFC7}010} &
			\cellcolor[HTML]{FFFFC7}011 &
			\multicolumn{1}{c|}{\cellcolor[HTML]{C4FF99}009} &
			\multicolumn{1}{c|}{\cellcolor[HTML]{C4FF99}008} &
			\multicolumn{1}{c|}{\cellcolor[HTML]{C4FF99}008} &
			\multicolumn{1}{c|}{\cellcolor[HTML]{C4FF99}012} &
			\multicolumn{1}{c|}{\cellcolor[HTML]{C4FF99}008} &
			\multicolumn{1}{c|}{\cellcolor[HTML]{C4FF99}007} &
			\multicolumn{1}{c|}{\cellcolor[HTML]{C4FF99}014} &
			\cellcolor[HTML]{C4FF99}025 \\ \cline{2-18} 
			\cellcolor[HTML]{C1C1C1} &
			\textbf{3} &
			\multicolumn{1}{c|}{\cellcolor[HTML]{FFFFC7}010} &
			\multicolumn{1}{c|}{\cellcolor[HTML]{FFFFC7}007} &
			\multicolumn{1}{c|}{\cellcolor[HTML]{FFFFC7}\textbf{005}} &
			\multicolumn{1}{c|}{\cellcolor[HTML]{FFFFC7}010} &
			\multicolumn{1}{c|}{\cellcolor[HTML]{FFFFC7}009} &
			\multicolumn{1}{c|}{\cellcolor[HTML]{FFFFC7}013} &
			\multicolumn{1}{c|}{\cellcolor[HTML]{FFFFC7}010} &
			\cellcolor[HTML]{FFFFC7}012 &
			\multicolumn{1}{c|}{\cellcolor[HTML]{C4FF99}010} &
			\multicolumn{1}{c|}{\cellcolor[HTML]{C4FF99}010} &
			\multicolumn{1}{c|}{\cellcolor[HTML]{C4FF99}006} &
			\multicolumn{1}{c|}{\cellcolor[HTML]{C4FF99}010} &
			\multicolumn{1}{c|}{\cellcolor[HTML]{C4FF99}008} &
			\multicolumn{1}{c|}{\cellcolor[HTML]{C4FF99}008} &
			\multicolumn{1}{c|}{\cellcolor[HTML]{C4FF99}015} &
			\cellcolor[HTML]{C4FF99}027 \\ \cline{2-18} 
			\cellcolor[HTML]{C1C1C1} &
			\textbf{4} &
			\multicolumn{1}{c|}{\cellcolor[HTML]{FFFFC7}007} &
			\multicolumn{1}{c|}{\cellcolor[HTML]{FFFFC7}011} &
			\multicolumn{1}{c|}{\cellcolor[HTML]{FFFFC7}010} &
			\multicolumn{1}{c|}{\cellcolor[HTML]{FFFFC7}\textbf{006}} &
			\multicolumn{1}{c|}{\cellcolor[HTML]{FFFFC7}012} &
			\multicolumn{1}{c|}{\cellcolor[HTML]{FFFFC7}019} &
			\multicolumn{1}{c|}{\cellcolor[HTML]{FFFFC7}011} &
			\cellcolor[HTML]{FFFFC7}016 &
			\multicolumn{1}{c|}{\cellcolor[HTML]{C4FF99}016} &
			\multicolumn{1}{c|}{\cellcolor[HTML]{C4FF99}016} &
			\multicolumn{1}{c|}{\cellcolor[HTML]{C4FF99}010} &
			\multicolumn{1}{c|}{\cellcolor[HTML]{C4FF99}007} &
			\multicolumn{1}{c|}{\cellcolor[HTML]{C4FF99}010} &
			\multicolumn{1}{c|}{\cellcolor[HTML]{C4FF99}010} &
			\multicolumn{1}{c|}{\cellcolor[HTML]{C4FF99}021} &
			\cellcolor[HTML]{C4FF99}033 \\ \cline{2-18} 
			\cellcolor[HTML]{C1C1C1} &
			\textbf{5} &
			\multicolumn{1}{c|}{\cellcolor[HTML]{FFFFC7}013} &
			\multicolumn{1}{c|}{\cellcolor[HTML]{FFFFC7}008} &
			\multicolumn{1}{c|}{\cellcolor[HTML]{FFFFC7}009} &
			\multicolumn{1}{c|}{\cellcolor[HTML]{FFFFC7}012} &
			\multicolumn{1}{c|}{\cellcolor[HTML]{FFFFC7}\textbf{006}} &
			\multicolumn{1}{c|}{\cellcolor[HTML]{FFFFC7}011} &
			\multicolumn{1}{c|}{\cellcolor[HTML]{FFFFC7}011} &
			\cellcolor[HTML]{FFFFC7}011 &
			\multicolumn{1}{c|}{\cellcolor[HTML]{C4FF99}010} &
			\multicolumn{1}{c|}{\cellcolor[HTML]{C4FF99}009} &
			\multicolumn{1}{c|}{\cellcolor[HTML]{C4FF99}010} &
			\multicolumn{1}{c|}{\cellcolor[HTML]{C4FF99}012} &
			\multicolumn{1}{c|}{\cellcolor[HTML]{C4FF99}007} &
			\multicolumn{1}{c|}{\cellcolor[HTML]{C4FF99}008} &
			\multicolumn{1}{c|}{\cellcolor[HTML]{C4FF99}014} &
			\cellcolor[HTML]{C4FF99}023 \\ \cline{2-18} 
			\cellcolor[HTML]{C1C1C1} &
			\textbf{6} &
			\multicolumn{1}{c|}{\cellcolor[HTML]{FFFFC7}021} &
			\multicolumn{1}{c|}{\cellcolor[HTML]{FFFFC7}012} &
			\multicolumn{1}{c|}{\cellcolor[HTML]{FFFFC7}013} &
			\multicolumn{1}{c|}{\cellcolor[HTML]{FFFFC7}019} &
			\multicolumn{1}{c|}{\cellcolor[HTML]{FFFFC7}011} &
			\multicolumn{1}{c|}{\cellcolor[HTML]{FFFFC7}\textbf{006}} &
			\multicolumn{1}{c|}{\cellcolor[HTML]{FFFFC7}015} &
			\cellcolor[HTML]{FFFFC7}010 &
			\multicolumn{1}{c|}{\cellcolor[HTML]{C4FF99}008} &
			\multicolumn{1}{c|}{\cellcolor[HTML]{C4FF99}009} &
			\multicolumn{1}{c|}{\cellcolor[HTML]{C4FF99}015} &
			\multicolumn{1}{c|}{\cellcolor[HTML]{C4FF99}019} &
			\multicolumn{1}{c|}{\cellcolor[HTML]{C4FF99}013} &
			\multicolumn{1}{c|}{\cellcolor[HTML]{C4FF99}014} &
			\multicolumn{1}{c|}{\cellcolor[HTML]{C4FF99}011} &
			\cellcolor[HTML]{C4FF99}015 \\ \cline{2-18} 
			\cellcolor[HTML]{C1C1C1} &
			\textbf{7} &
			\multicolumn{1}{c|}{\cellcolor[HTML]{FFFFC7}011} &
			\multicolumn{1}{c|}{\cellcolor[HTML]{FFFFC7}009} &
			\multicolumn{1}{c|}{\cellcolor[HTML]{FFFFC7}009} &
			\multicolumn{1}{c|}{\cellcolor[HTML]{FFFFC7}010} &
			\multicolumn{1}{c|}{\cellcolor[HTML]{FFFFC7}011} &
			\multicolumn{1}{c|}{\cellcolor[HTML]{FFFFC7}014} &
			\multicolumn{1}{c|}{\cellcolor[HTML]{FFFFC7}\textbf{007}} &
			\cellcolor[HTML]{FFFFC7}014 &
			\multicolumn{1}{c|}{\cellcolor[HTML]{C4FF99}012} &
			\multicolumn{1}{c|}{\cellcolor[HTML]{C4FF99}013} &
			\multicolumn{1}{c|}{\cellcolor[HTML]{C4FF99}010} &
			\multicolumn{1}{c|}{\cellcolor[HTML]{C4FF99}011} &
			\multicolumn{1}{c|}{\cellcolor[HTML]{C4FF99}010} &
			\multicolumn{1}{c|}{\cellcolor[HTML]{C4FF99}010} &
			\multicolumn{1}{c|}{\cellcolor[HTML]{C4FF99}017} &
			\cellcolor[HTML]{C4FF99}028 \\ \cline{2-18} 
			\multirow{-8}{*}{\rotatebox[origin=c]{90}{\cellcolor[HTML]{C1C1C1}\textbf{\racsi TRAINING}}} &
			\textbf{8} &
			\multicolumn{1}{c|}{\cellcolor[HTML]{FFFFC7}018} &
			\multicolumn{1}{c|}{\cellcolor[HTML]{FFFFC7}011} &
			\multicolumn{1}{c|}{\cellcolor[HTML]{FFFFC7}011} &
			\multicolumn{1}{c|}{\cellcolor[HTML]{FFFFC7}016} &
			\multicolumn{1}{c|}{\cellcolor[HTML]{FFFFC7}011} &
			\multicolumn{1}{c|}{\cellcolor[HTML]{FFFFC7}010} &
			\multicolumn{1}{c|}{\cellcolor[HTML]{FFFFC7}014} &
			\cellcolor[HTML]{FFFFC7}\textbf{007} &
			\multicolumn{1}{c|}{\cellcolor[HTML]{C4FF99}010} &
			\multicolumn{1}{c|}{\cellcolor[HTML]{C4FF99}010} &
			\multicolumn{1}{c|}{\cellcolor[HTML]{C4FF99}013} &
			\multicolumn{1}{c|}{\cellcolor[HTML]{C4FF99}017} &
			\multicolumn{1}{c|}{\cellcolor[HTML]{C4FF99}012} &
			\multicolumn{1}{c|}{\cellcolor[HTML]{C4FF99}013} &
			\multicolumn{1}{c|}{\cellcolor[HTML]{C4FF99}012} &
			\cellcolor[HTML]{C4FF99}018 \\ \hline
			\cellcolor[HTML]{C1C1C1} &
			\textbf{1} &
			\multicolumn{1}{c|}{\cellcolor[HTML]{BCFFFB}017} &
			\multicolumn{1}{c|}{\cellcolor[HTML]{BCFFFB}010} &
			\multicolumn{1}{c|}{\cellcolor[HTML]{BCFFFB}010} &
			\multicolumn{1}{c|}{\cellcolor[HTML]{BCFFFB}016} &
			\multicolumn{1}{c|}{\cellcolor[HTML]{BCFFFB}010} &
			\multicolumn{1}{c|}{\cellcolor[HTML]{BCFFFB}009} &
			\multicolumn{1}{c|}{\cellcolor[HTML]{BCFFFB}013} &
			\cellcolor[HTML]{BCFFFB}011 &
			\multicolumn{1}{c|}{\cellcolor[HTML]{FFFFC7}\textbf{005}} &
			\multicolumn{1}{c|}{\cellcolor[HTML]{FFFFC7}008} &
			\multicolumn{1}{c|}{\cellcolor[HTML]{FFFFC7}012} &
			\multicolumn{1}{c|}{\cellcolor[HTML]{FFFFC7}016} &
			\multicolumn{1}{c|}{\cellcolor[HTML]{FFFFC7}011} &
			\multicolumn{1}{c|}{\cellcolor[HTML]{FFFFC7}011} &
			\multicolumn{1}{c|}{\cellcolor[HTML]{FFFFC7}012} &
			\cellcolor[HTML]{FFFFC7}019 \\ \cline{2-18} 
			\cellcolor[HTML]{C1C1C1} &
			\textbf{2} &
			\multicolumn{1}{c|}{\cellcolor[HTML]{BCFFFB}017} &
			\multicolumn{1}{c|}{\cellcolor[HTML]{BCFFFB}009} &
			\multicolumn{1}{c|}{\cellcolor[HTML]{BCFFFB}010} &
			\multicolumn{1}{c|}{\cellcolor[HTML]{BCFFFB}016} &
			\multicolumn{1}{c|}{\cellcolor[HTML]{BCFFFB}009} &
			\multicolumn{1}{c|}{\cellcolor[HTML]{BCFFFB}009} &
			\multicolumn{1}{c|}{\cellcolor[HTML]{BCFFFB}013} &
			\cellcolor[HTML]{BCFFFB}011 &
			\multicolumn{1}{c|}{\cellcolor[HTML]{FFFFC7}008} &
			\multicolumn{1}{c|}{\cellcolor[HTML]{FFFFC7}\textbf{005}} &
			\multicolumn{1}{c|}{\cellcolor[HTML]{FFFFC7}011} &
			\multicolumn{1}{c|}{\cellcolor[HTML]{FFFFC7}016} &
			\multicolumn{1}{c|}{\cellcolor[HTML]{FFFFC7}011} &
			\multicolumn{1}{c|}{\cellcolor[HTML]{FFFFC7}011} &
			\multicolumn{1}{c|}{\cellcolor[HTML]{FFFFC7}011} &
			\cellcolor[HTML]{FFFFC7}019 \\ \cline{2-18} 
			\cellcolor[HTML]{C1C1C1} &
			\textbf{3} &
			\multicolumn{1}{c|}{\cellcolor[HTML]{BCFFFB}010} &
			\multicolumn{1}{c|}{\cellcolor[HTML]{BCFFFB}008} &
			\multicolumn{1}{c|}{\cellcolor[HTML]{BCFFFB}006} &
			\multicolumn{1}{c|}{\cellcolor[HTML]{BCFFFB}010} &
			\multicolumn{1}{c|}{\cellcolor[HTML]{BCFFFB}011} &
			\multicolumn{1}{c|}{\cellcolor[HTML]{BCFFFB}015} &
			\multicolumn{1}{c|}{\cellcolor[HTML]{BCFFFB}011} &
			\cellcolor[HTML]{BCFFFB}013 &
			\multicolumn{1}{c|}{\cellcolor[HTML]{FFFFC7}012} &
			\multicolumn{1}{c|}{\cellcolor[HTML]{FFFFC7}011} &
			\multicolumn{1}{c|}{\cellcolor[HTML]{FFFFC7}\textbf{006}} &
			\multicolumn{1}{c|}{\cellcolor[HTML]{FFFFC7}010} &
			\multicolumn{1}{c|}{\cellcolor[HTML]{FFFFC7}009} &
			\multicolumn{1}{c|}{\cellcolor[HTML]{FFFFC7}008} &
			\multicolumn{1}{c|}{\cellcolor[HTML]{FFFFC7}017} &
			\cellcolor[HTML]{FFFFC7}028 \\ \cline{2-18} 
			\cellcolor[HTML]{C1C1C1} &
			\textbf{4} &
			\multicolumn{1}{c|}{\cellcolor[HTML]{BCFFFB}008} &
			\multicolumn{1}{c|}{\cellcolor[HTML]{BCFFFB}011} &
			\multicolumn{1}{c|}{\cellcolor[HTML]{BCFFFB}010} &
			\multicolumn{1}{c|}{\cellcolor[HTML]{BCFFFB}007} &
			\multicolumn{1}{c|}{\cellcolor[HTML]{BCFFFB}012} &
			\multicolumn{1}{c|}{\cellcolor[HTML]{BCFFFB}019} &
			\multicolumn{1}{c|}{\cellcolor[HTML]{BCFFFB}011} &
			\cellcolor[HTML]{BCFFFB}017 &
			\multicolumn{1}{c|}{\cellcolor[HTML]{FFFFC7}015} &
			\multicolumn{1}{c|}{\cellcolor[HTML]{FFFFC7}016} &
			\multicolumn{1}{c|}{\cellcolor[HTML]{FFFFC7}010} &
			\multicolumn{1}{c|}{\cellcolor[HTML]{FFFFC7}\textbf{006}} &
			\multicolumn{1}{c|}{\cellcolor[HTML]{FFFFC7}011} &
			\multicolumn{1}{c|}{\cellcolor[HTML]{FFFFC7}010} &
			\multicolumn{1}{c|}{\cellcolor[HTML]{FFFFC7}021} &
			\cellcolor[HTML]{FFFFC7}033 \\ \cline{2-18} 
			\cellcolor[HTML]{C1C1C1} &
			\textbf{5} &
			\multicolumn{1}{c|}{\cellcolor[HTML]{BCFFFB}011} &
			\multicolumn{1}{c|}{\cellcolor[HTML]{BCFFFB}007} &
			\multicolumn{1}{c|}{\cellcolor[HTML]{BCFFFB}008} &
			\multicolumn{1}{c|}{\cellcolor[HTML]{BCFFFB}011} &
			\multicolumn{1}{c|}{\cellcolor[HTML]{BCFFFB}007} &
			\multicolumn{1}{c|}{\cellcolor[HTML]{BCFFFB}013} &
			\multicolumn{1}{c|}{\cellcolor[HTML]{BCFFFB}010} &
			\cellcolor[HTML]{BCFFFB}012 &
			\multicolumn{1}{c|}{\cellcolor[HTML]{FFFFC7}011} &
			\multicolumn{1}{c|}{\cellcolor[HTML]{FFFFC7}011} &
			\multicolumn{1}{c|}{\cellcolor[HTML]{FFFFC7}009} &
			\multicolumn{1}{c|}{\cellcolor[HTML]{FFFFC7}011} &
			\multicolumn{1}{c|}{\cellcolor[HTML]{FFFFC7}\textbf{006}} &
			\multicolumn{1}{c|}{\cellcolor[HTML]{FFFFC7}008} &
			\multicolumn{1}{c|}{\cellcolor[HTML]{FFFFC7}016} &
			\cellcolor[HTML]{FFFFC7}026 \\ \cline{2-18} 
			\cellcolor[HTML]{C1C1C1} &
			\textbf{6} &
			\multicolumn{1}{c|}{\cellcolor[HTML]{BCFFFB}010} &
			\multicolumn{1}{c|}{\cellcolor[HTML]{BCFFFB}008} &
			\multicolumn{1}{c|}{\cellcolor[HTML]{BCFFFB}008} &
			\multicolumn{1}{c|}{\cellcolor[HTML]{BCFFFB}010} &
			\multicolumn{1}{c|}{\cellcolor[HTML]{BCFFFB}009} &
			\multicolumn{1}{l|}{\cellcolor[HTML]{BCFFFB}014} &
			\multicolumn{1}{c|}{\cellcolor[HTML]{BCFFFB}011} &
			\cellcolor[HTML]{BCFFFB}013 &
			\multicolumn{1}{c|}{\cellcolor[HTML]{FFFFC7}011} &
			\multicolumn{1}{c|}{\cellcolor[HTML]{FFFFC7}011} &
			\multicolumn{1}{c|}{\cellcolor[HTML]{FFFFC7}008} &
			\multicolumn{1}{c|}{\cellcolor[HTML]{FFFFC7}010} &
			\multicolumn{1}{c|}{\cellcolor[HTML]{FFFFC7}008} &
			\multicolumn{1}{l|}{\cellcolor[HTML]{FFFFC7}\textbf{006}} &
			\multicolumn{1}{c|}{\cellcolor[HTML]{FFFFC7}016} &
			\cellcolor[HTML]{FFFFC7}027 \\ \cline{2-18} 
			\cellcolor[HTML]{C1C1C1} &
			\textbf{7} &
			\multicolumn{1}{c|}{\cellcolor[HTML]{BCFFFB}022} &
			\multicolumn{1}{c|}{\cellcolor[HTML]{BCFFFB}013} &
			\multicolumn{1}{c|}{\cellcolor[HTML]{BCFFFB}015} &
			\multicolumn{1}{c|}{\cellcolor[HTML]{BCFFFB}020} &
			\multicolumn{1}{c|}{\cellcolor[HTML]{BCFFFB}013} &
			\multicolumn{1}{l|}{\cellcolor[HTML]{BCFFFB}009} &
			\multicolumn{1}{c|}{\cellcolor[HTML]{BCFFFB}017} &
			\cellcolor[HTML]{BCFFFB}011 &
			\multicolumn{1}{c|}{\cellcolor[HTML]{FFFFC7}011} &
			\multicolumn{1}{c|}{\cellcolor[HTML]{FFFFC7}010} &
			\multicolumn{1}{c|}{\cellcolor[HTML]{FFFFC7}016} &
			\multicolumn{1}{c|}{\cellcolor[HTML]{FFFFC7}021} &
			\multicolumn{1}{c|}{\cellcolor[HTML]{FFFFC7}015} &
			\multicolumn{1}{c|}{\cellcolor[HTML]{FFFFC7}015} &
			\multicolumn{1}{c|}{\cellcolor[HTML]{FFFFC7}\textbf{008}} &
			\cellcolor[HTML]{FFFFC7}016 \\ \cline{2-18} 
			\multirow{-8}{*}{\rotatebox[origin=c]{90}{\cellcolor[HTML]{C1C1C1}\textbf{\racsi TESTING}}} &
			\textbf{8} &
			\multicolumn{1}{c|}{\cellcolor[HTML]{BCFFFB}035} &
			\multicolumn{1}{c|}{\cellcolor[HTML]{BCFFFB}025} &
			\multicolumn{1}{c|}{\cellcolor[HTML]{BCFFFB}026} &
			\multicolumn{1}{c|}{\cellcolor[HTML]{BCFFFB}033} &
			\multicolumn{1}{c|}{\cellcolor[HTML]{BCFFFB}023} &
			\multicolumn{1}{l|}{\cellcolor[HTML]{BCFFFB}015} &
			\multicolumn{1}{c|}{\cellcolor[HTML]{BCFFFB}028} &
			\cellcolor[HTML]{BCFFFB}018 &
			\multicolumn{1}{c|}{\cellcolor[HTML]{FFFFC7}019} &
			\multicolumn{1}{c|}{\cellcolor[HTML]{FFFFC7}019} &
			\multicolumn{1}{c|}{\cellcolor[HTML]{FFFFC7}028} &
			\multicolumn{1}{c|}{\cellcolor[HTML]{FFFFC7}033} &
			\multicolumn{1}{c|}{\cellcolor[HTML]{FFFFC7}026} &
			\multicolumn{1}{c|}{\cellcolor[HTML]{FFFFC7}027} &
			\multicolumn{1}{c|}{\cellcolor[HTML]{FFFFC7}016} &
			\cellcolor[HTML]{FFFFC7}\textbf{006} \\ \hline
		\end{tabular}%
	}
	\caption{Normalized average \whd computed on the partitions of the AntiSense dataset dedicated to training and testing with the receiver located in position 5 (rx5). The integer values are the first three digits after the comma, rounded to the nearest value. The POS parameter indicates the position of the person standing still within the experimental environment.}
	\label{tab:rx5whd}
\end{table}

These results confirm the behavior of the neural network, while simultaneously allowing to make preliminary predictions on its ability to succeed in locating a person within the chosen room. 
Hence, the $\overline{\whd}$ turns out to be a useful metric that can be exploited to make assumptions on the quality of the performance of neural networks in the positioning task. 
Nonetheless, it is still too coarse a metric to grasp the subtleties in \ac{CSI} values that a neural network is capable of identifying, making it hard to position a person within a room based solely on the \whd.

The three tables that have not been explicitly commented in this section can be found in \cref{app:ASwhdtables}, but the results of the analysis that was carried out on \cref{tab:rx1whd,,tab:rx5whd} remain consistent if extended to those tables as well.

%% file: future_work.tex
\chapter{Conclusions and Future Work}
\label{ch:conc}
\acresetall
This work took off from the findings of the Bachelor's Degree Thesis and focused on expanding the characterization of a \wifi channel through \ac{CSI} analysis. 
Given the complexity of the goal, the study was centered on the amplitude of the \acp{CSI}, temporarily discarding the phase values.

To favor a step-by-step approach, the work was subdivided into multiple tasks, the first consisting in identifying a representation of \ac{CSI} amplitudes that can simplify the comparison of the values across different experimental setups.
For each \ac{CSI}, other than normalizing its amplitudes by the integral of the energy to remove the effect of the \ac{AGC}, the values it takes on each \ac{OFDM} sub-carrier are normalized between 0 and 1 and then quantized on a finite number of bits.
The quantization allows for a more compact representation of the \acp{CSI}, introducing an upper bound to the amplitudes that was not implicitly present upon \ac{CSI} extraction. 
This approach lets us describe the traces on a closed set of finite values that remains unaltered across the different experiments, facilitating simultaneous analysis and comparison of the results relative to different captures, possibly even done with distinct technologies.  

Once a unique representation for the \acp{CSI} had been found, the second task consisted in considering each collection of traces as a source of information about the environment, which meant quantifying the amount of knowledge carried by each \ac{CSI}. 
This sub-goal was first approached through the computation of \ac{MI}, which should provide a measure of the information that can be gathered about `environment A', given a capture performed in `environment B'. 
Computing the \ac{MI} has proven less straightforward than expected, as each \ac{CSI} has one in $2^{\nsc\cdot\acsibits}$ chances of `happening', which of course results in unmanageable values, given that the number of useful sub-carriers is $\nsc = 256,~512,~1024$ and the quantization bits are $\acsibits=10$. 
Nonetheless, this approach would have allowed us to account for the probability of a \ac{CSI} belonging to a specific experiment, making \ac{MI} a possibly extremely useful metric for the environment classification task. 
Therefore, more research is needed to see whether the \ac{MI} could come in handy if its computation is implemented following an alternative path.

To avoid incurring problems linked to the numerical representation of the information content of \acp{CSI}, the study was redirected towards the measurement of the \ac{WHD} between two \ac{CSI} traces. 
This metric calculates the number of mismatching bits in the binary representation of two quantized \acp{CSI}, resulting in the determination of the number of differences between such traces. 
By switching to the integer representation of the quantized \ac{CSI}, the weight of the differing bits within the two traces is automatically accounted for, making mismatches in more significant bits more impactful on the resulting distance. 

As the \whd can only be computed between single \acp{CSI}, the comparison of whole experiments was carried out by measuring the \acp{WHD} between the reference \ac{CSI} of the first experiment and all \acp{CSI} of the second one and then averaging them. 
To ensure symmetry of the distances, the inverse was also computed. 
Results show that each capture is extremely self-similar (i.e., the average \whd between an experiment and its reference \ac{CSI} is close to zero), with increasingly larger values as the variability of the experimental setup grows. 
For instance, as the number of people in the room where the \acp{CSI} were collected increases, thus causing more changes in the environment, the distance between different experiments grows. 
As can be imagined, the smallest distances can be found in correspondence to the Empty Scenario (i.e., an empty room), whereas the largest ones can be obtained by comparing it with the Fully Dynamic Scenario with five people in the room. 

By only looking at the results relative to the \whd, we do not have enough information to tell environments apart based solely on the dispersion of the average distance from the reference \ac{CSI} representing an experiment. 
Specifically, this task can only be carried out for the Empty Scenario, as its variability is minimized and the distances from the reference \ac{CSI} are less dispersed, but, as the number of people in the room increases, the distribution of the average \whd loses such powerful meaning. 

Further research is needed to identify an ulterior metric that allows comparison of the distribution of \acp{CSI}: such a representation of the dispersion of \acp{CSI} within each capture would make it possible to find any overlap in the distributions relative to different experiments, enabling the computation of the probability of wrongly classifying an experimental setup. 
Reaching this goal would provide mathematical and probabilistic support to \ac{ML} classification algorithms, offering an insight into how they work and reducing their use as `black boxes'.

This work paves the road to such extension, having introduced a quantization mechanism to simplify manipulation of \ac{CSI} amplitudes, providing a method to view them as items within a finite set, which significantly simplifies theoretical reasoning and computations. 

%% file: appendixA.tex
\chapter{Detailed Classification of Collected Data}
\label{app:experiments}
The collection of the \ac{CSI} traces used in this work is obtained through multiple experiments, which are classified according to metadata that is specific to each set of captures.

Each experiment consists of a single capture of \acp{CSI} performed within a continuous period of time; stopping the traces capture and restarting it after a few minutes have passed without altering any configuration parameter still counts as the end of an experiment and the start of a new one. 
This leads to the possibility of having multiple experiments with the same configuration, hence the same metadata can be shared among different captures. 

The metadata are structured as fields of a \texttt{json} file, containing all information required to classify an experiment.
The content of the file is organized as follows:
\begin{itemize}
	\item Date: day, month, and year where the capture took place. All three sub-fields are integer values;
	\item Location ID: the unique identifier of the environment where the capture took place. The association of each ID with the corresponding description (e.g. the address or the geographical coordinates of the location) is contained in a separate file, as will be described further on;
	\item Experiment: a string that describes the type of experiment performed;
	\item \textit{Ad hoc} transmission: a boolean field that qualifies the traffic transmitted through the environment as artificially or user-generated;
	\item \texttt{usleep}: integer value indicating the interval --- in microseconds --- between each transmitted packet and the following one;
	\item Average duration: integer value indicating the duration (on average) of an experiment associated with the current metadata. The duration is expressed in seconds;
	\item Bandwidth: integer value indicating the bandwidth of the channel used to transmit traffic;
	\item Modulation: string indicating the type of 802.11 modulation used for transmission;
	\item Number of receivers: integer value indicating the number of receivers involved in the experiment;
	\item Number of transmitters: integer value indicating the number of transmitters involved in the experiment;
	\item Number of antennas used at the transmitter (integer);
	\item Number of antennas used at the receiver (integer);
	\item People: field used to identify the presence of people in the location where the experiment was performed. It is composed of four sub-fields: 
		\begin{itemize}
			\item Present: boolean field to state if anyone was within the environment where the \acp{CSI} were captured;
			\item Number: integer value indicating the number of people in the location;
			\item Moving: boolean field assessing whether the people in the room are walking around or standing still/sitting down. If no one is in the room, this field is set to \texttt{false};
			\item Names: list of the names of the people in the room (if any, empty list otherwise).
		\end{itemize}
	\item Notes: additional information that is deemed relevant.
\end{itemize}

To uniquely identify the locations of the experiments, an additional \texttt{json} file is generated, which contains all location IDs and corresponding descriptions.

This description of the classification of the captures corresponds to the latest version employed up to September 2024. 
Studies after this date may alter the structure of the \texttt{json} files containing the metadata of each experiment or even base the classification on entirely different mechanisms. 

All \ac{CSI} traces collected within this study will be published as open data according to the necessities of the projects listed in the acknowledgements of this work.

%% file: appendixB.tex
\chapter{Normalized Average \whd of the AntiSense Dataset}
\label{app:ASwhdtables}
The tables containing the normalized $\overline{\whd}$ computed on the AntiSense dataset that were not explicitly commented in \cref{ch:whdres} are here displayed. 
They reference the experiments identified as `rx2', `rx3', and `rx4' according to the position of the receiver, as shown in \cref{fig:tlclab}.

\begin{table}[]
	\centering
	\resizebox{\textwidth}{!}{%
		\begin{tabular}{|
				>{\columncolor[HTML]{C1C1C1}}c |
				>{\columncolor[HTML]{C1C1C1}}c |cccccccc|cccccccc|}
			\hline
			\textbf{} &
			\textbf{} &
			\multicolumn{8}{c|}{\cellcolor[HTML]{C1C1C1}\textbf{TRAINING}} &
			\multicolumn{8}{c|}{\cellcolor[HTML]{C1C1C1}\textbf{TESTING}} \\ \hline
			\textbf{} &
			\textbf{POS} &
			\multicolumn{1}{c|}{\cellcolor[HTML]{C1C1C1}\textbf{1}} &
			\multicolumn{1}{c|}{\cellcolor[HTML]{C1C1C1}\textbf{2}} &
			\multicolumn{1}{c|}{\cellcolor[HTML]{C1C1C1}\textbf{3}} &
			\multicolumn{1}{c|}{\cellcolor[HTML]{C1C1C1}\textbf{4}} &
			\multicolumn{1}{c|}{\cellcolor[HTML]{C1C1C1}\textbf{5}} &
			\multicolumn{1}{c|}{\cellcolor[HTML]{C1C1C1}\textbf{6}} &
			\multicolumn{1}{c|}{\cellcolor[HTML]{C1C1C1}\textbf{7}} &
			\cellcolor[HTML]{C1C1C1}\textbf{8} &
			\multicolumn{1}{c|}{\cellcolor[HTML]{C1C1C1}\textbf{1}} &
			\multicolumn{1}{c|}{\cellcolor[HTML]{C1C1C1}\textbf{2}} &
			\multicolumn{1}{c|}{\cellcolor[HTML]{C1C1C1}\textbf{3}} &
			\multicolumn{1}{c|}{\cellcolor[HTML]{C1C1C1}\textbf{4}} &
			\multicolumn{1}{c|}{\cellcolor[HTML]{C1C1C1}\textbf{5}} &
			\multicolumn{1}{c|}{\cellcolor[HTML]{C1C1C1}\textbf{6}} &
			\multicolumn{1}{c|}{\cellcolor[HTML]{C1C1C1}\textbf{7}} &
			\cellcolor[HTML]{C1C1C1}\textbf{8} \\ \hline
			\cellcolor[HTML]{C1C1C1} &
			\textbf{1} &
			\multicolumn{1}{c|}{\cellcolor[HTML]{FFFFC7}\textbf{009}} &
			\multicolumn{1}{c|}{\cellcolor[HTML]{FFFFC7}028} &
			\multicolumn{1}{c|}{\cellcolor[HTML]{FFFFC7}023} &
			\multicolumn{1}{c|}{\cellcolor[HTML]{FFFFC7}017} &
			\multicolumn{1}{c|}{\cellcolor[HTML]{FFFFC7}024} &
			\multicolumn{1}{c|}{\cellcolor[HTML]{FFFFC7}025} &
			\multicolumn{1}{c|}{\cellcolor[HTML]{FFFFC7}028} &
			\cellcolor[HTML]{FFFFC7}032 &
			\multicolumn{1}{c|}{\cellcolor[HTML]{C4FF99}028} &
			\multicolumn{1}{c|}{\cellcolor[HTML]{C4FF99}029} &
			\multicolumn{1}{c|}{\cellcolor[HTML]{C4FF99}023} &
			\multicolumn{1}{c|}{\cellcolor[HTML]{C4FF99}022} &
			\multicolumn{1}{c|}{\cellcolor[HTML]{C4FF99}020} &
			\multicolumn{1}{c|}{\cellcolor[HTML]{C4FF99}023} &
			\multicolumn{1}{c|}{\cellcolor[HTML]{C4FF99}028} &
			\cellcolor[HTML]{C4FF99}036 \\ \cline{2-18} 
			\cellcolor[HTML]{C1C1C1} &
			\textbf{2} &
			\multicolumn{1}{c|}{\cellcolor[HTML]{FFFFC7}028} &
			\multicolumn{1}{c|}{\cellcolor[HTML]{FFFFC7}\textbf{010}} &
			\multicolumn{1}{c|}{\cellcolor[HTML]{FFFFC7}024} &
			\multicolumn{1}{c|}{\cellcolor[HTML]{FFFFC7}021} &
			\multicolumn{1}{c|}{\cellcolor[HTML]{FFFFC7}028} &
			\multicolumn{1}{c|}{\cellcolor[HTML]{FFFFC7}022} &
			\multicolumn{1}{c|}{\cellcolor[HTML]{FFFFC7}026} &
			\cellcolor[HTML]{FFFFC7}028 &
			\multicolumn{1}{c|}{\cellcolor[HTML]{C4FF99}036} &
			\multicolumn{1}{c|}{\cellcolor[HTML]{C4FF99}015} &
			\multicolumn{1}{c|}{\cellcolor[HTML]{C4FF99}026} &
			\multicolumn{1}{c|}{\cellcolor[HTML]{C4FF99}025} &
			\multicolumn{1}{c|}{\cellcolor[HTML]{C4FF99}026} &
			\multicolumn{1}{c|}{\cellcolor[HTML]{C4FF99}023} &
			\multicolumn{1}{c|}{\cellcolor[HTML]{C4FF99}024} &
			\cellcolor[HTML]{C4FF99}029 \\ \cline{2-18} 
			\cellcolor[HTML]{C1C1C1} &
			\textbf{3} &
			\multicolumn{1}{c|}{\cellcolor[HTML]{FFFFC7}023} &
			\multicolumn{1}{c|}{\cellcolor[HTML]{FFFFC7}025} &
			\multicolumn{1}{c|}{\cellcolor[HTML]{FFFFC7}\textbf{008}} &
			\multicolumn{1}{c|}{\cellcolor[HTML]{FFFFC7}019} &
			\multicolumn{1}{c|}{\cellcolor[HTML]{FFFFC7}021} &
			\multicolumn{1}{c|}{\cellcolor[HTML]{FFFFC7}018} &
			\multicolumn{1}{c|}{\cellcolor[HTML]{FFFFC7}027} &
			\cellcolor[HTML]{FFFFC7}020 &
			\multicolumn{1}{c|}{\cellcolor[HTML]{C4FF99}026} &
			\multicolumn{1}{c|}{\cellcolor[HTML]{C4FF99}024} &
			\multicolumn{1}{c|}{\cellcolor[HTML]{C4FF99}015} &
			\multicolumn{1}{c|}{\cellcolor[HTML]{C4FF99}029} &
			\multicolumn{1}{c|}{\cellcolor[HTML]{C4FF99}022} &
			\multicolumn{1}{c|}{\cellcolor[HTML]{C4FF99}020} &
			\multicolumn{1}{c|}{\cellcolor[HTML]{C4FF99}023} &
			\cellcolor[HTML]{C4FF99}025 \\ \cline{2-18} 
			\cellcolor[HTML]{C1C1C1} &
			\textbf{4} &
			\multicolumn{1}{c|}{\cellcolor[HTML]{FFFFC7}018} &
			\multicolumn{1}{c|}{\cellcolor[HTML]{FFFFC7}022} &
			\multicolumn{1}{c|}{\cellcolor[HTML]{FFFFC7}019} &
			\multicolumn{1}{c|}{\cellcolor[HTML]{FFFFC7}\textbf{009}} &
			\multicolumn{1}{c|}{\cellcolor[HTML]{FFFFC7}020} &
			\multicolumn{1}{c|}{\cellcolor[HTML]{FFFFC7}021} &
			\multicolumn{1}{c|}{\cellcolor[HTML]{FFFFC7}024} &
			\cellcolor[HTML]{FFFFC7}025 &
			\multicolumn{1}{c|}{\cellcolor[HTML]{C4FF99}028} &
			\multicolumn{1}{c|}{\cellcolor[HTML]{C4FF99}021} &
			\multicolumn{1}{c|}{\cellcolor[HTML]{C4FF99}020} &
			\multicolumn{1}{c|}{\cellcolor[HTML]{C4FF99}016} &
			\multicolumn{1}{c|}{\cellcolor[HTML]{C4FF99}018} &
			\multicolumn{1}{c|}{\cellcolor[HTML]{C4FF99}019} &
			\multicolumn{1}{c|}{\cellcolor[HTML]{C4FF99}023} &
			\cellcolor[HTML]{C4FF99}030 \\ \cline{2-18} 
			\cellcolor[HTML]{C1C1C1} &
			\textbf{5} &
			\multicolumn{1}{c|}{\cellcolor[HTML]{FFFFC7}024} &
			\multicolumn{1}{c|}{\cellcolor[HTML]{FFFFC7}028} &
			\multicolumn{1}{c|}{\cellcolor[HTML]{FFFFC7}021} &
			\multicolumn{1}{c|}{\cellcolor[HTML]{FFFFC7}020} &
			\multicolumn{1}{c|}{\cellcolor[HTML]{FFFFC7}\textbf{010}} &
			\multicolumn{1}{c|}{\cellcolor[HTML]{FFFFC7}020} &
			\multicolumn{1}{c|}{\cellcolor[HTML]{FFFFC7}030} &
			\cellcolor[HTML]{FFFFC7}026 &
			\multicolumn{1}{c|}{\cellcolor[HTML]{C4FF99}023} &
			\multicolumn{1}{c|}{\cellcolor[HTML]{C4FF99}028} &
			\multicolumn{1}{c|}{\cellcolor[HTML]{C4FF99}019} &
			\multicolumn{1}{c|}{\cellcolor[HTML]{C4FF99}026} &
			\multicolumn{1}{c|}{\cellcolor[HTML]{C4FF99}012} &
			\multicolumn{1}{c|}{\cellcolor[HTML]{C4FF99}018} &
			\multicolumn{1}{c|}{\cellcolor[HTML]{C4FF99}028} &
			\cellcolor[HTML]{C4FF99}029 \\ \cline{2-18} 
			\cellcolor[HTML]{C1C1C1} &
			\textbf{6} &
			\multicolumn{1}{c|}{\cellcolor[HTML]{FFFFC7}025} &
			\multicolumn{1}{c|}{\cellcolor[HTML]{FFFFC7}021} &
			\multicolumn{1}{c|}{\cellcolor[HTML]{FFFFC7}017} &
			\multicolumn{1}{c|}{\cellcolor[HTML]{FFFFC7}020} &
			\multicolumn{1}{c|}{\cellcolor[HTML]{FFFFC7}019} &
			\multicolumn{1}{c|}{\cellcolor[HTML]{FFFFC7}\textbf{011}} &
			\multicolumn{1}{c|}{\cellcolor[HTML]{FFFFC7}024} &
			\cellcolor[HTML]{FFFFC7}023 &
			\multicolumn{1}{c|}{\cellcolor[HTML]{C4FF99}029} &
			\multicolumn{1}{c|}{\cellcolor[HTML]{C4FF99}021} &
			\multicolumn{1}{c|}{\cellcolor[HTML]{C4FF99}019} &
			\multicolumn{1}{c|}{\cellcolor[HTML]{C4FF99}026} &
			\multicolumn{1}{c|}{\cellcolor[HTML]{C4FF99}020} &
			\multicolumn{1}{c|}{\cellcolor[HTML]{C4FF99}015} &
			\multicolumn{1}{c|}{\cellcolor[HTML]{C4FF99}022} &
			\cellcolor[HTML]{C4FF99}028 \\ \cline{2-18} 
			\cellcolor[HTML]{C1C1C1} &
			\textbf{7} &
			\multicolumn{1}{c|}{\cellcolor[HTML]{FFFFC7}027} &
			\multicolumn{1}{c|}{\cellcolor[HTML]{FFFFC7}026} &
			\multicolumn{1}{c|}{\cellcolor[HTML]{FFFFC7}026} &
			\multicolumn{1}{c|}{\cellcolor[HTML]{FFFFC7}023} &
			\multicolumn{1}{c|}{\cellcolor[HTML]{FFFFC7}029} &
			\multicolumn{1}{c|}{\cellcolor[HTML]{FFFFC7}024} &
			\multicolumn{1}{c|}{\cellcolor[HTML]{FFFFC7}\textbf{012}} &
			\cellcolor[HTML]{FFFFC7}027 &
			\multicolumn{1}{c|}{\cellcolor[HTML]{C4FF99}041} &
			\multicolumn{1}{c|}{\cellcolor[HTML]{C4FF99}024} &
			\multicolumn{1}{c|}{\cellcolor[HTML]{C4FF99}028} &
			\multicolumn{1}{c|}{\cellcolor[HTML]{C4FF99}021} &
			\multicolumn{1}{c|}{\cellcolor[HTML]{C4FF99}030} &
			\multicolumn{1}{c|}{\cellcolor[HTML]{C4FF99}021} &
			\multicolumn{1}{c|}{\cellcolor[HTML]{C4FF99}021} &
			\cellcolor[HTML]{C4FF99}033 \\ \cline{2-18} 
			\multirow{-8}{*}{\rotatebox[origin=c]{90}{\cellcolor[HTML]{C1C1C1}\textbf{\racsi TRAINING}}} &
			\textbf{8} &
			\multicolumn{1}{c|}{\cellcolor[HTML]{FFFFC7}032} &
			\multicolumn{1}{c|}{\cellcolor[HTML]{FFFFC7}028} &
			\multicolumn{1}{c|}{\cellcolor[HTML]{FFFFC7}020} &
			\multicolumn{1}{c|}{\cellcolor[HTML]{FFFFC7}025} &
			\multicolumn{1}{c|}{\cellcolor[HTML]{FFFFC7}026} &
			\multicolumn{1}{c|}{\cellcolor[HTML]{FFFFC7}024} &
			\multicolumn{1}{c|}{\cellcolor[HTML]{FFFFC7}027} &
			\cellcolor[HTML]{FFFFC7}\textbf{010} &
			\multicolumn{1}{c|}{\cellcolor[HTML]{C4FF99}032} &
			\multicolumn{1}{c|}{\cellcolor[HTML]{C4FF99}029} &
			\multicolumn{1}{c|}{\cellcolor[HTML]{C4FF99}021} &
			\multicolumn{1}{c|}{\cellcolor[HTML]{C4FF99}030} &
			\multicolumn{1}{c|}{\cellcolor[HTML]{C4FF99}028} &
			\multicolumn{1}{c|}{\cellcolor[HTML]{C4FF99}023} &
			\multicolumn{1}{c|}{\cellcolor[HTML]{C4FF99}025} &
			\cellcolor[HTML]{C4FF99}017 \\ \hline
			\cellcolor[HTML]{C1C1C1} &
			\textbf{1} &
			\multicolumn{1}{c|}{\cellcolor[HTML]{BCFFFB}028} &
			\multicolumn{1}{c|}{\cellcolor[HTML]{BCFFFB}036} &
			\multicolumn{1}{c|}{\cellcolor[HTML]{BCFFFB}026} &
			\multicolumn{1}{c|}{\cellcolor[HTML]{BCFFFB}028} &
			\multicolumn{1}{c|}{\cellcolor[HTML]{BCFFFB}023} &
			\multicolumn{1}{c|}{\cellcolor[HTML]{BCFFFB}030} &
			\multicolumn{1}{c|}{\cellcolor[HTML]{BCFFFB}042} &
			\cellcolor[HTML]{BCFFFB}032 &
			\multicolumn{1}{c|}{\cellcolor[HTML]{FFFFC7}\textbf{009}} &
			\multicolumn{1}{c|}{\cellcolor[HTML]{FFFFC7}037} &
			\multicolumn{1}{c|}{\cellcolor[HTML]{FFFFC7}021} &
			\multicolumn{1}{c|}{\cellcolor[HTML]{FFFFC7}035} &
			\multicolumn{1}{c|}{\cellcolor[HTML]{FFFFC7}022} &
			\multicolumn{1}{c|}{\cellcolor[HTML]{FFFFC7}030} &
			\multicolumn{1}{c|}{\cellcolor[HTML]{FFFFC7}033} &
			\cellcolor[HTML]{FFFFC7}032 \\ \cline{2-18} 
			\cellcolor[HTML]{C1C1C1} &
			\textbf{2} &
			\multicolumn{1}{c|}{\cellcolor[HTML]{BCFFFB}029} &
			\multicolumn{1}{c|}{\cellcolor[HTML]{BCFFFB}015} &
			\multicolumn{1}{c|}{\cellcolor[HTML]{BCFFFB}024} &
			\multicolumn{1}{c|}{\cellcolor[HTML]{BCFFFB}021} &
			\multicolumn{1}{c|}{\cellcolor[HTML]{BCFFFB}028} &
			\multicolumn{1}{c|}{\cellcolor[HTML]{BCFFFB}022} &
			\multicolumn{1}{c|}{\cellcolor[HTML]{BCFFFB}025} &
			\cellcolor[HTML]{BCFFFB}029 &
			\multicolumn{1}{c|}{\cellcolor[HTML]{FFFFC7}037} &
			\multicolumn{1}{c|}{\cellcolor[HTML]{FFFFC7}\textbf{009}} &
			\multicolumn{1}{c|}{\cellcolor[HTML]{FFFFC7}026} &
			\multicolumn{1}{c|}{\cellcolor[HTML]{FFFFC7}023} &
			\multicolumn{1}{c|}{\cellcolor[HTML]{FFFFC7}027} &
			\multicolumn{1}{c|}{\cellcolor[HTML]{FFFFC7}022} &
			\multicolumn{1}{c|}{\cellcolor[HTML]{FFFFC7}024} &
			\cellcolor[HTML]{FFFFC7}032 \\ \cline{2-18} 
			\cellcolor[HTML]{C1C1C1} &
			\textbf{3} &
			\multicolumn{1}{c|}{\cellcolor[HTML]{BCFFFB}023} &
			\multicolumn{1}{c|}{\cellcolor[HTML]{BCFFFB}026} &
			\multicolumn{1}{c|}{\cellcolor[HTML]{BCFFFB}014} &
			\multicolumn{1}{c|}{\cellcolor[HTML]{BCFFFB}019} &
			\multicolumn{1}{c|}{\cellcolor[HTML]{BCFFFB}019} &
			\multicolumn{1}{c|}{\cellcolor[HTML]{BCFFFB}021} &
			\multicolumn{1}{c|}{\cellcolor[HTML]{BCFFFB}029} &
			\cellcolor[HTML]{BCFFFB}021 &
			\multicolumn{1}{c|}{\cellcolor[HTML]{FFFFC7}021} &
			\multicolumn{1}{c|}{\cellcolor[HTML]{FFFFC7}026} &
			\multicolumn{1}{c|}{\cellcolor[HTML]{FFFFC7}\textbf{009}} &
			\multicolumn{1}{c|}{\cellcolor[HTML]{FFFFC7}027} &
			\multicolumn{1}{c|}{\cellcolor[HTML]{FFFFC7}018} &
			\multicolumn{1}{c|}{\cellcolor[HTML]{FFFFC7}019} &
			\multicolumn{1}{c|}{\cellcolor[HTML]{FFFFC7}023} &
			\cellcolor[HTML]{FFFFC7}023 \\ \cline{2-18} 
			\cellcolor[HTML]{C1C1C1} &
			\textbf{4} &
			\multicolumn{1}{c|}{\cellcolor[HTML]{BCFFFB}022} &
			\multicolumn{1}{c|}{\cellcolor[HTML]{BCFFFB}025} &
			\multicolumn{1}{c|}{\cellcolor[HTML]{BCFFFB}029} &
			\multicolumn{1}{c|}{\cellcolor[HTML]{BCFFFB}016} &
			\multicolumn{1}{c|}{\cellcolor[HTML]{BCFFFB}027} &
			\multicolumn{1}{c|}{\cellcolor[HTML]{BCFFFB}027} &
			\multicolumn{1}{c|}{\cellcolor[HTML]{BCFFFB}023} &
			\cellcolor[HTML]{BCFFFB}030 &
			\multicolumn{1}{c|}{\cellcolor[HTML]{FFFFC7}035} &
			\multicolumn{1}{c|}{\cellcolor[HTML]{FFFFC7}023} &
			\multicolumn{1}{c|}{\cellcolor[HTML]{FFFFC7}027} &
			\multicolumn{1}{c|}{\cellcolor[HTML]{FFFFC7}\textbf{009}} &
			\multicolumn{1}{c|}{\cellcolor[HTML]{FFFFC7}022} &
			\multicolumn{1}{c|}{\cellcolor[HTML]{FFFFC7}024} &
			\multicolumn{1}{c|}{\cellcolor[HTML]{FFFFC7}028} &
			\cellcolor[HTML]{FFFFC7}034 \\ \cline{2-18} 
			\cellcolor[HTML]{C1C1C1} &
			\textbf{5} &
			\multicolumn{1}{c|}{\cellcolor[HTML]{BCFFFB}020} &
			\multicolumn{1}{c|}{\cellcolor[HTML]{BCFFFB}026} &
			\multicolumn{1}{c|}{\cellcolor[HTML]{BCFFFB}022} &
			\multicolumn{1}{c|}{\cellcolor[HTML]{BCFFFB}018} &
			\multicolumn{1}{c|}{\cellcolor[HTML]{BCFFFB}012} &
			\multicolumn{1}{c|}{\cellcolor[HTML]{BCFFFB}021} &
			\multicolumn{1}{c|}{\cellcolor[HTML]{BCFFFB}031} &
			\cellcolor[HTML]{BCFFFB}028 &
			\multicolumn{1}{c|}{\cellcolor[HTML]{FFFFC7}022} &
			\multicolumn{1}{c|}{\cellcolor[HTML]{FFFFC7}027} &
			\multicolumn{1}{c|}{\cellcolor[HTML]{FFFFC7}018} &
			\multicolumn{1}{c|}{\cellcolor[HTML]{FFFFC7}022} &
			\multicolumn{1}{c|}{\cellcolor[HTML]{FFFFC7}\textbf{009}} &
			\multicolumn{1}{c|}{\cellcolor[HTML]{FFFFC7}019} &
			\multicolumn{1}{c|}{\cellcolor[HTML]{FFFFC7}029} &
			\cellcolor[HTML]{FFFFC7}030 \\ \cline{2-18} 
			\cellcolor[HTML]{C1C1C1} &
			\textbf{6} &
			\multicolumn{1}{c|}{\cellcolor[HTML]{BCFFFB}024} &
			\multicolumn{1}{c|}{\cellcolor[HTML]{BCFFFB}024} &
			\multicolumn{1}{c|}{\cellcolor[HTML]{BCFFFB}019} &
			\multicolumn{1}{c|}{\cellcolor[HTML]{BCFFFB}019} &
			\multicolumn{1}{c|}{\cellcolor[HTML]{BCFFFB}018} &
			\multicolumn{1}{l|}{\cellcolor[HTML]{BCFFFB}016} &
			\multicolumn{1}{c|}{\cellcolor[HTML]{BCFFFB}022} &
			\cellcolor[HTML]{BCFFFB}023 &
			\multicolumn{1}{c|}{\cellcolor[HTML]{FFFFC7}030} &
			\multicolumn{1}{c|}{\cellcolor[HTML]{FFFFC7}022} &
			\multicolumn{1}{c|}{\cellcolor[HTML]{FFFFC7}019} &
			\multicolumn{1}{c|}{\cellcolor[HTML]{FFFFC7}023} &
			\multicolumn{1}{c|}{\cellcolor[HTML]{FFFFC7}019} &
			\multicolumn{1}{l|}{\cellcolor[HTML]{FFFFC7}\textbf{009}} &
			\multicolumn{1}{c|}{\cellcolor[HTML]{FFFFC7}022} &
			\cellcolor[HTML]{FFFFC7}028 \\ \cline{2-18} 
			\cellcolor[HTML]{C1C1C1} &
			\textbf{7} &
			\multicolumn{1}{c|}{\cellcolor[HTML]{BCFFFB}027} &
			\multicolumn{1}{c|}{\cellcolor[HTML]{BCFFFB}024} &
			\multicolumn{1}{c|}{\cellcolor[HTML]{BCFFFB}022} &
			\multicolumn{1}{c|}{\cellcolor[HTML]{BCFFFB}022} &
			\multicolumn{1}{c|}{\cellcolor[HTML]{BCFFFB}027} &
			\multicolumn{1}{l|}{\cellcolor[HTML]{BCFFFB}023} &
			\multicolumn{1}{c|}{\cellcolor[HTML]{BCFFFB}021} &
			\cellcolor[HTML]{BCFFFB}024 &
			\multicolumn{1}{c|}{\cellcolor[HTML]{FFFFC7}032} &
			\multicolumn{1}{c|}{\cellcolor[HTML]{FFFFC7}023} &
			\multicolumn{1}{c|}{\cellcolor[HTML]{FFFFC7}022} &
			\multicolumn{1}{c|}{\cellcolor[HTML]{FFFFC7}027} &
			\multicolumn{1}{c|}{\cellcolor[HTML]{FFFFC7}028} &
			\multicolumn{1}{c|}{\cellcolor[HTML]{FFFFC7}021} &
			\multicolumn{1}{c|}{\cellcolor[HTML]{FFFFC7}\textbf{011}} &
			\cellcolor[HTML]{FFFFC7}026 \\ \cline{2-18} 
			\multirow{-8}{*}{\rotatebox[origin=c]{90}{\cellcolor[HTML]{C1C1C1}\textbf{\racsi TESTING}}} &
			\textbf{8} &
			\multicolumn{1}{c|}{\cellcolor[HTML]{BCFFFB}036} &
			\multicolumn{1}{c|}{\cellcolor[HTML]{BCFFFB}029} &
			\multicolumn{1}{c|}{\cellcolor[HTML]{BCFFFB}024} &
			\multicolumn{1}{c|}{\cellcolor[HTML]{BCFFFB}029} &
			\multicolumn{1}{c|}{\cellcolor[HTML]{BCFFFB}028} &
			\multicolumn{1}{l|}{\cellcolor[HTML]{BCFFFB}028} &
			\multicolumn{1}{c|}{\cellcolor[HTML]{BCFFFB}033} &
			\cellcolor[HTML]{BCFFFB}017 &
			\multicolumn{1}{c|}{\cellcolor[HTML]{FFFFC7}032} &
			\multicolumn{1}{c|}{\cellcolor[HTML]{FFFFC7}031} &
			\multicolumn{1}{c|}{\cellcolor[HTML]{FFFFC7}023} &
			\multicolumn{1}{c|}{\cellcolor[HTML]{FFFFC7}034} &
			\multicolumn{1}{c|}{\cellcolor[HTML]{FFFFC7}030} &
			\multicolumn{1}{c|}{\cellcolor[HTML]{FFFFC7}028} &
			\multicolumn{1}{c|}{\cellcolor[HTML]{FFFFC7}027} &
			\cellcolor[HTML]{FFFFC7}\textbf{010} \\ \hline
		\end{tabular}%
	}
	\caption{Average normalized \whd computed on the partitions of the AntiSense dataset dedicated to training and testing with the receiver located in position 2 (rx2). The integer values are the first three digits after the comma, rounded to the nearest value. The POS parameter indicates the position of the person standing still within the experimental environment.}
	\label{tab:rx2whd}
\end{table}

\begin{table}[]
	\centering
	\resizebox{\textwidth}{!}{%
		\begin{tabular}{|
				>{\columncolor[HTML]{C1C1C1}}c |
				>{\columncolor[HTML]{C1C1C1}}c |cccccccc|cccccccc|}
			\hline
			\textbf{} &
			\textbf{} &
			\multicolumn{8}{c|}{\cellcolor[HTML]{C1C1C1}\textbf{TRAINING}} &
			\multicolumn{8}{c|}{\cellcolor[HTML]{C1C1C1}\textbf{TESTING}} \\ \hline
			\textbf{} &
			\textbf{POS} &
			\multicolumn{1}{c|}{\cellcolor[HTML]{C1C1C1}\textbf{1}} &
			\multicolumn{1}{c|}{\cellcolor[HTML]{C1C1C1}\textbf{2}} &
			\multicolumn{1}{c|}{\cellcolor[HTML]{C1C1C1}\textbf{3}} &
			\multicolumn{1}{c|}{\cellcolor[HTML]{C1C1C1}\textbf{4}} &
			\multicolumn{1}{c|}{\cellcolor[HTML]{C1C1C1}\textbf{5}} &
			\multicolumn{1}{c|}{\cellcolor[HTML]{C1C1C1}\textbf{6}} &
			\multicolumn{1}{c|}{\cellcolor[HTML]{C1C1C1}\textbf{7}} &
			\cellcolor[HTML]{C1C1C1}\textbf{8} &
			\multicolumn{1}{c|}{\cellcolor[HTML]{C1C1C1}\textbf{1}} &
			\multicolumn{1}{c|}{\cellcolor[HTML]{C1C1C1}\textbf{2}} &
			\multicolumn{1}{c|}{\cellcolor[HTML]{C1C1C1}\textbf{3}} &
			\multicolumn{1}{c|}{\cellcolor[HTML]{C1C1C1}\textbf{4}} &
			\multicolumn{1}{c|}{\cellcolor[HTML]{C1C1C1}\textbf{5}} &
			\multicolumn{1}{c|}{\cellcolor[HTML]{C1C1C1}\textbf{6}} &
			\multicolumn{1}{c|}{\cellcolor[HTML]{C1C1C1}\textbf{7}} &
			\cellcolor[HTML]{C1C1C1}\textbf{8} \\ \hline
			\cellcolor[HTML]{C1C1C1} &
			\textbf{1} &
			\multicolumn{1}{c|}{\cellcolor[HTML]{FFFFC7}\textbf{018}} &
			\multicolumn{1}{c|}{\cellcolor[HTML]{FFFFC7}051} &
			\multicolumn{1}{c|}{\cellcolor[HTML]{FFFFC7}065} &
			\multicolumn{1}{c|}{\cellcolor[HTML]{FFFFC7}076} &
			\multicolumn{1}{c|}{\cellcolor[HTML]{FFFFC7}071} &
			\multicolumn{1}{c|}{\cellcolor[HTML]{FFFFC7}057} &
			\multicolumn{1}{c|}{\cellcolor[HTML]{FFFFC7}086} &
			\cellcolor[HTML]{FFFFC7}116 &
			\multicolumn{1}{c|}{\cellcolor[HTML]{C4FF99}036} &
			\multicolumn{1}{c|}{\cellcolor[HTML]{C4FF99}049} &
			\multicolumn{1}{c|}{\cellcolor[HTML]{C4FF99}074} &
			\multicolumn{1}{c|}{\cellcolor[HTML]{C4FF99}226} &
			\multicolumn{1}{c|}{\cellcolor[HTML]{C4FF99}073} &
			\multicolumn{1}{c|}{\cellcolor[HTML]{C4FF99}064} &
			\multicolumn{1}{c|}{\cellcolor[HTML]{C4FF99}070} &
			\cellcolor[HTML]{C4FF99}131 \\ \cline{2-18} 
			\cellcolor[HTML]{C1C1C1} &
			\textbf{2} &
			\multicolumn{1}{c|}{\cellcolor[HTML]{FFFFC7}045} &
			\multicolumn{1}{c|}{\cellcolor[HTML]{FFFFC7}\textbf{027}} &
			\multicolumn{1}{c|}{\cellcolor[HTML]{FFFFC7}053} &
			\multicolumn{1}{c|}{\cellcolor[HTML]{FFFFC7}058} &
			\multicolumn{1}{c|}{\cellcolor[HTML]{FFFFC7}078} &
			\multicolumn{1}{c|}{\cellcolor[HTML]{FFFFC7}052} &
			\multicolumn{1}{c|}{\cellcolor[HTML]{FFFFC7}077} &
			\cellcolor[HTML]{FFFFC7}118 &
			\multicolumn{1}{c|}{\cellcolor[HTML]{C4FF99}052} &
			\multicolumn{1}{c|}{\cellcolor[HTML]{C4FF99}037} &
			\multicolumn{1}{c|}{\cellcolor[HTML]{C4FF99}063} &
			\multicolumn{1}{c|}{\cellcolor[HTML]{C4FF99}213} &
			\multicolumn{1}{c|}{\cellcolor[HTML]{C4FF99}070} &
			\multicolumn{1}{c|}{\cellcolor[HTML]{C4FF99}063} &
			\multicolumn{1}{c|}{\cellcolor[HTML]{C4FF99}060} &
			\cellcolor[HTML]{C4FF99}143 \\ \cline{2-18} 
			\cellcolor[HTML]{C1C1C1} &
			\textbf{3} &
			\multicolumn{1}{c|}{\cellcolor[HTML]{FFFFC7}064} &
			\multicolumn{1}{c|}{\cellcolor[HTML]{FFFFC7}055} &
			\multicolumn{1}{c|}{\cellcolor[HTML]{FFFFC7}\textbf{021}} &
			\multicolumn{1}{c|}{\cellcolor[HTML]{FFFFC7}072} &
			\multicolumn{1}{c|}{\cellcolor[HTML]{FFFFC7}089} &
			\multicolumn{1}{c|}{\cellcolor[HTML]{FFFFC7}073} &
			\multicolumn{1}{c|}{\cellcolor[HTML]{FFFFC7}086} &
			\cellcolor[HTML]{FFFFC7}117 &
			\multicolumn{1}{c|}{\cellcolor[HTML]{C4FF99}079} &
			\multicolumn{1}{c|}{\cellcolor[HTML]{C4FF99}062} &
			\multicolumn{1}{c|}{\cellcolor[HTML]{C4FF99}038} &
			\multicolumn{1}{c|}{\cellcolor[HTML]{C4FF99}190} &
			\multicolumn{1}{c|}{\cellcolor[HTML]{C4FF99}083} &
			\multicolumn{1}{c|}{\cellcolor[HTML]{C4FF99}084} &
			\multicolumn{1}{c|}{\cellcolor[HTML]{C4FF99}071} &
			\cellcolor[HTML]{C4FF99}139 \\ \cline{2-18} 
			\cellcolor[HTML]{C1C1C1} &
			\textbf{4} &
			\multicolumn{1}{c|}{\cellcolor[HTML]{FFFFC7}076} &
			\multicolumn{1}{c|}{\cellcolor[HTML]{FFFFC7}063} &
			\multicolumn{1}{c|}{\cellcolor[HTML]{FFFFC7}073} &
			\multicolumn{1}{c|}{\cellcolor[HTML]{FFFFC7}\textbf{016}} &
			\multicolumn{1}{c|}{\cellcolor[HTML]{FFFFC7}068} &
			\multicolumn{1}{c|}{\cellcolor[HTML]{FFFFC7}064} &
			\multicolumn{1}{c|}{\cellcolor[HTML]{FFFFC7}055} &
			\cellcolor[HTML]{FFFFC7}080 &
			\multicolumn{1}{c|}{\cellcolor[HTML]{C4FF99}072} &
			\multicolumn{1}{c|}{\cellcolor[HTML]{C4FF99}073} &
			\multicolumn{1}{c|}{\cellcolor[HTML]{C4FF99}088} &
			\multicolumn{1}{c|}{\cellcolor[HTML]{C4FF99}203} &
			\multicolumn{1}{c|}{\cellcolor[HTML]{C4FF99}064} &
			\multicolumn{1}{c|}{\cellcolor[HTML]{C4FF99}066} &
			\multicolumn{1}{c|}{\cellcolor[HTML]{C4FF99}053} &
			\cellcolor[HTML]{C4FF99}107 \\ \cline{2-18} 
			\cellcolor[HTML]{C1C1C1} &
			\textbf{5} &
			\multicolumn{1}{c|}{\cellcolor[HTML]{FFFFC7}069} &
			\multicolumn{1}{c|}{\cellcolor[HTML]{FFFFC7}080} &
			\multicolumn{1}{c|}{\cellcolor[HTML]{FFFFC7}088} &
			\multicolumn{1}{c|}{\cellcolor[HTML]{FFFFC7}065} &
			\multicolumn{1}{c|}{\cellcolor[HTML]{FFFFC7}\textbf{021}} &
			\multicolumn{1}{c|}{\cellcolor[HTML]{FFFFC7}057} &
			\multicolumn{1}{c|}{\cellcolor[HTML]{FFFFC7}053} &
			\cellcolor[HTML]{FFFFC7}077 &
			\multicolumn{1}{c|}{\cellcolor[HTML]{C4FF99}067} &
			\multicolumn{1}{c|}{\cellcolor[HTML]{C4FF99}088} &
			\multicolumn{1}{c|}{\cellcolor[HTML]{C4FF99}112} &
			\multicolumn{1}{c|}{\cellcolor[HTML]{C4FF99}217} &
			\multicolumn{1}{c|}{\cellcolor[HTML]{C4FF99}035} &
			\multicolumn{1}{c|}{\cellcolor[HTML]{C4FF99}050} &
			\multicolumn{1}{c|}{\cellcolor[HTML]{C4FF99}046} &
			\cellcolor[HTML]{C4FF99}092 \\ \cline{2-18} 
			\cellcolor[HTML]{C1C1C1} &
			\textbf{6} &
			\multicolumn{1}{c|}{\cellcolor[HTML]{FFFFC7}056} &
			\multicolumn{1}{c|}{\cellcolor[HTML]{FFFFC7}055} &
			\multicolumn{1}{c|}{\cellcolor[HTML]{FFFFC7}075} &
			\multicolumn{1}{c|}{\cellcolor[HTML]{FFFFC7}062} &
			\multicolumn{1}{c|}{\cellcolor[HTML]{FFFFC7}058} &
			\multicolumn{1}{c|}{\cellcolor[HTML]{FFFFC7}\textbf{020}} &
			\multicolumn{1}{c|}{\cellcolor[HTML]{FFFFC7}057} &
			\cellcolor[HTML]{FFFFC7}088 &
			\multicolumn{1}{c|}{\cellcolor[HTML]{C4FF99}050} &
			\multicolumn{1}{c|}{\cellcolor[HTML]{C4FF99}066} &
			\multicolumn{1}{c|}{\cellcolor[HTML]{C4FF99}090} &
			\multicolumn{1}{c|}{\cellcolor[HTML]{C4FF99}232} &
			\multicolumn{1}{c|}{\cellcolor[HTML]{C4FF99}048} &
			\multicolumn{1}{c|}{\cellcolor[HTML]{C4FF99}030} &
			\multicolumn{1}{c|}{\cellcolor[HTML]{C4FF99}048} &
			\cellcolor[HTML]{C4FF99}113 \\ \cline{2-18} 
			\cellcolor[HTML]{C1C1C1} &
			\textbf{7} &
			\multicolumn{1}{c|}{\cellcolor[HTML]{FFFFC7}086} &
			\multicolumn{1}{c|}{\cellcolor[HTML]{FFFFC7}080} &
			\multicolumn{1}{c|}{\cellcolor[HTML]{FFFFC7}087} &
			\multicolumn{1}{c|}{\cellcolor[HTML]{FFFFC7}053} &
			\multicolumn{1}{c|}{\cellcolor[HTML]{FFFFC7}054} &
			\multicolumn{1}{c|}{\cellcolor[HTML]{FFFFC7}058} &
			\multicolumn{1}{c|}{\cellcolor[HTML]{FFFFC7}\textbf{020}} &
			\cellcolor[HTML]{FFFFC7}066 &
			\multicolumn{1}{c|}{\cellcolor[HTML]{C4FF99}082} &
			\multicolumn{1}{c|}{\cellcolor[HTML]{C4FF99}091} &
			\multicolumn{1}{c|}{\cellcolor[HTML]{C4FF99}107} &
			\multicolumn{1}{c|}{\cellcolor[HTML]{C4FF99}206} &
			\multicolumn{1}{c|}{\cellcolor[HTML]{C4FF99}050} &
			\multicolumn{1}{c|}{\cellcolor[HTML]{C4FF99}056} &
			\multicolumn{1}{c|}{\cellcolor[HTML]{C4FF99}033} &
			\cellcolor[HTML]{C4FF99}097 \\ \cline{2-18} 
			\multirow{-8}{*}{\rotatebox[origin=c]{90}{\cellcolor[HTML]{C1C1C1}\textbf{\racsi TRAINING}}} &
			\textbf{8} &
			\multicolumn{1}{c|}{\cellcolor[HTML]{FFFFC7}114} &
			\multicolumn{1}{c|}{\cellcolor[HTML]{FFFFC7}119} &
			\multicolumn{1}{c|}{\cellcolor[HTML]{FFFFC7}116} &
			\multicolumn{1}{c|}{\cellcolor[HTML]{FFFFC7}079} &
			\multicolumn{1}{c|}{\cellcolor[HTML]{FFFFC7}075} &
			\multicolumn{1}{c|}{\cellcolor[HTML]{FFFFC7}085} &
			\multicolumn{1}{c|}{\cellcolor[HTML]{FFFFC7}065} &
			\cellcolor[HTML]{FFFFC7}\textbf{026} &
			\multicolumn{1}{c|}{\cellcolor[HTML]{C4FF99}111} &
			\multicolumn{1}{c|}{\cellcolor[HTML]{C4FF99}129} &
			\multicolumn{1}{c|}{\cellcolor[HTML]{C4FF99}140} &
			\multicolumn{1}{c|}{\cellcolor[HTML]{C4FF99}215} &
			\multicolumn{1}{c|}{\cellcolor[HTML]{C4FF99}081} &
			\multicolumn{1}{c|}{\cellcolor[HTML]{C4FF99}083} &
			\multicolumn{1}{c|}{\cellcolor[HTML]{C4FF99}076} &
			\cellcolor[HTML]{C4FF99}043 \\ \hline
			\cellcolor[HTML]{C1C1C1} &
			\textbf{1} &
			\multicolumn{1}{c|}{\cellcolor[HTML]{BCFFFB}036} &
			\multicolumn{1}{c|}{\cellcolor[HTML]{BCFFFB}057} &
			\multicolumn{1}{c|}{\cellcolor[HTML]{BCFFFB}080} &
			\multicolumn{1}{c|}{\cellcolor[HTML]{BCFFFB}072} &
			\multicolumn{1}{c|}{\cellcolor[HTML]{BCFFFB}068} &
			\multicolumn{1}{c|}{\cellcolor[HTML]{BCFFFB}051} &
			\multicolumn{1}{c|}{\cellcolor[HTML]{BCFFFB}083} &
			\cellcolor[HTML]{BCFFFB}113 &
			\multicolumn{1}{c|}{\cellcolor[HTML]{FFFFC7}\textbf{017}} &
			\multicolumn{1}{c|}{\cellcolor[HTML]{FFFFC7}061} &
			\multicolumn{1}{c|}{\cellcolor[HTML]{FFFFC7}090} &
			\multicolumn{1}{c|}{\cellcolor[HTML]{FFFFC7}245} &
			\multicolumn{1}{c|}{\cellcolor[HTML]{FFFFC7}069} &
			\multicolumn{1}{c|}{\cellcolor[HTML]{FFFFC7}051} &
			\multicolumn{1}{c|}{\cellcolor[HTML]{FFFFC7}070} &
			\cellcolor[HTML]{FFFFC7}132 \\ \cline{2-18} 
			\cellcolor[HTML]{C1C1C1} &
			\textbf{2} &
			\multicolumn{1}{c|}{\cellcolor[HTML]{BCFFFB}050} &
			\multicolumn{1}{c|}{\cellcolor[HTML]{BCFFFB}042} &
			\multicolumn{1}{c|}{\cellcolor[HTML]{BCFFFB}063} &
			\multicolumn{1}{c|}{\cellcolor[HTML]{BCFFFB}073} &
			\multicolumn{1}{c|}{\cellcolor[HTML]{BCFFFB}089} &
			\multicolumn{1}{c|}{\cellcolor[HTML]{BCFFFB}067} &
			\multicolumn{1}{c|}{\cellcolor[HTML]{BCFFFB}092} &
			\cellcolor[HTML]{BCFFFB}130 &
			\multicolumn{1}{c|}{\cellcolor[HTML]{FFFFC7}061} &
			\multicolumn{1}{c|}{\cellcolor[HTML]{FFFFC7}\textbf{017}} &
			\multicolumn{1}{c|}{\cellcolor[HTML]{FFFFC7}073} &
			\multicolumn{1}{c|}{\cellcolor[HTML]{FFFFC7}222} &
			\multicolumn{1}{c|}{\cellcolor[HTML]{FFFFC7}082} &
			\multicolumn{1}{c|}{\cellcolor[HTML]{FFFFC7}080} &
			\multicolumn{1}{c|}{\cellcolor[HTML]{FFFFC7}076} &
			\cellcolor[HTML]{FFFFC7}156 \\ \cline{2-18} 
			\cellcolor[HTML]{C1C1C1} &
			\textbf{3} &
			\multicolumn{1}{c|}{\cellcolor[HTML]{BCFFFB}073} &
			\multicolumn{1}{c|}{\cellcolor[HTML]{BCFFFB}067} &
			\multicolumn{1}{c|}{\cellcolor[HTML]{BCFFFB}040} &
			\multicolumn{1}{c|}{\cellcolor[HTML]{BCFFFB}088} &
			\multicolumn{1}{c|}{\cellcolor[HTML]{BCFFFB}113} &
			\multicolumn{1}{c|}{\cellcolor[HTML]{BCFFFB}091} &
			\multicolumn{1}{c|}{\cellcolor[HTML]{BCFFFB}107} &
			\cellcolor[HTML]{BCFFFB}141 &
			\multicolumn{1}{c|}{\cellcolor[HTML]{FFFFC7}089} &
			\multicolumn{1}{c|}{\cellcolor[HTML]{FFFFC7}073} &
			\multicolumn{1}{c|}{\cellcolor[HTML]{FFFFC7}\textbf{019}} &
			\multicolumn{1}{c|}{\cellcolor[HTML]{FFFFC7}179} &
			\multicolumn{1}{c|}{\cellcolor[HTML]{FFFFC7}105} &
			\multicolumn{1}{c|}{\cellcolor[HTML]{FFFFC7}102} &
			\multicolumn{1}{c|}{\cellcolor[HTML]{FFFFC7}091} &
			\cellcolor[HTML]{FFFFC7}164 \\ \cline{2-18} 
			\cellcolor[HTML]{C1C1C1} &
			\textbf{4} &
			\multicolumn{1}{c|}{\cellcolor[HTML]{BCFFFB}226} &
			\multicolumn{1}{c|}{\cellcolor[HTML]{BCFFFB}213} &
			\multicolumn{1}{c|}{\cellcolor[HTML]{BCFFFB}190} &
			\multicolumn{1}{c|}{\cellcolor[HTML]{BCFFFB}202} &
			\multicolumn{1}{c|}{\cellcolor[HTML]{BCFFFB}217} &
			\multicolumn{1}{c|}{\cellcolor[HTML]{BCFFFB}232} &
			\multicolumn{1}{c|}{\cellcolor[HTML]{BCFFFB}206} &
			\cellcolor[HTML]{BCFFFB}215 &
			\multicolumn{1}{c|}{\cellcolor[HTML]{FFFFC7}245} &
			\multicolumn{1}{c|}{\cellcolor[HTML]{FFFFC7}222} &
			\multicolumn{1}{c|}{\cellcolor[HTML]{FFFFC7}179} &
			\multicolumn{1}{c|}{\cellcolor[HTML]{FFFFC7}\textbf{010}} &
			\multicolumn{1}{c|}{\cellcolor[HTML]{FFFFC7}222} &
			\multicolumn{1}{c|}{\cellcolor[HTML]{FFFFC7}228} &
			\multicolumn{1}{c|}{\cellcolor[HTML]{FFFFC7}197} &
			\cellcolor[HTML]{FFFFC7}207 \\ \cline{2-18} 
			\cellcolor[HTML]{C1C1C1} &
			\textbf{5} &
			\multicolumn{1}{c|}{\cellcolor[HTML]{BCFFFB}072} &
			\multicolumn{1}{c|}{\cellcolor[HTML]{BCFFFB}073} &
			\multicolumn{1}{c|}{\cellcolor[HTML]{BCFFFB}083} &
			\multicolumn{1}{c|}{\cellcolor[HTML]{BCFFFB}062} &
			\multicolumn{1}{c|}{\cellcolor[HTML]{BCFFFB}036} &
			\multicolumn{1}{c|}{\cellcolor[HTML]{BCFFFB}049} &
			\multicolumn{1}{c|}{\cellcolor[HTML]{BCFFFB}050} &
			\cellcolor[HTML]{BCFFFB}083 &
			\multicolumn{1}{c|}{\cellcolor[HTML]{FFFFC7}068} &
			\multicolumn{1}{c|}{\cellcolor[HTML]{FFFFC7}082} &
			\multicolumn{1}{c|}{\cellcolor[HTML]{FFFFC7}105} &
			\multicolumn{1}{c|}{\cellcolor[HTML]{FFFFC7}222} &
			\multicolumn{1}{c|}{\cellcolor[HTML]{FFFFC7}\textbf{019}} &
			\multicolumn{1}{c|}{\cellcolor[HTML]{FFFFC7}041} &
			\multicolumn{1}{c|}{\cellcolor[HTML]{FFFFC7}047} &
			\cellcolor[HTML]{FFFFC7}106 \\ \cline{2-18} 
			\cellcolor[HTML]{C1C1C1} &
			\textbf{6} &
			\multicolumn{1}{c|}{\cellcolor[HTML]{BCFFFB}064} &
			\multicolumn{1}{c|}{\cellcolor[HTML]{BCFFFB}066} &
			\multicolumn{1}{c|}{\cellcolor[HTML]{BCFFFB}085} &
			\multicolumn{1}{c|}{\cellcolor[HTML]{BCFFFB}066} &
			\multicolumn{1}{c|}{\cellcolor[HTML]{BCFFFB}050} &
			\multicolumn{1}{l|}{\cellcolor[HTML]{BCFFFB}031} &
			\multicolumn{1}{c|}{\cellcolor[HTML]{BCFFFB}056} &
			\cellcolor[HTML]{BCFFFB}086 &
			\multicolumn{1}{c|}{\cellcolor[HTML]{FFFFC7}051} &
			\multicolumn{1}{c|}{\cellcolor[HTML]{FFFFC7}080} &
			\multicolumn{1}{c|}{\cellcolor[HTML]{FFFFC7}102} &
			\multicolumn{1}{c|}{\cellcolor[HTML]{FFFFC7}228} &
			\multicolumn{1}{c|}{\cellcolor[HTML]{FFFFC7}042} &
			\multicolumn{1}{l|}{\cellcolor[HTML]{FFFFC7}\textbf{018}} &
			\multicolumn{1}{c|}{\cellcolor[HTML]{FFFFC7}049} &
			\cellcolor[HTML]{FFFFC7}109 \\ \cline{2-18} 
			\cellcolor[HTML]{C1C1C1} &
			\textbf{7} &
			\multicolumn{1}{c|}{\cellcolor[HTML]{BCFFFB}070} &
			\multicolumn{1}{c|}{\cellcolor[HTML]{BCFFFB}064} &
			\multicolumn{1}{c|}{\cellcolor[HTML]{BCFFFB}072} &
			\multicolumn{1}{c|}{\cellcolor[HTML]{BCFFFB}052} &
			\multicolumn{1}{c|}{\cellcolor[HTML]{BCFFFB}047} &
			\multicolumn{1}{l|}{\cellcolor[HTML]{BCFFFB}049} &
			\multicolumn{1}{c|}{\cellcolor[HTML]{BCFFFB}033} &
			\cellcolor[HTML]{BCFFFB}078 &
			\multicolumn{1}{c|}{\cellcolor[HTML]{FFFFC7}069} &
			\multicolumn{1}{c|}{\cellcolor[HTML]{FFFFC7}076} &
			\multicolumn{1}{c|}{\cellcolor[HTML]{FFFFC7}091} &
			\multicolumn{1}{c|}{\cellcolor[HTML]{FFFFC7}197} &
			\multicolumn{1}{c|}{\cellcolor[HTML]{FFFFC7}048} &
			\multicolumn{1}{c|}{\cellcolor[HTML]{FFFFC7}048} &
			\multicolumn{1}{c|}{\cellcolor[HTML]{FFFFC7}\textbf{018}} &
			\cellcolor[HTML]{FFFFC7}106 \\ \cline{2-18} 
			\multirow{-8}{*}{\rotatebox[origin=c]{90}{\cellcolor[HTML]{C1C1C1}\textbf{\racsi TESTING}}} &
			\textbf{8} &
			\multicolumn{1}{c|}{\cellcolor[HTML]{BCFFFB}131} &
			\multicolumn{1}{c|}{\cellcolor[HTML]{BCFFFB}145} &
			\multicolumn{1}{c|}{\cellcolor[HTML]{BCFFFB}139} &
			\multicolumn{1}{c|}{\cellcolor[HTML]{BCFFFB}107} &
			\multicolumn{1}{c|}{\cellcolor[HTML]{BCFFFB}092} &
			\multicolumn{1}{l|}{\cellcolor[HTML]{BCFFFB}113} &
			\multicolumn{1}{c|}{\cellcolor[HTML]{BCFFFB}096} &
			\cellcolor[HTML]{BCFFFB}046 &
			\multicolumn{1}{c|}{\cellcolor[HTML]{FFFFC7}132} &
			\multicolumn{1}{c|}{\cellcolor[HTML]{FFFFC7}156} &
			\multicolumn{1}{c|}{\cellcolor[HTML]{FFFFC7}164} &
			\multicolumn{1}{c|}{\cellcolor[HTML]{FFFFC7}207} &
			\multicolumn{1}{c|}{\cellcolor[HTML]{FFFFC7}106} &
			\multicolumn{1}{c|}{\cellcolor[HTML]{FFFFC7}109} &
			\multicolumn{1}{c|}{\cellcolor[HTML]{FFFFC7}106} &
			\cellcolor[HTML]{FFFFC7}\textbf{020} \\ \hline
		\end{tabular}%
	}
	\caption{Average normalized \whd computed on the partitions of the AntiSense dataset dedicated to training and testing with the receiver located in position 3 (rx3). The integer values are the first three digits after the comma, rounded to the nearest value. The POS parameter indicates the position of the person standing still within the experimental environment.}
	\label{tab:rx3whd}
\end{table}

\begin{table}[]
	\centering
	\resizebox{\textwidth}{!}{%
		\begin{tabular}{|
				>{\columncolor[HTML]{C1C1C1}}c |
				>{\columncolor[HTML]{C1C1C1}}c |cccccccc|cccccccc|}
			\hline
			\textbf{} &
			\textbf{} &
			\multicolumn{8}{c|}{\cellcolor[HTML]{C1C1C1}\textbf{TRAINING}} &
			\multicolumn{8}{c|}{\cellcolor[HTML]{C1C1C1}\textbf{TESTING}} \\ \hline
			\textbf{} &
			\textbf{POS} &
			\multicolumn{1}{c|}{\cellcolor[HTML]{C1C1C1}\textbf{1}} &
			\multicolumn{1}{c|}{\cellcolor[HTML]{C1C1C1}\textbf{2}} &
			\multicolumn{1}{c|}{\cellcolor[HTML]{C1C1C1}\textbf{3}} &
			\multicolumn{1}{c|}{\cellcolor[HTML]{C1C1C1}\textbf{4}} &
			\multicolumn{1}{c|}{\cellcolor[HTML]{C1C1C1}\textbf{5}} &
			\multicolumn{1}{c|}{\cellcolor[HTML]{C1C1C1}\textbf{6}} &
			\multicolumn{1}{c|}{\cellcolor[HTML]{C1C1C1}\textbf{7}} &
			\cellcolor[HTML]{C1C1C1}\textbf{8} &
			\multicolumn{1}{c|}{\cellcolor[HTML]{C1C1C1}\textbf{1}} &
			\multicolumn{1}{c|}{\cellcolor[HTML]{C1C1C1}\textbf{2}} &
			\multicolumn{1}{c|}{\cellcolor[HTML]{C1C1C1}\textbf{3}} &
			\multicolumn{1}{c|}{\cellcolor[HTML]{C1C1C1}\textbf{4}} &
			\multicolumn{1}{c|}{\cellcolor[HTML]{C1C1C1}\textbf{5}} &
			\multicolumn{1}{c|}{\cellcolor[HTML]{C1C1C1}\textbf{6}} &
			\multicolumn{1}{c|}{\cellcolor[HTML]{C1C1C1}\textbf{7}} &
			\cellcolor[HTML]{C1C1C1}\textbf{8} \\ \hline
			\cellcolor[HTML]{C1C1C1} &
			\textbf{1} &
			\multicolumn{1}{c|}{\cellcolor[HTML]{FFFFC7}\textbf{021}} &
			\multicolumn{1}{c|}{\cellcolor[HTML]{FFFFC7}052} &
			\multicolumn{1}{c|}{\cellcolor[HTML]{FFFFC7}201} &
			\multicolumn{1}{c|}{\cellcolor[HTML]{FFFFC7}057} &
			\multicolumn{1}{c|}{\cellcolor[HTML]{FFFFC7}225} &
			\multicolumn{1}{c|}{\cellcolor[HTML]{FFFFC7}052} &
			\multicolumn{1}{c|}{\cellcolor[HTML]{FFFFC7}089} &
			\cellcolor[HTML]{FFFFC7}169 &
			\multicolumn{1}{c|}{\cellcolor[HTML]{C4FF99}033} &
			\multicolumn{1}{c|}{\cellcolor[HTML]{C4FF99}063} &
			\multicolumn{1}{c|}{\cellcolor[HTML]{C4FF99}038} &
			\multicolumn{1}{c|}{\cellcolor[HTML]{C4FF99}063} &
			\multicolumn{1}{c|}{\cellcolor[HTML]{C4FF99}260} &
			\multicolumn{1}{c|}{\cellcolor[HTML]{C4FF99}047} &
			\multicolumn{1}{c|}{\cellcolor[HTML]{C4FF99}069} &
			\cellcolor[HTML]{C4FF99}087 \\ \cline{2-18} 
			\cellcolor[HTML]{C1C1C1} &
			\textbf{2} &
			\multicolumn{1}{c|}{\cellcolor[HTML]{FFFFC7}053} &
			\multicolumn{1}{c|}{\cellcolor[HTML]{FFFFC7}\textbf{020}} &
			\multicolumn{1}{c|}{\cellcolor[HTML]{FFFFC7}221} &
			\multicolumn{1}{c|}{\cellcolor[HTML]{FFFFC7}061} &
			\multicolumn{1}{c|}{\cellcolor[HTML]{FFFFC7}232} &
			\multicolumn{1}{c|}{\cellcolor[HTML]{FFFFC7}052} &
			\multicolumn{1}{c|}{\cellcolor[HTML]{FFFFC7}085} &
			\cellcolor[HTML]{FFFFC7}186 &
			\multicolumn{1}{c|}{\cellcolor[HTML]{C4FF99}048} &
			\multicolumn{1}{c|}{\cellcolor[HTML]{C4FF99}033} &
			\multicolumn{1}{c|}{\cellcolor[HTML]{C4FF99}044} &
			\multicolumn{1}{c|}{\cellcolor[HTML]{C4FF99}057} &
			\multicolumn{1}{c|}{\cellcolor[HTML]{C4FF99}275} &
			\multicolumn{1}{c|}{\cellcolor[HTML]{C4FF99}054} &
			\multicolumn{1}{c|}{\cellcolor[HTML]{C4FF99}064} &
			\cellcolor[HTML]{C4FF99}083 \\ \cline{2-18} 
			\cellcolor[HTML]{C1C1C1} &
			\textbf{3} &
			\multicolumn{1}{c|}{\cellcolor[HTML]{FFFFC7}200} &
			\multicolumn{1}{c|}{\cellcolor[HTML]{FFFFC7}220} &
			\multicolumn{1}{c|}{\cellcolor[HTML]{FFFFC7}\textbf{011}} &
			\multicolumn{1}{c|}{\cellcolor[HTML]{FFFFC7}236} &
			\multicolumn{1}{c|}{\cellcolor[HTML]{FFFFC7}211} &
			\multicolumn{1}{c|}{\cellcolor[HTML]{FFFFC7}213} &
			\multicolumn{1}{c|}{\cellcolor[HTML]{FFFFC7}249} &
			\cellcolor[HTML]{FFFFC7}054 &
			\multicolumn{1}{c|}{\cellcolor[HTML]{C4FF99}226} &
			\multicolumn{1}{c|}{\cellcolor[HTML]{C4FF99}230} &
			\multicolumn{1}{c|}{\cellcolor[HTML]{C4FF99}212} &
			\multicolumn{1}{c|}{\cellcolor[HTML]{C4FF99}242} &
			\multicolumn{1}{c|}{\cellcolor[HTML]{C4FF99}111} &
			\multicolumn{1}{c|}{\cellcolor[HTML]{C4FF99}214} &
			\multicolumn{1}{c|}{\cellcolor[HTML]{C4FF99}229} &
			\cellcolor[HTML]{C4FF99}252 \\ \cline{2-18} 
			\cellcolor[HTML]{C1C1C1} &
			\textbf{4} &
			\multicolumn{1}{c|}{\cellcolor[HTML]{FFFFC7}059} &
			\multicolumn{1}{c|}{\cellcolor[HTML]{FFFFC7}061} &
			\multicolumn{1}{c|}{\cellcolor[HTML]{FFFFC7}236} &
			\multicolumn{1}{c|}{\cellcolor[HTML]{FFFFC7}\textbf{019}} &
			\multicolumn{1}{c|}{\cellcolor[HTML]{FFFFC7}260} &
			\multicolumn{1}{c|}{\cellcolor[HTML]{FFFFC7}049} &
			\multicolumn{1}{c|}{\cellcolor[HTML]{FFFFC7}068} &
			\cellcolor[HTML]{FFFFC7}201 &
			\multicolumn{1}{c|}{\cellcolor[HTML]{C4FF99}043} &
			\multicolumn{1}{c|}{\cellcolor[HTML]{C4FF99}063} &
			\multicolumn{1}{c|}{\cellcolor[HTML]{C4FF99}046} &
			\multicolumn{1}{c|}{\cellcolor[HTML]{C4FF99}037} &
			\multicolumn{1}{c|}{\cellcolor[HTML]{C4FF99}300} &
			\multicolumn{1}{c|}{\cellcolor[HTML]{C4FF99}050} &
			\multicolumn{1}{c|}{\cellcolor[HTML]{C4FF99}058} &
			\cellcolor[HTML]{C4FF99}066 \\ \cline{2-18} 
			\cellcolor[HTML]{C1C1C1} &
			\textbf{5} &
			\multicolumn{1}{c|}{\cellcolor[HTML]{FFFFC7}224} &
			\multicolumn{1}{c|}{\cellcolor[HTML]{FFFFC7}231} &
			\multicolumn{1}{c|}{\cellcolor[HTML]{FFFFC7}208} &
			\multicolumn{1}{c|}{\cellcolor[HTML]{FFFFC7}259} &
			\multicolumn{1}{c|}{\cellcolor[HTML]{FFFFC7}\textbf{035}} &
			\multicolumn{1}{c|}{\cellcolor[HTML]{FFFFC7}246} &
			\multicolumn{1}{c|}{\cellcolor[HTML]{FFFFC7}254} &
			\cellcolor[HTML]{FFFFC7}211 &
			\multicolumn{1}{c|}{\cellcolor[HTML]{C4FF99}234} &
			\multicolumn{1}{c|}{\cellcolor[HTML]{C4FF99}240} &
			\multicolumn{1}{c|}{\cellcolor[HTML]{C4FF99}230} &
			\multicolumn{1}{c|}{\cellcolor[HTML]{C4FF99}248} &
			\multicolumn{1}{c|}{\cellcolor[HTML]{C4FF99}196} &
			\multicolumn{1}{c|}{\cellcolor[HTML]{C4FF99}237} &
			\multicolumn{1}{c|}{\cellcolor[HTML]{C4FF99}244} &
			\cellcolor[HTML]{C4FF99}253 \\ \cline{2-18} 
			\cellcolor[HTML]{C1C1C1} &
			\textbf{6} &
			\multicolumn{1}{c|}{\cellcolor[HTML]{FFFFC7}051} &
			\multicolumn{1}{c|}{\cellcolor[HTML]{FFFFC7}053} &
			\multicolumn{1}{c|}{\cellcolor[HTML]{FFFFC7}213} &
			\multicolumn{1}{c|}{\cellcolor[HTML]{FFFFC7}048} &
			\multicolumn{1}{c|}{\cellcolor[HTML]{FFFFC7}247} &
			\multicolumn{1}{c|}{\cellcolor[HTML]{FFFFC7}\textbf{020}} &
			\multicolumn{1}{c|}{\cellcolor[HTML]{FFFFC7}073} &
			\cellcolor[HTML]{FFFFC7}177 &
			\multicolumn{1}{c|}{\cellcolor[HTML]{C4FF99}043} &
			\multicolumn{1}{c|}{\cellcolor[HTML]{C4FF99}059} &
			\multicolumn{1}{c|}{\cellcolor[HTML]{C4FF99}050} &
			\multicolumn{1}{c|}{\cellcolor[HTML]{C4FF99}047} &
			\multicolumn{1}{c|}{\cellcolor[HTML]{C4FF99}275} &
			\multicolumn{1}{c|}{\cellcolor[HTML]{C4FF99}037} &
			\multicolumn{1}{c|}{\cellcolor[HTML]{C4FF99}058} &
			\cellcolor[HTML]{C4FF99}065 \\ \cline{2-18} 
			\cellcolor[HTML]{C1C1C1} &
			\textbf{7} &
			\multicolumn{1}{c|}{\cellcolor[HTML]{FFFFC7}089} &
			\multicolumn{1}{c|}{\cellcolor[HTML]{FFFFC7}085} &
			\multicolumn{1}{c|}{\cellcolor[HTML]{FFFFC7}249} &
			\multicolumn{1}{c|}{\cellcolor[HTML]{FFFFC7}067} &
			\multicolumn{1}{c|}{\cellcolor[HTML]{FFFFC7}255} &
			\multicolumn{1}{c|}{\cellcolor[HTML]{FFFFC7}071} &
			\multicolumn{1}{c|}{\cellcolor[HTML]{FFFFC7}\textbf{022}} &
			\cellcolor[HTML]{FFFFC7}213 &
			\multicolumn{1}{c|}{\cellcolor[HTML]{C4FF99}075} &
			\multicolumn{1}{c|}{\cellcolor[HTML]{C4FF99}073} &
			\multicolumn{1}{c|}{\cellcolor[HTML]{C4FF99}079} &
			\multicolumn{1}{c|}{\cellcolor[HTML]{C4FF99}067} &
			\multicolumn{1}{c|}{\cellcolor[HTML]{C4FF99}305} &
			\multicolumn{1}{c|}{\cellcolor[HTML]{C4FF99}068} &
			\multicolumn{1}{c|}{\cellcolor[HTML]{C4FF99}039} &
			\cellcolor[HTML]{C4FF99}052 \\ \cline{2-18} 
			\multirow{-8}{*}{\rotatebox[origin=c]{90}{\cellcolor[HTML]{C1C1C1}\textbf{\racsi TRAINING}}} &
			\textbf{8} &
			\multicolumn{1}{c|}{\cellcolor[HTML]{FFFFC7}169} &
			\multicolumn{1}{c|}{\cellcolor[HTML]{FFFFC7}187} &
			\multicolumn{1}{c|}{\cellcolor[HTML]{FFFFC7}052} &
			\multicolumn{1}{c|}{\cellcolor[HTML]{FFFFC7}201} &
			\multicolumn{1}{c|}{\cellcolor[HTML]{FFFFC7}217} &
			\multicolumn{1}{c|}{\cellcolor[HTML]{FFFFC7}177} &
			\multicolumn{1}{c|}{\cellcolor[HTML]{FFFFC7}213} &
			\cellcolor[HTML]{FFFFC7}\textbf{016} &
			\multicolumn{1}{c|}{\cellcolor[HTML]{C4FF99}193} &
			\multicolumn{1}{c|}{\cellcolor[HTML]{C4FF99}199} &
			\multicolumn{1}{c|}{\cellcolor[HTML]{C4FF99}180} &
			\multicolumn{1}{c|}{\cellcolor[HTML]{C4FF99}206} &
			\multicolumn{1}{c|}{\cellcolor[HTML]{C4FF99}137} &
			\multicolumn{1}{c|}{\cellcolor[HTML]{C4FF99}180} &
			\multicolumn{1}{c|}{\cellcolor[HTML]{C4FF99}194} &
			\cellcolor[HTML]{C4FF99}217 \\ \hline
			\cellcolor[HTML]{C1C1C1} &
			\textbf{1} &
			\multicolumn{1}{c|}{\cellcolor[HTML]{BCFFFB}035} &
			\multicolumn{1}{c|}{\cellcolor[HTML]{BCFFFB}050} &
			\multicolumn{1}{c|}{\cellcolor[HTML]{BCFFFB}227} &
			\multicolumn{1}{c|}{\cellcolor[HTML]{BCFFFB}043} &
			\multicolumn{1}{c|}{\cellcolor[HTML]{BCFFFB}236} &
			\multicolumn{1}{c|}{\cellcolor[HTML]{BCFFFB}045} &
			\multicolumn{1}{c|}{\cellcolor[HTML]{BCFFFB}076} &
			\cellcolor[HTML]{BCFFFB}193 &
			\multicolumn{1}{c|}{\cellcolor[HTML]{FFFFC7}\textbf{017}} &
			\multicolumn{1}{c|}{\cellcolor[HTML]{FFFFC7}059} &
			\multicolumn{1}{c|}{\cellcolor[HTML]{FFFFC7}034} &
			\multicolumn{1}{c|}{\cellcolor[HTML]{FFFFC7}050} &
			\multicolumn{1}{c|}{\cellcolor[HTML]{FFFFC7}282} &
			\multicolumn{1}{c|}{\cellcolor[HTML]{FFFFC7}042} &
			\multicolumn{1}{c|}{\cellcolor[HTML]{FFFFC7}062} &
			\cellcolor[HTML]{FFFFC7}071 \\ \cline{2-18} 
			\cellcolor[HTML]{C1C1C1} &
			\textbf{2} &
			\multicolumn{1}{c|}{\cellcolor[HTML]{BCFFFB}065} &
			\multicolumn{1}{c|}{\cellcolor[HTML]{BCFFFB}035} &
			\multicolumn{1}{c|}{\cellcolor[HTML]{BCFFFB}231} &
			\multicolumn{1}{c|}{\cellcolor[HTML]{BCFFFB}064} &
			\multicolumn{1}{c|}{\cellcolor[HTML]{BCFFFB}242} &
			\multicolumn{1}{c|}{\cellcolor[HTML]{BCFFFB}061} &
			\multicolumn{1}{c|}{\cellcolor[HTML]{BCFFFB}074} &
			\cellcolor[HTML]{BCFFFB}199 &
			\multicolumn{1}{c|}{\cellcolor[HTML]{FFFFC7}059} &
			\multicolumn{1}{c|}{\cellcolor[HTML]{FFFFC7}\textbf{016}} &
			\multicolumn{1}{c|}{\cellcolor[HTML]{FFFFC7}053} &
			\multicolumn{1}{c|}{\cellcolor[HTML]{FFFFC7}063} &
			\multicolumn{1}{c|}{\cellcolor[HTML]{FFFFC7}293} &
			\multicolumn{1}{c|}{\cellcolor[HTML]{FFFFC7}061} &
			\multicolumn{1}{c|}{\cellcolor[HTML]{FFFFC7}058} &
			\cellcolor[HTML]{FFFFC7}077 \\ \cline{2-18} 
			\cellcolor[HTML]{C1C1C1} &
			\textbf{3} &
			\multicolumn{1}{c|}{\cellcolor[HTML]{BCFFFB}041} &
			\multicolumn{1}{c|}{\cellcolor[HTML]{BCFFFB}046} &
			\multicolumn{1}{c|}{\cellcolor[HTML]{BCFFFB}213} &
			\multicolumn{1}{c|}{\cellcolor[HTML]{BCFFFB}046} &
			\multicolumn{1}{c|}{\cellcolor[HTML]{BCFFFB}231} &
			\multicolumn{1}{c|}{\cellcolor[HTML]{BCFFFB}053} &
			\multicolumn{1}{c|}{\cellcolor[HTML]{BCFFFB}080} &
			\cellcolor[HTML]{BCFFFB}180 &
			\multicolumn{1}{c|}{\cellcolor[HTML]{FFFFC7}035} &
			\multicolumn{1}{c|}{\cellcolor[HTML]{FFFFC7}052} &
			\multicolumn{1}{c|}{\cellcolor[HTML]{FFFFC7}\textbf{016}} &
			\multicolumn{1}{c|}{\cellcolor[HTML]{FFFFC7}055} &
			\multicolumn{1}{c|}{\cellcolor[HTML]{FFFFC7}274} &
			\multicolumn{1}{c|}{\cellcolor[HTML]{FFFFC7}050} &
			\multicolumn{1}{c|}{\cellcolor[HTML]{FFFFC7}065} &
			\cellcolor[HTML]{FFFFC7}077 \\ \cline{2-18} 
			\cellcolor[HTML]{C1C1C1} &
			\textbf{4} &
			\multicolumn{1}{c|}{\cellcolor[HTML]{BCFFFB}063} &
			\multicolumn{1}{c|}{\cellcolor[HTML]{BCFFFB}057} &
			\multicolumn{1}{c|}{\cellcolor[HTML]{BCFFFB}243} &
			\multicolumn{1}{c|}{\cellcolor[HTML]{BCFFFB}035} &
			\multicolumn{1}{c|}{\cellcolor[HTML]{BCFFFB}249} &
			\multicolumn{1}{c|}{\cellcolor[HTML]{BCFFFB}046} &
			\multicolumn{1}{c|}{\cellcolor[HTML]{BCFFFB}067} &
			\cellcolor[HTML]{BCFFFB}206 &
			\multicolumn{1}{c|}{\cellcolor[HTML]{FFFFC7}048} &
			\multicolumn{1}{c|}{\cellcolor[HTML]{FFFFC7}061} &
			\multicolumn{1}{c|}{\cellcolor[HTML]{FFFFC7}053} &
			\multicolumn{1}{c|}{\cellcolor[HTML]{FFFFC7}\textbf{021}} &
			\multicolumn{1}{c|}{\cellcolor[HTML]{FFFFC7}300} &
			\multicolumn{1}{c|}{\cellcolor[HTML]{FFFFC7}052} &
			\multicolumn{1}{c|}{\cellcolor[HTML]{FFFFC7}056} &
			\cellcolor[HTML]{FFFFC7}060 \\ \cline{2-18} 
			\cellcolor[HTML]{C1C1C1} &
			\textbf{5} &
			\multicolumn{1}{c|}{\cellcolor[HTML]{BCFFFB}258} &
			\multicolumn{1}{c|}{\cellcolor[HTML]{BCFFFB}273} &
			\multicolumn{1}{c|}{\cellcolor[HTML]{BCFFFB}105} &
			\multicolumn{1}{c|}{\cellcolor[HTML]{BCFFFB}299} &
			\multicolumn{1}{c|}{\cellcolor[HTML]{BCFFFB}194} &
			\multicolumn{1}{c|}{\cellcolor[HTML]{BCFFFB}274} &
			\multicolumn{1}{c|}{\cellcolor[HTML]{BCFFFB}304} &
			\cellcolor[HTML]{BCFFFB}132 &
			\multicolumn{1}{c|}{\cellcolor[HTML]{FFFFC7}281} &
			\multicolumn{1}{c|}{\cellcolor[HTML]{FFFFC7}290} &
			\multicolumn{1}{c|}{\cellcolor[HTML]{FFFFC7}273} &
			\multicolumn{1}{c|}{\cellcolor[HTML]{FFFFC7}299} &
			\multicolumn{1}{c|}{\cellcolor[HTML]{FFFFC7}\textbf{023}} &
			\multicolumn{1}{c|}{\cellcolor[HTML]{FFFFC7}271} &
			\multicolumn{1}{c|}{\cellcolor[HTML]{FFFFC7}284} &
			\cellcolor[HTML]{FFFFC7}306 \\ \cline{2-18} 
			\cellcolor[HTML]{C1C1C1} &
			\textbf{6} &
			\multicolumn{1}{c|}{\cellcolor[HTML]{BCFFFB}046} &
			\multicolumn{1}{c|}{\cellcolor[HTML]{BCFFFB}053} &
			\multicolumn{1}{c|}{\cellcolor[HTML]{BCFFFB}215} &
			\multicolumn{1}{c|}{\cellcolor[HTML]{BCFFFB}048} &
			\multicolumn{1}{c|}{\cellcolor[HTML]{BCFFFB}239} &
			\multicolumn{1}{l|}{\cellcolor[HTML]{BCFFFB}035} &
			\multicolumn{1}{c|}{\cellcolor[HTML]{BCFFFB}067} &
			\cellcolor[HTML]{BCFFFB}180 &
			\multicolumn{1}{c|}{\cellcolor[HTML]{FFFFC7}038} &
			\multicolumn{1}{c|}{\cellcolor[HTML]{FFFFC7}057} &
			\multicolumn{1}{c|}{\cellcolor[HTML]{FFFFC7}046} &
			\multicolumn{1}{c|}{\cellcolor[HTML]{FFFFC7}051} &
			\multicolumn{1}{c|}{\cellcolor[HTML]{FFFFC7}273} &
			\multicolumn{1}{l|}{\cellcolor[HTML]{FFFFC7}\textbf{023}} &
			\multicolumn{1}{c|}{\cellcolor[HTML]{FFFFC7}055} &
			\cellcolor[HTML]{FFFFC7}062 \\ \cline{2-18} 
			\cellcolor[HTML]{C1C1C1} &
			\textbf{7} &
			\multicolumn{1}{c|}{\cellcolor[HTML]{BCFFFB}068} &
			\multicolumn{1}{c|}{\cellcolor[HTML]{BCFFFB}063} &
			\multicolumn{1}{c|}{\cellcolor[HTML]{BCFFFB}230} &
			\multicolumn{1}{c|}{\cellcolor[HTML]{BCFFFB}055} &
			\multicolumn{1}{c|}{\cellcolor[HTML]{BCFFFB}245} &
			\multicolumn{1}{l|}{\cellcolor[HTML]{BCFFFB}055} &
			\multicolumn{1}{c|}{\cellcolor[HTML]{BCFFFB}038} &
			\cellcolor[HTML]{BCFFFB}194 &
			\multicolumn{1}{c|}{\cellcolor[HTML]{FFFFC7}059} &
			\multicolumn{1}{c|}{\cellcolor[HTML]{FFFFC7}055} &
			\multicolumn{1}{c|}{\cellcolor[HTML]{FFFFC7}062} &
			\multicolumn{1}{c|}{\cellcolor[HTML]{FFFFC7}054} &
			\multicolumn{1}{c|}{\cellcolor[HTML]{FFFFC7}285} &
			\multicolumn{1}{c|}{\cellcolor[HTML]{FFFFC7}055} &
			\multicolumn{1}{c|}{\cellcolor[HTML]{FFFFC7}\textbf{024}} &
			\cellcolor[HTML]{FFFFC7}054 \\ \cline{2-18} 
			\multirow{-8}{*}{\rotatebox[origin=c]{90}{\cellcolor[HTML]{C1C1C1}\textbf{\racsi TESTING}}} &
			\textbf{8} &
			\multicolumn{1}{c|}{\cellcolor[HTML]{BCFFFB}085} &
			\multicolumn{1}{c|}{\cellcolor[HTML]{BCFFFB}081} &
			\multicolumn{1}{c|}{\cellcolor[HTML]{BCFFFB}252} &
			\multicolumn{1}{c|}{\cellcolor[HTML]{BCFFFB}063} &
			\multicolumn{1}{c|}{\cellcolor[HTML]{BCFFFB}254} &
			\multicolumn{1}{l|}{\cellcolor[HTML]{BCFFFB}062} &
			\multicolumn{1}{c|}{\cellcolor[HTML]{BCFFFB}050} &
			\cellcolor[HTML]{BCFFFB}217 &
			\multicolumn{1}{c|}{\cellcolor[HTML]{FFFFC7}068} &
			\multicolumn{1}{c|}{\cellcolor[HTML]{FFFFC7}074} &
			\multicolumn{1}{c|}{\cellcolor[HTML]{FFFFC7}073} &
			\multicolumn{1}{c|}{\cellcolor[HTML]{FFFFC7}059} &
			\multicolumn{1}{c|}{\cellcolor[HTML]{FFFFC7}308} &
			\multicolumn{1}{c|}{\cellcolor[HTML]{FFFFC7}059} &
			\multicolumn{1}{c|}{\cellcolor[HTML]{FFFFC7}054} &
			\cellcolor[HTML]{FFFFC7}\textbf{027} \\ \hline
		\end{tabular}%
	}
	\caption{Average normalized \whd computed on the partitions of the AntiSense dataset dedicated to training and testing with the receiver located in position 4 (rx4). The integer values are the first three digits after the comma, rounded to the nearest value. The POS parameter indicates the position of the person standing still within the experimental environment.}
	\label{tab:rx4whd}
\end{table}

%% file: acknowledgements.tex
\section*{Ringraziamenti}
\selectlanguage{italian}
\pagenumbering{gobble}
\setstretch{1.2}


\selectlanguage{american}